\documentclass[useAMS,usenatbib]{mn2e}   

\usepackage{xspace,graphicx,epsfig, rotating,amsmath,amssymb}
\usepackage[usenames,dvipsnames]{xcolor}
\usepackage{subfigure}
\usepackage{times}
\usepackage{float,lscape}
\usepackage{rotating}
\usepackage{isotope}
\usepackage{color}

\newcommand\ion[2]{#1$\;${\small \uppercase\expandafter{\romannumeral #2\relax}}\relax}%
\newcommand{\nodata}[1]{$\dots$}

\newcommand{\iher}{{\rm $\iota$~Her}}
\newcommand{\kms}{{\rm km/s}}
\newcommand{\vsi}{{ $v \sin i$}}

\title[UV Abundance Analysis of Iota Her]{Ultraviolet Spectral Synthesis of Iota Herculis}

\author[Golriz \& Landstreet]{S.S.~Golriz$^{1}$, J. D.~Landstreet$^{1,2}$\\
$^{1}$ Department of Physics \& Astronomy, The University of
  Western Ontario, London N6A 3K7, Canada \\
$^{2}$ Armagh Observatory, College Hill, Armagh BT91 6DG, Northern Ireland} %

\begin{document}

\date{January 30 2015}

\pagerange{\pageref{firstpage}--\pageref{lastpage}} \pubyear{2014}

\maketitle

\label{firstpage}

\begin{abstract}
The atmospheric abundances of elements provide essential insights into
the formation and evolution history of stars. The visible wavelength
window has been used almost exclusively in the past to determine the
abundances of chemical elements in B-type stars.

However, some elements do not have useful spectral lines in the
visible.  A high resolution spectrum of $\iota$~Herculis is available
from 999 to 1400\,\AA.\ In this project, we investigate the abundance
determination in the UV. We identify the elements whose abundances can
be tested, and search for elements whose abundances can be determined
in the UV to add to those in the current literature. We also
investigate the completeness of the VALD line list in this region, and
the adequacy of LTE modeling in the UV for this star. We have used the
LTE spectrum synthesis code {\sc zeeman} to model the UV spectrum of
$\iota$~Herculis for elements with 5 $\le$ Z $\le$ 80. Abundances or
upper limits are derived for 24 elements.

We find that most of our results are in reasonable agreement
with previous results. We estimate a value or an upper limit for the
abundance of nine elements in this star that were not detected in the
visible. 

LTE UV spectral synthesis is found to be a
useful tool for abundance determination, even though limitations such
as incomplete and uncertain atomic data, uncertain continuum
normalization and scattered light, and severe blending can introduce
difficulties.  The high abundance of two heavy elements may be a sign
of radiative levitation.
\end{abstract}


\section{Introduction}
\label{intro}
The star $\iota$~Herculis (HD 160762, HR 6588; B3 IV) is a
slowly-rotating early B-type star. Such stars are of great interest for
elemental abundance analysis, since their initial chemical abundances
remain fairly intact throughout their short lifetimes,  and thus these
abundances also constrain the chemistry of the local interstellar
medium. Although
B-type stars are usually rapid rotators ($v \sin i \gtrsim 100$\,\kms),
$\iota$~Herculis is an exception. Its very low rotational velocity
makes this star particularly suitable for the measurement of weak
and/or blended spectral lines.

Over the past four decades, the optical spectrum of $\iota$~Herculis
has been exhaustively studied; the literature now approximately agrees
on the values of effective temperature, surface gravity and rotational
velocity \citep[see][]{Grigsby1996}. 

The literature contains numerous studies regarding the chemical
composition of $\iota$~Herculis \citep[see
  e.g.][]{Peters1970,Kane1980,Dufton1981:Nitrogen,Lennon1983:Cii,Balona1984,Peters1985,Lester1986,Barnett1988,Grigsby1991,Pintado1993,Grigsby1996,
  Nieva2012}.  Table~\ref{Table:literature_abundance} summarizes these
abundance determinations.

The measurements listed in Table~\ref{Table:literature_abundance} have
mostly been carried out at visible wavelengths except for those of
\citet{Peters1985} who studied the abundances of $\iota$~Herculis in
the ultraviolet, using the same spectrum as this project, and reported
near solar abundance for all the elements they identified.

Since then, however, the literature has not yet converged into a fully
consistent set of abundances, and the number of elements included in
the analyses is rather limited. Therefore, a complementary study of
this target may be very useful and potentially improve the
situation. 

In this paper we model a high resolution, high signal-to-noise ratio
(S/N) spectrum of $\iota$~Herculis covering almost the full spectral window
from 999 to 1400\AA,~ in an attempt to improve our knowledge of the
elemental abundances in this star.  In spite of being publicly
available data, this ultraviolet spectrum has not previously been used
for modern analysis. We will further investigate: (1) the usefulness
of full spectrum synthesis of small windows as a tool to measure
abundances in crowded spectral regions; (2) the accuracy and
completeness of the line-list in the UV for spectrum synthesis in this
wavelength region; (3)
the accuracy of abundance measurements that can be achieved in the UV
in comparison to abundances determined in the optical in the previous
literature; (4) abundance values of elements that are accessible in
the UV but have not been previously measured in the literature. Finally (5)
we try to understand how well LTE synthesis works at this
temperature in the UV. 

Section~\ref{Sect:observation} describes the observation and reduction
of the data in more detail. In Section~\ref{Sect:modeling} we discuss
the modeling methods and tools used in this work.
Section~\ref{Sect:Abundances} provides  an introduction to the
  comprehensive element-by-element discussion of our results (found in
  the on-line Appendix) as well as a brief discussion of the "missing
lines". Finally, in Section~\ref{discussion}, we compare our results
to the literature and summarize our new results.

\section{Observation and Data Reduction} 
\label{Sect:observation}
The results presented in this work are based on the
high-spectral-resolution ultraviolet observations obtained from the
{\em Copernicus} Spectral Atlas of $\iota$~Herculis \citep[see][for
  details]{Upson1980}. The {\em Copernicus} satellite was the third
Orbiting Astronomical Observatory (OAO-3), launched in August 1972; it
operated until 1982.




The spectra used in this work contain two orders, covering wavelengths
from 999.3-1422.2\AA~(U1, second order) and 1417.9-1467.7\AA~(U1,
first order).  The data have reasonably high spectral resolution of
$\lambda/\Delta\lambda\sim 14,000$ in first order, and
$\lambda/\Delta\lambda\sim 24,000$ in second order, and the co-added
data reach S/N$>$100 \citep{Kling1996}. 

The data acquired from the {\em Copernicus} spectrometer have been reduced
with considerable care, including careful wavelength calibration,
corrections for effects due to changes in
spectrometer temperature, terrestrial and spacecraft Doppler shifts,
background corrections for cosmic rays and charged particles, 
correction for (estimated) scattered light, and for stellar image
drift on the spectrometer slit \citep[see][for more
  details]{Upson1980}.

 The total uncertainty of the relative flux presented in the atlas is
 of the order of 2 or 3\% of the local continuum flux for wavelengths
 below 1200\,\AA\ and rises gradually at longer wavelengths to about
 15\% at 1400\,\AA\ in the second order spectrum; it is about 10\%
 throughout the first order spectrum \citep{Upson1980}. Most of this
 uncertainty is due to statistical fluctuations in the photon counts.



  A normalized version of the spectrum is provided in the {\em
    Copernicus} atlas, using a smooth envelope defined by maxima in
  the Fourier-smoothed spectra. However,  we have found that this
  normalization is not a very accurate representation of the real
  stellar continuum.

  We originally used the normalized atlas spectrum in our modeling
  routine, but we found a clear mismatch between models and the
  observations in the sense that the computed high points in the
  spectrum often lie above the atlas continuum (see \S
  \ref{Sect:modeling} for details).  We resolved this by
  re-normalizing the  observed spectrum in three steps; (1)
  manually specifying continuum points over intervals of
  $\sim$20\,\AA, (2) fitting the continuum with the exponential of a
  cubic spline, and (3) dividing the spectrum by our smooth fit.  The
  re-normalization process was iterated until the least mismatch
  between the observation and model was achieved (see section \S
  \ref{Sect:modeling}). However, this process  still leaves some
    uncertainty in the correct location of the continuum, which will
  be discussed in section \S \ref{uncertainty}.

\section{Methods}


\subsection{Abundance Determination Method}
\label{Sect:modeling}

 To carry out an analysis of a spectrum as crowded as that of \iher, in
which essentially every spectral feature is blended at least to a
minor extent with other features, usually mixing individual spectral
lines of more than one element, it is essential to use the methods of
spectrum synthesis. A model of the spectrum is computed assuming
various basic parameters of the star (effective temperature, etc) and
an initial abundance table. The spectrum is computed using a line list
that one hopes contains all the significant spectral lines of all
elements, including the best atomic data (excitation level, oscillator
strength $\log gf$, damping constants) available for each
line. Assuming that the basic parameters of the star have already been
established, the abundance table is varied until the best computed fit
to the observed spectrum is obtained. This may be done by varying a
single element at a time, iterating through the elements that
contribute to a spectral window, or by varying several elements at
once. To the extent that the underlying atmospheric model is
appropriate, and the atomic data are accurate, this procedure allows
one to establish the abundances of a number of elements even in the
presence of severe blending in almost every feature.

In this work, we use the {\sc FORTRAN} spectral synthesis program {\sc zeeman.f}
\citep{Landstreet1988,Landstreet1989,Wade2001}. This program was
originally designed for  optical spectropolarimetry of magnetic
stellar atmospheres, but works well (if a little slowly) for
non-magnetic stellar spectra as well. (Note that the literature does
not contain any conclusive evidence for the presence of magnetic
fields in the atmosphere of Iota Herculis. Recent attempts to search
for weak or complex magnetic fields in this star have not detected any
non-zero Zeeman signatures \citep{Wade2014}.)

 To use a program like {\sc zeeman} that was originally designed
  for use on spectra obtained in visible light, the main concern is
  whether all necessary continuous opacity sources are included. The
  continuum opacity subroutine used by {\sc zeeman} includes the
  essential H and He bound-free and free-free ultraviolet opacities of
  H$^-$, H, and H$^+$, and of He$^-$, He, He$^+$ and
  He$^{+2}$. Continuous opacity due to electron scattering and
  Rayleigh scattering from neutral H (longward of Lyman~$\alpha$) are
  treated as pure absorption. {\sc zeeman} also includes a small
  number of ground state and low-lying bound-free transitions of
  common neutral elements, but none of ions.

{\sc zeeman} uses stellar atmospheric models produced with the code of
\citet{Kurucz1970} with solar composition, precomputed for us by
\citet{Piskunov2001}, and a set of atomic line data from the Vienna
Atomic Line Database(VALD;
\citet{Piskunov1995,Ryabchikova1997,Kupka1999,Kupka2000}). As input
parameters, the program requires the effective temperature (T$_e$),
gravitational acceleration ($\log g$),  estimated rotational
velocity ($v \sin i$), microturbulence parameter ($\xi$), and 
  initial individualized abundances, in order to compute the emergent
atmospheric spectrum. For our synthesis we used T$_e$ = 17500K, $\log
g$ = 3.8, $\xi = 1$\,\kms, and $v \sin i = 6.0$\,\kms\ \citep[see
Table~5;][]{Nieva2012}. These parameters were not varied. {\sc zeeman}
is able to synthesise a spectral window of some tens of \AA\ at once,
utilizing a line list of up to about 2000 lines. {\sc zeeman} can then
adjust the abundance of any element of choice through an iterative
process until a least squares best fit with (automatically selected)
spectral features in the window under study is achieved. This process
of course works best for unblended lines for which the abundance can
be adjusted without any contamination from other elements.

{\sc zeeman} was designed for LTE conditions; this is marginally appropriate
for $\iota$~Herculis, whose effective temperature is near the boundary
at which non-LTE effects start to become important. However,
\citet{Przybilla2011} have shown that for the main sequence stars with
effective temperatures (T $\lesssim$ 22000K) and carefully selected
spectral lines, pure LTE modeling can yield meaningful results. They
also show that despite the conceptual superiority of NLTE modeling,
inadequate model atoms can potentially result in larger
systematic errors than LTE modeling.

In this work, we use LTE modeling, even though an ideal global
match between synthetic and observed spectra can only be achieved when
the NLTE effects on the lines are fully taken
into consideration. We used a spectral line list retrieved from VALD
covering wavelengths 900 to 1500\,\AA, selected for
significant depth at the effective temperature and gravity of
$\iota$~Her. The elements included in the resulting line list in
this spectral range, assuming solar abundance, are all listed in
Table~\ref{Tab:final_results}.

We extracted line data from the VALD database using the {\it Extract
  Stellar} option. The criteria we used to select our line-list were
the following: starting wavelength: 900\,\AA,\ ending wavelength:
1500\,\AA,\ detection threshold: 0.01  of the continuum with $v \sin i
= 0$, microturbulence: 1\,\kms, T$_{\rm eff}$: 17500\,K, and $\log g$: 4
(in cgs units). We requested long extraction format in October,
2014. 

As discussed in general above, one of the main challenges in using a
crowded spectral region for abundance analysis is to identify spectral
features that respond primarily to the abundance of a single
element. {\sc zeeman} has a mode for determination of abundance of
elements one at time, in which the program first tries to identify
features that respond primarily to the abundance of the element being
fit, and then iteratively modifies the abundance until a best least
squares fit is achieved to the features, or parts of features,
identified as sensitive primarily to that element. The identification
of such features is carried out by computing two trial synthetic
spectra of the observational window being fit, one with an assumed
abundance of the element being fit that is slightly enhanced compared
to the final value expected (or guessed), and a second computed with
the element entirely absent.  {\sc zeeman} compares these two spectra
and then selects individual wavelengths in the model spectrum to use
for fitting based on simple criteria (which can be adjusted by the
user by modifying a couple of lines of code). Two examples of criteria
used would be to require (1) that any blending feature (i.e. still present
in the spectrum with the fitted element removed) be no
deeper than 0.95 of the continuum, and that the depth of the feature
with the desired element present be deeper than 0.7 of the continuum,
or (2) that the line be deeper than 0.7 of the continuum, and more
than 4 times deeper than blending features.

After selecting a list of wavelength grid points in the computed
spectrum based on such criteria, {\sc zeeman} iterates the abundance
of the desired element to find a least squares best fit to the
selected wavelengths, using a downhill simplex method. At the end of
the process, the value of the abundance which produces the best fit is
reported, along with a plot file of the current best fit spectrum for
the window, the initial spectrum without the element being fit, and a
map of wavelengths used in the fit. Graphical visual examination of
these spectra enables the user to decide if the program has actually
found useful spectral features or not, to decide if the fit found is
based on strong enough evidence to consider that the reported value is
useful, and to see whether the best fit is consistent over the various
features available in the window, whether explicitly fitted or not.
Many such fits to small windows, comparing the observed spectrum to
both the model spectrum with a best fitting abundance, and also to a
spectrum with the fitted element absent (e.g. Figure~1), will be used
in this paper to illustrate our results. The examination of such
comparison spectra is an extremely useful technique for studying
synthetic spectral models of crowded spectral regions.

Using the techniques outlined above, we find that several elements
exhibit a few apparently clean and unblended lines. After identifying
those lines, we used them to adjust the abundances of those elements
(see \S \ref{Sect:Abundances} ). Table~\ref{Tab:final_results}, which
presents the abundance values (ion by ion) that are found to best fit
the observed spectrum, lists the lines that we found to be most
useful. 

For some elements of interest, reasonably strong lines are present in
our spectrum, but most are not really free of blends.  The analysis of
blended lines is slightly more challenging since they can have more
than one element contributing to the overall shape of the line
profile. In these cases, a few more steps are required.  We first
iterated over the abundance of each element, assuming solar abundance
for the rest in order to identify which wavelength window could best
be used to adjust the abundance of that particular element. This first
approximate model gives us a first estimate of the abundance even
though it is not the final result.  Once the first iterations are
carried out, we are left with an initial set of model
parameters. Having selected the most useful window(s) for each
element, we then iterated the abundances of both the element of
interest and also the blending element(s) until we obtained the best
possible global fit to the observation.

The computed model spectrum, using the final set of abundances,
results in a reasonably good fit to the observed spectrum (see
Appendix B), although it is quite clear that there are
lines in the observed spectrum that are missing from the VALD list.
One significant source of such lines is due to absorption by the Lyman
bands of interstellar H$_2$. 

\subsection{Line Broadening}
 Because we are trying to fit individual spectral features as
  accurately as possible in spectra of fairly high resolution, we must
  take into account both local line broadening (natural broadening,
  van der Waals and Stark broadening, which are included in the
  computation of the local line profile opacity coefficients as a
  function of depth in the atmosphere), and broadening due to various
  velocity fields, including micro- and macroturbulence, pulsation,
  and rotation. 

The microturbulent velocity $\xi$ is the non-thermal component of the
local gas velocity in the spectral line formation region of the
stellar atmosphere \citep{Cowley1996}. The validity of this concept
has been the subject of debate for decades. For instance, it has been
suggested that the use of non-LTE in spectral analysis should
eliminate the need for microturbulence.  However, it is found by
\citet{Nieva2011} that consistent fitting of many spectral lines in
early B stars with an advanced non-LTE code still requires the
introduction of microturbulence.
It is suggested that the origin of the photospheric microturbulence in
hot stars is the sub-surface convection. Convective regions in the
envelope of a hot massive star are generated by the opacity peaks of
ionized iron and (nearer the surface) of the He~{\sc ii} to He~{\sc
  iii} ionization zone \citep[see][for more
  details]{Cantiello2009}. (There may also be very weak convection at
the He~{\sc i} to He~{\sc ii} ionization zone, which is within the
visible atmosphere.)

\citet{Nieva2012} have carefully carried out an extensive non-LTE
analysis of high-resolution optical spectra for a sample of early
B-type stars including $\iota$~Herculis. Through an iterative process,
they have constrained the stellar parameters including the
microturbulence. They find a microturbulence of $\xi \sim 1$\,\kms for
$\iota$~Herculis. We have included their result in our calculations.

In contrast to microturbulence, macroturbulence is the photospheric
velocity field with scales much longer than the mean free path of the
photons. This broadening component changes the overall shape of
spectral lines from roughly Gaussian (when thermal and microturbulent
broadening dominate) or U-shaped (when rotational velocity broadening
dominates) to a more triangular form. It has been proposed that
macroturbulent broadening in hot stars may be the result of many
low-amplitude pulsation modes \citep{Aerts2009}.  Since \iher\ is
  known to be a slowly pulsating B star \citep{Chapellier2000}, we
  might expect to find some indication of macroturbulence in the line
  profiles.  However, \citet{Nieva2012} showed that in case of
$\iota$~Herculis, macroturbulence is negligible, and the spectral 
profiles can be reproduced with only rotational broadening and
microturbulence.  The small (a few \kms) and quite slow excursions
  in radial velocity due to the pulsations do not seem to have led to
  obvious radial velocity inhomogeneity in the available Copernicus
  spectra, perhaps because each short scan was corrected for radial
  velocity individually. 

 The projected rotational velocity of the star around its axis,
  $v \sin i$, is difficult to measure accurately because it is not
  much larger than typical values of other velocity fields.  It is
  necessary to take into account the impact of microturbulence as
well as macroturbulence when measuring the $v \sin i$ values
\citep{Simon2014}. The $v \sin i$ value of 6\,km\,s$^{-1}$, determined
by \citet{Nieva2012},  is used here.. 

 In fact, in the Copernicus spectrum, with a resolving power of at
  most only about 24\,000, which is not sufficient to fully resolve
  the rotational velocity, instrumental broadening is the dominant
  line broadening mechanism. We find that combining the various
  broadening mechanisms above with the instrumental profile (assumed
  Gaussian in {\sc zeeman}) yields a
  satisfactory fit to observed quasi-isolated lines in the spectrum.

\subsection{Sources of Uncertainty}
\label{uncertainty}

The stellar parameters used here (e.g. $T_{\rm eff}$ and $\log g$), of
course, are not exact but have relatively small uncertainties.  The
uncertainties associated with $T_{\rm eff}$ and $\log g$ are about
$\pm 200$\,K and $\pm 0.05$\,(cgs), respectively \citep[see
  Table~5;][]{Nieva2012}. We find that an increase of 0.05 in surface
gravity would result in an increase of typically about 0.02 in the
abundance, and the rise in temperature by 200K will cause the
abundance to increase by 0.035. Therefore, the uncertainty arising
from uncertain stellar parameters is roughly $\pm 0.04$.

Another, probably more important, source of error is the
re-normalization of the spectrum (\S \ref{Sect:observation}). In order
to evaluate this uncertainty, we have used multiple spectral lines of
one atom and attempted to determine abundance for each of them
separately (see e.g. nickel \S Appendix A).

A similar source of error is the correction for background and
scattered light, which sets the zero level of the spectrum. In some
parts of the spectrum, the strongest computed lines are deeper than
the observed lines (for example in the cores of the C~{\sc ii}
resonance doublet of UV multiplet~(1) at 1335~\AA), suggesting that
the background correction should be a little larger than the one used
by \citet{Upson1980}. We have not tried to correct for this
effect. The uncertainty actually introduced by this effect is small
because in lines deep enough for this to be a significant source of
error, we frequently have fitted the (sometimes rather broad) damping
wings rather than the core depth.

In some cases, the best-fitting models of several individual lines or
multiplets will not result in the same abundance value for that
element. This may be produced by the effect of  incorrect
  continuum normalization, lines missing from our
VALD line list, inaccurate atomic data, and/or non-LTE effects of
under- or over-ionization of minority ionization stages
\citep{Lanz2007}.

The effects of missing blending lines can sometimes be identified by
careful comparison of the observed and calculated line profiles. It is
not possible to generalize about the uncertainty resulting from this
problem, but in a few cases we can include an estimate of the effect in
our uncertainty estimates.

The atomic data uncertainties for individual lines can be estimated in
some cases from the literature, especially for resonance lines for
which very accurate theoretical and/or experimental oscillator
strengths are sometimes available (for example, according to the NIST
database \citep{Kramida2014} the accuracy of the oscillator strengths
of the C~{\sc ii} resonance lines of multiplets (1) (at 1335~\AA) and
(2) (at 1036~\AA) is "A", with uncertainties of about $\pm 3$\% or
0.01~dex). 

For most of the line data in VALD, very accurate oscillator strengths
are not available (typical uncertainties for logarithmic oscillator
strengths of a light ion such as Si~{\sc ii} are of the order of $\pm
0.1$~dex, while those of Fe~{\sc ii} lines are of the order of $\pm
0.2$~dex, for example). These uncertainties are similar to those for
the optical lines often used for abundance analysis
\citep[e.g.][]{Ryabchikova1994}, and probably set an important lower limit
to uncertainty. Except when we use lines of very well determined
oscillator strengths, we will assume that atomic data introduce
uncertainties of this order.

Non-LTE effects can alter the relative populations of various levels
of a single ionization stage, and also the relative numbers of
different ionization stages. We do not have any very useful method for
identifying such effects within a single ionization stage except by
looking for large discrepancies between different "reliable" lines of
that ionization stage, and comments in the literature about individual
lines and multiplets. In contrast, we expect the non-LTE effects on
relative numbers of different ionization stages to affect primarily the
ionization stages with relatively small populations. In order to
identify the sensitive stages, we have computed (in LTE)
the relative populations of various ions of interest for temperature
-- electron density values appropriate to about continuum optical
depth (evaluated at 5000~\AA) $\tau_{\rm cont} \sim 0.2$ and $\sim
0.001$. The results are given in Table \ref{Tab:saha_ratios}. We
believe that abundances derived for ionization stages that in LTE
account for at least roughly 10\% of the total population of an element
are not very sensitive to under- or over-ionization effects, but that
less populated ionization stages may be. Thus we may be able to identify a few
cases of non-LTE ionization effects by finding systematic
discrepancies between abundances derived from dominant ionization
stages and from weakly populated stages.


In a general way, we expect that the effects discussed above will lead
to uncertainties in the derived (logarithmic) abundances of the order
of $\pm 0.3$~dex or perhaps somewhat more (or less) in some cases. We
are sometimes able to estimate the abundance uncertainties of
individual atoms by looking at the dispersion of results obtained from
different multiplets, and the scatter is indeed often of this
amplitude.


\section{Summary of Individual Elements}
\label{Sect:Abundances}  In this section, we present sample
  results for a few element identified in the spectrum of $\iota$~Her.
  We have used several spectral lines for abundance determination, in
  order to reduce the bias due to the kinds of uncertainties discussed
  above. The final abundances determined for individual ionization
  stages of these elements are listed in Table
  \ref{Tab:final_results}.  We have correspondingly omitted from our
  modeling a few \AA\ on either side of Lyman $\alpha$ (1215.67~\AA)
  and Lyman $\beta$ (1025.72~\AA).  In fact the wings of these two
  lines extend to approximately $\pm 20$ and $\pm 15$~\AA\ from line
  center respectively. {\sc zeeman} has not been enabled to compute
  hydrogen lines correctly, although there would be no difficulty in
  principle with this. The modification of the code and the proper
  analysis of these lines will be the focus of future work. We have
  modeled strong lines in the farther wings of these two lines in
  spite of the omission of Lyman line wing opacity, because the
  opacity of strong lines so completely overwhelms the opacity of the
  Lyman line wings (important mainly in the deeper atmosphere layers
  in any case) that the line strengths are largely independent of the
  neglect of the Lyman line wings. In fact the syntheses of the strong
  metallic lines in the Lyman line wings are surprisingly successful.

\begin{figure}
\resizebox{\hsize}{!}{\includegraphics{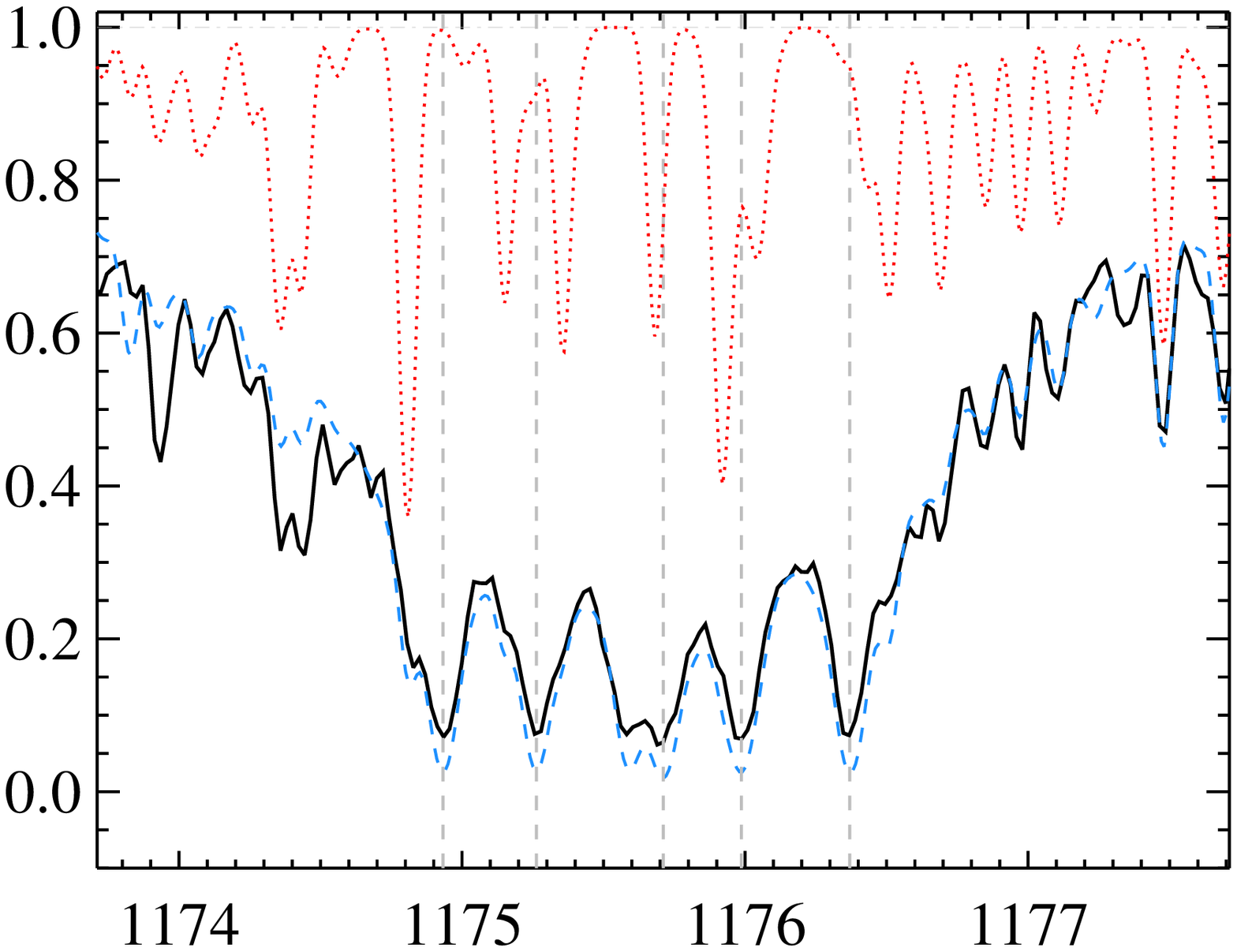}}\\
\resizebox{\hsize}{!}{\includegraphics{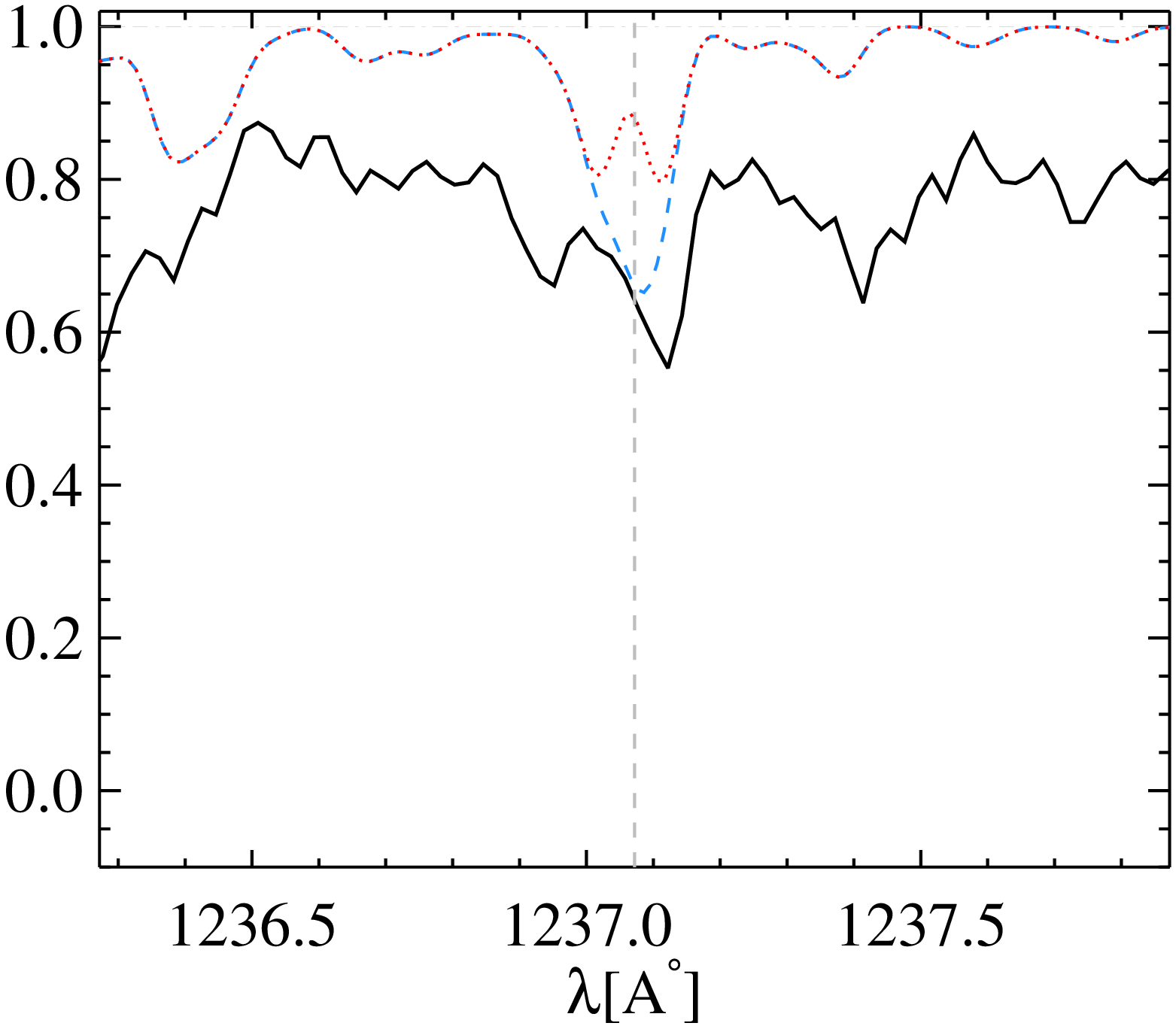}%
\includegraphics{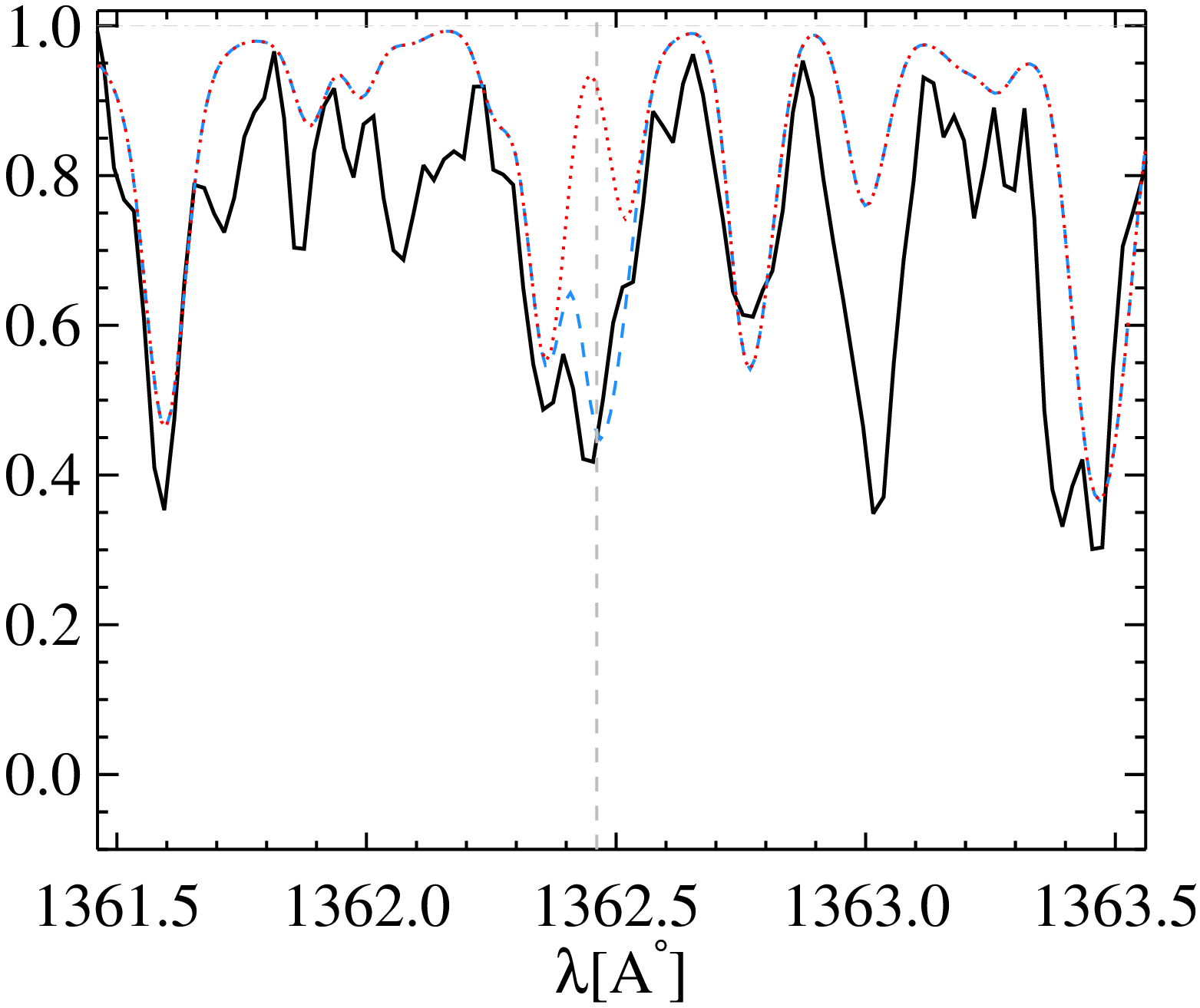}}
\caption{\label{Fig:sample_figures} In this figure the y-axis
  represents the normalized flux and the x-axis represents the
  wavelength in units of angstrom. The black curve is the observed
  spectrum of $\iota$-Herculis. The blue dashed line is the calculated
  model spectrum using the specified abundance and the red dotted line
  is the calculated model in the absence of that element. {\it Top:
    The C\,{\sc iii} lines of multiplet (4) around 1174\,\AA, with
    prominent wings extending to several \AA\ from the deepest part of
    the feature.}  {\it Bottom left:}A strong line Ge\,{\sc ii} at
  1237.072\AA,\ {\it Bottom right:} B\,{\sc ii} at 1362.461\AA.}
\end{figure}

  It is worthwhile to mention that iron provides strong line opacity
  over the entire wavelength range studied here. Probably most of the
  lines of other elements are at least slightly blended with lines of
  iron. Therefore, iron was the first element for which we determined
  the abundance (see section \S 1.17 in Appendix A). However, in what
  follows we only provide a few examples of these results.  Figure
  \ref{Fig:sample_figures} shows a sample of our results for three
  elements carbon, boron, and germanium. In case of carbon, we have a
  relatively clean and unblended set of multiplets which allows a
  clear determination of the abundance value. In contrast, in the case of
  germanium or boron, the lines are not unblended and thus are more
  suitable for determination of an upper limit for the abundance. We
  refer the reader to Appendix A for a detailed discussion of each
  element studied. Table \ref{Tab_line_list} contains a list of the
  spectral lines that we found particularly useful for this project,
  with the adopted $gf$ values and lower ionisation potentials. Table
  \ref{Tab:final_results} provides the derived abundance and upper
  limits for all the elements studied here.

Notice in Figure 1 how our technique of visually (and within {\sc
  zeeman}, numerically) comparing a spectral region with and without
the element of interest allows u to easily assess how useful a region
is for abundance determination. This method relies simply on the
availablity of a relatively complete line list, such as those provided
by VALD. In fact, such comparison allow us to test line lists and in
some cases to identify lines in the list with seriously incorrect
atomic data (see the example of Si~{\sc i} in Sirius~A
\citep{Landstreet2011}).  

\subsection{The Missing Lines}
We have done the spectrum synthesis from which we have derived
abundances using the atomic data provided in the VALD database.
However, the spectrum of $\iota$~Her still contains some fairly strong
lines that are partially or entirely missing from the calculated model
(see e.g. 1001.85\,\AA, 1003.33\, \AA, 1051.05\,\AA, 1062.9\,\AA,
1077.72\,\AA, 1078.95\, \AA, etc. in Appendix B).

One source of lines missing from our synthesis is lines due to
interstellar absorption by the Lyman bands of the H$_2$ molecule. The
transitions that affect our data are the resonance lines of electronic
transitions between the ground electronic-vibrational state (the normal
state for interstellar H$_2$ below about 30~K) and the low vibrational
states of the first electronic state. The relevant wavelengths are
provided for example in Table 4 of \citet{Krishna1972}.  These absorption
lines are all present in $\iota$~Her, in some cases clearly coinciding
with strong observed lines for which our synthesis has no match
whatever (1001.8~\AA, 1062.9~\AA, 1077.1~\AA, in other cases also
coinciding with strong absorption lines, but with ones for which
stellar lines in our synthesis fill a part of the profile (1012.8~\AA,
1092.2~\AA). It is quite clear that H$_2$ Lyman band interstellar
absorption contributes a number of significant lines to the observed
spectrum of $\iota$~Her below 1108~\AA.

We also investigated the possibility that a few of the missing lines
might be in the VALD database with incorrect $\log gf$ values. Using
the ``Extract all'' option, we requested all lines in the database
close to some of the strongest unidentified features, and examined the
resulting short lists for good wavelength coincidence. We then
considered how large a change in oscillator strength would be
required to fit the observations, and whether such a change would be
plausible. Although we found a few good wavelength coincidences, in all cases no
reasonable change in oscillator strength would lead to a good fit. This
line of investigation was eventually abandoned. 

We suspect that most of the remaining strong unidentified features are
high-excitation lines of abundant ions, perhaps usually of iron peak
elements, that simply lack oscillator strengths.


\section{Discussion and Conclusions}
\label{discussion}

\subsection{Comparison with Literature}

\begin{figure*}
\resizebox{\hsize}{!}{\includegraphics{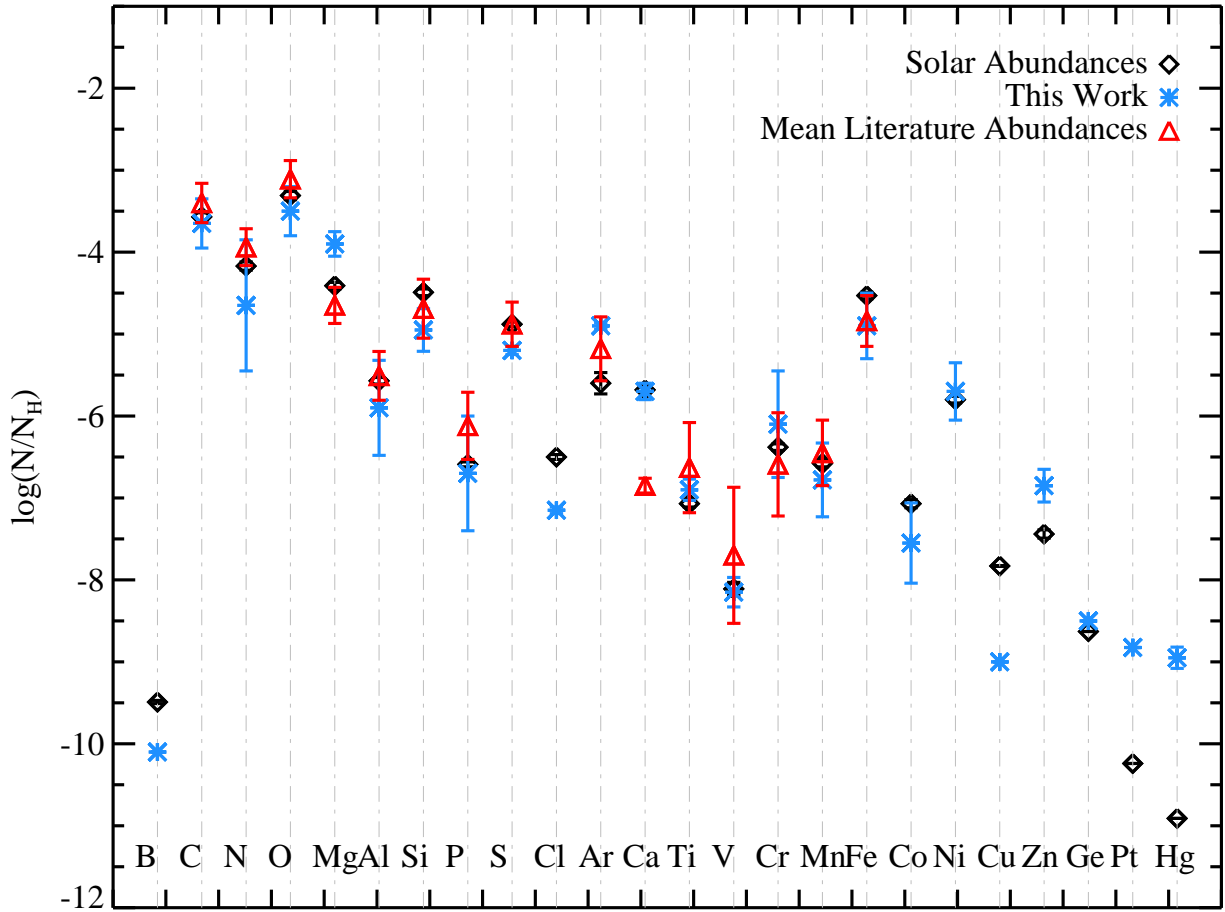}}
\caption{\label{fig:results} This figure shows a comparison between
  the solar abundances of 24 elements and the abundances derived from
  this work and the literature. Note that the literature values are
  the mean taken from Table~\ref{Table:literature_abundance}.}
\end{figure*}

This work presents a detailed UV spectral synthesis for Iota
Herculis. It is important to compare the outcome of this new approach
with previous work. In this project, we investigated the abundance of
27 elements starting from Z=5 up to Z=80. We could not determine the
abundance of a few elements that we tried to detect, specifically
neon (Z=10), gold (Z=79), and tin (Z=50) because their spectral features were
too weak and heavily blended to be useful for abundance
analysis. Among the remaining 24 elements, there are 16 that have been
studied in previous works.

The results of this study are listed in Table~\ref{Tab:final_results},
and shown graphically in Figure~\ref{fig:results}.  All 16 elements,
except for one, are in good agreement with previous work (see
Table~\ref{Table:literature_abundance}) within the estimated
uncertainties. This indicates that the UV spectral synthesis is indeed
a reasonable tool for abundance determination, at least for sharp-line
stars with fundamental parameters similar to $\iota$~Her.

The only exception to this generalization is magnesium. We find a
higher abundance of --3.90$\pm$0.19 for magnesium that even within the
measured uncertainties, is still higher than previous estimates. This
difference is probably due to the fact, discussed in Sec.~4.5, that
the dominant ionization state for magnesium at this temperature is
Mg\,{\sc iii}, but the line list contains only lines of Mg\,{\sc ii}.
Thus abundance determinations for this element are likely to be
affected by non-LTE effects which may well vary from line to line.

The line list used here does contain a few lines of gold (Z=79),
tin (Z=50), and neon (Z=10). These lines are found to be very weak and
heavily blended. Thus, unfortunately, they do not enable us to
determine the abundance of these elements. However, we have been able
to estimate a value or an upper limit for the abundance of nine
elements that were not previously studied, presumably because of a
lack of useful lines in the optical wavelength window usually used for
analysis; boron (--10.10$\pm$0.10), chlorine (-7.15$\pm$0.05),
cobalt (--7.55$\pm$0.49), nickel (--5.70$\pm$0.36), copper (--9.00),
zinc (--6.85$\pm$0.24), germanium (--8.5), platinum (--8.825$\pm$0.100),
and mercury (--8.95$\pm$0.17).  In the atmosphere of $\iota$~Herculis
the abundance of boron is sub-solar, nickel, cobalt and germanium
are at most almost solar, and platinum and mercury could even be
higher than solar by 1-2 orders of magnitude.

Overall, the aims of this project, as described in \S \ref{intro},
have been fulfilled. In addition to identifying previously unreported
elements, we find a good agreement with previous work. This conclusion
probably only applies to quite sharp-line stars because of severe line
blending even at very low $v \sin i$ values. We find that LTE modeling
in the UV is a useful tool for determining elemental abundances, at
least in stars with $T_{\rm eff}$ not much higher than that of
$\iota$~Her. Finally, in agreement with the results reported by
\citet{Landstreet2011}, we find that the available VALD line lists
seem to be missing a significant number of moderately strong lines
(quite possibly many are high-excitation lines of iron peak elements),
and that the incompleteness of the line lists appears to grow with
decreasing wavelength, even over the limited wavelength range of this
study.

\subsection{Further Considerations and future work}

 We have now carried out abundance analyses using using
  ultraviolet spectra analyzed with synthesis methods for three A or B
  stars: Sirius~A = HD~48915 from HST GHRS spectra of $R \approx
  25000$ \citep{Landstreet2011}, HD~72660 = HR~3383 using STIS spectra
  of $R \approx 114\,000$ \citep{Golriz2015}, and
  \iher\ using Copernicus spectra of $R \approx 24\,000$ or less (the
  present work). We have concluded that the technique of full spectrum
  synthesis of selected spectral windows makes it possible, in spite
  of the extreme line crowding in the UV in early-type stars, to carry
  out useful abundance analysis on the basis of such spectra. It is of
  interest to step back and look more broadly at what the limitations
  of this technique are for other spectral material and other stars.

  We can first note an important feature of the available atomic data,
  as assembled by VALD. The line lists appear to become both less
  complete and to have less accurate atomic data as analysis moves
  father into the UV from the edge of the visible spectrum at about
  3000~\AA. This effect is strongly noticeable below about 1700~\AA.
  In part this is a product of the limitations of the work done by the
  atomic physics community, but it also reflects the so far quite
  limited use of spectrum synthesis in the UV, resulting in line lists that
  have been far less carefully vetted for accuracy than those used in
  the visible window.

  However, the accuracy of results also depends strongly on various
  characteristics of the stars studied and of the available spectra.
  The model spectra are more likely to be successful in fitting the
  observations for spectra of high S/N (say well above 100) and for
  high resolution (say 40\,000 or above so that individual lines are
  nearly or completely resolved), and for stars of very low \vsi. The
  three stars analyzed by us all have \vsi\ of between 5 and 16\,\kms.
  Thus IUE high-resolution spectra, obtained with resolving power of
  around 12--13\,000 and S/N of typically 30 or less, will necessarily
  be much more difficult to model with confidence. It will be very
  important to test line lists on suitable high-resolution UV spectra
  of sharp-line standard stars covering a range of 
  effective temperature in order to empirically improve the quality of
  uncertain $gf$ values. 

  Improvement to the accuracy of much atomic data will certainly also be
  required if accurate determination of stellar parameters using
  ultraviolet spectra is to be attempted. The best determinations of
  basic parameters using visible spectra \citep[e.g.][]{Nieva2012}
  have been carried out on stars with sufficiently sparse line spectra
  that individual lines can easily be studied. This is simply not
  possible in the UV spectra of A and B stars of roughly solar
  abundance, and parameter determination will depend not only on
  accurate atomic data for selected lines, but also for the many
  possible blending lines.

We have searched our data for possible indications of a stellar wind.
Numerous studies have shown that early type stars reveal the presence
of a stellar wind primarily through blue-shifted excess absorption in
the wings of resonance lines \citep{Lamers1981,Lamers1982}. We have
modeled resonance lines of C~{\sc ii} at 1334-35~\AA, of N~{\sc ii}
at 1083-85~\AA, and of Si~{\sc iv} at 1393-1402~\AA. These appear to be
the strongest such lines present in regions for which we have data and
that we can model confidently (i.e.  they are not in the wings of
Lyman~$\alpha$ or Lyman~$\beta$. All these lines were modeled
reasonably well (e.g. Figure~1), and in none of them did we find any
hint of excess blue wing absorption.  Modeling with a code able to
model the Lyman lines, and/or data covering a larger wavelength
window, could further clarify this situation. In any case, the absence
of a detectable (mixed) stellar wind in this star is consistent with
the wind computations of \citet{Babel1996}.

Diffusion is also another process that can occur in the stable
atmosphere of slow rotating B stars. Under the action of gravity
alone, heavier atoms would sink through the ambient hydrogen down into
the stellar interior under the influence of gravity.

However, sometimes observation shows a large abundance of heavy
elements (e.g. Mn, Sr, Pt, Hg, etc.) in the atmosphere, and an
under-abundance of lighter elements (e.g. He, Ne, O, etc.). This kind
of abundance peculiarity occurs when radiation pressure from the
radiation flowing out of the stellar interior reverses the downward
drift of low-abundance elements. This phenomenon is known as
``radiative levitation''. Radiative levitation can operate
particularly effectively in slowly rotating stars, and although we do
not know the actual rotation rate of $\iota$~Her, its very low $v \sin
i$ value suggests that it may well be a slow rotator.

In the atmosphere of $\iota$~Herculis, we find,in agreement with other
studies, that most of our light elements have almost solar abundances.
For heavier elements such as zinc (Z=30), platinum (Z=78) and mercury
(Z=80),  our upper limits are unfortunately higher than the solar
  values. This is unfortunate, as it prevents us from looking for the
phenomenon observed by \citet{Hill93}, that even most normal but
slowly rotating A0V stars seem to have overabundance of some heavy
elements such as Zr, Ba and La.

Chemical abundances of stars provide probably the most powerful probes
of their structure and of the complex mixing and separation processes
that occur in the stellar interior and envelope and that influence
evolution on the main sequence and in later stages. Obtaining accurate
measurements of the relative abundances of as many elements as
possible is an essential first step to interpreting the information
content of stellar chemistry. This work has shown that the ultraviolet
wavelength window can make a very useful contribution to this step 
towards a better understanding of such sharp-line mid-B stars as
$\iota$~Her. 

\section{Acknowledgments}
\label{acknowledgement}
SSG acknowledges support from the Department of Physics and Astronomy
of the University of Western Ontario. SSG and JDL acknowledge
financial support for this work from the Natural Sciences and
Engineering Council of Canada.  This work is based on observations
made with {\it Copernicus} satellite or Orbiting Astronomical
Observatory 3 (OAO-3), a collaborative effort between the USA(NASA)
and the UK(SERC). The UV telescope on board belonged to the Princeton
University. Some of the data presented in this paper were obtained
from the Mikulski Archive for Space Telescopes (MAST). STScI is
operated by the Association of Universities for Research in Astronomy,
Inc., under NASA contract NAS5-26555. Support for MAST for non-HST
data is provided by the NASA Office of Space Science via grant
NNX13AC07G and by other grants and contracts.

\begin{table*}
 \centering
 \caption{\label{Table:literature_abundance} Iota Herculis- Summary of
   literature abundances. (1) \citet{Pintado1993}; (2)
   \citet{Schmidt1979}; (3) \citet{Balona1984}; (4)
   \citet{Lester1986}; (5) \citet{Kane1980}; (6) \citet{Barnett1988};
   (7) \citet{Grigsby1991}; (8) \citet{Dufton1981:Nitrogen}; (9)
   \citet{Peters1985} ; (10) \citet{Peters1970}; (11)
   \citet{Nieva2012}. Note: The first eleven columns represent
   abundances from difference sources and the last column represents
   the solar abundances which are taken from \citet[][I \&
     II]{scott2014I}, \citet{Grevesse2014} and \citep{Asplund2009}}
\begin{tabular}{ccccccccccccc}
  \hline
Species &  1 & 2  & 3 &4 & 5 & 6 & 7 &8 &9 &10& 11& $\log (N/N_{\rm H})_{\odot}$ \\  \hline 
\ion{He}{1} & --1.07 &  &  &  &  & &  --0.042 &  &  &  & & --1.07$\pm$0.01 \\   
\ion{C}{2} & --3.39$\pm$0.28 &--3.50&--3.47&--3.30&--3.87&--3.87& --3.48 & &
&--3.23$\pm$0.38& --3.60$\pm$0.07 & --3.57$\pm$0.05\\   
\ion{N}{2}  & --3.84$\pm$0.17 & &&&--4.10& & --3.99 &--4.10 &-4.11$\pm$0.39&--3.72$\pm$0.22 & --4.11$\pm$0.12 &--4.17$\pm$0.05 \\   
\ion{O}{1} & & & & & & &  & & --3.01$\pm$0.15 & & --3.20$\pm$0.09 & --3.31$\pm$0.05 \\   
\ion{O}{2}  & --2.91$\pm$0.23 & & & & --3.30 & & --3.11 & & --3.33$\pm$0.44 & & &  --3.31$\pm$0.05 \\   
\ion{Ne}{1} & & & & & & & & & & --3.35$\pm$0.19  & --3.95$\pm$0.07 & --4.07$\pm$0.10 \\  
\ion{Ne}{2} & & & & & & & --3.97 &  & --3.36$\pm$0.26 & & &--4.07$\pm$0.10 \\  
\ion{Mg}{2} & --4.78$\pm$0.26 & & & & & & --4.46 & &--4.66$\pm$0.09 & --4.72$\pm$0.09 & --4.44$\pm$0.06 &--4.41$\pm$0.04 \\   
\ion{Al}{2} & --6.03 &  & & & & & --5.57   & & & & &--5.57$\pm$0.04\\  
\ion{Al}{3} & --5.49$\pm$0.25  & & & & & & & & --5.58$\pm$0.24  & --5.47$\pm$0.17  && \\  
\ion{Si}{2} & --5.16$\pm$0.23  & & & & & & --4.49 & &--4.96$\pm$0.47 &--4.86$\pm$0.34 & --4.49$\pm$0.05 & --4.49$\pm$0.03 \\  
\ion{Si}{3} & --4.45$\pm$0.34  &  & & & & & & & --4.59$\pm$0.44 & --4.45$\pm$0.25 &&\\  
\ion{Si}{4} & & & & & & & & & --4.61$\pm$0.06   &  & &\\  
\ion{P}{2}  & --5.90 & & & & & & & & --5.63$\pm$0.55 & --5.65$\pm$0.36& & --6.59$\pm$0.03\\   
\ion{P}{3}  & --6.25$\pm$0.13  & & &&&&&&--6.98$\pm$0.6&& &\\  
\ion{S}{2}  & --4.91$\pm$0.20 &  & & & & & & &  --4.83$\pm$0.35& --4.90$\pm$0.18 & & --4.88$\pm$0.03 \\  
\ion{S}{3}  & --4.64$\pm$0.10  &  & & & & & & & --5.05$\pm$0.36 &--4.99$\pm$0.43 & & \\  
\ion{Ar}{2} & --5.18$\pm$0.25  &  & & & & & & &--5.14$\pm$0.59 & --5.22$\pm$0.33& & --5.60$\pm$0.13\\   
\ion{Ca}{2} & --6.03 & & & & & & & & --5.84 &  --6.85$\pm$0.09  & & --5.68$\pm$0.03\\  
\ion{Ti}{3} & &&&&&&&& --6.63$\pm$0.55  & & & --7.07$\pm$0.04 \\  
\ion{V}{3}  &&&&&&&&&  --7.70$\pm$0.83   &  & & --8.11$\pm$0.08\\  
\ion{Cr}{3} & &&&&&&&& --6.59$\pm$0.63   & & & --6.38$\pm$0.04\\  
\ion{Mn}{3} &&&&&&&&& --6.45$\pm$0.40   & & & --6.58$\pm$0.04\\  
\ion{Fe}{2} &--5.14$\pm$0.24&&&&&& --4.37 &&-5.18$\pm$0.41 &--5.87$\pm$0.43& --4.49$\pm$0.08 & --4.53$\pm$0.04 \\   
\ion{Fe}{3} & --4.35$\pm$0.24 & & & & & & & & --4.50$\pm$0.31 & --4.37$\pm$0.23 & & --4.53$\pm$0.04 \\      
\hline
\end{tabular}
\end{table*}
 
\begin{table*}
 \centering
 \caption{\label{Tab:final_results} This table shows the abundances
   determined in this work for every ion and the recommended value for
   every element, the solar abundances from \citet[][I \&
     II]{scott2014I}, \citet{Grevesse2014} and \citet{Asplund2009},
   and the wavelengths used to determine these values. Note that the
   recommended value is the variance of all the available literature
   values. The subset of VALD list used in our models can be found
   here: www.astro.uwo.ca/$\sim$jlandstr/iota\_her\_linelist/linelist.txt}
\begin{tabular}{cccccc}
\hline
Element & Abundance &  $\log (N/N_{\rm H})$ & $\log (N/N_{\rm H})_{\odot}$  \\ 
or ion & (this work) & &  \\ \hline 
B\,{\sc ii}   & --10.100$\pm$0.100 & --10.100$\pm$0.100  &--9.300$\pm$0.020  \\
C\,{\sc ii}   & &--4.450$\pm$0.300  &  \\
C\,{\sc iii}  & &--3.750$\pm$0.250  &  \\
C & --3.55$\pm$0.300 & & --3.570$\pm$0.05  \\
N\,{\sc i}    & &--4.350$\pm$0.340  & \\ 
N\,{\sc ii}   & &--4.860$\pm$0.340  & \\ 
N & --4.650$\pm$0.800& & --4.170$\pm$0.050 \\ 
O\,{\sc i}    & &--3.500$\pm$0.300  &  \\ 
O\,{\sc ii}   & &--3.550$\pm$ 0.300 &  \\
O & --3.500$\pm$ 0.300&& --3.310$\pm$0.050  \\ 
Mg\,{\sc ii}  & --3.900$\pm$0.150&--3.900$\pm$0.190  & --4.410$\pm$0.040 \\
Al\,{\sc iii} & --5.900$\pm$0.580&--5.900$\pm$0.580  & --5.570$\pm$0.040 \\
Si\,{\sc ii}& &--4.650$\pm$0.27 & \\ 
Si\,{\sc iii} & &--4.950$\pm$0.20 & \\
Si\,{\sc iv}& &--4.750$\pm$0.15 &  \\ 
Si & --4.780$\pm$0.260&  & --4.490$\pm$0.030 \\
P\,{\sc iii}  & &--6.750$\pm$0.680  & \\
P\,{\sc ii}   & &--6.400$\pm$0.660  & \\
P & --6.700$\pm$ 0.700 && --6.590$\pm$0.030 \\ 
S\,{\sc ii}  &  &--5.300$\pm$0.050  & \\
S\,{\sc iii}  & &--5.200$\pm$0.100  & \\
S\,{\sc iv}   & &--5.200$\pm$0.100  & \\ 
S & --5.200$\pm$ 0.100&&--4.880$\pm$0.030 \\ 
Cl\,{\sc ii}  & --7.150$\pm$0.100&--7.150$\pm$0.041 & --6.500$\pm$0.030 \\
Ar\,{\sc i}   & --4.900$\pm$0.100&--4.900$\pm$0.041 & --5.600$\pm$0.13 \\
Ca\,{\sc ii}  & --5.700$\pm$0.100&--5.700$\pm$0.010 & --5.680$\pm$0.030 \\
Ti\,{\sc iii} & --6.900$\pm$0.170&--6.900$\pm$0.130 & --7.070$\pm$0.040\\
V\,{\sc iii}  & --8.150$\pm$0.180&--8.150$\pm$0.180 &--8.110$\pm$0.080\\
Cr\,{\sc iii} & --6.100$\pm$0.650&--6.100$\pm$0.650 &--6.380$\pm$0.040  \\
Mn\,{\sc iii} & --6.780$\pm$0.450&--6.780$\pm$0.450 & --6.580$\pm$0.040 \\
Fe\,{\sc ii} & &--4.950$\pm$0.390 \\
Fe\,{\sc iii} & &--4.750$\pm$0.400  & \\ 
Fe & --4.900$\pm$0.400&& --4.530$\pm$0.040 \\ 
Co\,{\sc ii}  & &--7.550$\pm$0.100  & \\
Co\,{\sc iii} & &--7.450$\pm$0.490  & \\
Co & --7.55$\pm$0.490& &  --7.070$\pm$0.050 \\ 
Ni\,{\sc ii}  & &--5.500 $\pm$0.360  & \\
Ni\,{\sc iii} & &--5.800 $\pm$0.300  & \\
Ni & --5.700$\pm$0.350 && --5.800$\pm$0.040 \\ 
Cu\,{\sc ii} & --9.000 & & --7.830  \\
Zn\,{\sc iii} & --6.850$\pm$0.200&--6.850$\pm$0.200  &--7.440 $\pm$0.050 \\
Ge\,{\sc ii} & --8.500 && --8.630  \\
Pt\,{\sc iii} & --8.825&--8.825  & --10.240   \\
Hg\,{\sc iii} & --8.950$\pm$0.130&--8.950$\pm$0.130  &  --10.91\\
\hline 
\end{tabular}
\end{table*}

\begin{table*}
\centering
\caption{\label{Tab_line_list} This table contains a list of adopted
  atomic data for the most useful and important spectral lines used
  in the modelling of various chemical elements in the UV spectrum of
$\iota$~Her. }
\begin{tabular}{cccccc}
\hline
Spec Ion & $WL_{\rm vac}$ & $\log gf$ &  $E_{\rm low}$ & Ref  \\
        &       (A)      &           &      (eV)      &      \\
\hline
  'B 2' &  1362.4610     & -0.076    &  0.000         & VALD  \\ \\ 
  'C 3' &  1174.9330     & -0.468    &  6.496         & VALD  \\  
  'C 3' &  1175.2630     & -0.565    &  6.493         & VALD  \\  
  'C 3' &  1175.5900     & -0.690    &  6.496         & VALD  \\  
  'C 3' &  1175.7110     &  0.009    &  6.503         & VALD  \\  
  'C 3' &  1175.9870     & -0.565    &  6.496         & VALD  \\  
  'C 3' &  1176.3700     & -0.468    &  6.503         & VALD  \\  
  'C 3' &  1247.3830     & -0.314    & 12.690         & VALD  \\  
  'C 2' &  1323.8617     & -1.284    &  9.290         & VALD  \\  
  'C 2' &  1323.9059     & -0.337    &  9.291         & VALD  \\  
  'C 2' &  1323.9510     & -0.144    &  9.290         & VALD  \\  
  'C 2' &  1323.9950     & -1.288    &  9.291         & VALD  \\  
  'C 1' &  1328.8333     & -1.236    &  0.000         & VALD  \\  
  'C 1' &  1329.0849     & -1.231    &  0.002         & VALD  \\  
  'C 1' &  1329.1000     & -1.147    &  0.002         & VALD  \\  
  'C 1' &  1329.1230     & -1.355    &  0.002         & VALD  \\  
  'C 1' &  1329.5775     & -0.662    &  0.005         & VALD  \\  
  'C 1' &  1329.6000     & -1.136    &  0.005         & VALD  \\  
  'C 2' &  1334.5320     & -0.589    &  0.000         & VALD  \\  
  'C 2' &  1335.6627     & -1.293    &  0.008         & VALD  \\  
  'C 2' &  1335.7077     & -0.335    &  0.008         & VALD  \\  \\
  'N 2' &  1085.5290     & -2.277    &  0.016         & VALD  \\  
  'N 2' &  1085.5460     & -1.095    &  0.016         & VALD  \\  
  'N 2' &  1085.7010     & -0.337    &  0.016         & VALD  \\  
  'N 1' &  1199.5500     & -0.278    &  0.000         & VALD  \\  
  'N 1' &  1200.2230     & -0.459    &  0.000         & VALD  \\  
  'N 1' &  1200.7100     & -0.762    &  0.000         & VALD  \\  
  'N 1' &  1243.1710     & -1.511    &  2.384         & VALD  \\  
  'N 1' &  1243.1790     & -0.349    &  2.384         & VALD  \\  
  'N 1' &  1243.3060     & -0.541    &  2.385         & VALD  \\  
  'N 1' &  1243.3130     & -1.484    &  2.385         & VALD  \\  
  'N 2' &  1275.0380     & -1.206    & 11.436         & VALD  \\  
  'N 2' &  1275.2510     & -1.944    & 11.438         & VALD  \\  
  'N 2' &  1276.2010     & -1.478    & 11.438         & VALD  \\  
  'N 2' &  1276.2250     & -1.948    & 11.438         & VALD  \\  
  'N 2' &  1276.8000     & -1.828    & 11.438         & VALD  \\ \\
  'O 1' &  1039.2300     & -1.339    &  0.000         & VALD  \\  
  'O 1' &  1040.9430     & -1.561    &  0.020         & VALD  \\  
  'O 1' &  1041.6880     & -2.037    &  0.028         & VALD  \\  
  'O 2' &  1132.3890     & -1.854    & 14.889         & VALD  \\  
  'O 1' &  1302.1680     & -0.585    &  0.000         & VALD  \\  
  'O 1' &  1304.8580     & -0.808    &  0.020         & VALD  \\  
  'O 1' &  1306.0290     & -1.285    &  0.028         & VALD  \\ \\
 'Mg 2' &  1239.9250     & -3.530    &  0.000         & VALD  \\  
 'Mg 2' &  1240.3950     & -3.830    &  0.000         & VALD  \\  
 'Mg 2' &  1367.2570     & -2.080    &  4.434         & VALD  \\  
 'Mg 2' &  1367.7080     & -2.240    &  4.422         & VALD  \\  \\
  'Al 2' &  1190.0460     & -1.260    &  4.644         & VALD  \\  
  'Al 2' &  1190.0510     & -0.780    &  4.644         & VALD  \\  
  'Al 2' &  1191.8080     & -1.260    &  4.659         & VALD  \\  
  'Al 2' &  1191.8140     & -0.510    &  4.659         & VALD  \\  
  'Al 3' &  1352.8100     & -0.020    & 14.377         & VALD  \\  
  'Al 3' &  1352.8580     & -0.180    & 14.377         & VALD  \\  
  'Al 3' &  1379.6700     & -0.600    &  6.656         & VALD  \\  \\
  'Si 4' &  1128.3250     & -0.480    &  8.896         & VALD  \\  
  'Si 4' &  1128.3400     &  0.470    &  8.896         & VALD  \\  
  'Si 2' &  1190.4158     & -0.245    &  0.000         & VALD  \\  
  'Si 2' &  1193.2897     &  0.075    &  0.000         & VALD  \\  
  'Si 2' &  1260.4221     &  0.462    &  0.000         & VALD  \\  
  'Si 2' &  1264.7377     &  0.710    &  0.036         & VALD  \\  
  'Si 2' &  1265.0020     & -0.273    &  0.036         & VALD  \\  
  'Si 3' &  1296.7260     & -0.127    &  6.537         & VALD  \\  
  'Si 3' &  1298.8920     & -0.257    &  6.553         & VALD  \\  
  'Si 3' &  1298.9460     &  0.443    &  6.585         & VALD  \\  
  'Si 3' &  1301.1490     & -0.127    &  6.553         & VALD  \\  
  'Si 3' &  1303.3230     & -0.037    &  6.585         & VALD  \\  \\
   'P 3' &  1003.6000     & -0.400    &  0.069         & VALD  \\  
   'P 2' &  1249.8296     &  0.102    &  1.101         & VALD  \\  
   'P 3' &  1334.8130     & -1.300    &  0.000         & VALD  \\  
   'P 3' &  1344.3260     & -1.040    &  0.069         & VALD  \\  
   'P 3' &  1344.8500     & -2.000    &  0.069         & VALD  \\  \\
   'S 1' &  1070.5492     & -1.096    &  0.000         & VALD  \\  
   'S 4' &  1072.9960     & -0.829    &  0.118         & VALD  \\  
   'S 4' &  1073.5280     & -1.789    &  0.118         & VALD  \\  
   'S 3' &  1077.1340     & -1.078    &  1.404         & VALD  \\  
   'S 3' &  1200.9560     & -1.030    &  0.103         & VALD  \\  
   'S 3' &  1201.7220     & -1.780    &  0.103         & VALD  \\  
   'S 3' &  1202.1200     & -2.949    &  0.103         & VALD  \\  
   'S 2' &  1250.5840     & -1.670    &  0.000         & VALD  \\  
   'S 2' &  1253.8110     & -1.400    &  0.000         & VALD  \\  
   'S 2' &  1259.5190     & -1.320    &  0.000         & VALD  \\  \\
  'Cl 2' &  1071.0360     & -1.140    &  0.000         & VALD  \\  
  'Cl 2' &  1075.2290     & -1.710    &  0.124         & VALD  \\  \\
  'Ar 1' &  1048.2200     & -0.590    &  0.000         & VALD  \\  
  'Ar 1' &  1066.6600     & -1.190    &  0.000         & VALD  \\  \\
  'Ca 2' &  1369.5595     & -0.866    &  1.700         & VALD  \\  
  'Ca 2' &  1432.5028     & -0.793    &  1.692         & VALD  \\  \\
  'Ti 3' &  1282.4835     & -1.425    &  0.023         & VALD  \\  
  'Ti 3' &  1286.2325     & -2.659    &  0.000         & VALD  \\  
  'Ti 3' &  1286.3687     & -0.360    &  0.052         & VALD  \\  
  'Ti 3' &  1295.8835     & -0.439    &  0.000         & VALD  \\  
  'Ti 3' &  1298.6330     & -0.906    &  0.052         & VALD  \\  
  'Ti 3' &  1298.6974     & -0.271    &  0.000         & VALD  \\  \\
   'V 3' &  1154.2254     &  0.030    &  1.511         & VALD  \\  
   'V 3' &  1154.2663     & -1.030    &  0.072         & VALD  \\  
   'V 3' &  1160.7610     &  0.093    &  2.084         & VALD  \\  
   'V 3' &  1252.1043     & -0.183    &  1.511         & VALD  \\  
   'V 3' &  1332.0022     &  0.035    &  2.105         & VALD  \\  \\
  'Cr 3' &  1038.1569     & -0.483    &  0.008         & VALD  \\  
  'Cr 3' &  1038.7815     & -1.580    &  2.567         & VALD  \\  
  'Cr 3' &  1038.7913     & -1.178    &  2.288         & VALD  \\  
  'Cr 3' &  1038.9642     & -1.242    &  2.585         & VALD  \\  
  'Cr 3' &  1039.6145     & -2.115    &  2.129         & VALD  \\  
  'Cr 3' &  1040.0594     & -0.925    &  0.000         & VALD  \\  
  'Cr 3' &  1040.1681     & -0.940    &  0.071         & VALD  \\  
  'Cr 3' &  1040.3963     & -0.461    &  2.585         & VALD  \\  
  'Cr 3' &  1040.5132     & -0.100    &  2.603         & VALD  \\  
  'Cr 3' &  1040.7280     & -1.742    &  0.008         & VALD  \\  
  'Cr 3' &  1040.8037     & -0.896    &  2.567         & VALD  \\  
  'Cr 3' &  1041.1433     & -1.036    &  2.213         & VALD  \\  
  'Cr 3' &  1041.3379     & -1.170    &  0.044         & VALD  \\  
  'Cr 3' &  1051.5224     & -1.566    &  2.142         & VALD  \\  
  'Cr 3' &  1051.8971     & -0.786    &  4.588         & VALD  \\  
  'Cr 3' &  1052.3525     & -1.116    &  3.394         & VALD  \\  
  'Cr 3' &  1052.8860     & -1.638    &  2.157         & VALD  \\  
  'Cr 3' &  1063.6493     & -1.778    &  2.288         & VALD  \\  
  'Cr 3' &  1064.3176     & -0.275    &  2.295         & VALD  \\  
  'Cr 3' &  1064.4098     & -0.370    &  2.288         & VALD  \\  
  'Cr 3' &  1065.0790     & -1.164    &  2.295         & VALD  \\  
  'Cr 3' &  1065.1372     & -1.222    &  2.304         & VALD  \\  
  'Cr 3' &  1098.6094     & -2.112    &  2.295         & VALD  \\  
  'Cr 3' &  1098.8730     & -1.209    &  3.196         & VALD  \\  
  'Cr 3' &  1099.4410     & -0.815    &  3.986         & VALD  \\  \\
  'Mn 3' &  1046.1717     & -0.899    &  5.344         & VALD  \\  
  'Mn 3' &  1046.1809     & -1.345    &  5.113         & VALD  \\  
  'Mn 3' &  1088.7056     & -1.039    &  5.952         & VALD  \\  
  'Mn 3' &  1088.7354     & -1.742    &  3.626         & VALD  \\  
  'Mn 3' &  1111.1038     & -0.351    &  3.329         & VALD  \\  
  'Mn 3' &  1111.2125     & -1.308    &  3.330         & VALD  \\  
  'Mn 3' &  1112.2778     & -0.799    &  5.767         & VALD  \\  
  'Mn 3' &  1239.2397     & -1.434    &  4.015         & VALD  \\  
  'Mn 3' &  1239.2550     & -1.554    &  4.015         & VALD  \\  \\
  'Fe 2' &  1129.6209     & -1.264    &  0.232         & VALD  \\  
  'Fe 2' &  1129.7656     & -1.828    &  0.232         & VALD  \\  
  'Fe 2' &  1129.8143     & -1.290    &  3.230         & VALD  \\  
  'Fe 3' &  1129.8552     & -1.806    &  6.222         & VALD  \\  
  'Fe 2' &  1130.3427     & -1.848    &  0.048         & VALD  \\  
  'Fe 2' &  1130.3447     & -2.021    &  0.107         & VALD  \\  
  'Fe 3' &  1130.3969     & -1.120    &  0.127         & VALD  \\  
  'Fe 3' &  1130.4104     & -3.670    &  3.117         & VALD  \\  
  'Fe 2' &  1130.4431     & -1.327    &  0.083         & VALD  \\  
  'Fe 2' &  1130.5596     & -1.792    &  0.083         & VALD  \\  
  'Fe 3' &  1130.6841     & -2.958    &  3.117         & VALD  \\  
  'Fe 2' &  1130.8632     & -1.900    &  0.232         & VALD  \\  
  'Fe 3' &  1131.1888     & -1.244    &  0.116         & VALD  \\  
  'Fe 2' &  1131.2969     & -1.473    &  2.676         & VALD  \\  
  'Fe 2' &  1135.3020     & -1.751    &  0.387         & VALD  \\  
  'Fe 2' &  1135.5488     & -2.364    &  0.352         & VALD  \\  
  'Fe 2' &  1135.5778     & -1.760    &  0.352         & VALD  \\  
  'Fe 3' &  1136.3235     & -2.105    &  6.222         & VALD  \\  
  'Fe 3' &  1136.3707     & -3.302    &  3.764         & VALD  \\  
  'Fe 3' &  1142.4549     & -0.861    &  3.808         & VALD  \\  
  'Fe 2' &  1142.4687     & -1.249    &  2.844         & VALD  \\  
  'Fe 2' &  1142.4741     & -0.753    &  2.807         & VALD  \\  
  'Fe 2' &  1142.7573     & -0.878    &  2.807         & VALD  \\  
  'Fe 3' &  1142.9501     & -0.646    &  3.826         & VALD  \\  
  'Fe 2' &  1143.1133     & -1.009    &  2.828         & VALD  \\  
  'Fe 2' &  1143.1501     & -0.945    &  2.635         & VALD  \\  
  'Fe 2' &  1143.2257     & -0.716    &  0.000         & VALD  \\  
  'Fe 2' &  1143.2814     & -0.983    &  2.844         & VALD  \\  
  'Fe 3' &  1143.3251     & -3.115    &  3.829         & VALD  \\  
  'Fe 3' &  1143.5399     & -1.747    &  3.808         & VALD  \\  
  'Fe 3' &  1143.6662     & -0.966    &  3.809         & VALD  \\  
  'Fe 2' &  1144.0493     & -0.735    &  2.583         & VALD  \\  
  'Fe 3' &  1153.8769     & -3.124    &  3.826         & VALD  \\  
  'Fe 3' &  1153.9014     & -1.908    &  7.095         & VALD  \\  
  'Fe 2' &  1153.9449     & -1.975    &  0.121         & VALD  \\  
  'Fe 3' &  1153.9497     & -2.385    &  6.147         & VALD  \\  
  'Fe 2' &  1154.3448     & -1.181    &  2.583         & VALD  \\  
  'Fe 2' &  1154.3512     & -1.149    &  3.153         & VALD  \\  
  'Fe 2' &  1154.3974     & -1.438    &  0.121         & VALD  \\  
  'Fe 3' &  1154.5736     & -2.730    &  6.222         & VALD  \\  
  'Fe 3' &  1154.6153     & -1.928    &  6.222         & VALD  \\  \\
  'Co 3' &  1043.2424     & -0.698    &  1.913         & VALD  \\  
  'Co 3' &  1046.7565     & -1.172    &  1.913         & VALD  \\  
  'Co 3' &  1088.5088     & -1.412    &  1.913         & VALD  \\  
  'Co 2' &  1466.2110     & -0.112    &  0.000         & VALD  \\  \\
 'Ni 2' &  1308.8657     & -1.485    &  0.000         & VALD  \\ 
 'Ni 3' &  1321.3248     & -1.924    &  7.605         & VALD  \\  
 'Ni 3' &  1321.7990     & -1.663    &  6.777         & VALD  \\  
 'Ni 3' &  1322.2543     & -1.799    &  7.605         & VALD  \\  
 'Ni 2' &  1345.8782     & -1.411    &  0.000         & VALD  \\  
 'Ni 2' &  1370.1323     & -0.090    &  0.000         & VALD  \\ \\
  'Cu 2' &  1358.7730     & -0.174    &  0.000         & VALD  \\  \\
  'Zn 3' &  1456.7090     & -0.557    &  9.829         & VALD  \\  
  'Zn 3' &  1464.1800     & -0.698    &  9.829         & VALD  \\  
  'Zn 3' &  1359.7990     &  0.525    & 17.916         & VALD  \\  \\
 'Ge 2' &  1237.0720     &  0.145    &  0.000         & VALD  \\ \\
  'Pt 3' &   999.6650     & -0.474    &  0.656         & VALD  \\  \\
  'Hg 3' &  1330.7700     &  0.658    &  5.313         & VALD  \\  
  'Hg 3' &  1377.8300     &  0.065    &  5.707         & VALD  \\

\hline 
\end{tabular}
\end{table*}

\begin{table*}
 \centering
 \caption{\label{Tab:saha_ratios} The percentage ionization ratios
   derived from the Saha equation. Here $n_{tot}$ is
   the total number density in every state ($n_{tot}$ =
   $n_0$+$n_1$+$n_2$+$n_3$). This table shows the
   ionization balances in two different regions; {\it left column: } the
   region where the continuum and weak lines are formed ($\tau_{5000} \sim 0.7$, $\tau_{1110} \sim \tau_{1320} \sim 0.2$, 
   where $T \approx 17000$~K, $n_e \approx 3.84\,10^{14}$) and {\it right
     column: } the region where line cores of strong lines are formed
   ($\tau_{5000} \sim  \tau_{1110} \sim \tau_{1320} \sim 7\,10^{-4}$, where $T \approx 11800$~K, $n_e \approx 5.13\,10^{12}$) respectively \citep{Kurucz1979}.}
\begin{tabular}{ccccc|ccccc}
\hline
    &   & \ \ \ $\tau$ = 0.2 & &  & & &   \ \ \ \ \ $\tau$ = 7$\times$10$^{-4}$&  &  \\

Species & [I]/ $ n_{tot}$ & [II] /$ n_{tot}$ & [III] /$ n_{tot}$ & [IV] /$ n_{tot}$  && [I]/$ n_{tot}$ & [II] /$ n_{tot}$ & [III] /$ n_{tot}$ & [IV] /$ n_{tot}$  \\ 
\hline 
B &   0.002&27.30&72.69&0.009 &\vline & 2.07$\times$10$^{-3}$ &97.52 & 2.50 & 4.25$\times$10$^{-8}$ \\
C &   0.007&72.84&27.14&1.71$\times$10$^{-5}$&\vline&0.01&99.53&0.46&0.00 \\
N &   0.03&95.68&4.29&3.03$\times$10$^{-7}$ & \vline&0.08&99.91&9.76$\times$10$^{-3}$&  0.00 \\
O &   0.073&99.55&0.37&0.00&\vline&0.17&99.83& 1.3$\times$10$^{-4}$&  0.00 \\
Ne &  1.11&98.88&0.005&0.00&\vline&26.31&73.70&1.81$\times$10$^{-7}$&0.00\\
Mg &  4.98$\times$10$^{-7}$&0.17&99.83&0.00&\vline& 5.60$\times$10$^{-7}$&0.60&99.40&0.00\\
Al &  5.80$\times$10$^{-6}$&0.49&92.73&6.77&\vline&1.34$\times$10$^{-5}$&6.42&93.56& 2.20$\times$10$^{-2}$ \\
Si &  1.49$\times$10$^{-5}$ & 1.19  &97.84&0.97& \vline&2.93$\times$10$^{-5}$ &6.39&93.61&5.53$\times$10$^{-4}$\\
P &   5.02$\times$10$^{-5}$ & 2.84  &96.42&0.73& \vline&4.75$\times$10$^{-4}$&33.66&66.34&9.06$\times$10$^{-4}$\\
S &   7.30$\times$10$^{-4}$ & 8.92  &90.96&0.12&\vline&5.35$\times$10$^{-3}$&84.85 & 15.15&7.78$\times$10$^{-6}$ \\
Cl &  5.70$\times$10$^{-3}$ & 40.64 &59.34&0.01& \vline&2.50$\times$10$^{-2}$&97.84&2.13&3.05$\times$10$^{-8}$\\
Ar &  0.02&72.46&27.52&4.00$\times$10$^{-4}$&\vline&0.11&99.74&0.16&0.00\\
Ca &  2.13$\times$10$^{-8}$&0.02&99.98&2.48$\times$10$^{-5}$&\vline&0.00&2.55$\times$10$^{-2}$ &99.97 & 0.00\\
Ti &  9.53$\times$10$^{-8}$&0.04&88.28&11.68&\vline&5.68$\times$10$^{-8}$&9.44$\times$10$^{-2}$&99.85&5.80$\times$10$^{-2}$  \\
V &  1.98$\times$10$^{-7}$&0.05&94.16&5.78&\vline&1.53$\times$10$^{-7}$& 0.18&99.80&1.47 $\times$10$^{-2}$\\
Cr &  2.01$\times$10$^{-7}$&0.05&97.01&2.94&\vline&2.82$\times$10$^{-7}$&0.32&99.68&4.20$\times$10$^{-3}$\\
Mn &  6.37$\times$10$^{-7}$&0.14&98.06&1.79&\vline&8.30$\times$10$^{-7}$&0.65&99.35&1.02$\times$10$^{-3}$ \\
Fe &  1.27$\times$10$^{-6}$&0.22&98.99&0.79&\vline&2.30$\times$10$^{-6}$&1.15&98.85&1.21$\times$10$^{-3}$ \\
Co &  2.23$\times$10$^{-6}$&0.24&99.32&0.43&\vline&5.22$\times$10$^{-6}$&1.72&98.28&2.60$\times$10$^{-4}$ \\
Ni &  4.03$\times$10$^{-6}$&0.33&99.46&0.21&\vline&1.25$\times$10$^{-5}$&3.27&96.73&7.07$\times$10$^{-5}$\\
Cu &  3.57$\times$10$^{-6}$&0.29&99.60&0.11&\vline&2.25$\times$10$^{-5}$&5.64&94.36&2.05$\times$10$^{-5}$\\
Zn &  1.14$\times$10$^{-5}$&1.19&98.73&0.07&\vline&5.50$\times$10$^{-5}$&10.36&89.64&5.10$\times$10$^{-6}$\\
Ge &  9.60$\times$10$^{-6}$&0.91&98.49&0.60&\vline&1.54$\times$10$^{-5}$&4.32&95.67&2.72 $\times$10$^{-4}$  \\
Sn &  2.60$\times$10$^{-6}$&0.36&92.80&6.84&\vline&2.46$\times$10$^{-6}$&1.21&98.78&1.17$\times$10$^{-2}$ \\
Pt &  1.07$\times$10$^{-5}$&0.75&82.89&16.36&\vline&6.54$\times$10$^{-5}$&9.53&90.41&6.61$\times$10$^{-2}$ \\
Au &  5.20$\times$10$^{-5}$&3.04&92.20&4.75&\vline&3.81$\times$10$^{-4}$&42.40&57.60&5.65$\times$10$^{-3}$ \\
Hg &  1.95$\times$10$^{-5}$&1.01&98.68&0.30&\vline&1.72$\times$10$^{-4}$&11.34&88.66&1.30$\times$10$^{-4}$ \\
\hline 
\end{tabular}
\end{table*}

\bibliographystyle{mn2e}


\bsp
\label{lastpage}

\end{document}


\section{Appendix A: Individual Abundances}

In this Appendix, we present separately our detailed results for every
element clearly identified in the spectrum of $\iota$~Her. The
following sections are presented in ascending order of atomic number.
The figures shows the observation in black (solid lines), the
calculated model using the assigned abundance in blue
(dash-dot-dot-dot lines) and calculated model in the absence of the
specified element in red (long dash lines). In all figures, the
vertical axis represents the normalized flux and the horizontal axis
is the wavelength in Angstrom units, in vacuum.

\subsection{Boron, Z=5}
\label{boron}

\begin{figure}
\resizebox{\hsize}{!}{\includegraphics{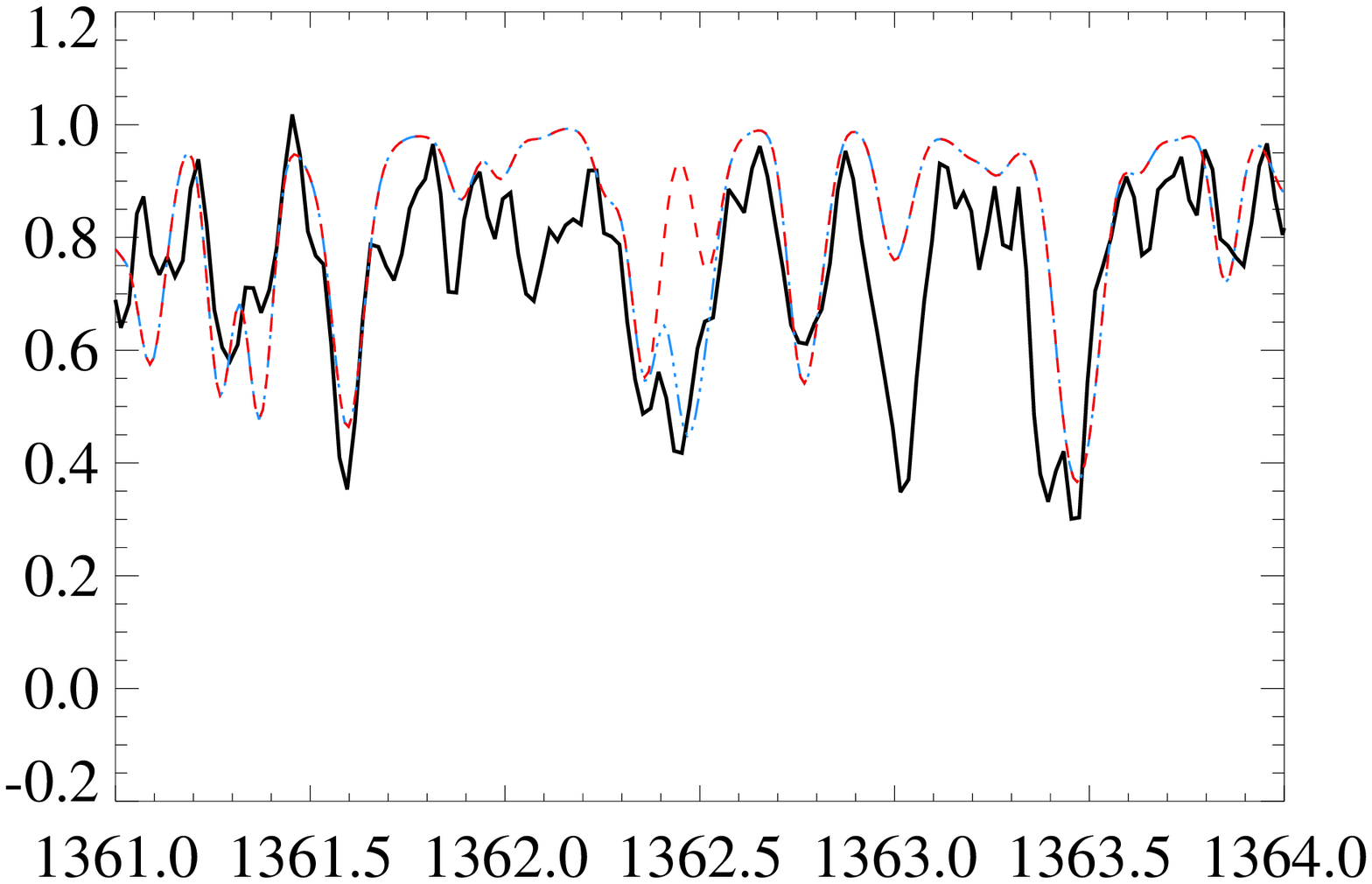}}
\caption{\label{Fig:boron_clean_line} This figure shows the
  observation (in black), calculated model with $\log(n_{\rm B}/n_{\rm
    H})= $--10.1 (in blue) and without boron (in red).}
\end{figure}

There are only a few boron lines in our VALD line list.  According to
the Saha equation (see Table~3), the dominant states of ionization in
the atmosphere are B\,{\sc ii} and {\sc iii}. We have used the B\,{\sc
  ii} resonance line at 1362.461\AA\ (Ryabtsev, 2005) for abundance
determination. This is the only boron line expected to be detectable
in our spectrum, and fortunately it is not heavily blended.  The
oscillator strength is very precisely known (it has quality ``A'' on
the NIST Atomic Spectra Database scale, corresponding to an
uncertainty of the transition probability of only about 3\,\% (Kramida
et al., 2014), and since this is a resonance line of a dominant
ionization state, our LTE treatment should introduce no important
errors. We estimate the uncertainty from fitting to be about
$\pm$0.1. Therefore the abundance is $\log(n_{\rm B}/n_{\rm H})$=
--10.1$\pm$0.1 for this element. Figure~\ref{Fig:boron_clean_line}
shows the fit to the observed profile.

Boron is a rare and fragile light element that is not produced through
stellar nucleosynthesis (except for the very minor PP~{\sc iii}
chain), but is destroyed in the interiors of stars.  There are several
suggestion regarding the formation of this element, such as cosmic ray
interactions with the ISM, or supernova neutrino spallation on
particularly \isotope[4][]{He} or \isotope[12][]{C} (see Vangioni-Flam
et al., 2000 for extensive details).

In comparison with the solar abundance of $\log(n_{\rm B}/n_{\rm H})=
-9.30 \pm 0.03$ (Asplund et al., 2009), the abundance value found
here is about a factor of 6 smaller. The presence of boron may
indicate that outer layers of $\iota$~Her have at most been partially
mixed into the stellar interior since the period of star
formation. (In contrast, for instance, in Sirius~A, the absence of
detectable boron is consistent with the hypothesis that the outer
layers were earlier inside its companion (Landstreet, 2011).)

\subsection{Carbon, Z=6}
\label{carbon}

Carbon is a very abundant element in the spectrum of $\iota$~Herculis.
In the VALD line list used here, most of the carbon lines are in
neutral form. The rest are lines of C\,{\sc ii}, C\,{\sc iii},
and C\,{\sc iv}. The distribution of carbon over ionization states
in LTE in the atmosphere of $\iota$~Her is shown in
Table~3.  Most carbon is in the form of C\,{\sc
  ii} and C\,{\sc iii}, and both contribute numerous lines. However,
even though it is a very minor constituent, weaker C\,{\sc i} lines
are evident throughout the spectrum. No C\,{\sc iv} lines are
detected. 

We determined the abundance of carbon using a simultaneous fit to
lines of C~{\sc i} UV multiplet (4) at 1329~\AA, C~{\sc ii} UV
multiplet (11) at 1324~\AA\ and UV muliplet (1) at 1335~\AA, the C~{\sc
  iii} UV multiplet (4) between 1172 and 1180~\AA, and UV multiplet
(9) at 1247~\AA. Although most of these lines have saturated line
cores, several have strong wings that are reasonably sensitive to
abundance. A good global fit is found for a logarithmic abundance of
$-3.55$, quite close to the solar value.

The 5 C~{\sc iii} triplet lines at 1174-76~\AA\ and the singlet line
at 1247.38~\AA\ have ``A+'' $gf$ values according to NIST. The
ionisation energies for the 1175~\AA\ (Moore, 1970) triplets and the
1247~\AA\ line are 6.5 and 12.7 eV, respectively. The 4 C~{\sc ii}
lines at 1323.9~\AA\ (Tachiev, 2000) have ``A'' $gf$s, and are at
9.3 eV. The C~{\sc ii} resonance lines at 1334-35~\AA\ (Moore, 1970)
also have "A" quality $gf$s. Non-LTE effects may play a role but the
$gf$ values are very accurate, and contribute negligibly to the
overall uncertainty.

It is notable that the worst fit in these windows is to the C~{\sc i}
lines at 1329~\AA. These lines are considered by Kramida et al. (2014)
to have accuracy "B", with uncertainties of about $\pm 0.04$ in $\log
gf$, so the atomic data make an unimportant contribution to the total
error budget. The abundance derived from the best fit of these lines
is about --4.65, about one dex lower than that derived from lines of
C~{\sc ii} and C~{\sc iii}, the two dominant ionization stages. This
illustrates the dangers of relying on a (very) minor ionization stage,
which in this case seems to be subject to quite significant
overionisation.

\begin{figure}
\resizebox{\hsize}{!}{\includegraphics{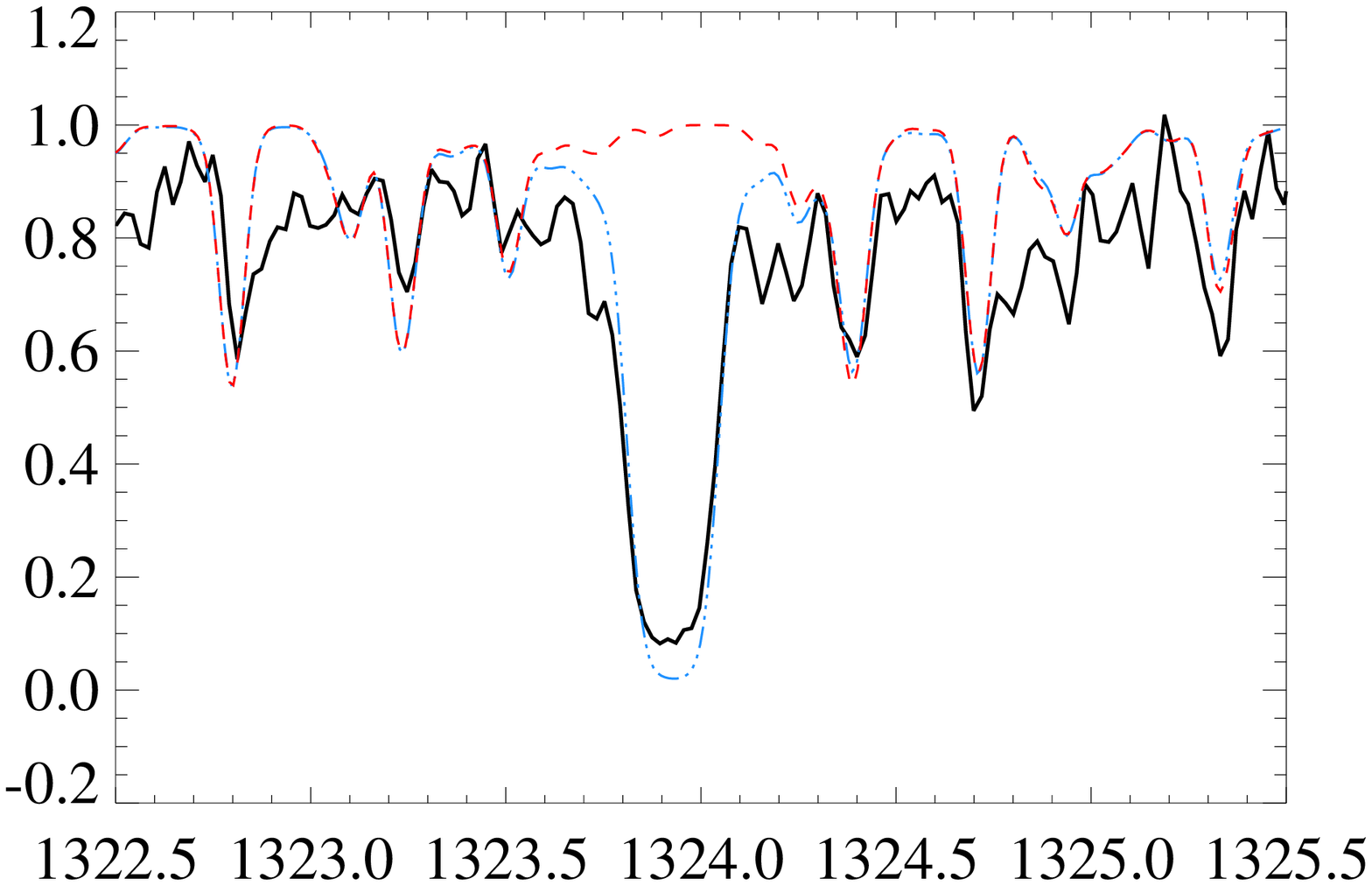}%
\includegraphics{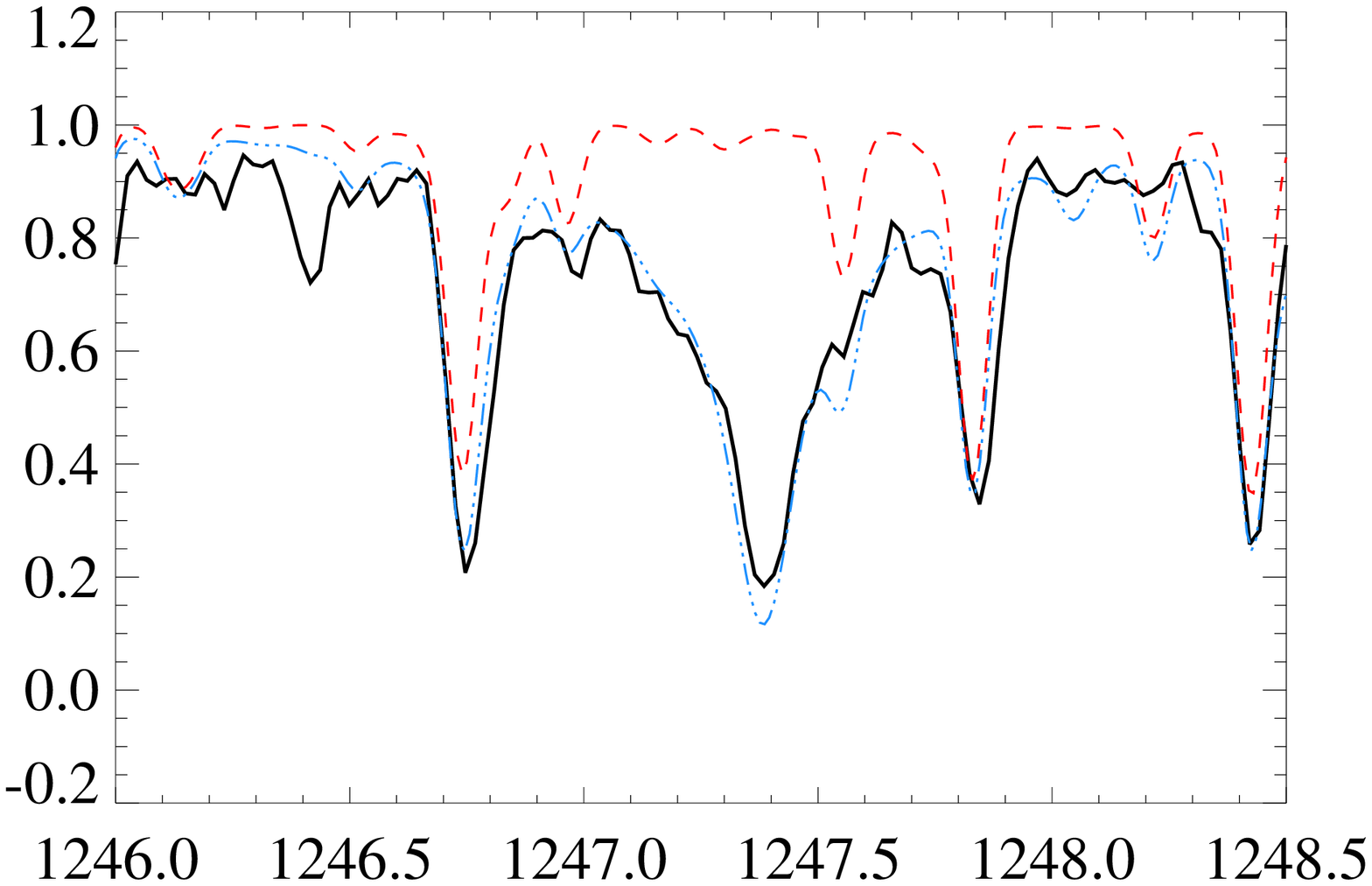}}\\ 
\resizebox{\hsize}{!}{\includegraphics{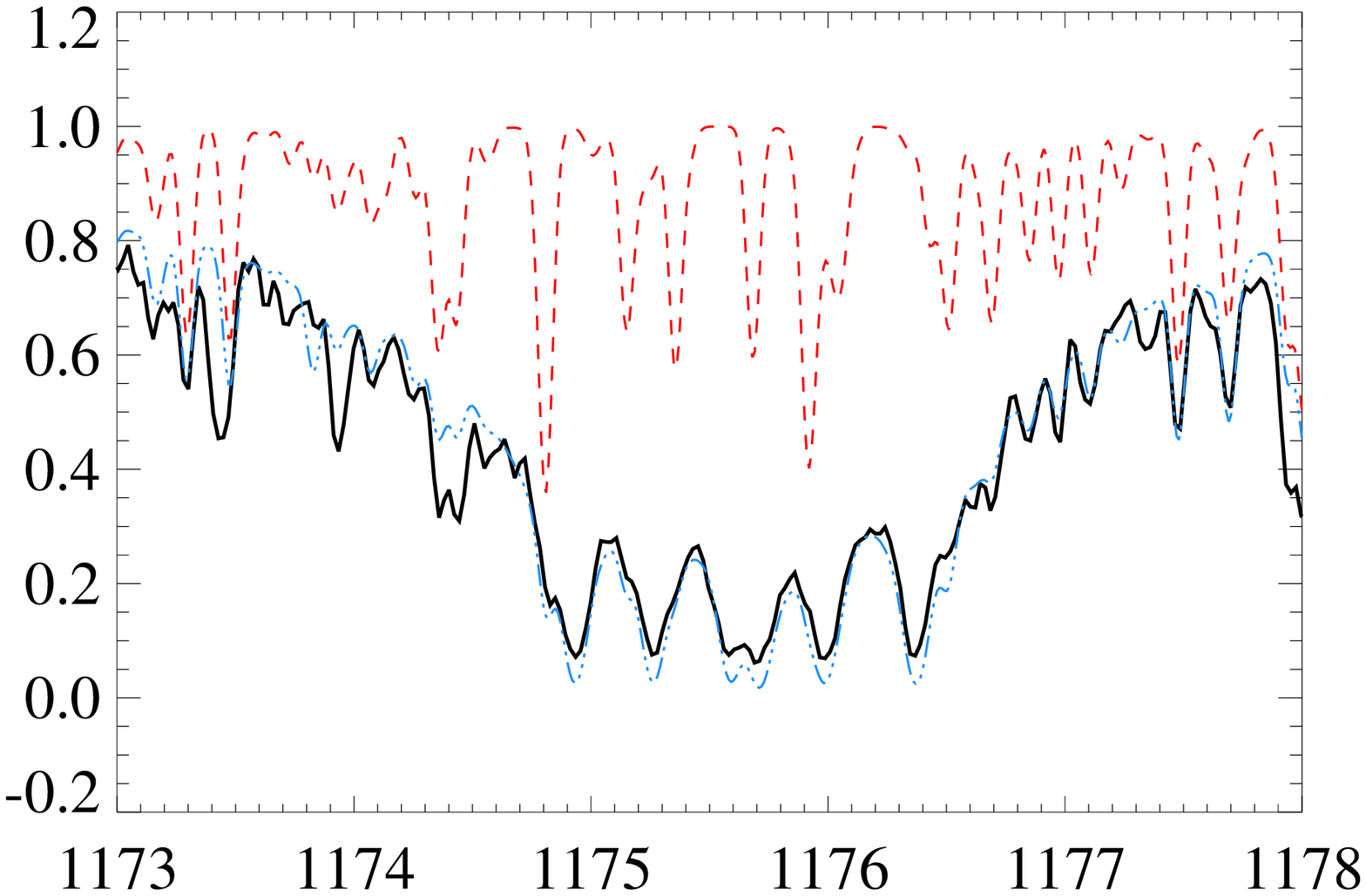}}
\caption{\label{Fig:carbon_clean_line} The figure shows the
  observation (in black), calculated model with $\log(n_{\rm C}/n_{\rm H})$=
  --3.55 $\pm$ 0.30 (in blue) and without carbon (in red). {\it Top
    left:} The unblended multiplet (11) triplet of C\,{\sc ii} at
  1324~\AA.  {\it Top right:} The single line of multiplet (9) of
  C\,{\sc iii} at 1247\,\AA.\ {\it Bottom:} The C\,{\sc iii} lines of
  multiplet (4) around 1174\,\AA, with prominent wings extending to
  several \AA\ from the deepest part of the feature.}
\end{figure}

It is worthwhile to mention that we removed from our VALD3 linelist,
six lines of C\,{\sc i} at the following wavelengths; 1253.410,
1253.467, 1253.539, 1253.541, 1254.489, 1254.511\AA,\ due to
apparently inaccurate values of oscillator strength and/or damping
constants. There are large inconsistencies between the values reported
in VALD2 and VALD3, and neither set resulted in a good fit with our
observation. Figure~\ref{Fig:carbon_clean_line} shows the fit to the
observed profile at these wavelengths.

\subsection{Nitrogen, Z=7}
\label{nitrogen}
The selected subset of the VALD database contains overall 233 lines of
nitrogen. Almost half of these lines are for the  neutral form (N\,{\sc i}),
and the rest are N\,{\sc ii} and N\,{\sc iii}, together with
a few N\,{\sc iv} lines. The strongest nitrogen contribution to the UV
spectrum seems to be primarily due to N\,{\sc ii}. There are only a
few strong lines of of N\,{\sc i} and N\,{\sc iii}. All the N\,{\sc
  iv} lines are much weaker and less prominent. This distribution is
consistent with the expected ionization ratios in the atmosphere according to
the Saha equation (see Table 3).

With a few exceptions, most nitrogen lines are rather blended in the UV
spectrum of this target. Although some N~{\sc i} lines appear largely unblended, the tiny fraction of N in this form makes abundance determination using such lines rather unreliable. Instead we have used the weakly blended N~{\sc ii} line of UV multiplet (1) at 1083.99~\AA\, and lines of N~{\sc ii} at 1275.04, 1276.20 and 1276.22~\AA\, to estimate the abundance of nitrogen. Uncertainty in both the zero point and continuum normalization, together with the fairly weak dependence of line strength on abundance, lead to a large uncertainty on the best compromise abundance of --4.65 $\pm$ 0.8~dex. 

\begin{figure}
\resizebox{\hsize}{!}{\includegraphics{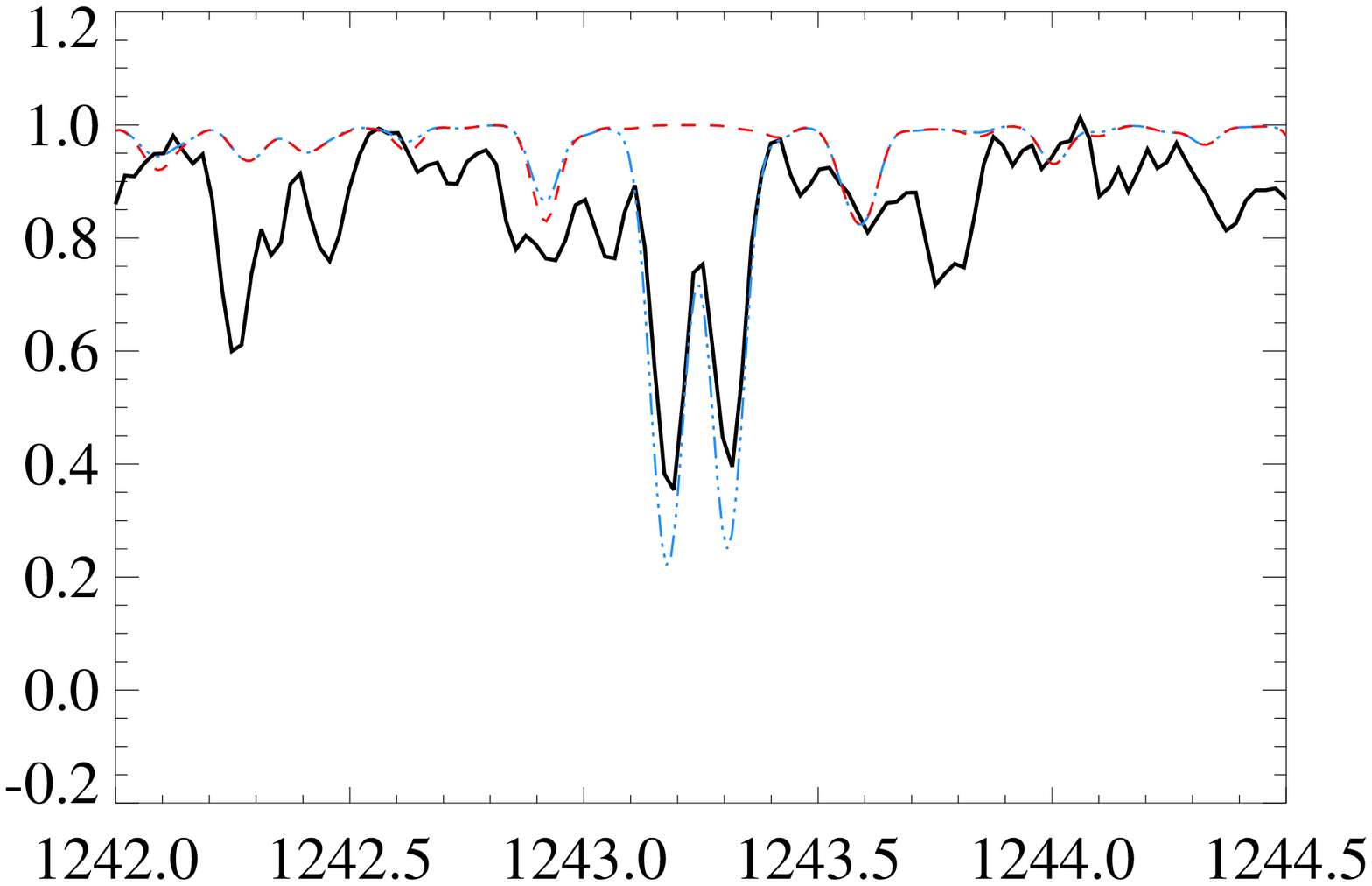}%
\includegraphics{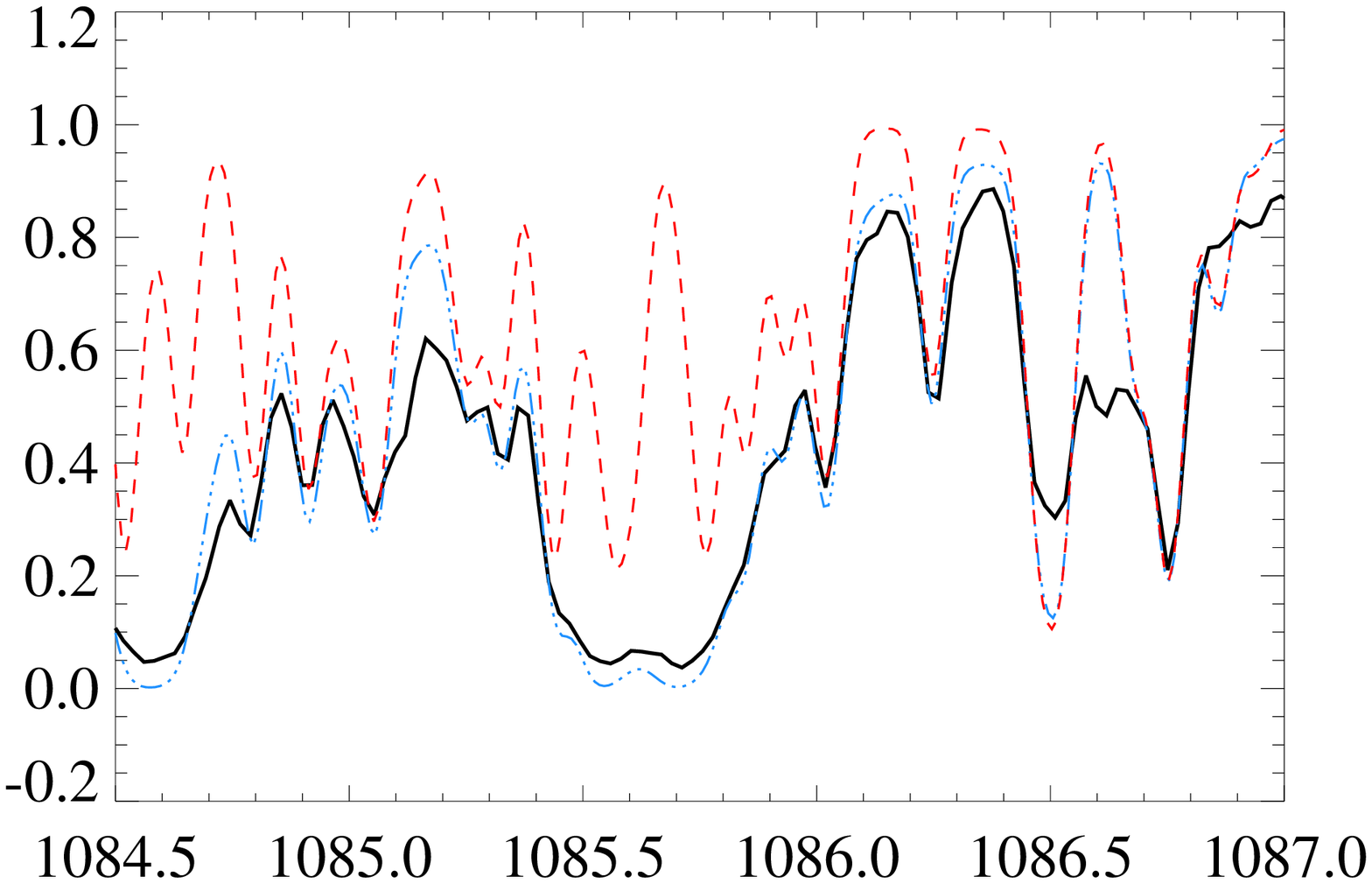}}\\
\resizebox{\hsize}{!}{\includegraphics{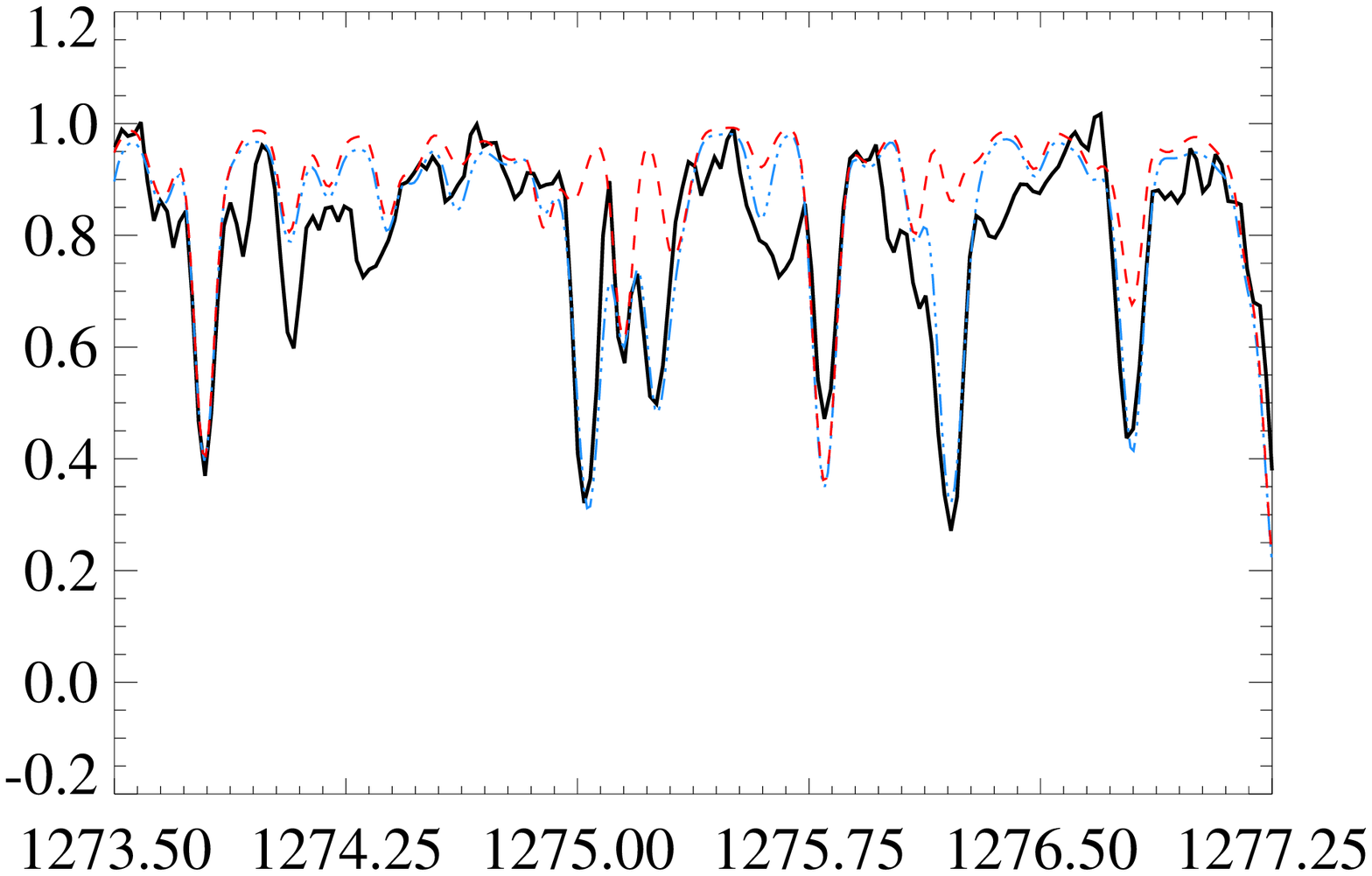}%
\includegraphics{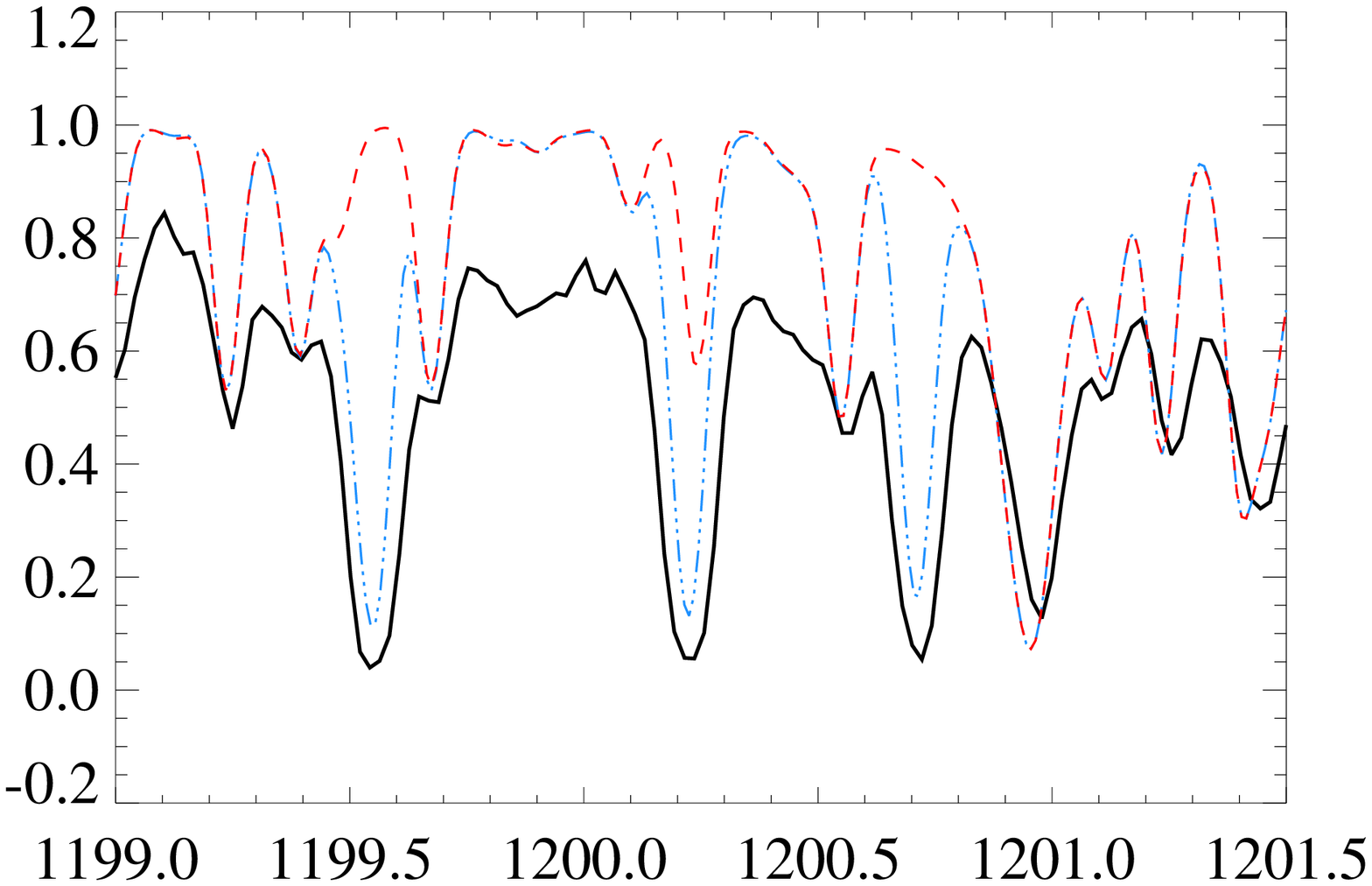}}
\caption{\label{Fig:nitrogen_clean_line} This figure shows the
  observation (in black), the best-fitting model calculated with
  $\log(n_{\rm N}/n_{\rm H})=$--4.65$\pm$0.80 (in blue) and without nitrogen (in
  red).  {\it Top left: }N\,{\sc i} UV multiplet (5) at 1243\,\AA,\ 
  {\it Top right: }N\,{\sc ii} multiplet (1) at
  1085\,\AA,\ {\it Bottom left: }N\,{\sc ii} multiplet
  at 1275--1276\,\AA,\ {\it Bottom right: }N\,{\sc i} multiplet (1) at 
  1199--1200\,\AA}
\end{figure}

Several observed lines as well as the model are shown in Figure
\ref{Fig:nitrogen_clean_line}.

\subsection{Oxygen, Z=8}

Oxygen- has a few very strong but blended lines in the spectrum of
$\iota$~Herculis. The selection of VALD lines used here contains 218
oxygen lines, of which half are in neutral form and the rest are in
the form of O\,{\sc ii} and O\,{\sc iii}, together with a few O\,{\sc
  iv} lines. The UV spectrum of $\iota$~Herculis shows a strong
presence of O\,{\sc ii} and O\,{\sc iii} with a moderate contribution
from O\,{\sc i} and absolutely no trace of O\,{\sc iv}, which is
expected given the effective temperature of this target. This
distribution is consistent with the expected ionization ratios
measured from the Saha equation (see Table~3).

\begin{figure}
\resizebox{\hsize}{!}{\includegraphics{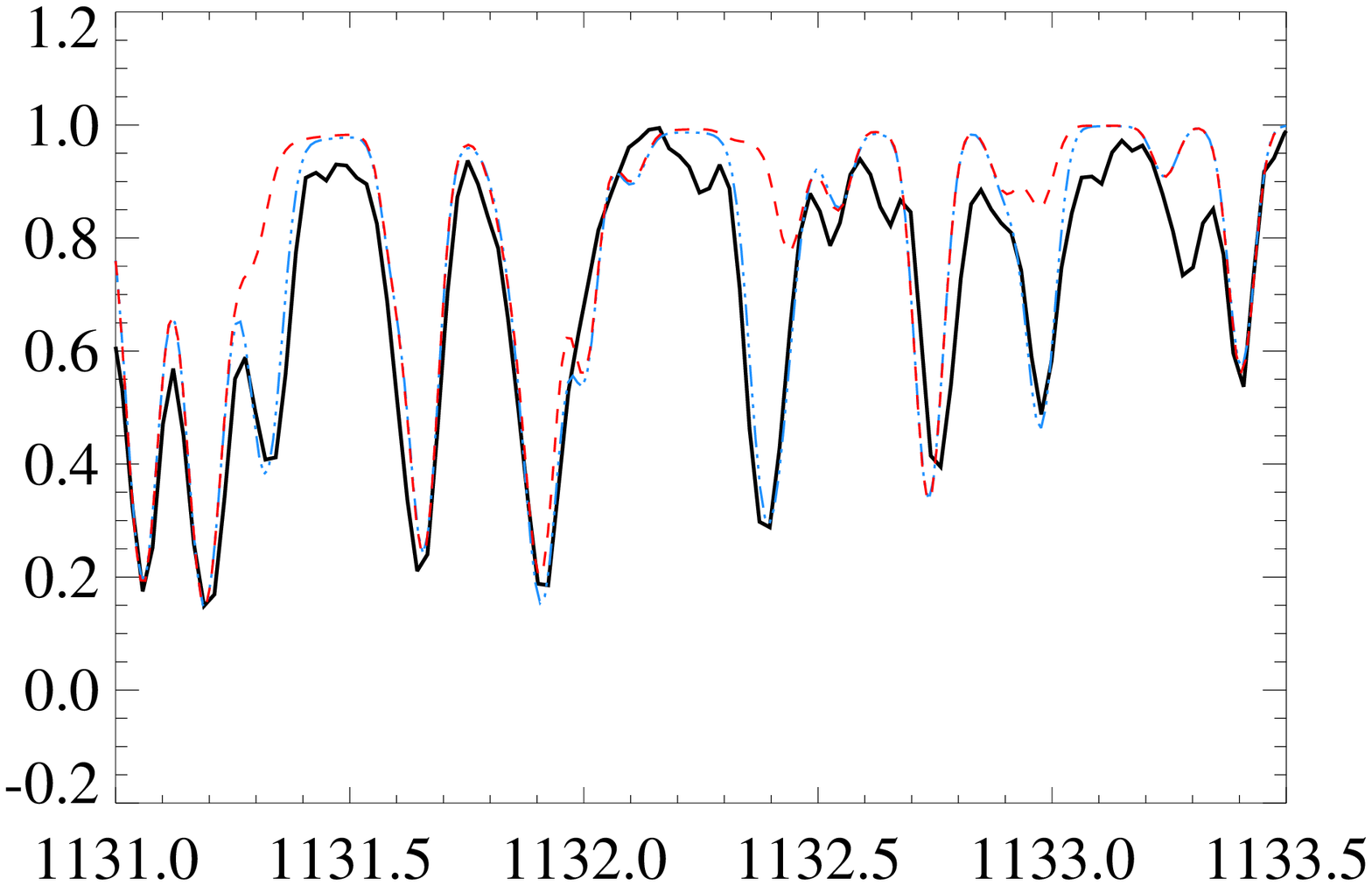}%
  \includegraphics{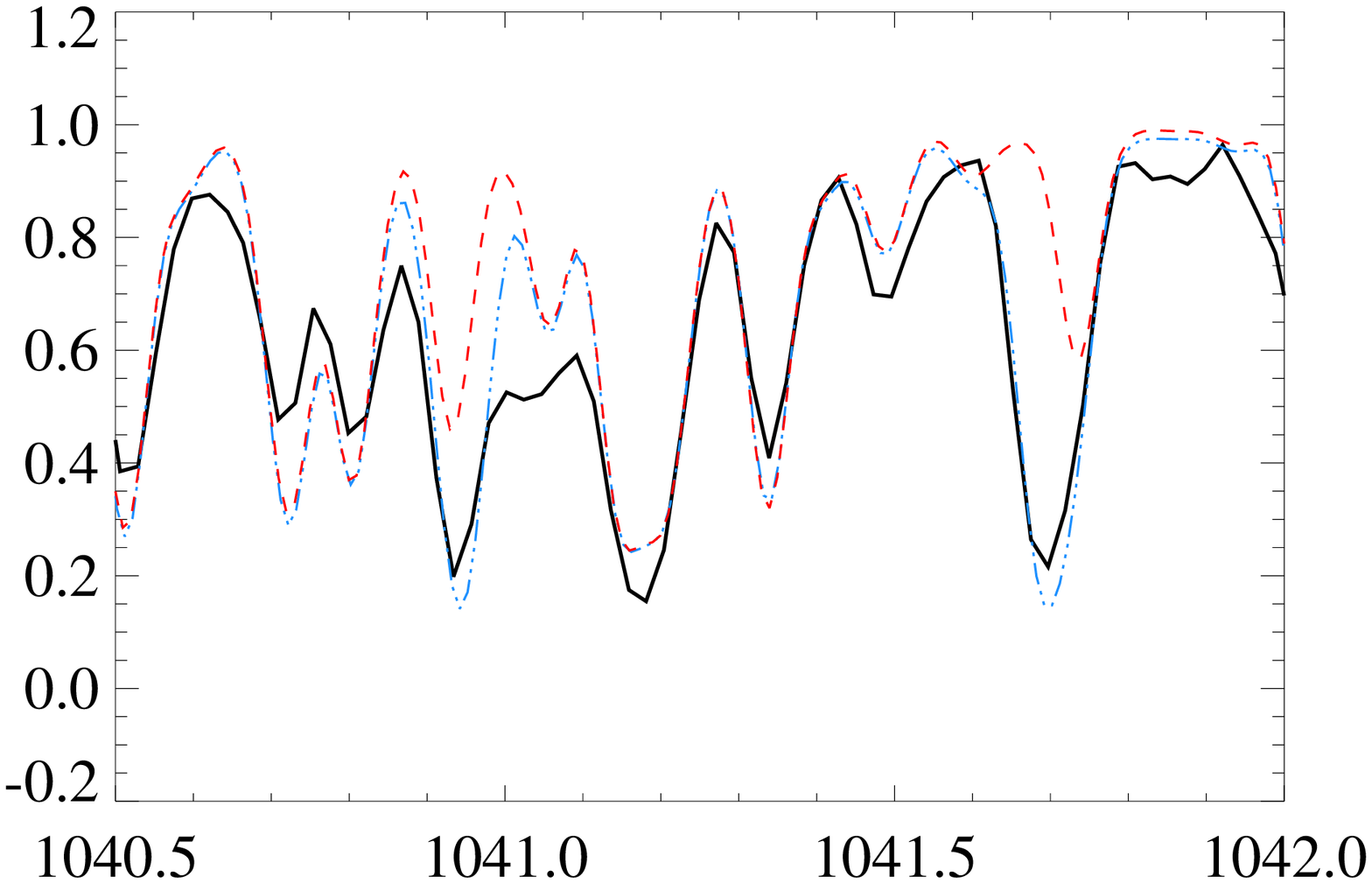}}\\ 
\resizebox{\hsize}{!}{\includegraphics{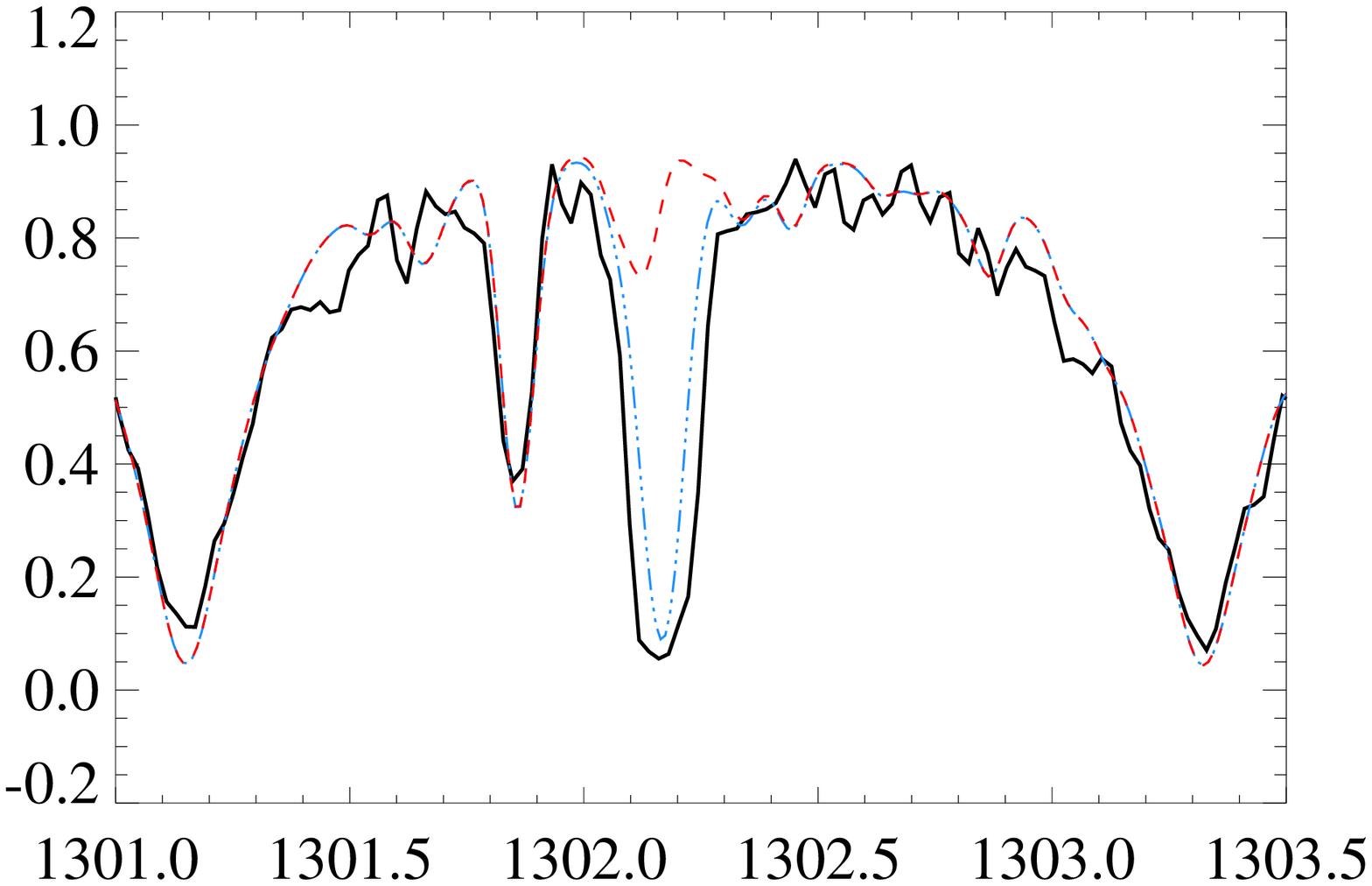}%
  \includegraphics{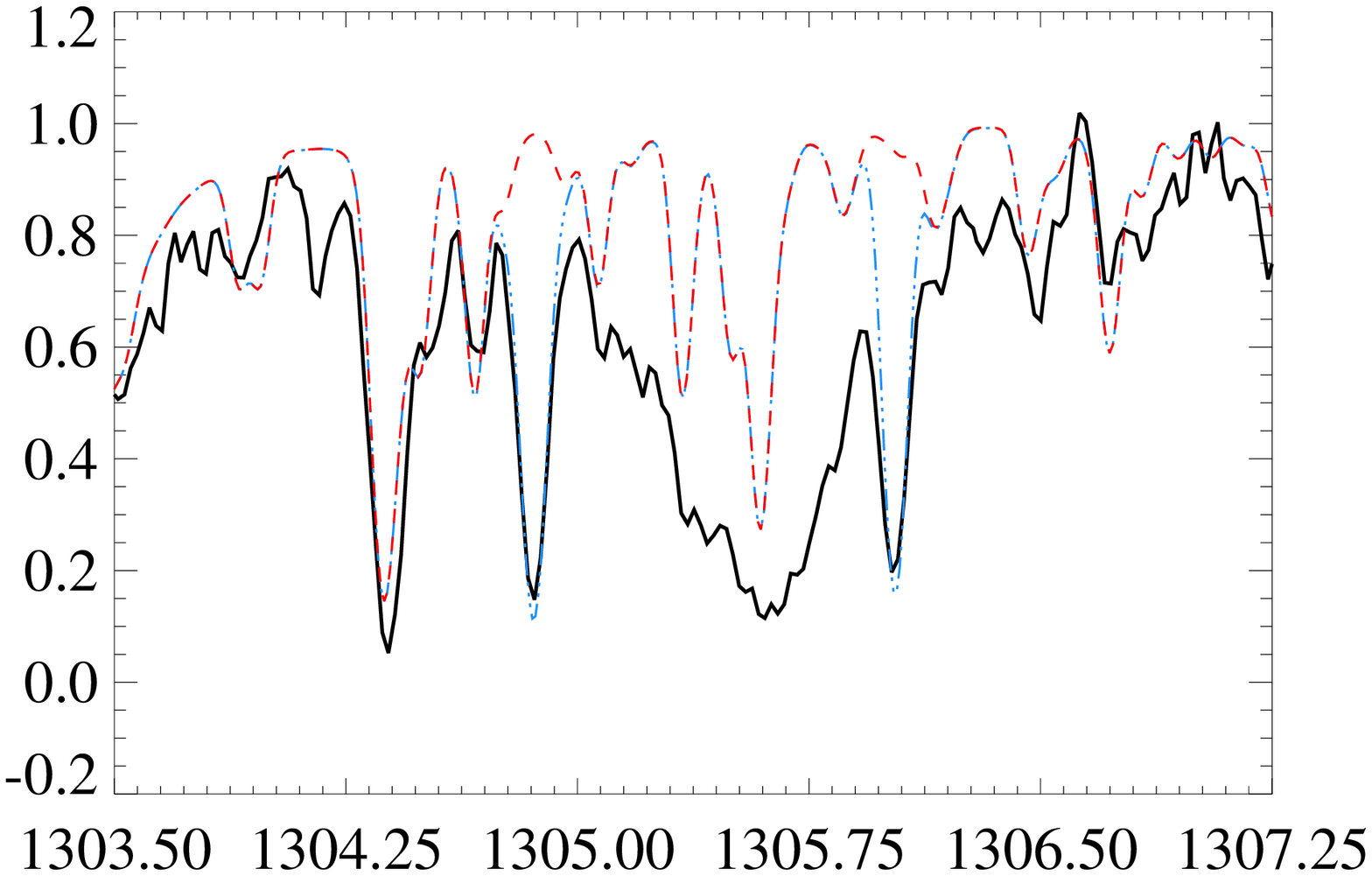}}
\caption{\label{Fig:oxygen_line} The figure shows the observed line
  (in black), the best-fitting model calculated with
  $\log(n_{\rm O}/n_{\rm H})$=--3.5$\pm$0.3 (in blue) and without oxygen (in
  red). The lines are blended with iron. {\it Top left:} O\,{\sc ii}
  quartet lines at 1132\,\AA.\ {\it Top right:}
  O\,{\sc i} triplet lines at 1041\,\AA,\ {\it Bottom left: }
  Strong O\,{\sc i} at 1302\,\AA,\ {\it Bottom Right:} O\,{\sc i}
  triplet lines at 1305--1306\,\AA.}
\end{figure}

We used the strong O\,{\sc ii} quartet lines at 1131\,\AA.\ These lines are
slightly blended with nickel and iron, for which we have reasonable
abundance estimates (see \S \ref{iron} \& \S \ref{nickel}). The
best-fitting model yields $\log(n_{\rm O}/n_{\rm
H})=$--3.5$\pm$0.3. We have used the strong O\,{\sc i} UV multiplet
(3) at 1039--1041\,\AA\ to confirm this value;
Figure~\ref{Fig:oxygen_line} shows only the longer wavelength end of
the feature due to a long gap between strong lines.  We also used the
strong 1302--1306\,\AA\ lines of O\,{\sc i} of UV multiplet (2); in
Figure~\ref{Fig:oxygen_line}, we have shown only the short
wavelength end. These lines are blended with iron lines
(see \S \ref{iron}). The best-fitting model results in the same value
of $\log(n_{\rm O}/n_{\rm H})=$--3.5$\pm$0.3, thus we adopted this
value for the oxygen abundance. Figure~\ref{Fig:oxygen_line} shows the
observation and the models.

\subsection{Magnesium, Z=12}
In our subset of VALD database, there are only a few magnesium lines.
The dominant ionization stage is predicted by the Saha equation to be
Mg~{\sc iii} (see Table~3). However, because the
lowest excited states are more than 50~eV above the ground state, no
lines of Mg~{\sc iii} are detectable in our window. We are therefore
obliged to use lines of the weakly populated Mg~{\sc ii}. There are
numerous lines in our window that arise from low excitation levels
about 4.5~eV above ground, and even a couple of resonance lines.
However, all these lines have very small $\log gf$ values, around -3.

\begin{figure}
\resizebox{\hsize}{!}{\includegraphics{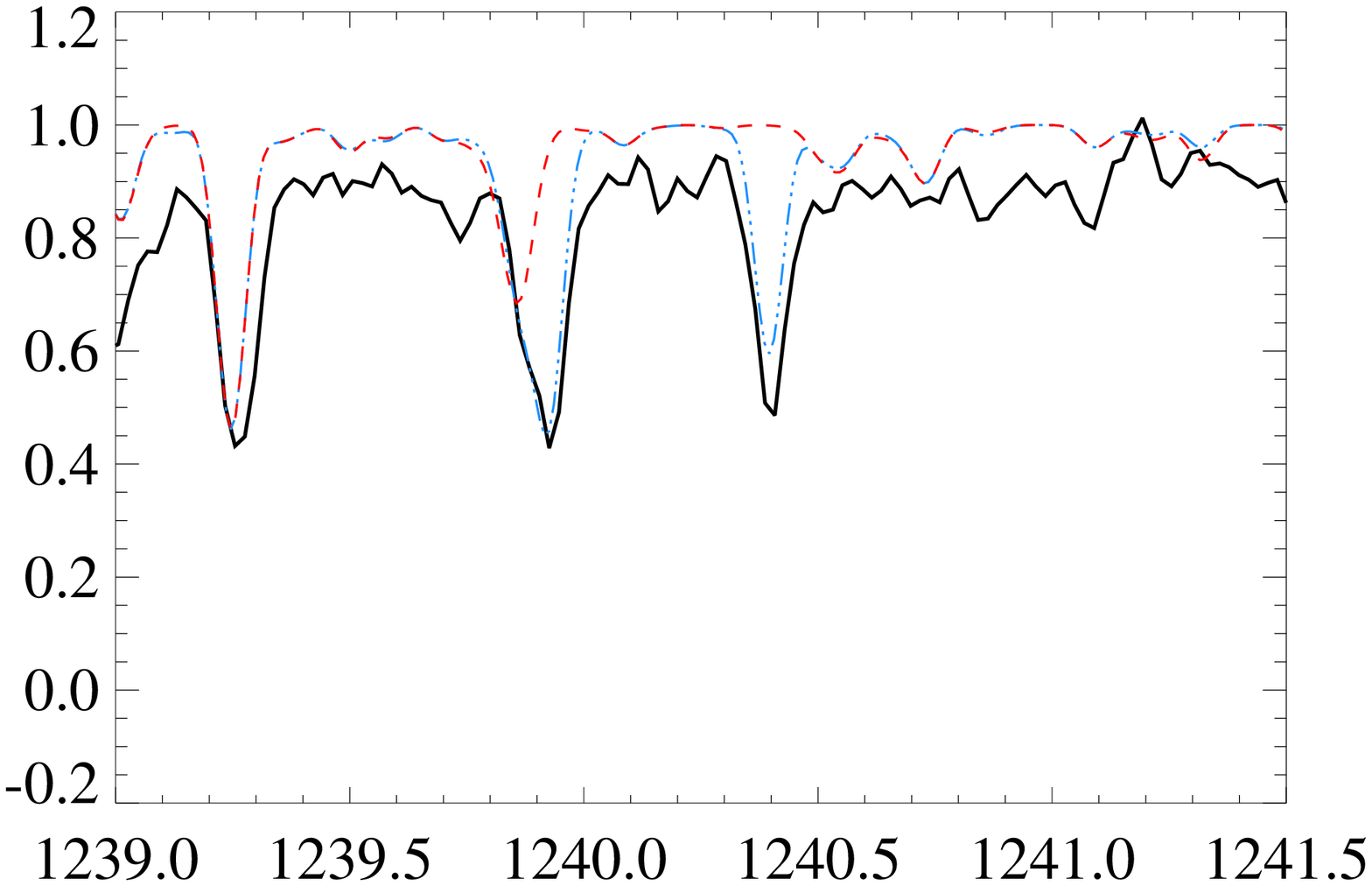}%
  \includegraphics{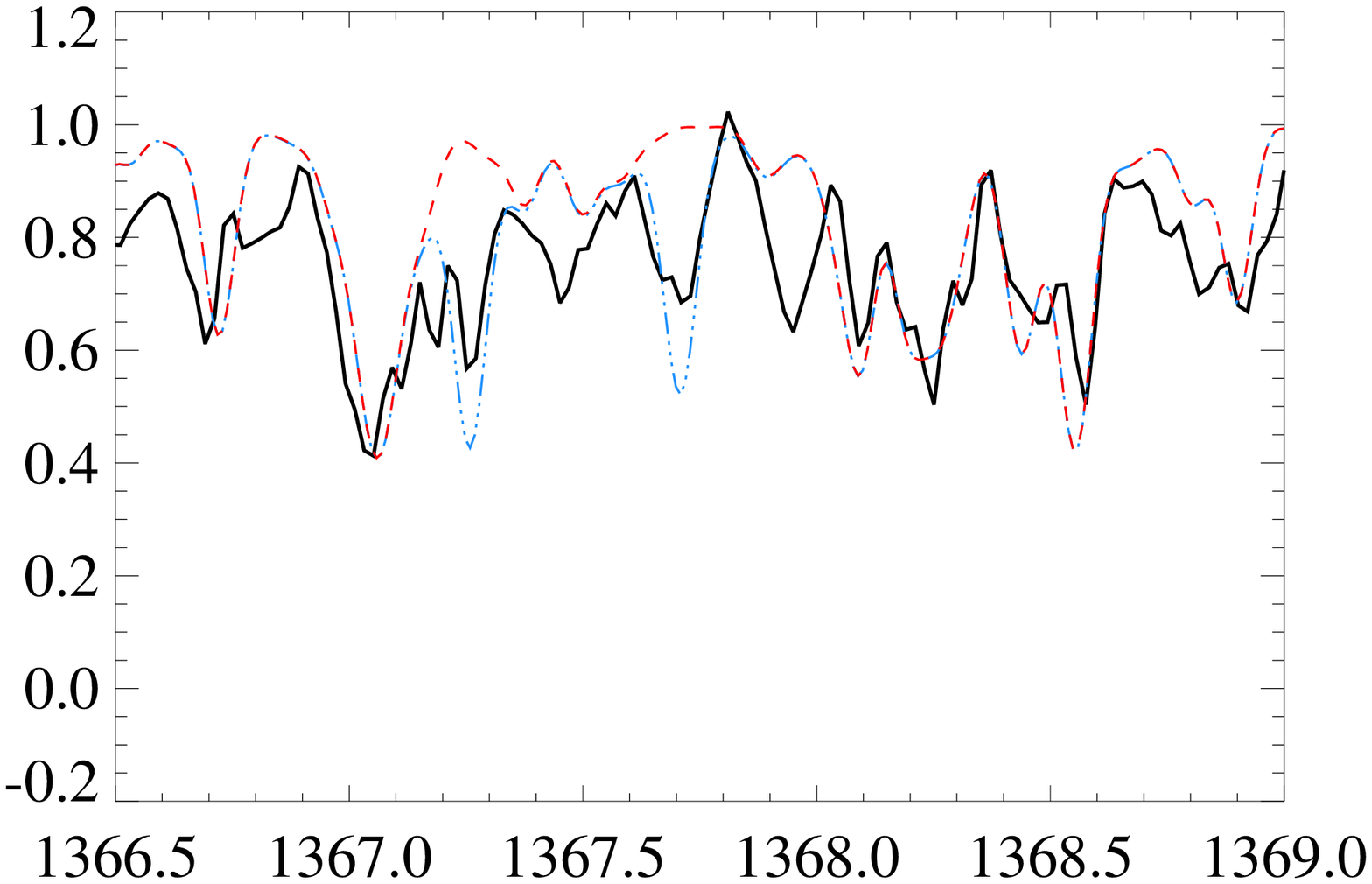}}
\caption{\label{Fig:Mg_lines} This figure shows the observation (in
  black), the calculated model with $\log(n_{\rm Mg}/n_{\rm H})$= --3.9$\pm$0.19
  (in blue) and without magnesium (in red). {\it left:} Mg\,{\sc ii}
  doublet at 1367\,\AA,\ {\it right:} strong line at
  Mg\,{\sc ii} 1240\,\AA}
\end{figure}

We used the Mg\,{\sc ii} doublet at 1367\,\AA\ to
determine the abundance of magnesium. We find $\log(n_{\rm Mg}/n_{\rm H})$=
--3.90$\pm$0.19. We confirmed this value using the strong line at
Mg\,{\sc ii} 1240\,\AA. Figure~\ref{Fig:Mg_lines} shows the fit to
the observed profile at these wavelengths.

\subsection{Aluminum, Z=13}
\label{aluminum}
Aluminum is most prominent at longer wavelengths. Our selection of the
VALD lines contains 88 aluminum lines in total; more than half of them
are in the form of Al\,{\sc ii} and the rest are in Al\,{\sc iii}. At
this temperature, almost all of aluminum is expected to be in Al\,{\sc
  iii} with a moderate Al\,{\sc iv} contribution (see table 3).


In the spectrum of $\iota$~Herculis, most of the aluminum lines
present are contributed by Al\,{\sc ii}, with a moderate number of
Al\,{\sc iii} lines and no trace of Al\,{\sc iv}. The lack of lines of
Al~{\sc iv} is clearly due to the atomic energy structure, which has
no excited states below 75~eV.

\begin{figure}
\resizebox{\hsize}{!}{\includegraphics{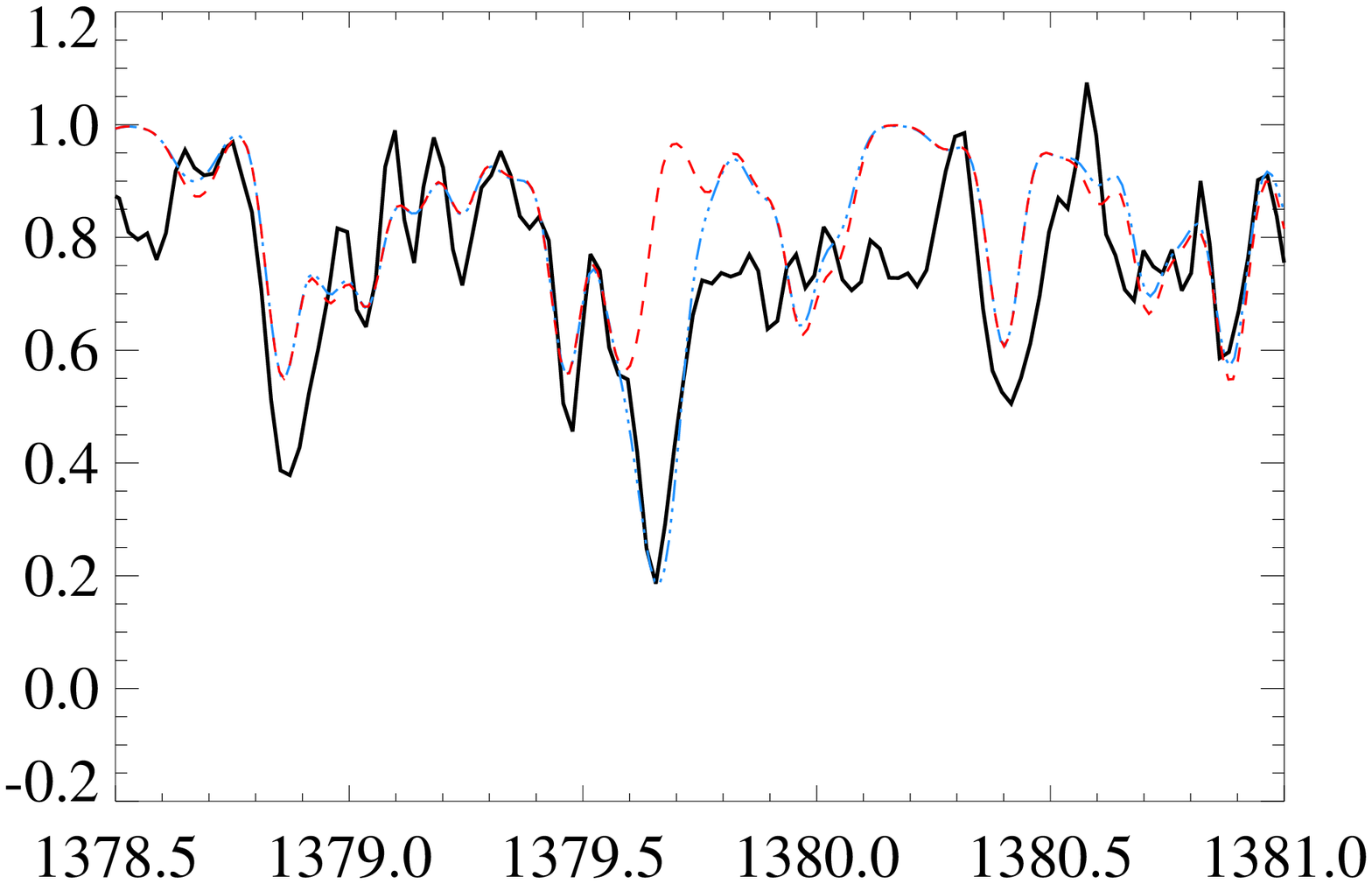}%
\includegraphics{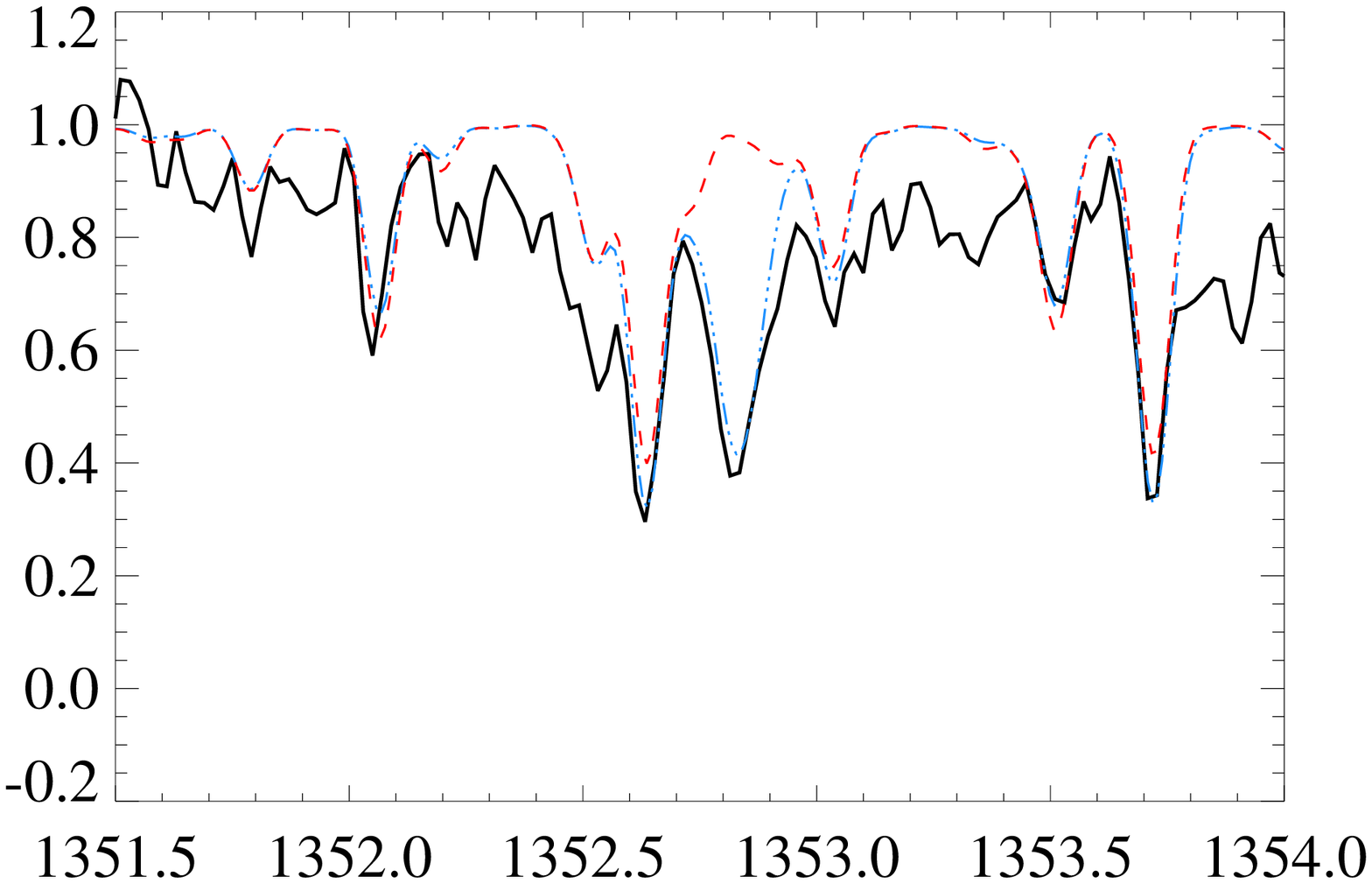}}\\
\resizebox{\hsize}{!}{\includegraphics{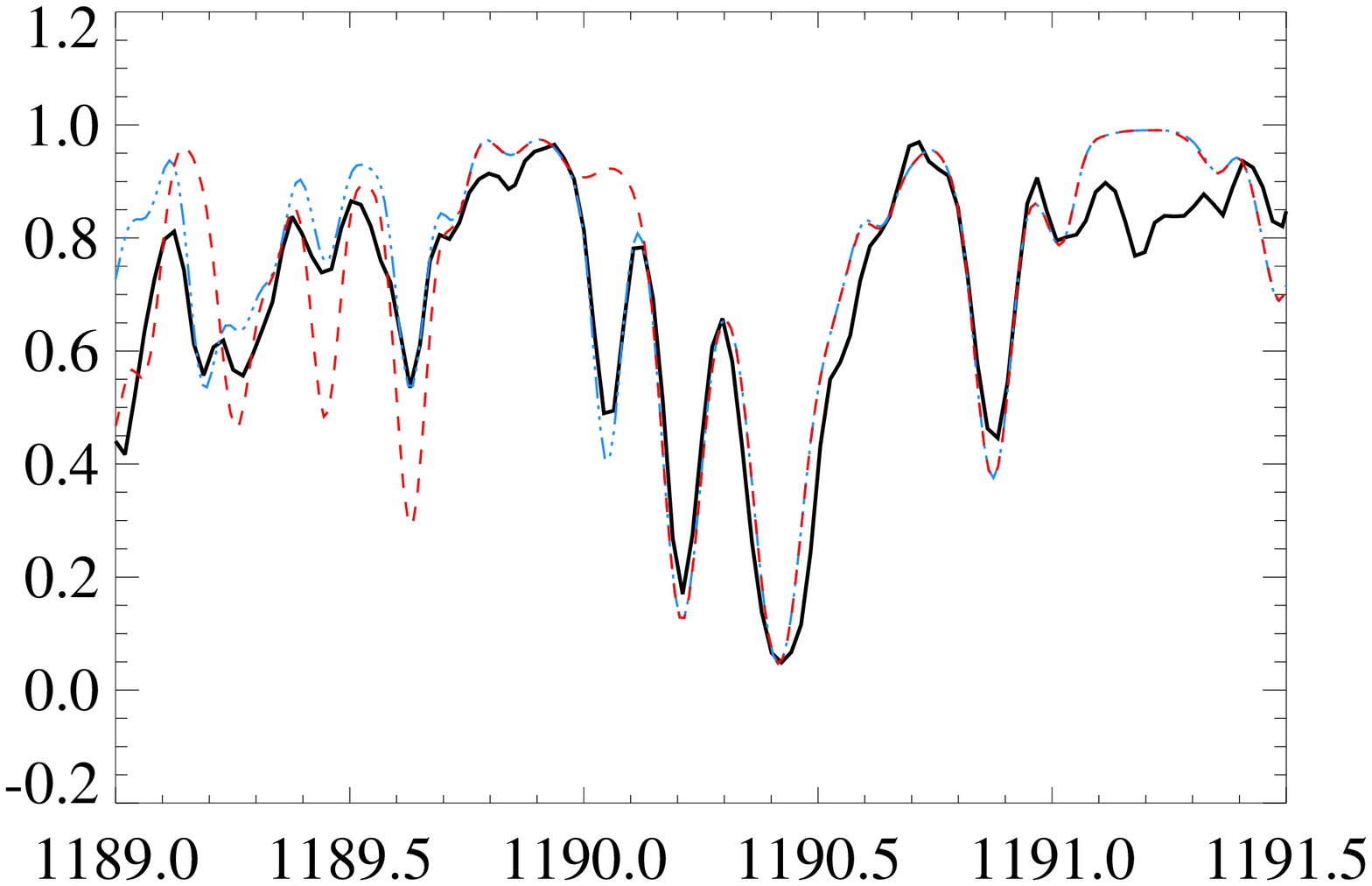}%
\includegraphics{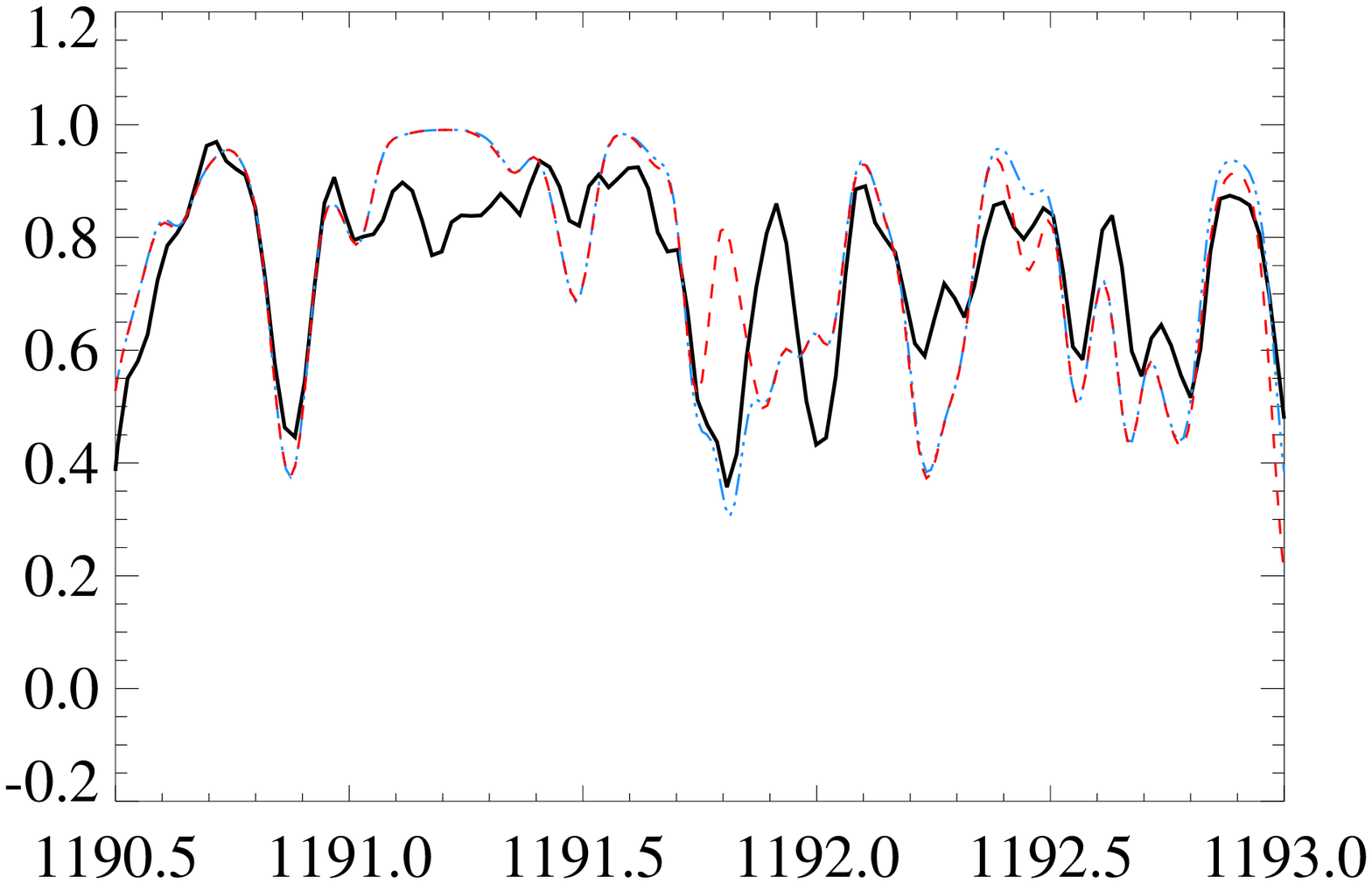}}
\caption{\label{Fig:Al_lines} This figure shows the observation (in
  black), the calculated model with $\log(n_{\rm Al}/n_{\rm H})$= --5.90$\pm$0.58 (in
  blue) and without aluminum (in red). {\it Top left:} Al\,{\sc iii}
  line at 1379\,\AA.\ {\it Top right:} Al\,{\sc iii} doublet line at
  1352\,\AA,\ {\it Bottom left:} Al\,{\sc ii}
  triplet line at 1189\,\AA,\ {\it Bottom right:}
  Al\,{\sc ii} triplet line at 1191\,\AA.\ }
\end{figure}

The available lines of Al are heavily blended throughout the
spectrum. We used the Al\,{\sc iii} line at 1379\,\AA\ which is only
blended with iron (see \S \ref{iron}). The best-fitting model results
in the abundance of $\log(n_{\rm Al}/n_{\rm H})$= --5.90$\pm$0.58 for
aluminum. We confirmed this value by fitting the following lines;
Al\,{\sc ii} triplet lines at 1189--1190\,\AA,\ a strong Al\,{\sc iii}
doublet line at 1352\,\AA,\ and another Al\,{\sc ii} triplet line at
1191\,\AA. Figure~\ref{Fig:Al_lines} shows the observation and the
model fits.


\subsection{Silicon, Z=14}

The VALD lines selected here include 361 silicon lines. The majority
of these lines are in the form of Si\,{\sc ii} and Si\,{\sc iii} and a
small fraction of them are in the form of Si\,{\sc i} and Si\,{\sc
  iv}. In the UV spectrum of $\iota$~Herculis, silicon is largely
observed in the form of Si\,{\sc ii} and Si\,{\sc iii} with a very
minor contribution from Si\,{\sc iv} and no trace of Si\,{\sc i}. This
is generally consistent with the expectations from the Saha equation.
The fact that lines of Si~{\sc ii}, Si~{\sc iii} and Si~{\sc iv} are
all found in the spectrum is because all these ionization stages have
lines arising from the ground state or low lying levels that occur in
the window we are modeling.

\begin{figure}
\resizebox{\hsize}{!}{\includegraphics{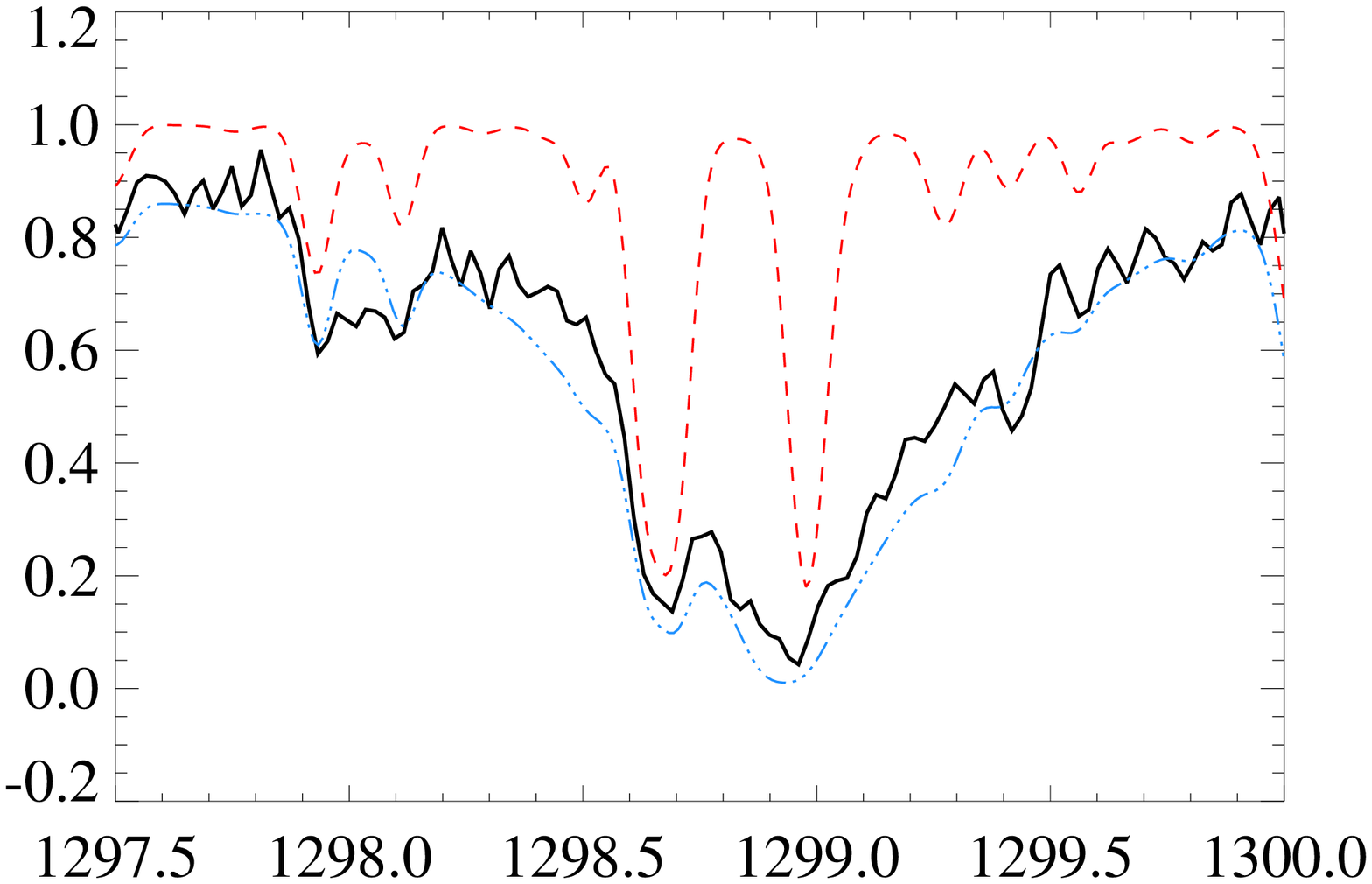}%
\includegraphics{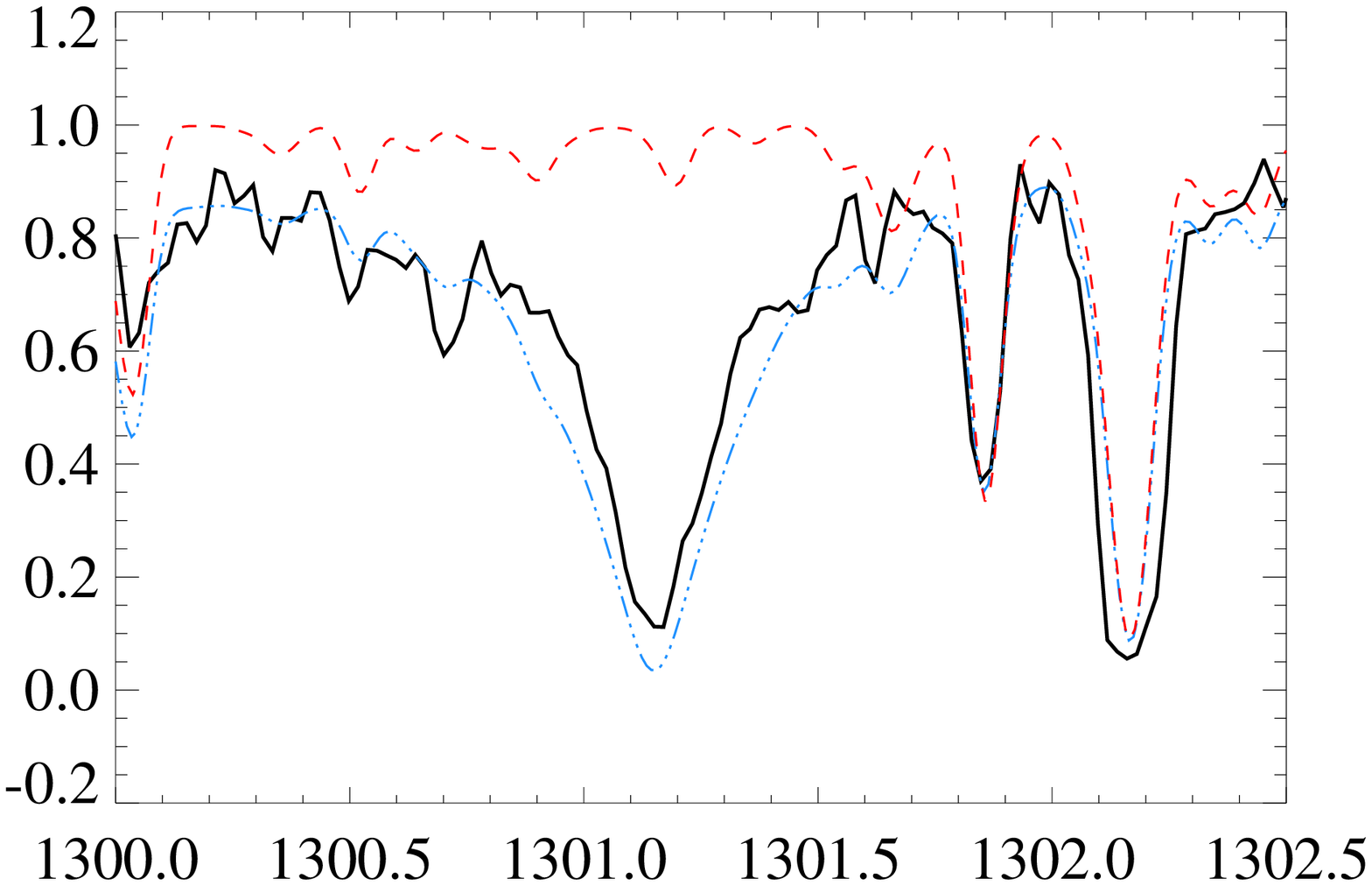}}\\
\resizebox{\hsize}{!}{\includegraphics{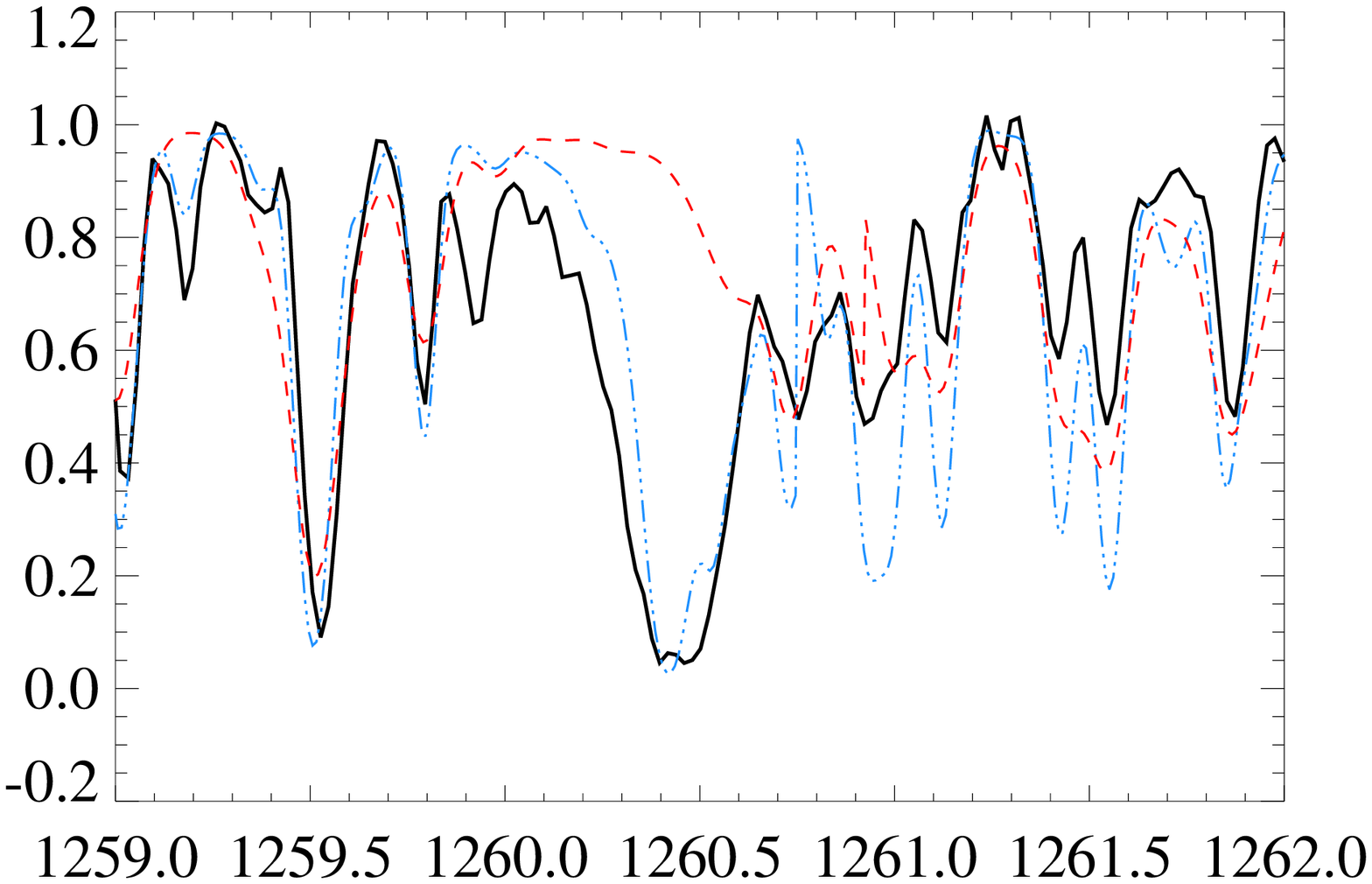}%
\includegraphics{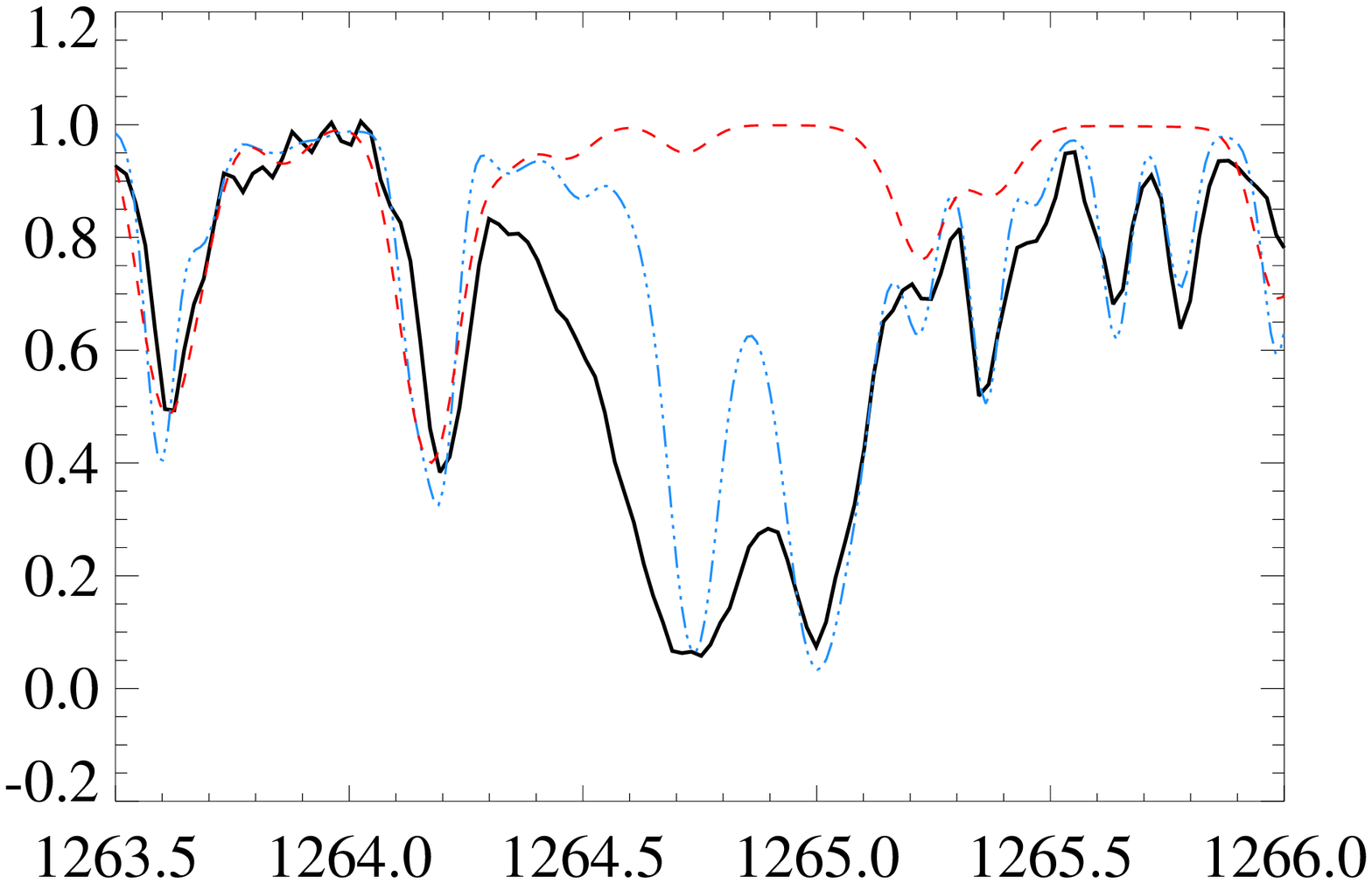}}\\
\resizebox{\hsize}{!}{\includegraphics{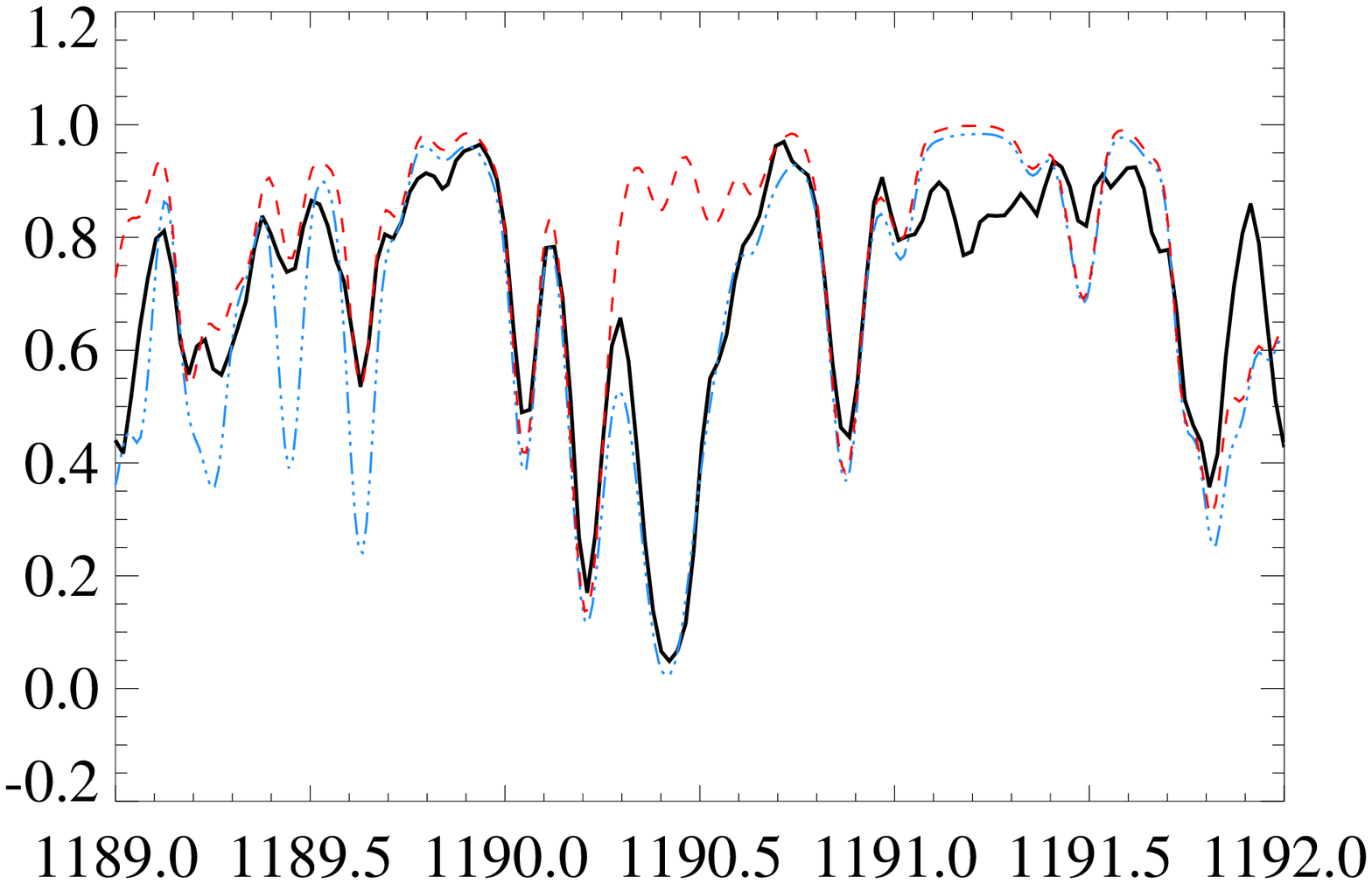}%
\includegraphics{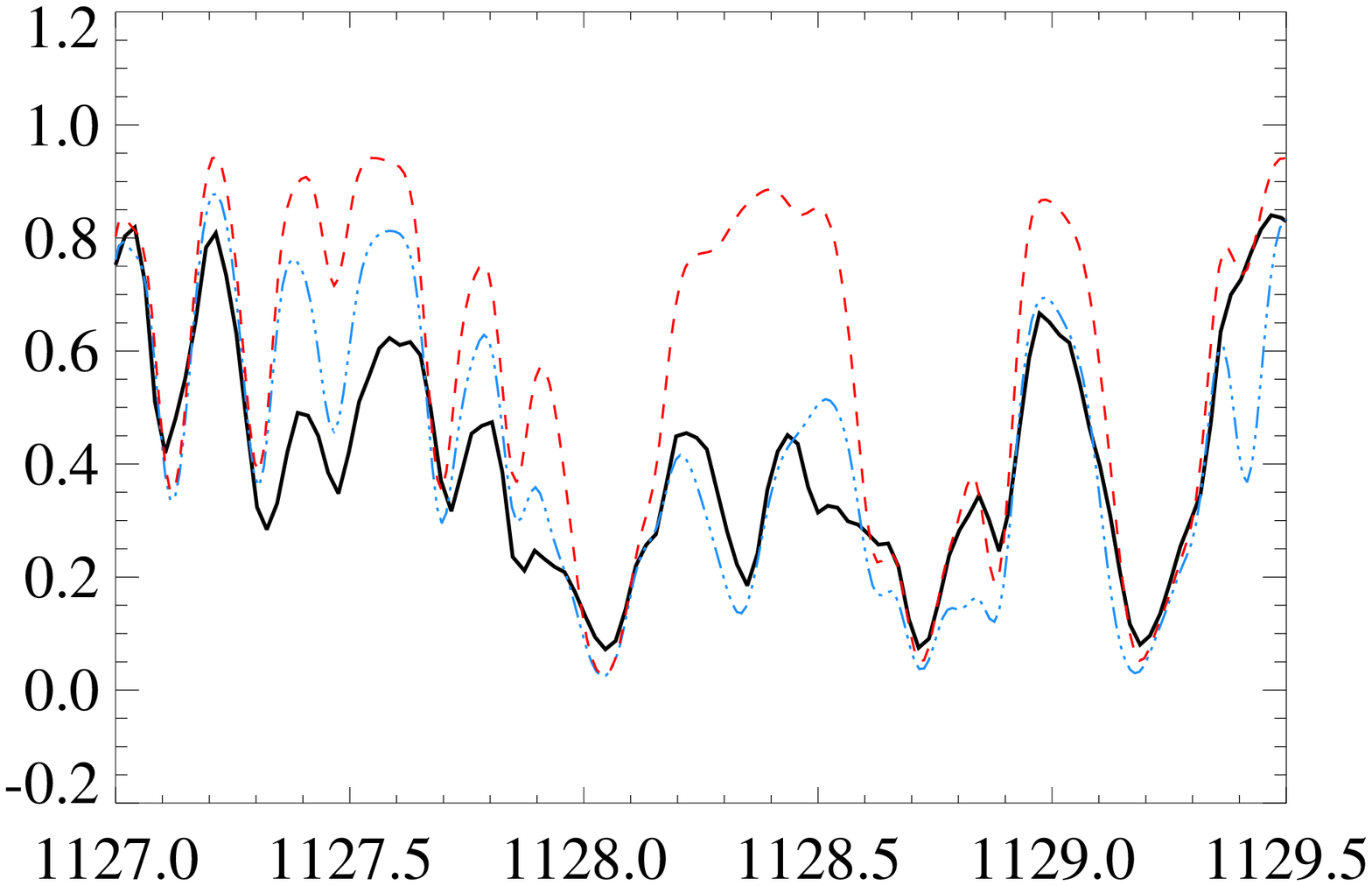}}
\caption{\label{Fig:Si_clean_lines} This figure shows the observation
  (in black), the best-fitting model spectrum with
  $\log(n_{\rm Si}/n_{\rm H})=$--4.65$\pm$0.26 (in blue) and without silicon (in
  red). {\it Top: } Strong Si\,{\sc iii} lines of multiplet (4)
  between 1294--1310\,\AA,\ we only show 1298\,\AA\ ({\it left}) and
  1301\,\AA\ ({\it right}), {\it Middle: } Multiplet (5) of
  Si\,{\sc ii} between 1260--64\,\AA\ around 1260\,\AA\ ({\it left}) and
  1264\,\AA\ ({\it right}), {\it Bottom left: } Si\,{\sc ii}
  multiplet (4) between 1190--93\,\AA,\ {\it Bottom
    right:} A strong Si,{\sc iv} line at 1128\,\AA.}
\end{figure}

We determined the abundance of Si using the relatively unblended and
strong Si\,{\sc iii} lines of multiplet (4) between 1294--1310\,\AA,
and the lines of multiplet (5) between 1108--1114\,\AA. We find the
abundance of $\log(n_{\rm Si}/n_{\rm H})$= --4.65$\pm$0.26 for silicon.  For
confirmation, we have used this abundance to model the blended
multiplet (4) and (5) lines of Si\,{\sc ii} at 1190--93\,\AA\ and
1260--64\,\AA, and a strong Si\,{\sc iv} multiplet (3) line at
1128\,\AA.\ Figure~\ref{Fig:Si_clean_lines} shows the best-fitting
models and the observed spectrum. 

Note that the fit to the lines of Si~{\sc ii} in the middle panel is
poor, and suggests an abundance of Si larger than we have
chosen. Major discrepancies between abundances deduced with Si~{\sc
  ii} and Si~{\sc iii} are common in hot stars, and may represent
important departures from LTE or possibly stratification of Si in the
atmosphere (Bailey \& Landstreet, 2013), but in this case the
discrepancies may be due to unidentified blends.


\subsection{Phosphorus, Z=15}
There are overall 133 phosphorus lines in the subset of the VALD
database selected here; mostly in the form of P\,{\sc ii}, a smaller
subset are in the form of P\,{\sc iv}, P\,{\sc iii}, and there are
only a few P\,{\sc i} and P\,{\sc v} lines. In the UV spectrum of
$\iota$~Herculis, phosphorus lines are mostly observed in the form of
P\,{\sc ii} and P\,{\sc iii} and there is a very minor contribution
from other ionization states (P\,{\sc iv}, P\,{\sc v} and P\,{\sc
  i}). This is consistent with the ionization ratios obtained from the
Saha equation at this temperature (see Table~3).
All the P\,{\sc ii} and P\,{\sc iii} lines in the wavelength range
studied here arise from low lying energy levels (0--1.1~eV) which
explains their clear appearance in the observed spectrum.

\begin{figure}
\resizebox{\hsize}{!}{\includegraphics{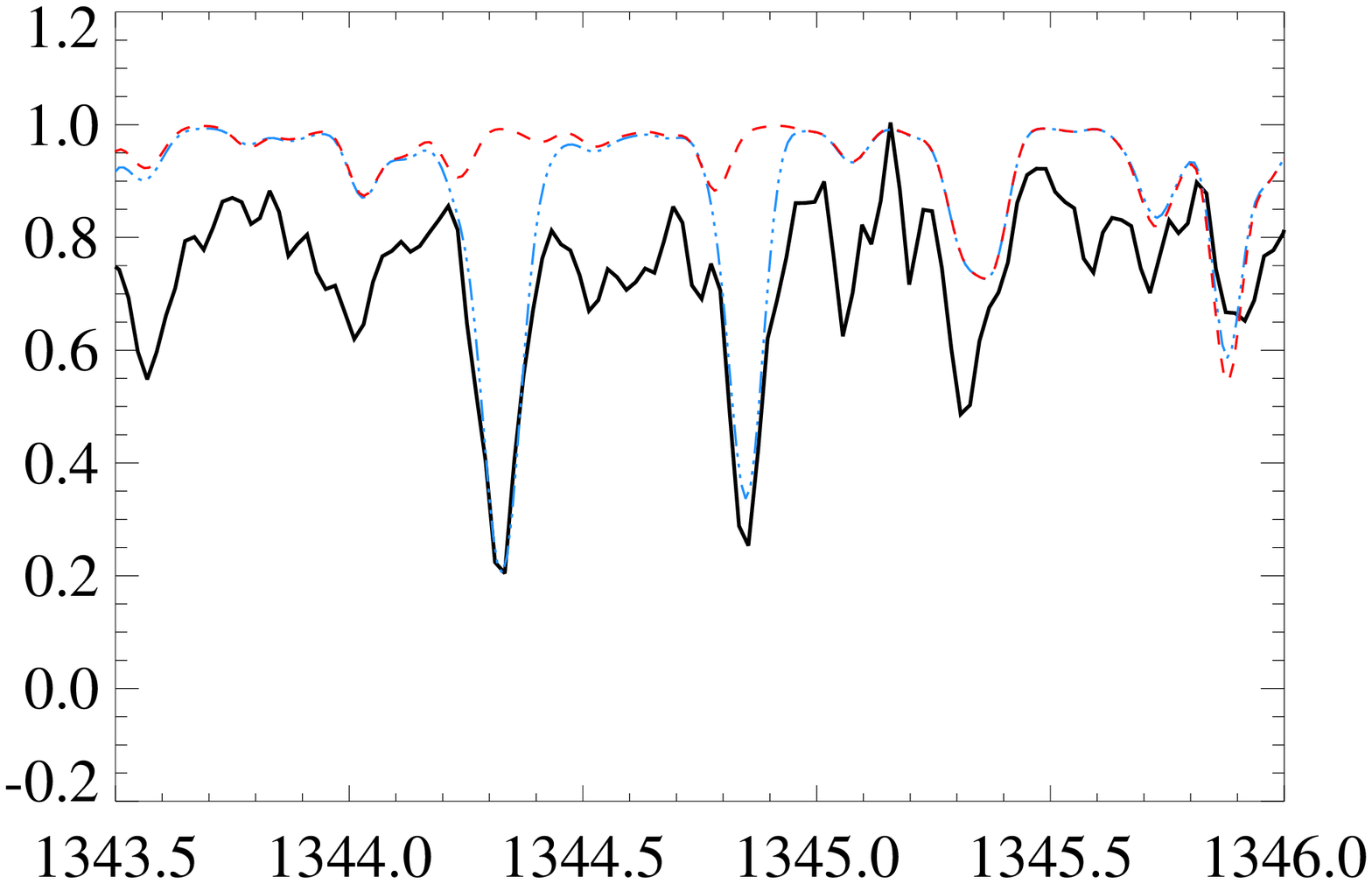}%
\includegraphics{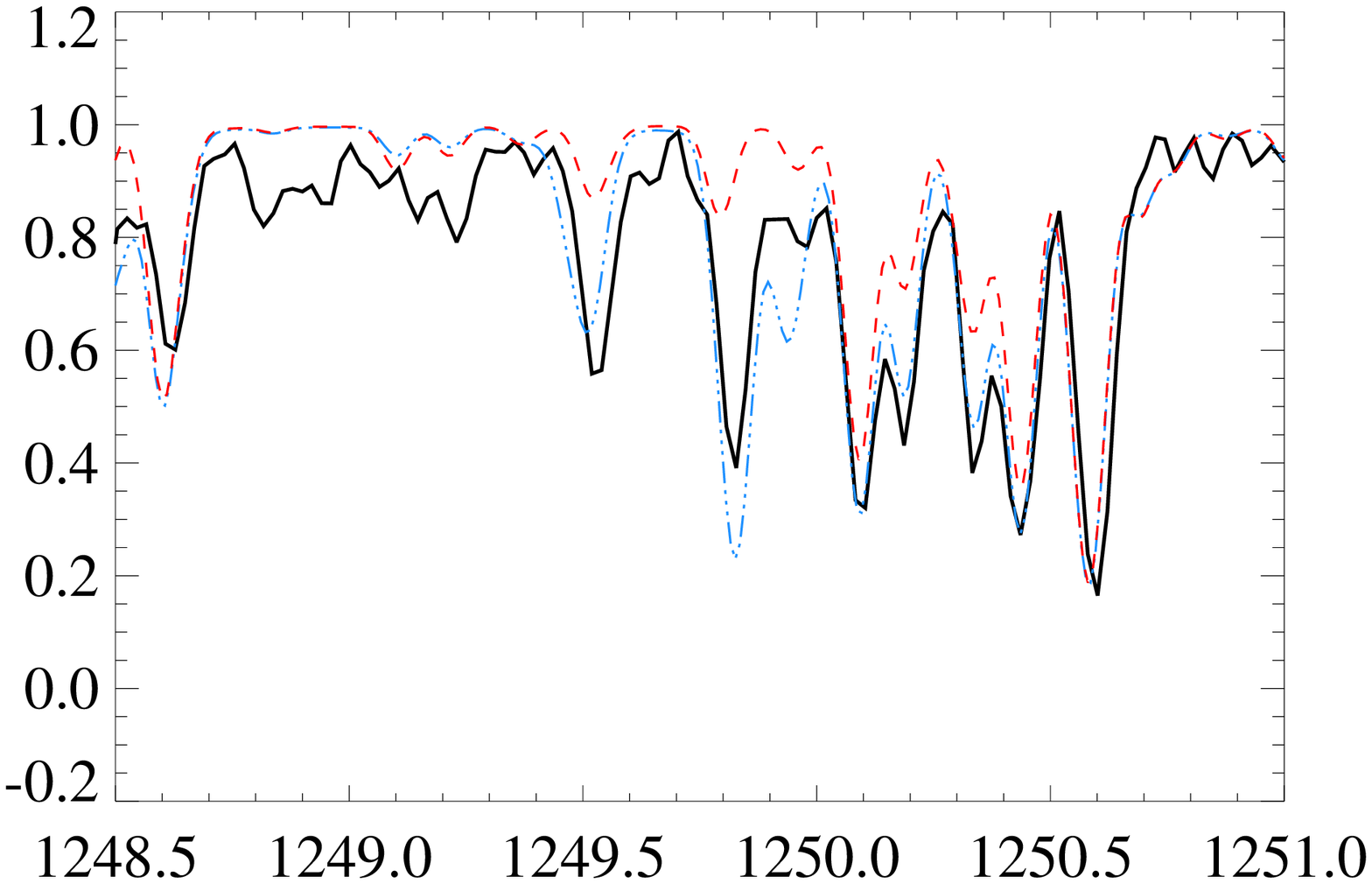}}\\
\resizebox{\hsize}{!}{\includegraphics{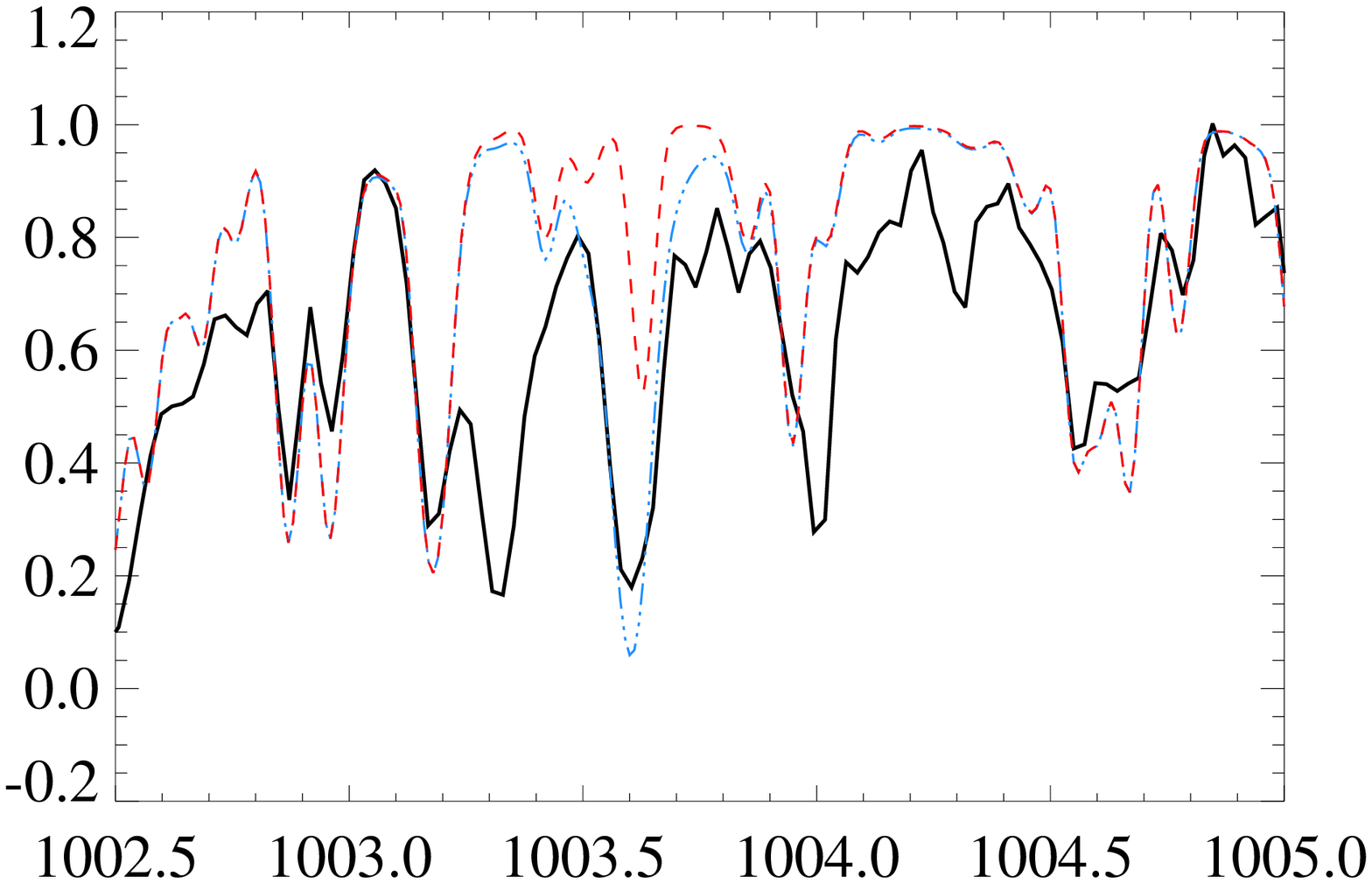}%
\includegraphics{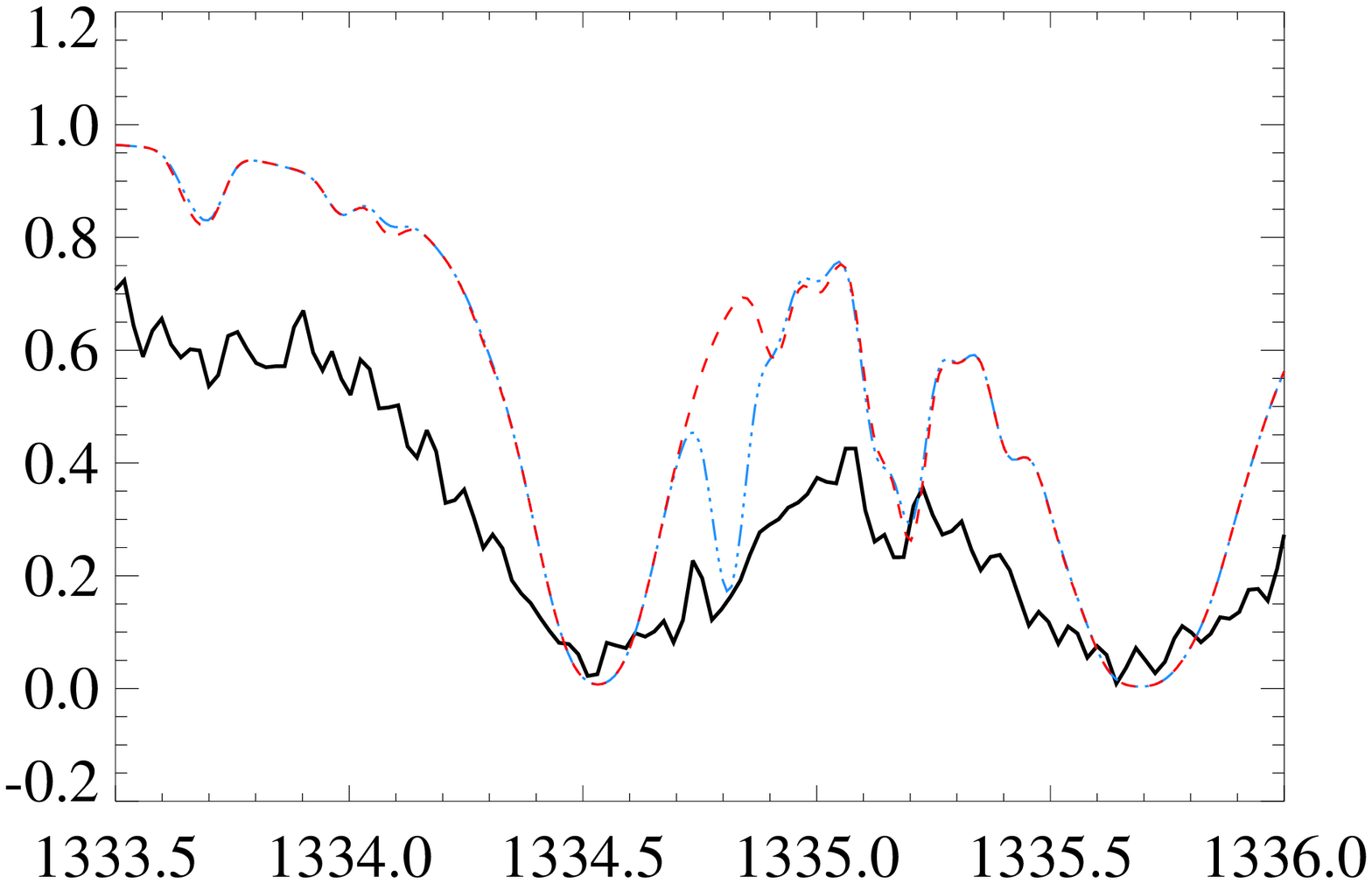}}
\caption{\label{Fig:P_clean_lines} This figure shows the observation
  (in black), the calculated model with $\log(n_{\rm P}/n_{\rm
    H})=$--6.7$\pm$0.7 (in blue), and without phosphorus (in red).
  {\it Top left:} A P\,{\sc iii} doublet at 1344\,\AA,\ {\it Top
    right:} A P\,{\sc ii} line at 1249\,\AA,\ {\it Bottom left:}
  A P\,{\sc iii} at 1003\,\AA\ {\it Bottom right:} A resonance P\,{\sc
    iii} line at 1334\,\AA.}
\end{figure}

We have used a strong P\,{\sc iii} tripet of UV multiplet (1) at
1344-45\,\AA.\ These lines both arise from low initial energy level
(~0.07eV) with a fairly accurate (Kramida et al., 2014) oscillator
strength value of ``D'' $\log gf \sim$ -1.5. They are slightly blended
with iron for which we have a reasonable abundance determination (see
\S \ref{iron}). Modeling these lines yields $\log(n_{\rm P}/n_{\rm
  H})=$--6.7$\pm$0.7. We confirmed this value using two strong lines
of P\,{\sc ii} at 1249.8\AA\ and P\,{\sc iii} at 1003.6\AA\ with $\log
gf$ values close to zero and a P\,{\sc iii} resonance line at
1334.8\AA\ with ``D'' $\log gf$ values close to -1. Figure
\ref{Fig:P_clean_lines} shows the observation and the best-fitting
models.

\subsection{Sulfur, Z=16}
\label{sulfur}
There are 474 lines in the subset of the VALD database used here. In
the observed UV spectrum of $\iota$~Herculis as well as the VALD
database, sulfur has a few strong lines in the form of S\,{\sc ii},
S\,{\sc iii} and S\,{\sc iv} and a large number of weaker S\,{\sc i}
lines. This is consistent with the expected ionization ratios predicted
by the Saha equation.

\begin{figure}
\resizebox{\hsize}{!}{\includegraphics{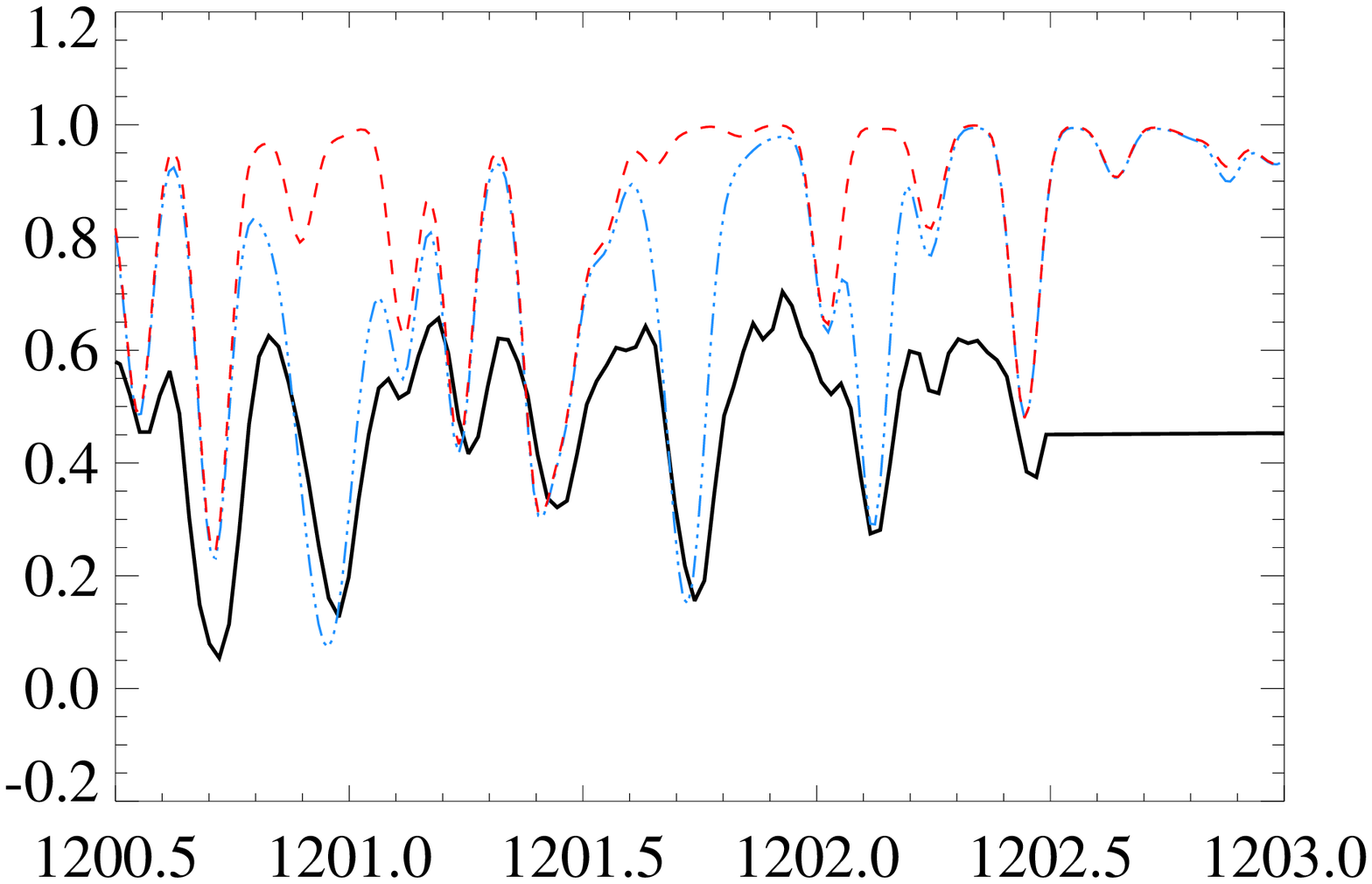}%
  \includegraphics{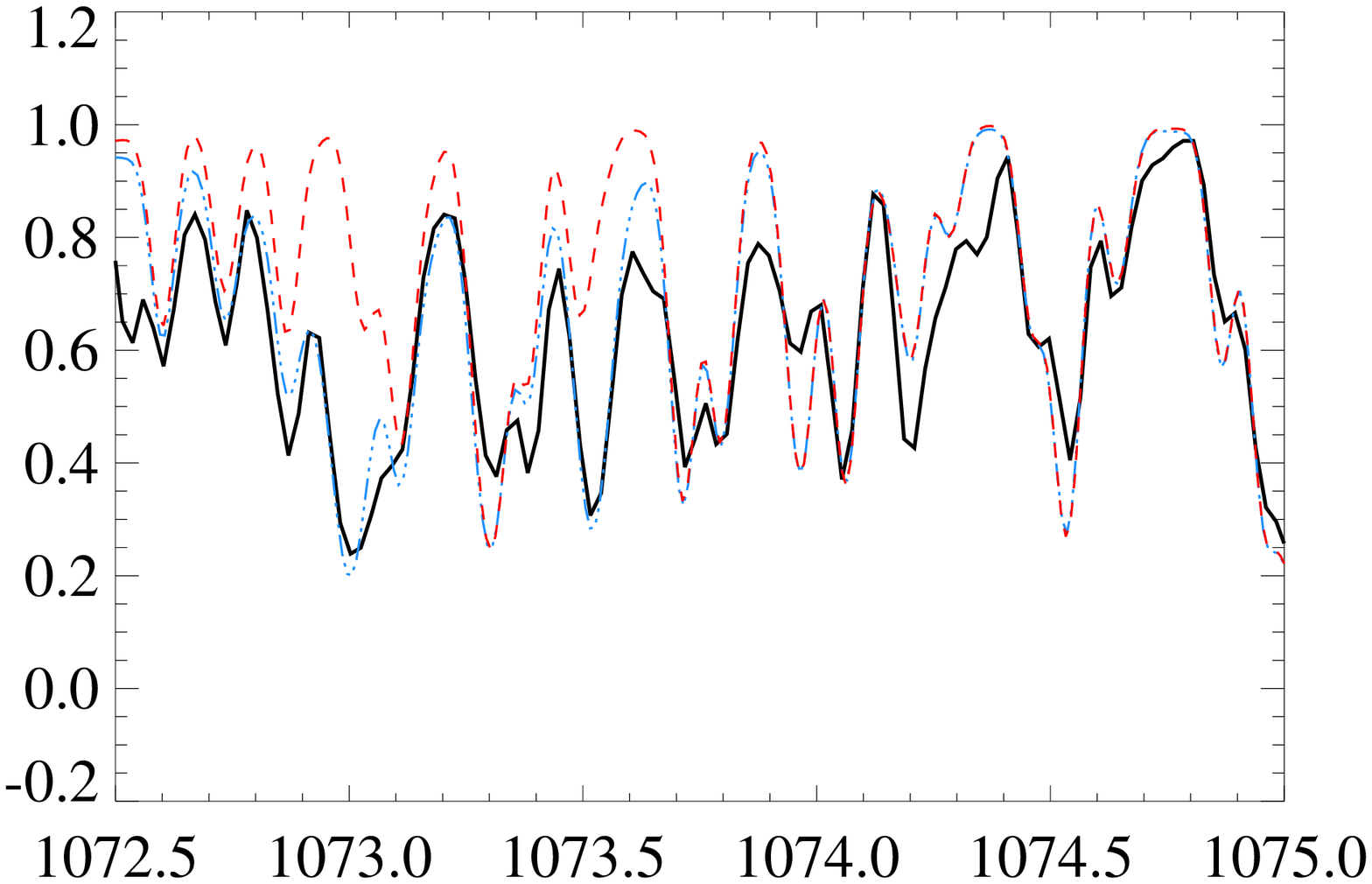}}\\ 
\resizebox{\hsize}{!}{\includegraphics{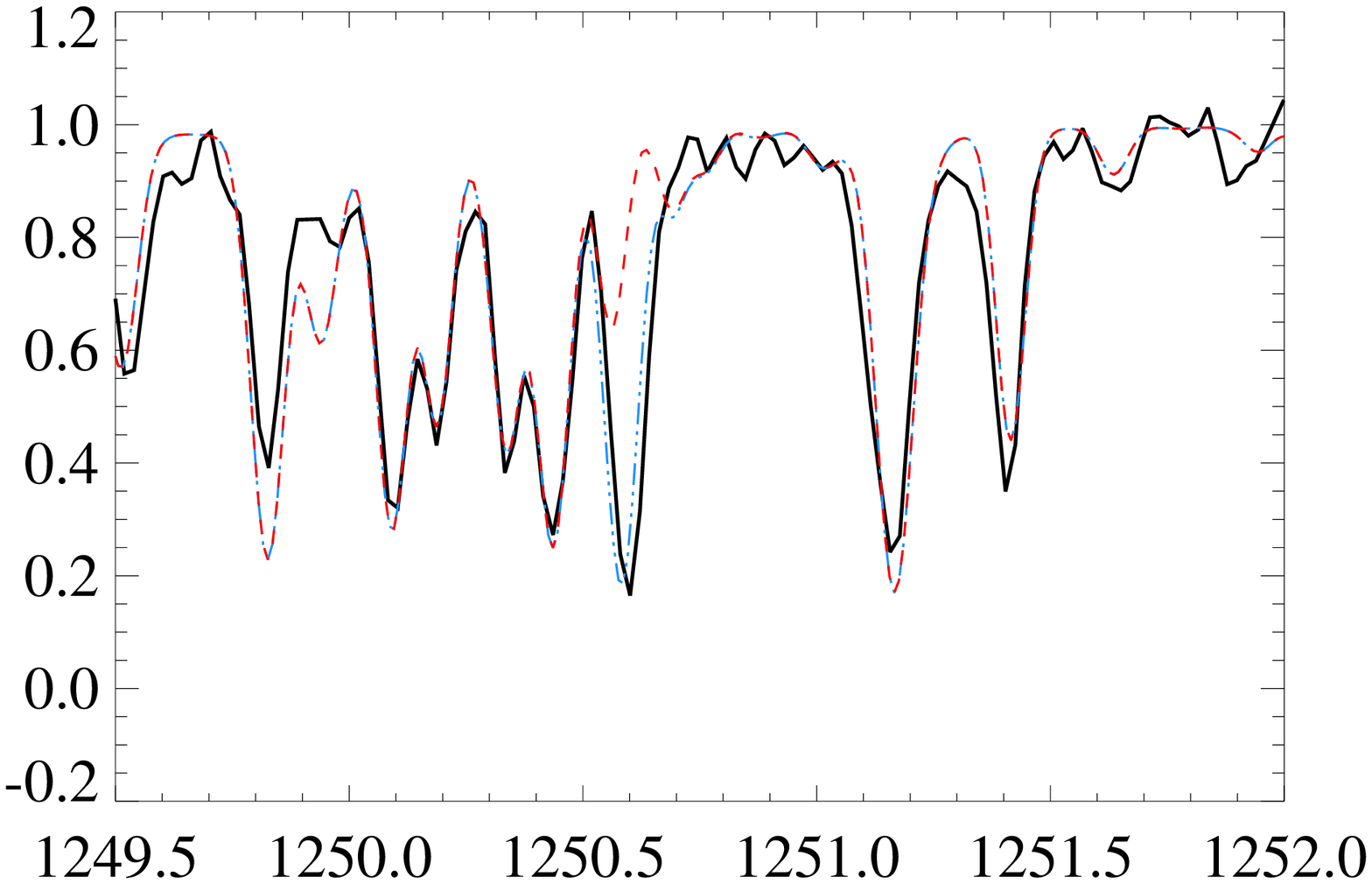}%
  \includegraphics{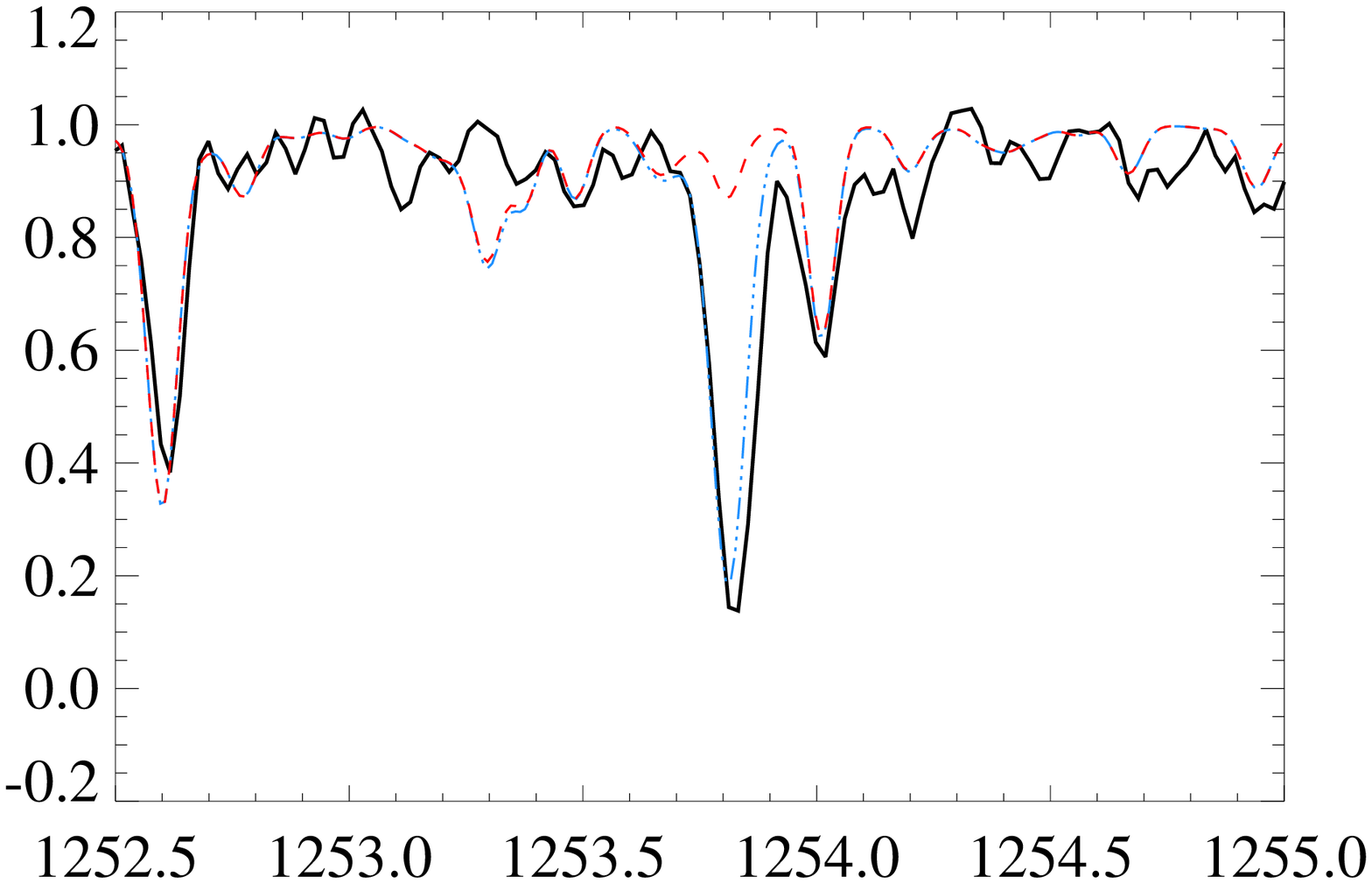}}\\
\resizebox{\hsize}{!}{\includegraphics{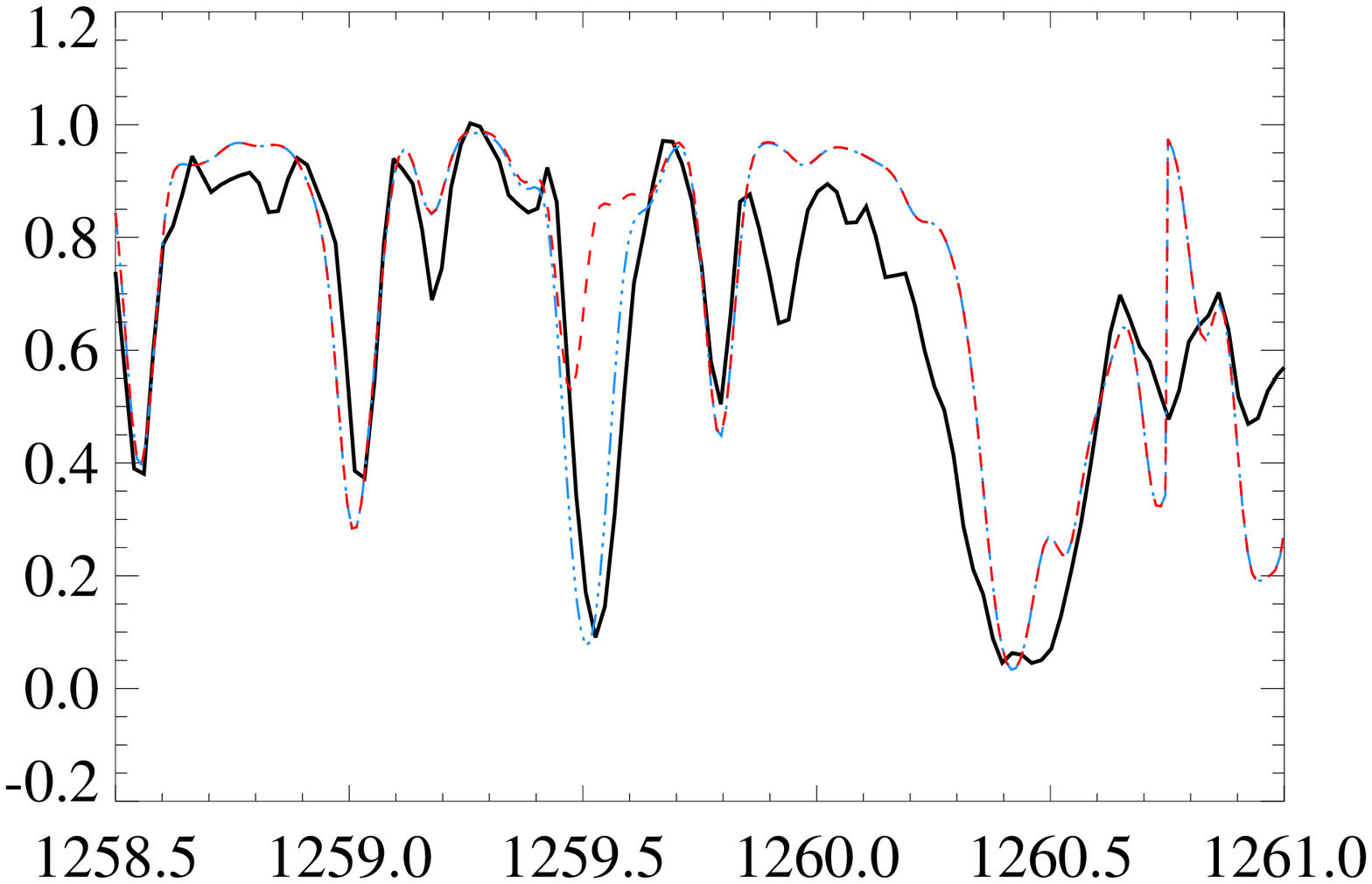}%
  \includegraphics{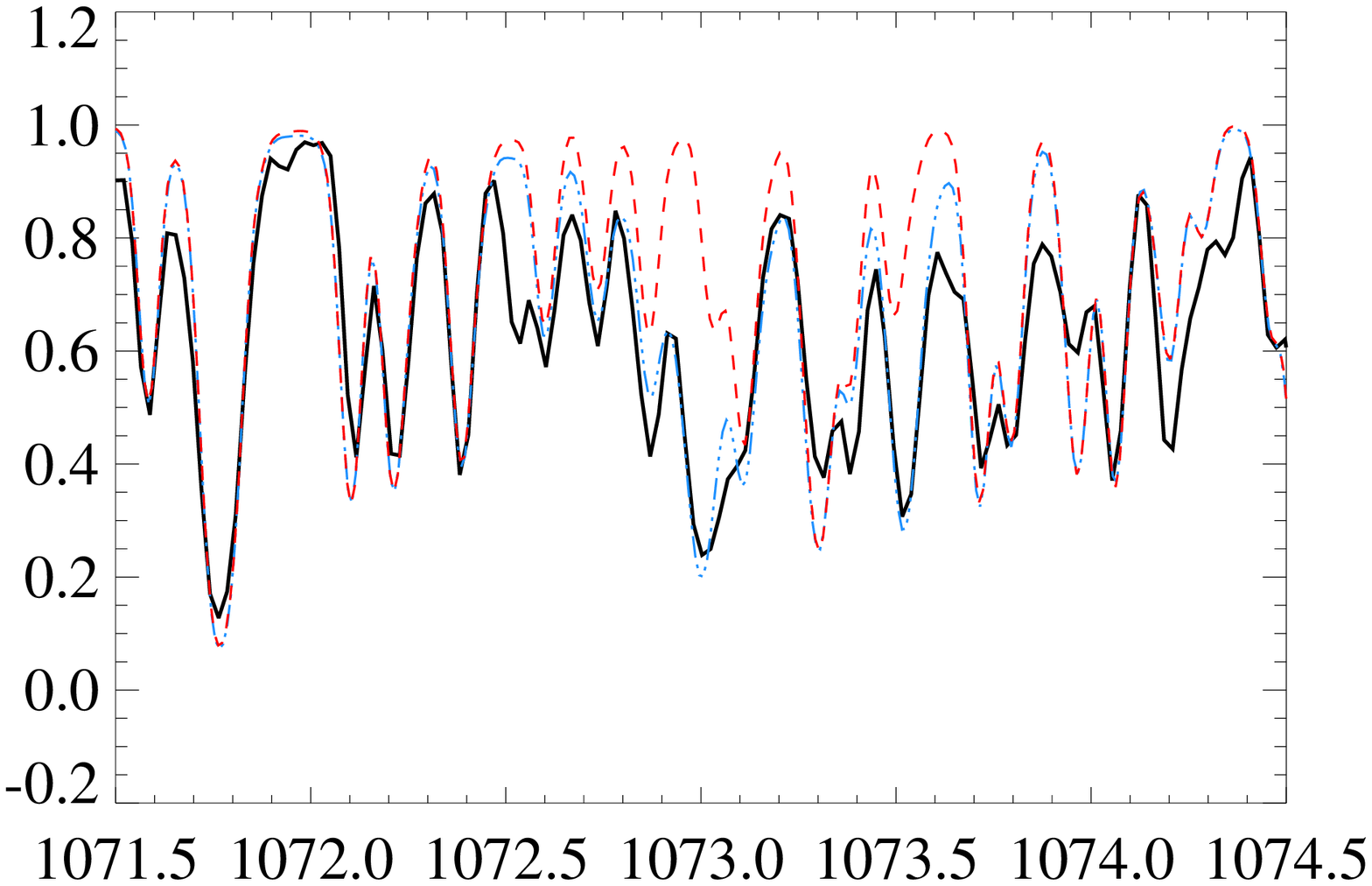}}\\
\caption{\label{Fig:sulfur_lines} This figure shows the observation
  (in black), the calculated model with $\log(n_{\rm S}/n_{\rm
    H})=$--5.20$\pm$0.10 (in blue), and without sulfur (in red). {\it
    Top left:} The strong S\,{\sc iii} triplet lines at 1200\,\AA,\
  {\it Top right:} The S\,{\sc iv} doublet at 1072\,\AA,\ {\it Middle:
  } A strong S\,{\sc ii} resonance doublet at 1250\,\AA,\ {\it Bottom
    left: } A strong S\,{\sc ii} resonance line at 1259\,\AA,\ {\it
    Bottom right: } A strong S\,{\sc ii} line at 1072\,\AA}
\end{figure}

We have modeled the strong S\,{\sc iii} resonance lines of UV
multiplet (1) at 1200\AA. The best-fitting model results in the value
of $\log(n_{\rm S}/n_{\rm H})=$--5.20$\pm$0.10. We confirmed this
value using a line of S\,{\sc iv} multiplet (1) at 1072\,\AA.\ This line is
blended with iron, nickel and manganese, but can still be used for our
purposes (see \S \ref{iron}, \S \ref{nickel}, \S \ref{manganese}). We
also used S\,{\sc ii} resonance lines of multiplet (1) in the window
between 1250-59\AA,\ and a S\,{\sc ii} resonance line at 1072\AA. The
oscillator strength of all the lines used here are around -1 with a
``D'' accuracy (Kramida et al., 2014) and they all arise from low
energy levels. The observation and models are shown in Figure
\ref{Fig:sulfur_lines}.

\subsection{Chlorine, Z=17}
Chlorine has 117 lines in our selected database. The majority of them
are in neutral form. In our spectral window of $\iota$~Herculis,
chlorine is observed mostly in shorter wavelengths. The observed
spectrum is mostly dominated by Cl\,{\sc i} lines with minor
contribution from other ionization states (Cl\,{\sc ii}, Cl\,{\sc
  iii}, and Cl\,{\sc iv}). As predicted by the Saha equation (see
Table 3), the dominant ionization state at this temperature should be
Cl\,{\sc iii}. However, this ionization state contributes only very
weakly. The lowest excited states of Cl\,{\sc iii} are 18~eV above the
ground state (Kramida et al., 2014), which explains why there are
virtually no Cl\,{\sc iii} lines seen in the spectral window studied
here.

\begin{figure}
\resizebox{\hsize}{!}{\includegraphics{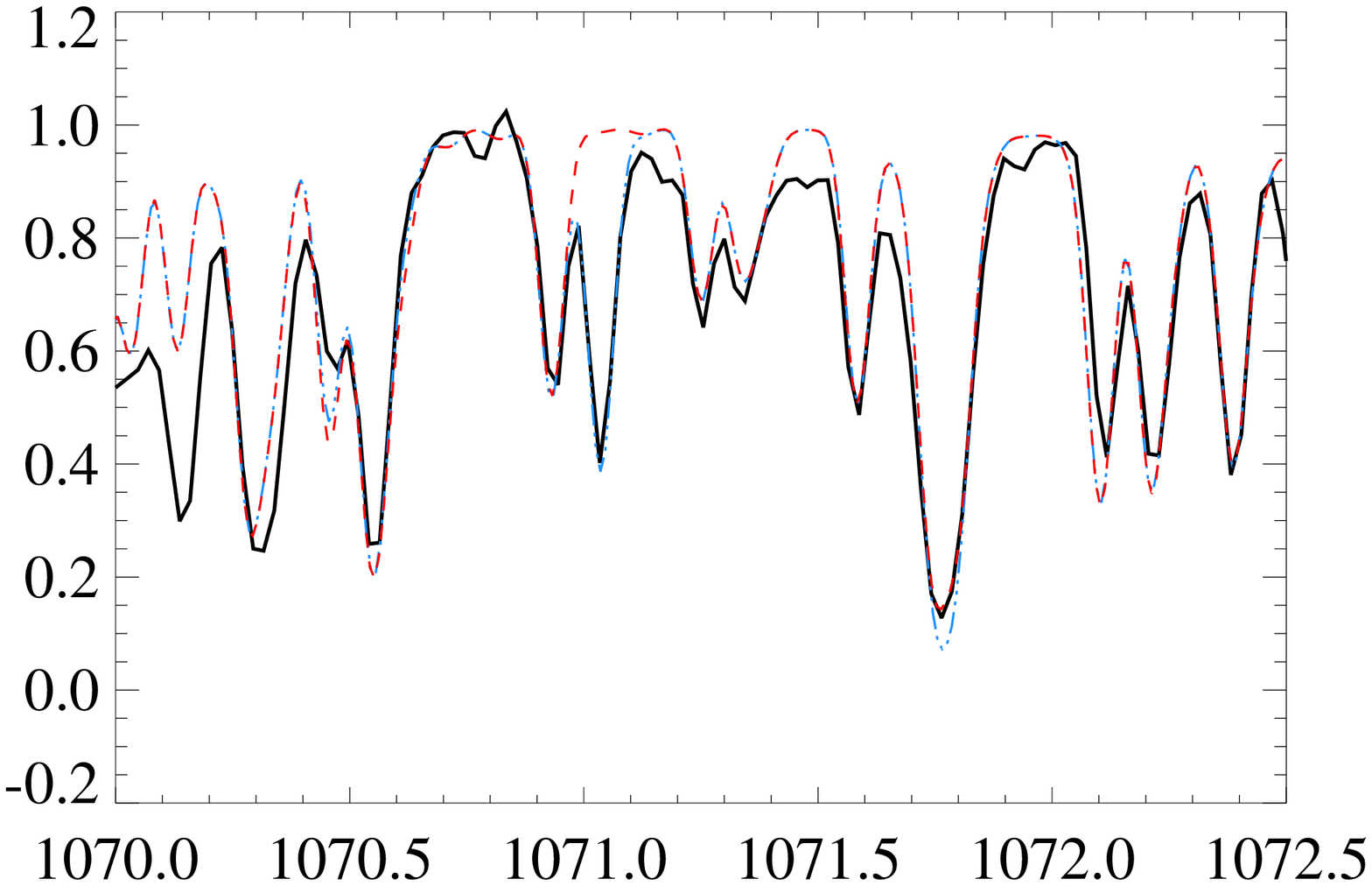}%
\includegraphics{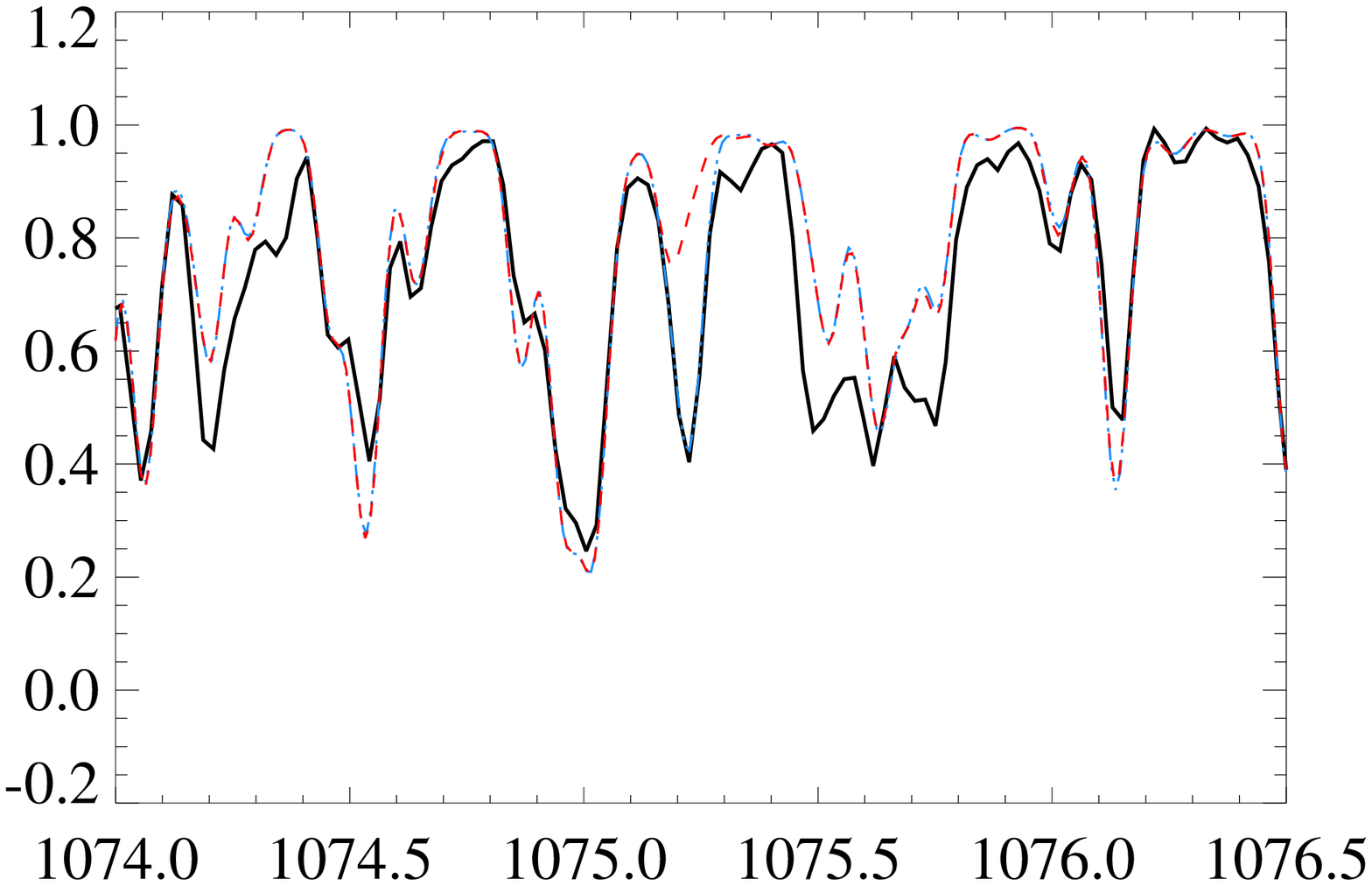}}
\caption{\label{Fig:cl_line} This figure shows the observation
  (in black), the calculated model with $\log(n_{\rm Cl}/n_{\rm H})=$--7.150$\pm$0.100 (in
  blue), and without chlorine (in red). {\it Left:} A Cl\,{\sc ii} line
  at 1071\,\AA,\ {\it Right:} A Cl\,{\sc ii} line at 1075\,\AA.}
\end{figure}

We determined the abundance of chlorine using the unblended Cl\,{\sc
  ii} resonance line of multiplet (1) at 1071\,\AA\ shown in Figure
\ref{Fig:cl_line}. This line has a ``C'' $\log gf$ value of roughly
--1. The best-fitting model to this line results in the value of
$\log(n_{\rm Cl}/n_{\rm H})=$--7.15$\pm$0.10. We confirmed this value
by also modeling the Cl\,{\sc ii} line at 1075\,\AA.\ This line also
arises from a very low energy level of 0.1eV and has a ``C'' $\log gf$
of --1.7 (Kramida et al., 2014). The observation and models are shown
in Figure~\ref{Fig:cl_line}.

\subsection{Argon, Z=18}
Argon appears with 66 lines in the subset of VALD database selected
here. In the spectrum of $\iota$~Herculis, argon is observed mostly in
the form of weak Ar\,{\sc ii} lines with a minor contribution from
stronger Ar\,{\sc i} and Ar\,{\sc iii} lines. As measured from the
Saha equation, the dominant ionization state at this temperature is
Ar\,{\sc ii} (see Table 3).  The reason for the
under-population of Ar\,{\sc ii} lines in the observed spectrum is
that in this wavelength window, almost all of the Ar\,{\sc ii} lines
arise from high excitation energy levels which are at least
$\sim$16~eV above the ground state.

\begin{figure}
\resizebox{\hsize}{!}{\includegraphics{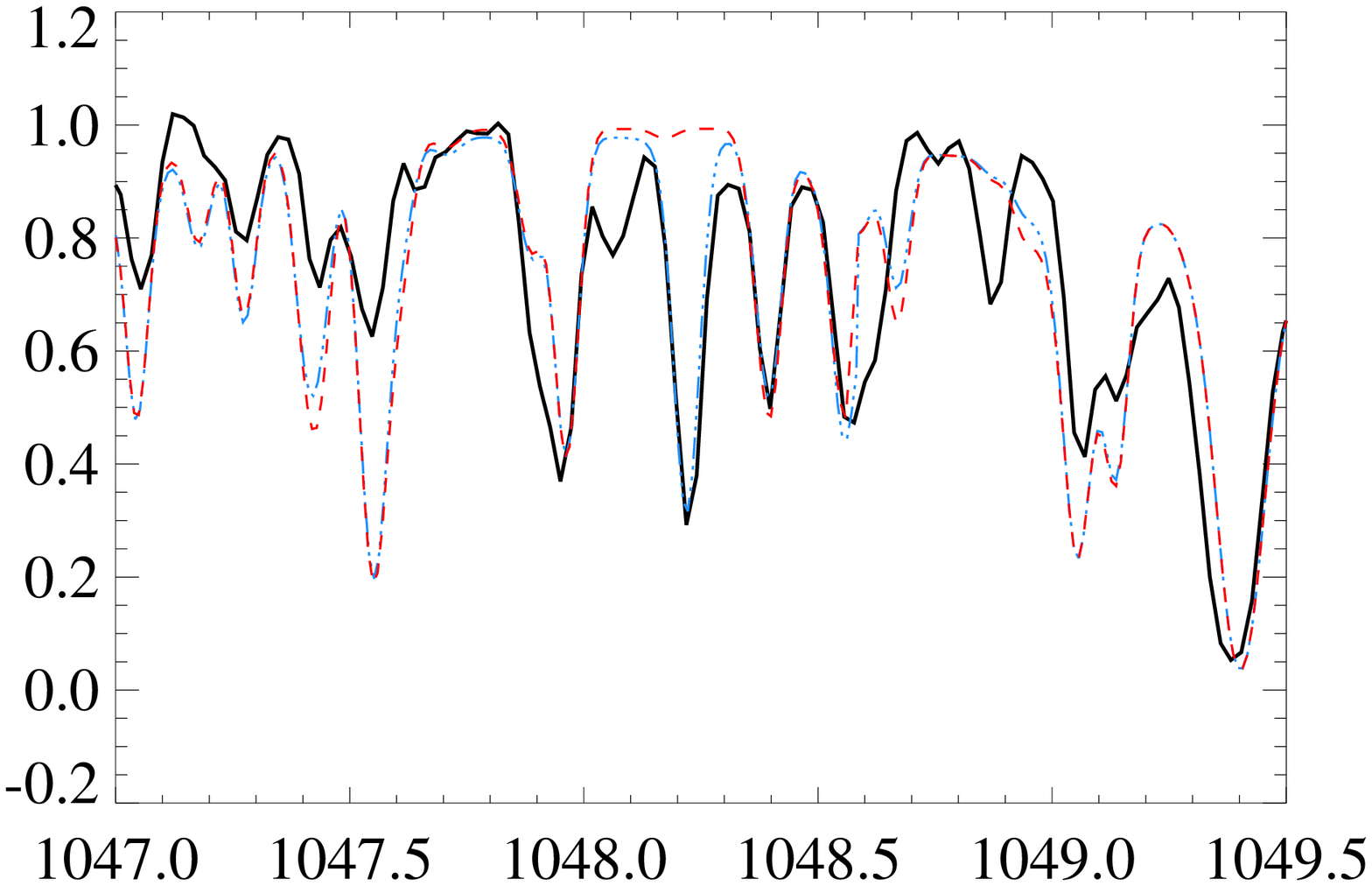}%
\includegraphics{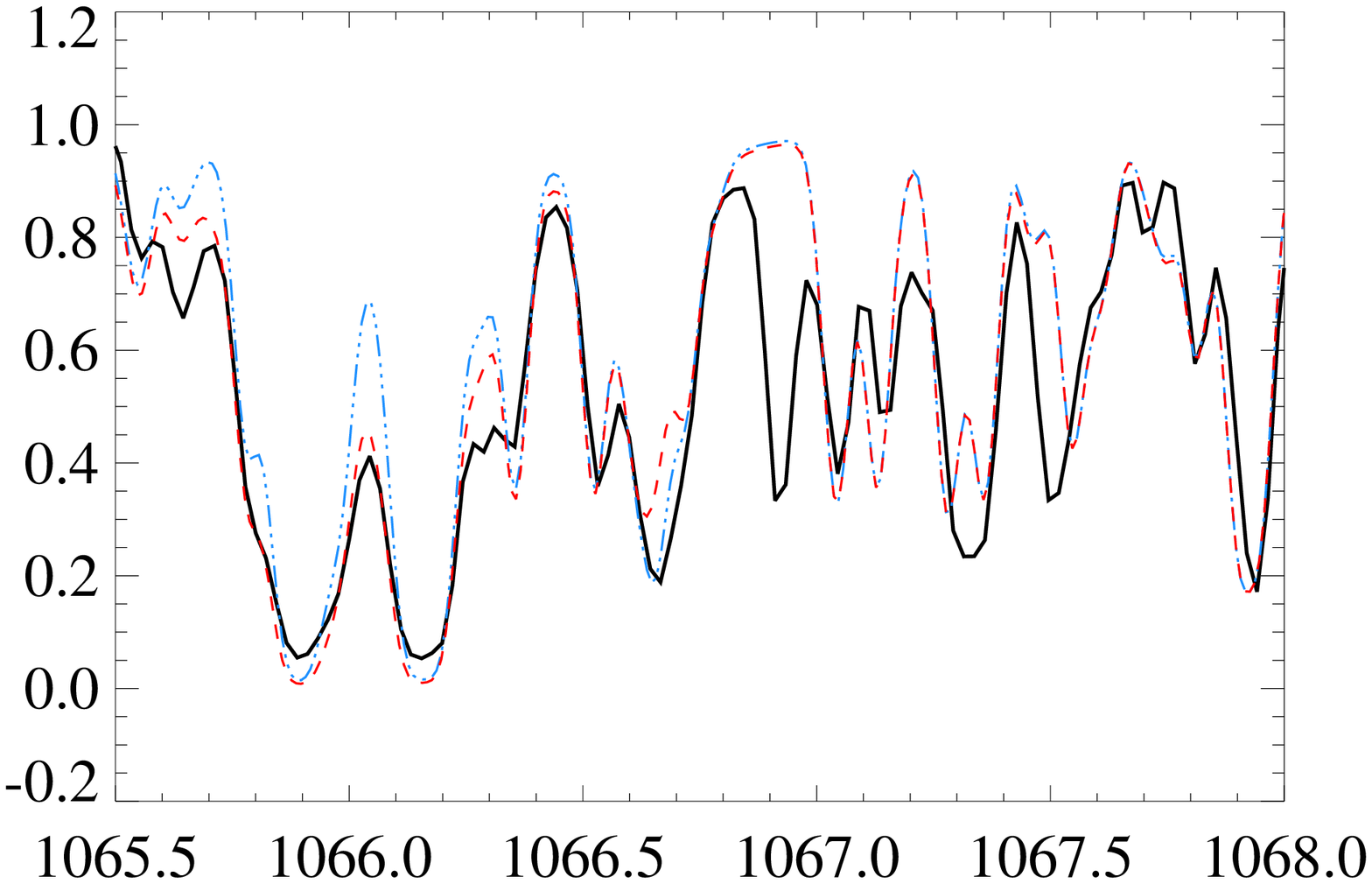}}
\caption{\label{Fig:ar_clean_line} This figure shows the observation
  (in black), the calculated model with $\log(n_{\rm Ar}/n_{\rm H})=$--4.900$\pm$0.100 (in
  blue), and without argon (in red). {\it Left:} Ar\,{\sc i} line at
  1048\,\AA\ {\it Right:} Ar\,{\sc i} line at 1066\,\AA.}
\end{figure}

For the abundance determination, we modeled two suitable lines of
Ar\,{\sc i}; first a UV multiplet (1) line at 1048\AA\ as well as
another UV multiplet (1) line at 1066\,\AA.\ These two lines are both
resonance lines and have relatively well determined ``C+'' $\log gf$
values of around -0.5 and -1.0, respectively (Kramida et al., 2014),
and the line at 1048~\AA\ appears unblended. The best-fitting model
yields $\log(n_{\rm Ar}/n_{\rm H})=$--4.90$\pm$0.10 in both cases.
Figure~\ref{Fig:ar_clean_line} shows the observation and the models.

\subsection{Calcium, Z=20}
In the UV spectrum of $\iota$~Herculis, calcium is not strongly
observed. The subset of the VALD database selected here contains
overall 43 calcium lines which are almost equally distributed between
Ca\,{\sc ii} and Ca\,{\sc iii}. The prediction of the Saha equation
shows that Ca\,{\sc iii} must be the dominant state of ionization at
this temperature (see Table 3).  However, because the lowest excited
states are more than 30~eV above ground, there is almost no
contribution from the Ca\,{\sc iii} state in the observed spectrum of
$\iota$~Herculis.

\begin{figure}
\resizebox{\hsize}{!}{\includegraphics{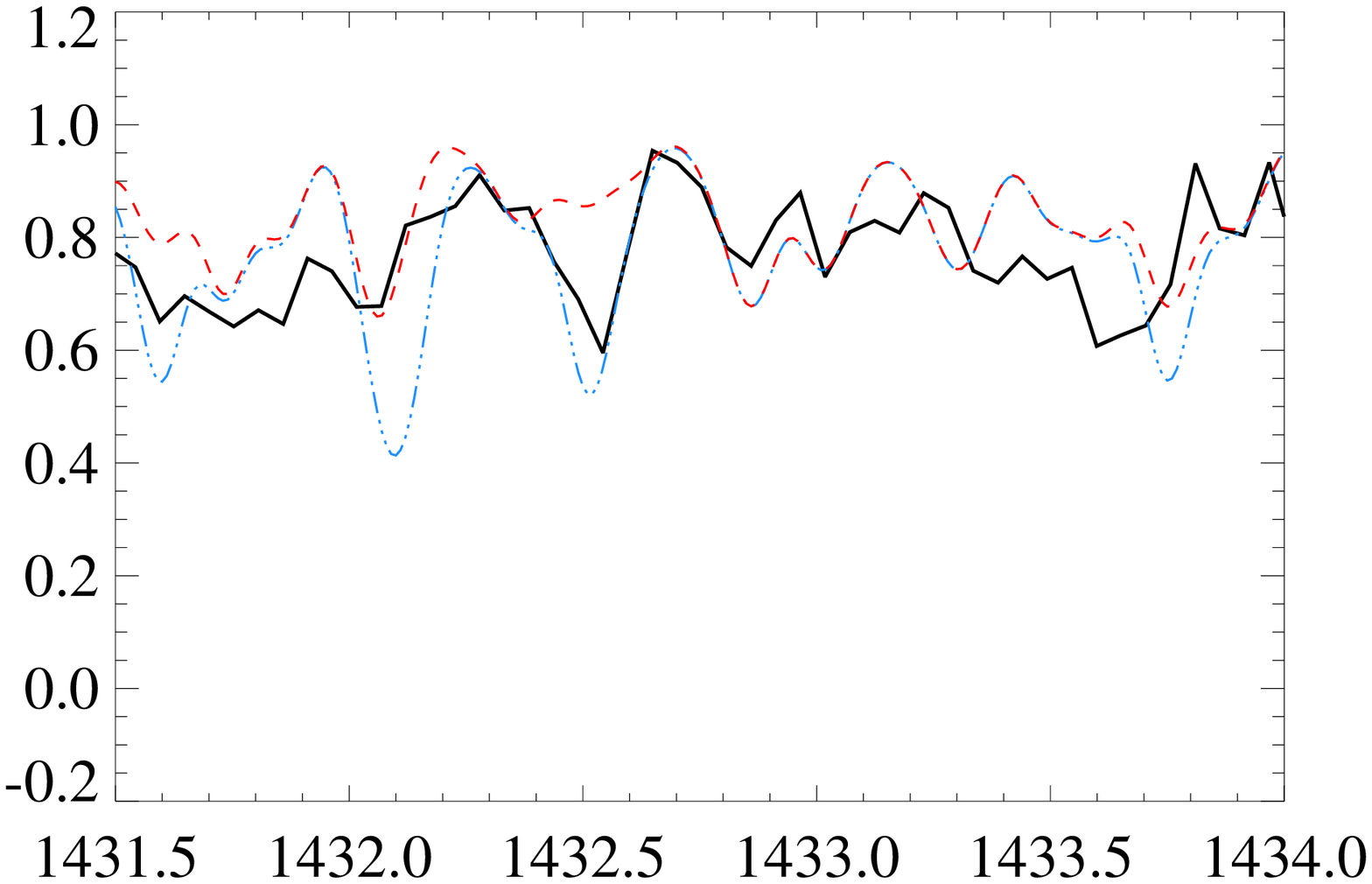}%
\includegraphics{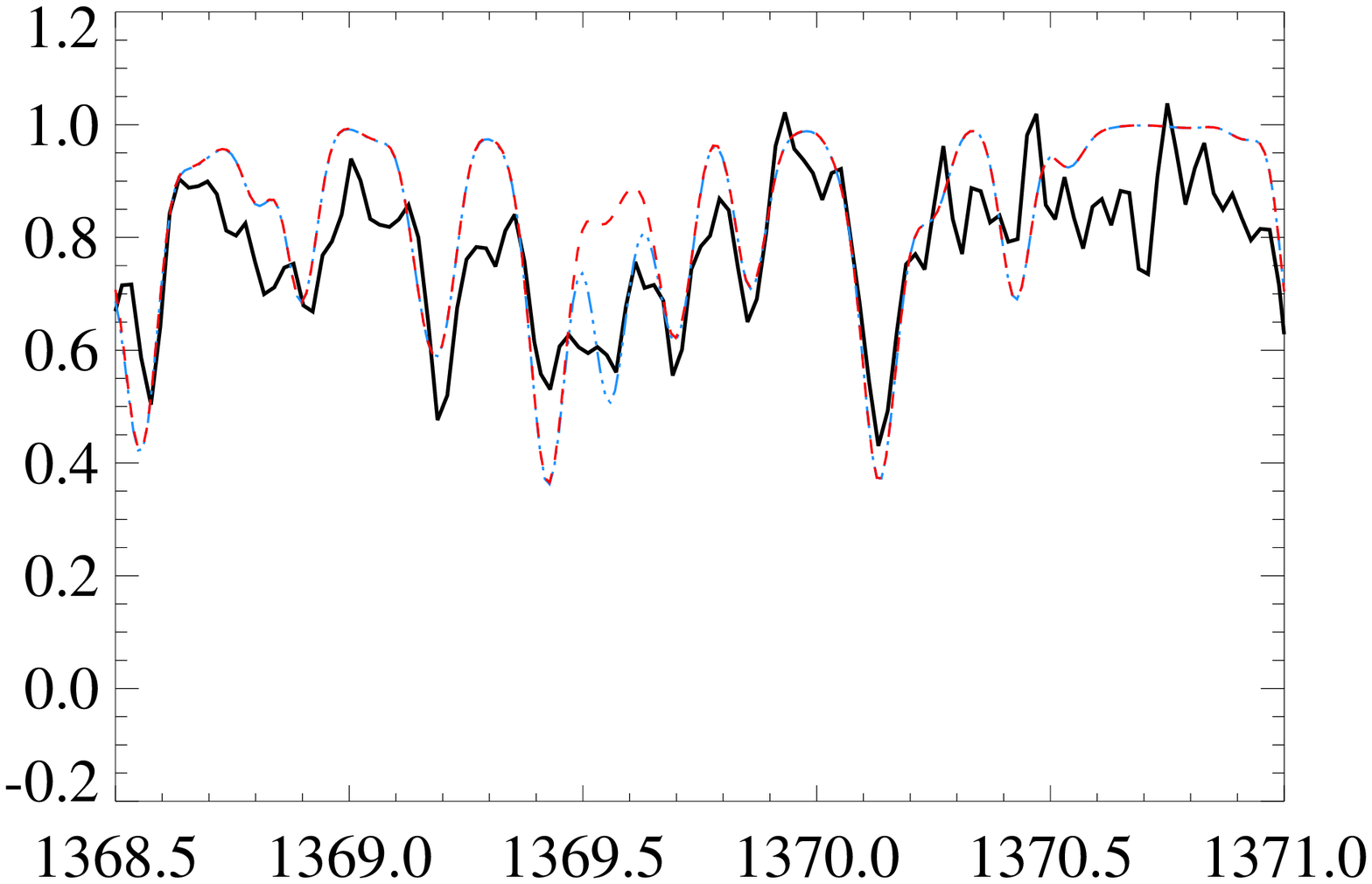}}
\caption{\label{Fig:cl_clean_line} This figure shows the observation
  (in black), the calculated model using
  $\log(n_{\rm Ca}/n_{\rm H})=$--5.30$\pm$0.10 (in blue), and without calcium
  (in red). {\it Left:} Ca\,{\sc ii} lines at 1432\,\AA, {\it
    Right:} Ca\,{\sc ii} line at 1369\,\AA.}
\end{figure}

Despite the lack of clean and unblended lines, we could determine the
abundance using blended lines. Modeling the two Ca\,{\sc ii} lines in the 
1369 and 1432\,\AA\ windows yields $\log(n_{\rm Ca}/n_{\rm
  H})=$--5.30$\pm$0.10. These lines are suited for our purpose since
they both arise from a low energy state ($\sim$1.6eV). They both have
relatively well determined ``C'' $\log gf$ values of --0.8 (Kramida et
al., 2014). Figure~\ref{Fig:cl_clean_line} shows the observations and
models.

\subsection{Titanium, Z=22}
Titanium has 156 titanium lines in our database. Most of them
are in the form of Ti\,{\sc iii}, and a smaller fraction of them are in
Ti\,{\sc ii} with little to contribution from Ti\,{\sc iv}. In the UV
spectrum of $\iota$~Herculis, Ti\,{\sc iii} is dominantly observed and
there is a minor contribution from Ti\,{\sc ii} lines at longer
wavelengths. This is consistent with the ionization ratios calculated
from the Saha equation for this temperature (see
Table~3).

\begin{figure}
\resizebox{\hsize}{!}{\includegraphics{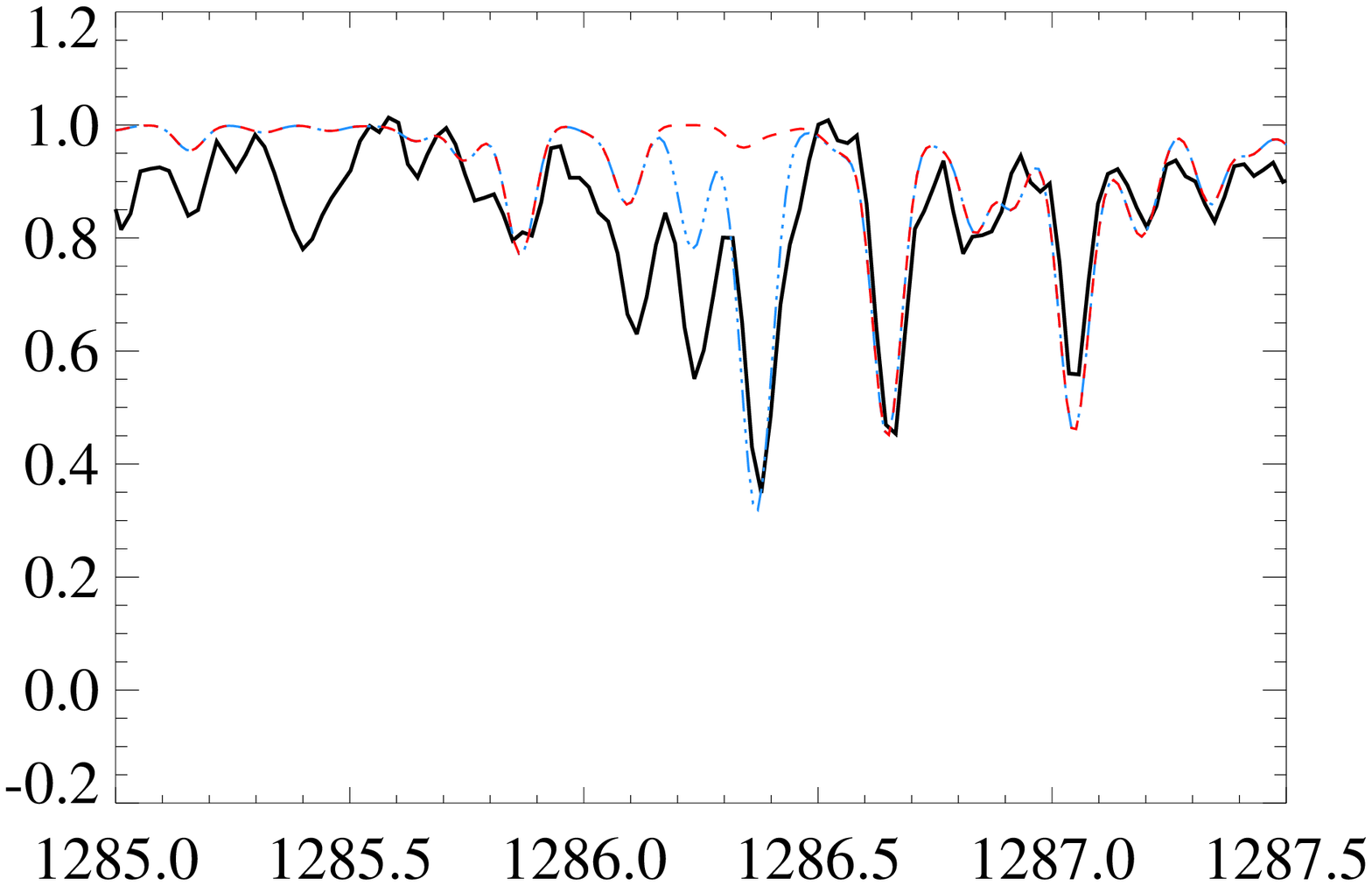}%
\includegraphics{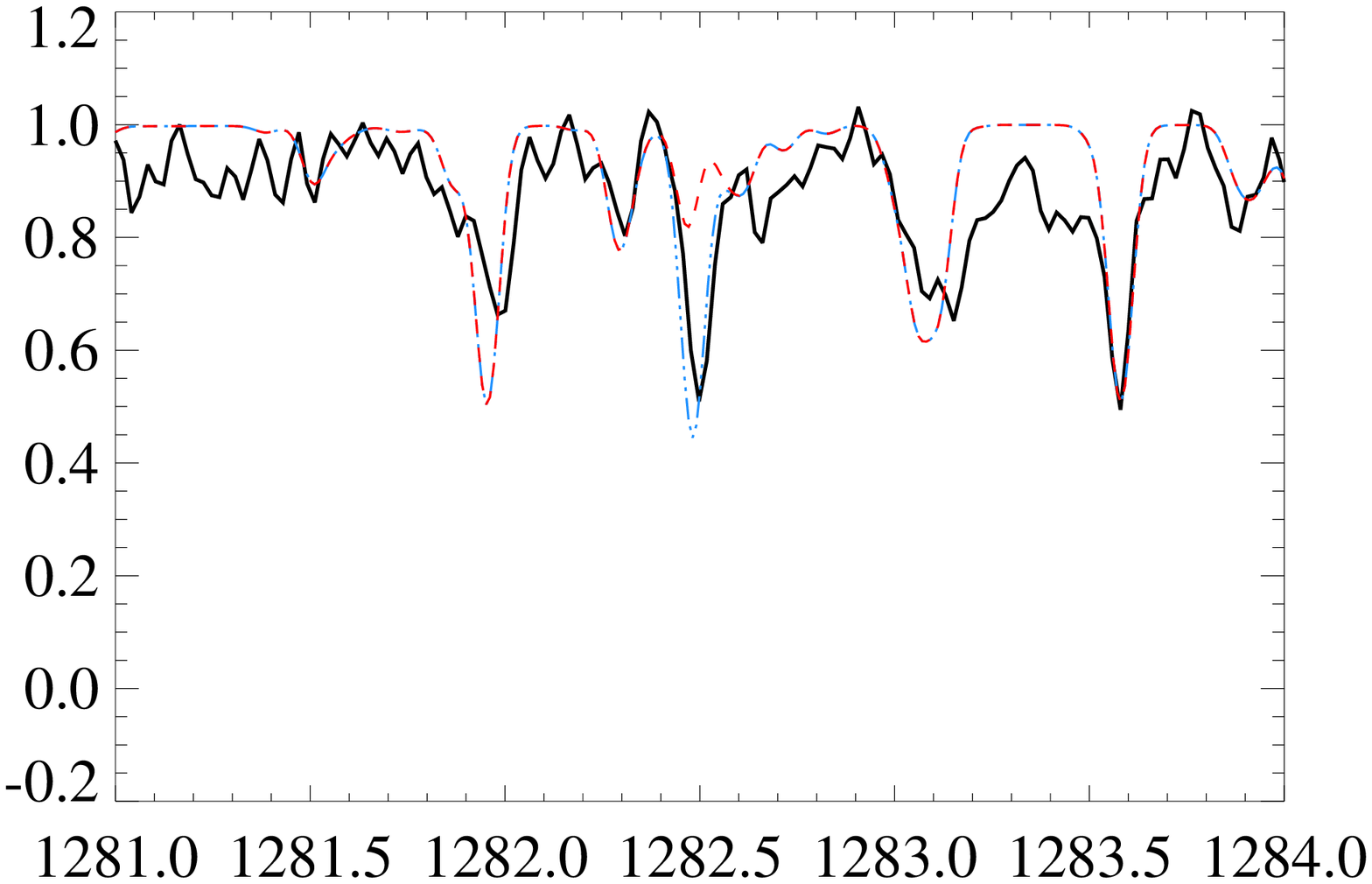}}\\
\resizebox{\hsize}{!}{\includegraphics{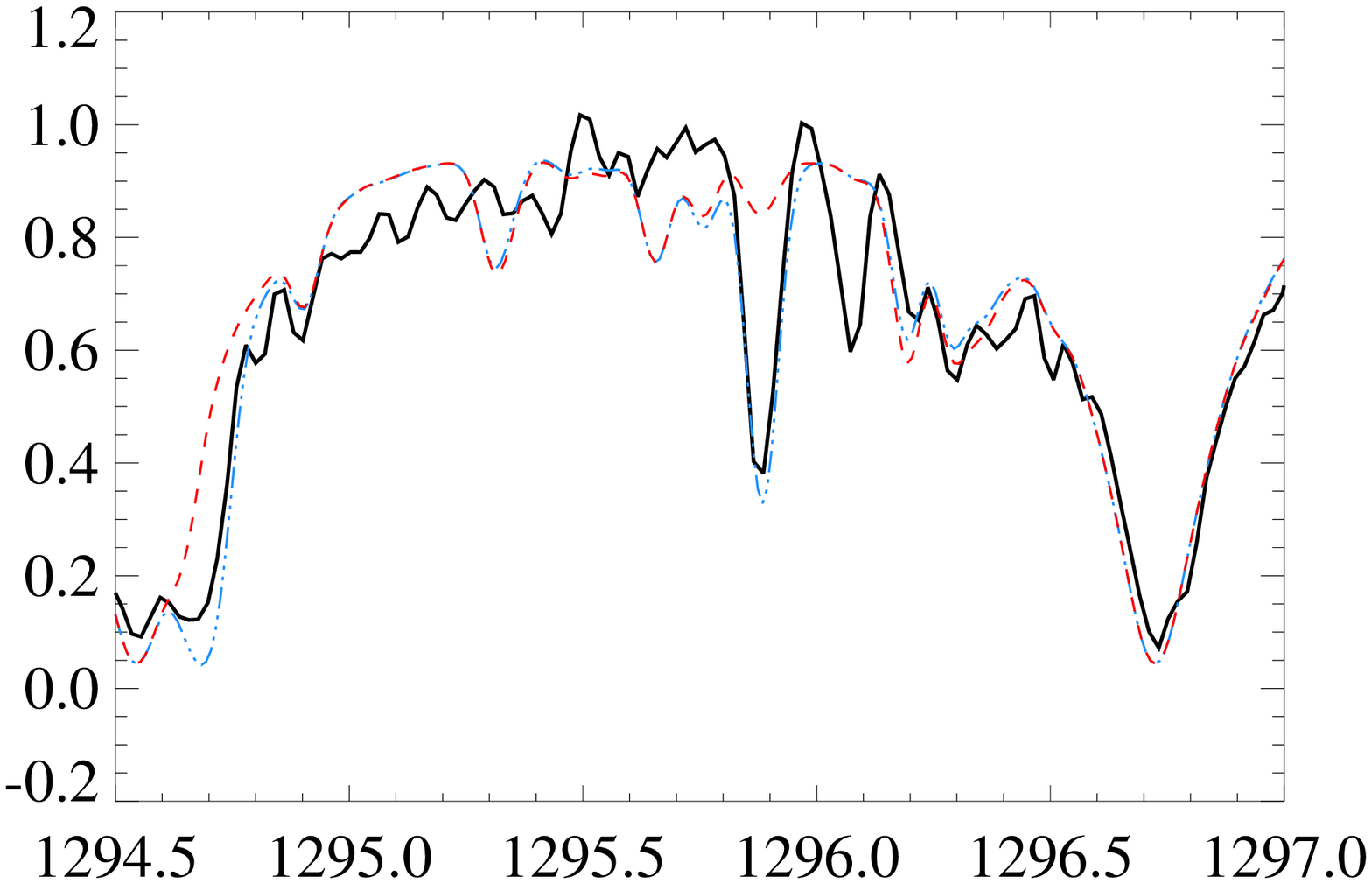}%
\includegraphics{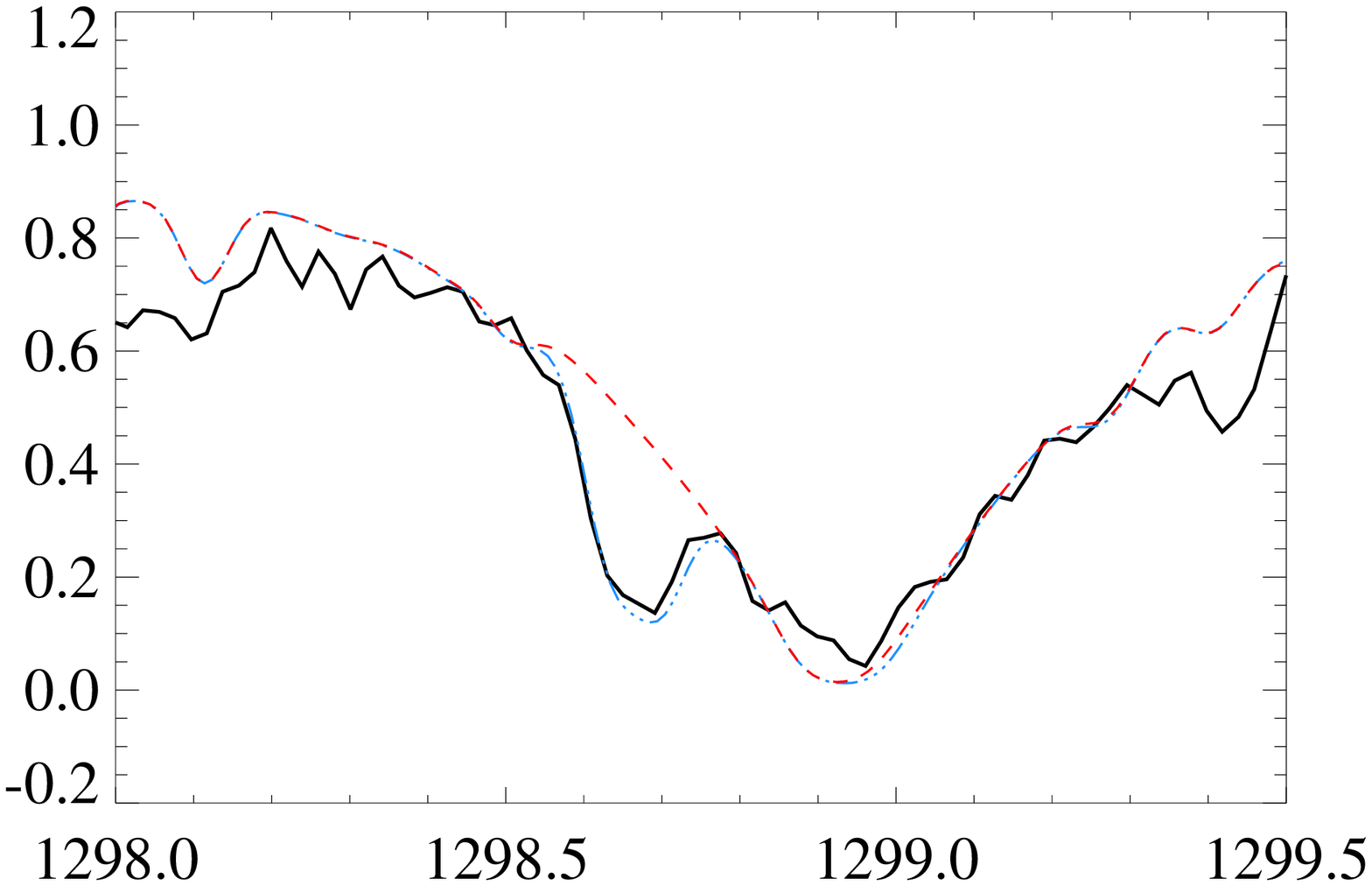}}
\caption{\label{Fig:Ti_line} This figure shows the observation (in
  black), the calculated model with $\log(n_{\rm Ti}/n_{\rm H})=$--6.90$\pm$0.17
  (in blue), and without titanium (in red). All the plots belong to
  the same ionization state Ti\,{\sc iii}; {\it Top left: } A
  resonance line at 1286\,\AA,\ {\it Top right:} at 1282\,\AA,\ {\it Bottom
    left: } at 1295\AA,\ {\it Bottom right: } at 1298\,\AA}.
\end{figure}

We determined the abundance of titanium using lines of the two strong
Ti\,{\sc iii} multiplets (1) and (2) in the 1286-93\,\AA\ window. They
are suitable choices because they arise from low lying energy levels
($\sim$0.00-0.05) and have relatively well determined (``D'') $\log gf$
values of around --0.5. Only parts of these multiplets are shown in
Figure~\ref{Fig:Ti_line}.

 The best-fitting model results in the abundance $\log(n_{\rm
 Ti}/n_{\rm H})= -6.90 \pm 0.17$. We confirmed this abundance by
 modeling the Ti\,{\sc iii} UV resonance multiplet (2) at 1282\AA\ 
 and Ti\,{\sc iii} resonance lines at 1295-98\,\AA. These lines all
 have a ``D'' $\log gf$ values of around --0.50 and despite a slight
 blending with nickel (see \S \ref{nickel}) they still serve our
 purpose. Figure~\ref{Fig:Ti_line} shows the observed lines as well as
 the calculated models.

\subsection{Vanadium, Z=23}
Our line database contains 234 vanadium lines. All are in the
form of V\,{\sc iii}. In the spectrum of $\iota$~Herculis vanadium is
also observed only as V\,{\sc iii}. The predictions of the Saha
equation shows (see Table 3) that V\,{\sc iii} is
the dominant state of ionization and V\,{\sc i}, V\,{\sc ii} are
severely underpopulated in this temperature and wavelength range. The
lines of V\,{\sc iv} are absent because the allowed transitions in our
spectral region arise from excited states that are higher than
$\sim$18~eV.

We have used the four strong and fairly clean V\,{\sc iii} lines at
1154-1160, 1252, 1332\,\AA.\ They all arise from low lying energy
levels ($\sim$1.5~eV) and $\log gf$ values around zero. Despite slight
blending with Ni\,{\sc ii} line (see \S \ref{nickel}), they still
serve our purpose.

\begin{figure}
 \resizebox{\hsize}{!}{\includegraphics{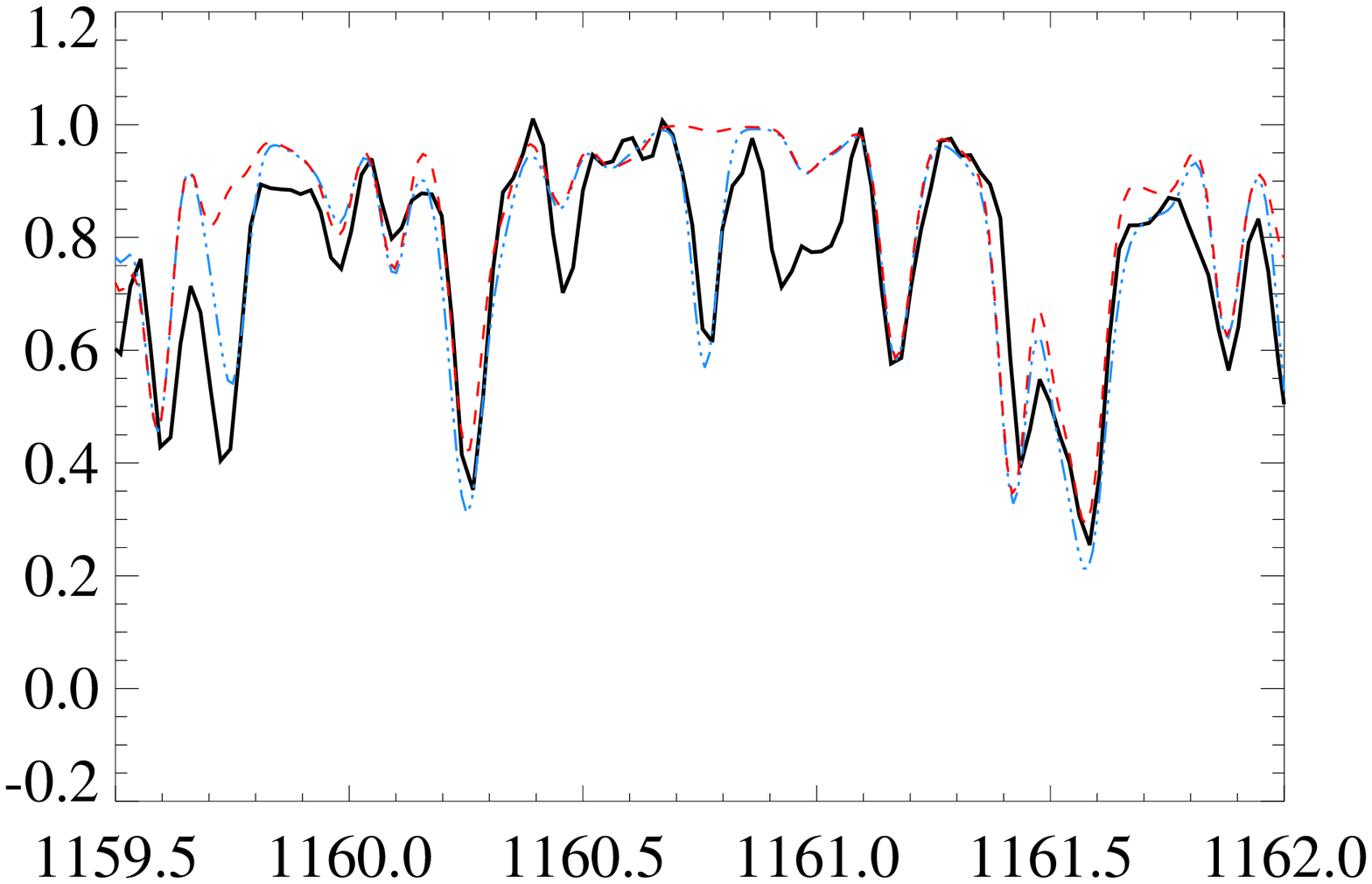}%
 \includegraphics{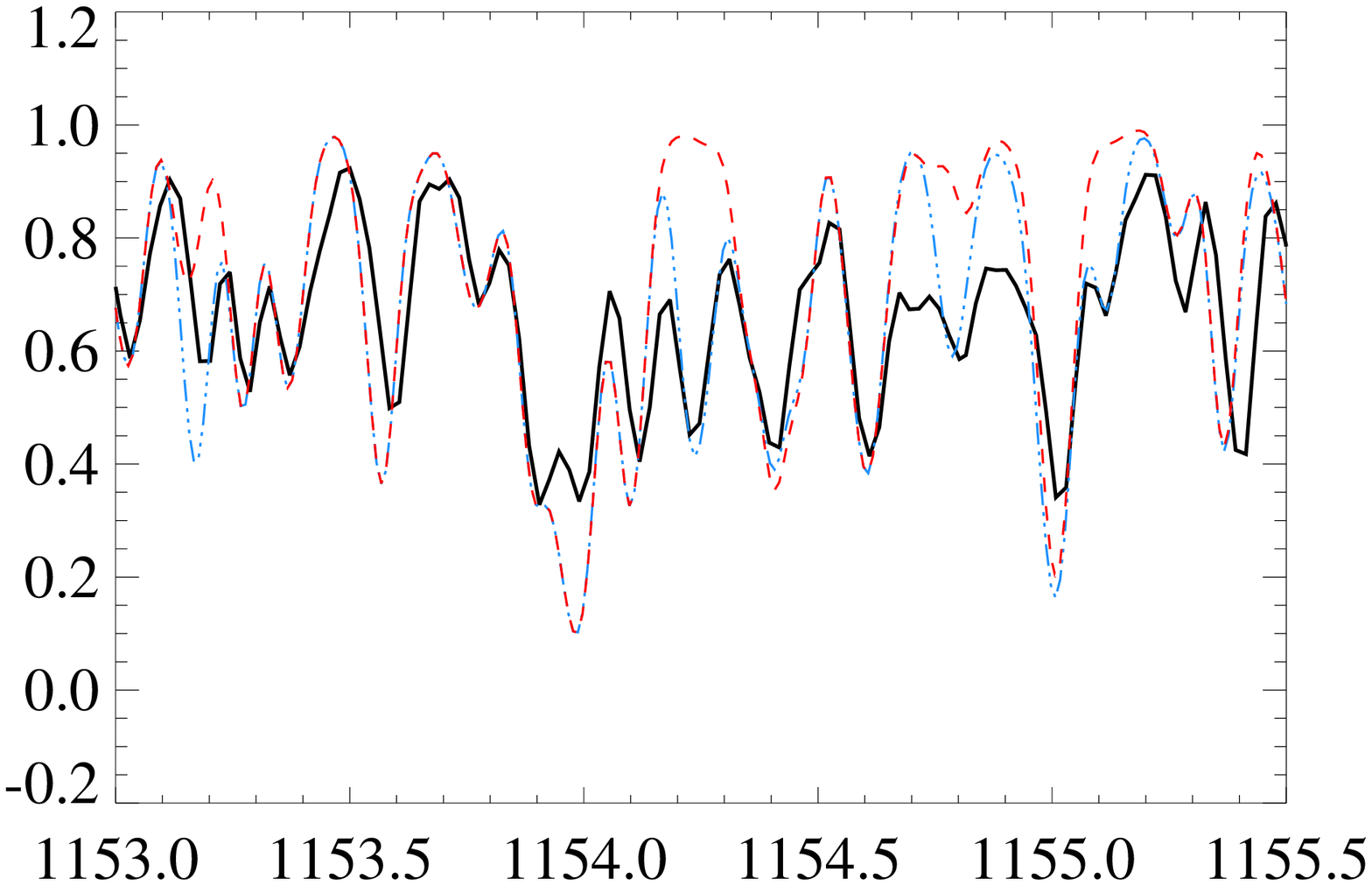}}\\ 
 \resizebox{\hsize}{!}{\includegraphics{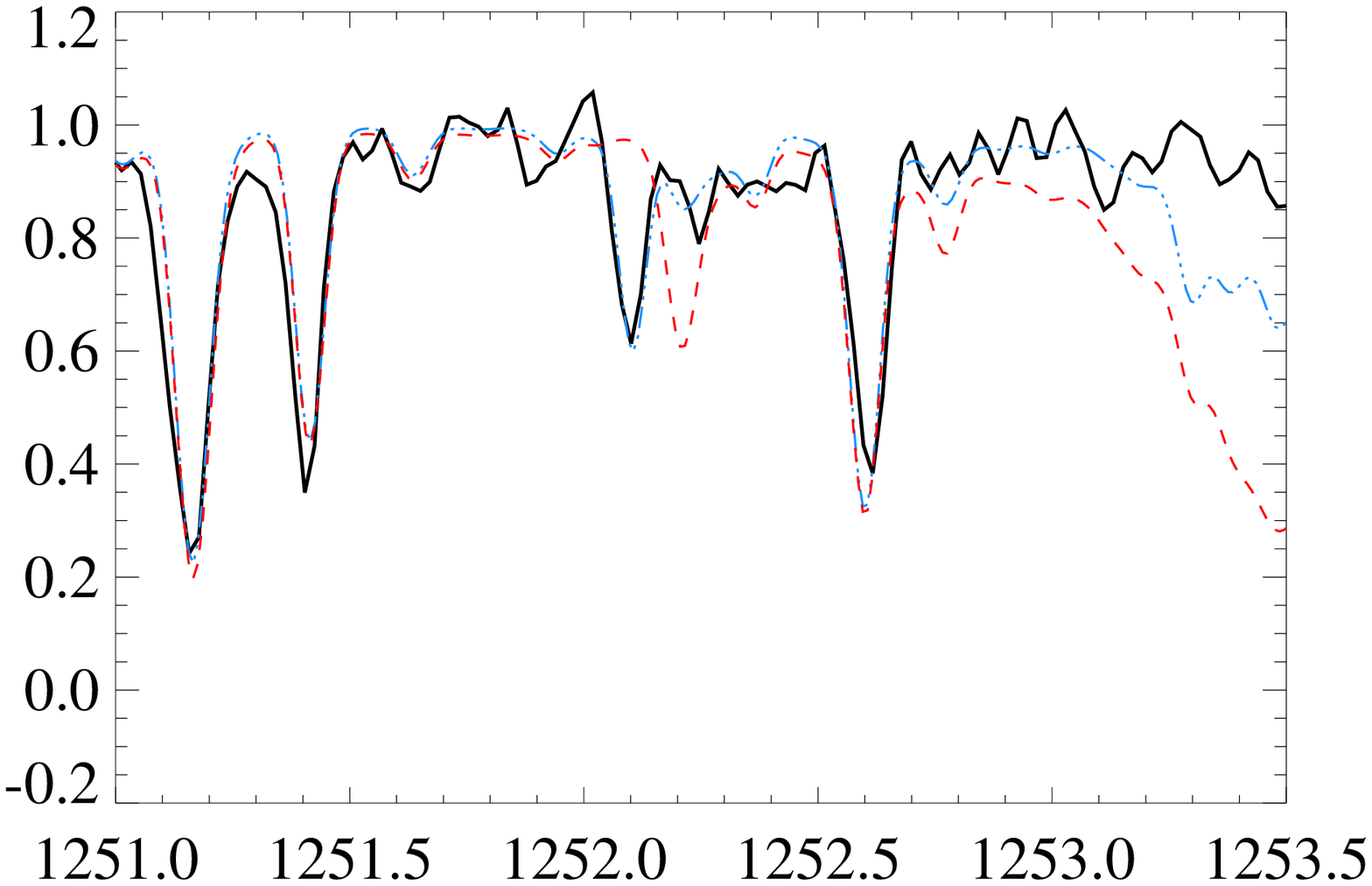}%
 \includegraphics{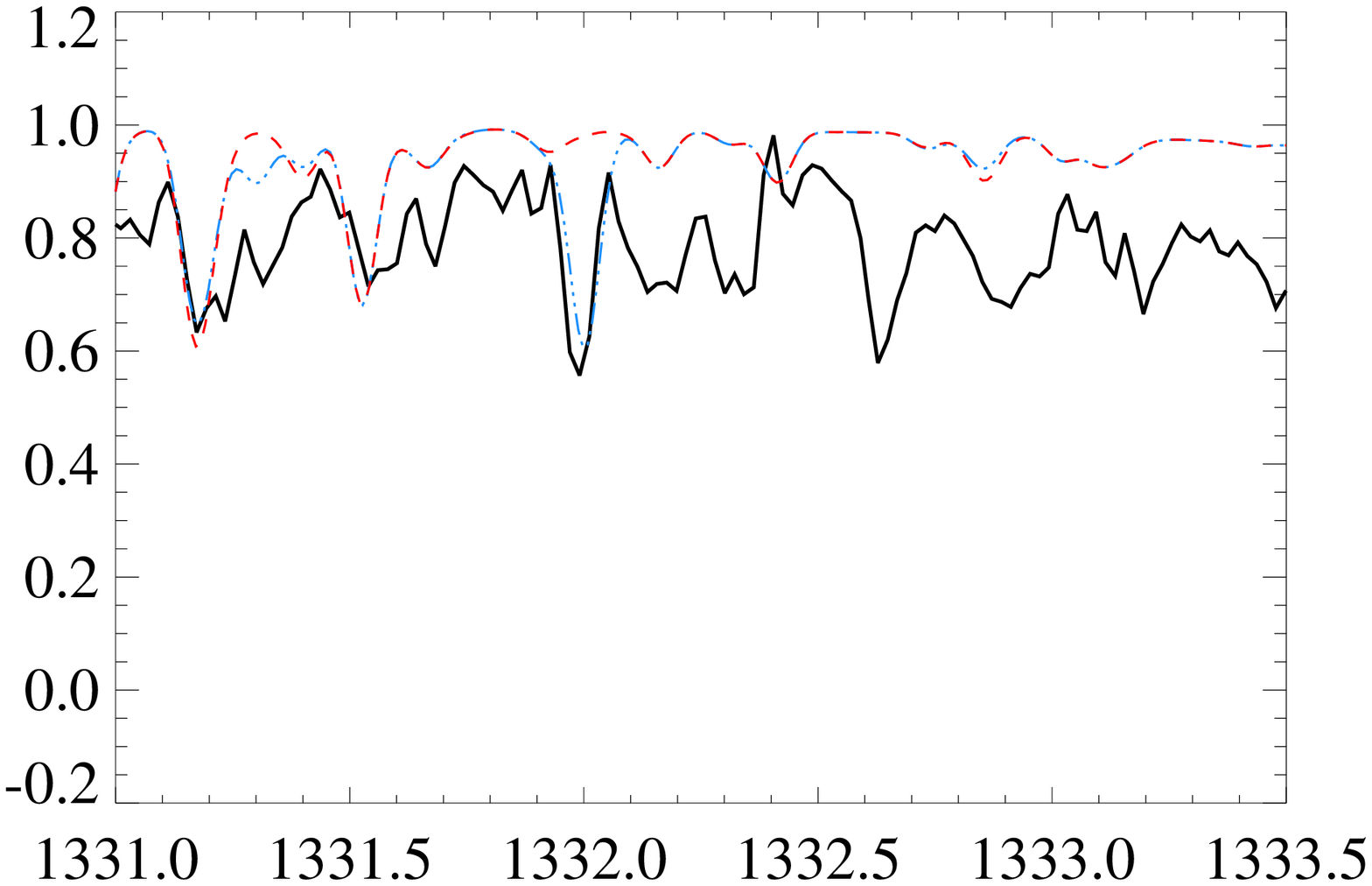}}
\caption{\label{Fig:clean_V_line} This figure shows the observation
  (in black), the calculated model with $\log(n_{\rm V}/n_{\rm
    H})$=--8.15$\pm$0.18 (in blue), and without vanadium (in red). All
  of the plots belong to the same state of ionization; V\,{\sc iii}.
  {\it Top left: } 1160\,\AA,\ {\it Top right:
  }1154\,\AA,\ {\it Bottom left: } 1252\,\AA,\ {\it Bottom
    right: } 1332\,\AA}
\end{figure}

The best-fitting model to these lines results from a vanadium
abundance of $\log(n_{\rm V}/n_{\rm H}) =$--8.15$\pm$0.18. Figure
\ref{Fig:clean_V_line} shows the best-fitting model to the
observation. The blending of nickel and vanadium provides a reasonable
fit to the observation.

\subsection{Chromium, Z=24}
\label{chromium}
The selected line-list contains 1645 chromium lines. In the UV
  spectrum of $\iota$~Herculis, chromium is mostly observed as
  Cr\,{\sc iii}, with smaller contributions from Cr\,{\sc ii} and
  Cr\,{\sc iv}. This is consistent with the ionization ratios predicted
  from the Saha equation (see Table~3.

 We used the Cr\,{\sc iii} triplet line in the 1098\,\AA\ window to
 determine the abundance. This line is unblended and it arises from a
 fairly low lying energy level of 2~eV and $\log$gf of $\sim$--2. The
 accuracy on the oscillator strengths of chromium lines are not known
 very well. The best-fitting model results in $\log(n_{\rm Cr}/n_{\rm
   H})$=--6.10$\pm$0.65.

\begin{figure}
\resizebox{\hsize}{!}{\includegraphics{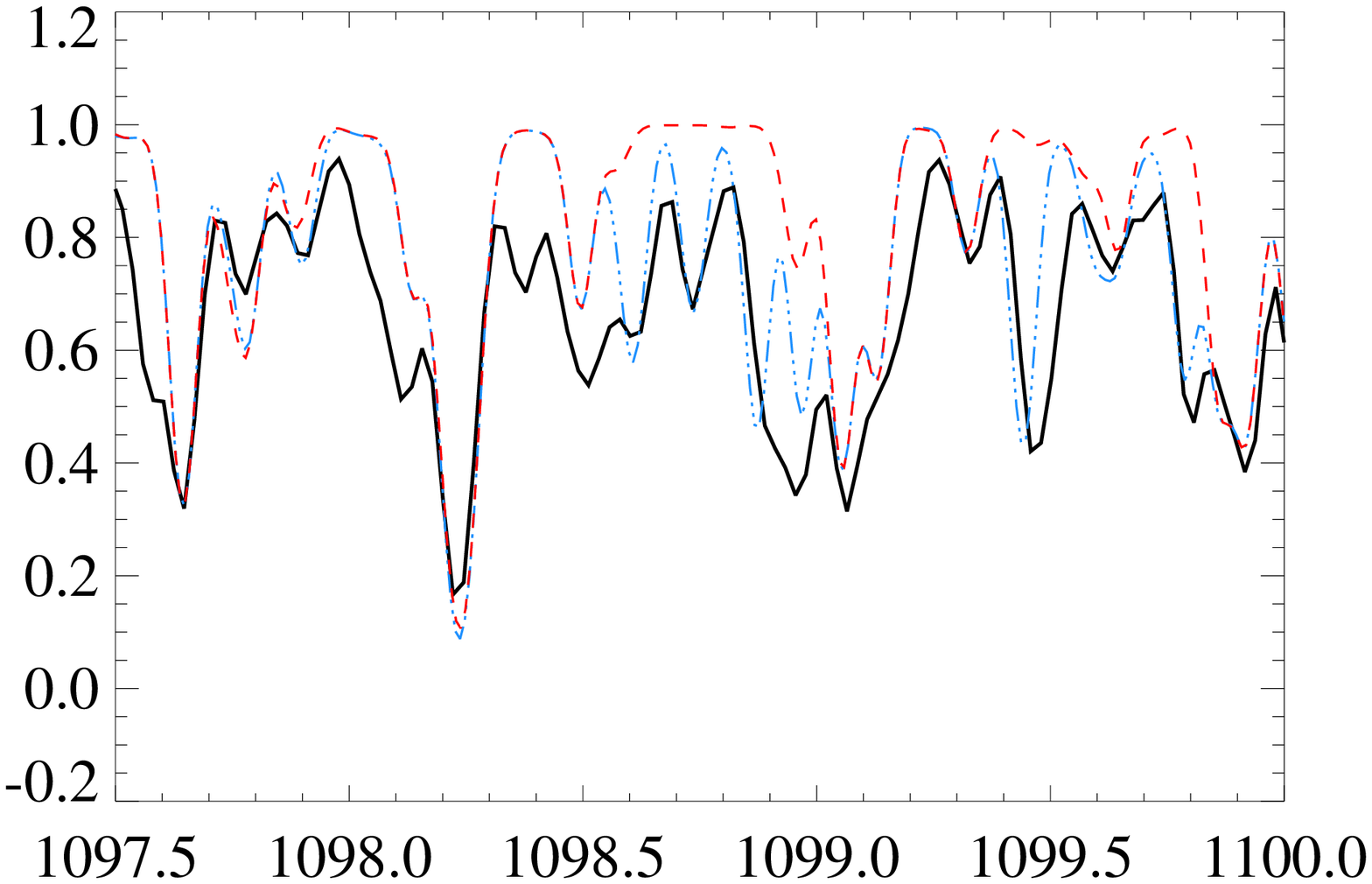}%
\includegraphics{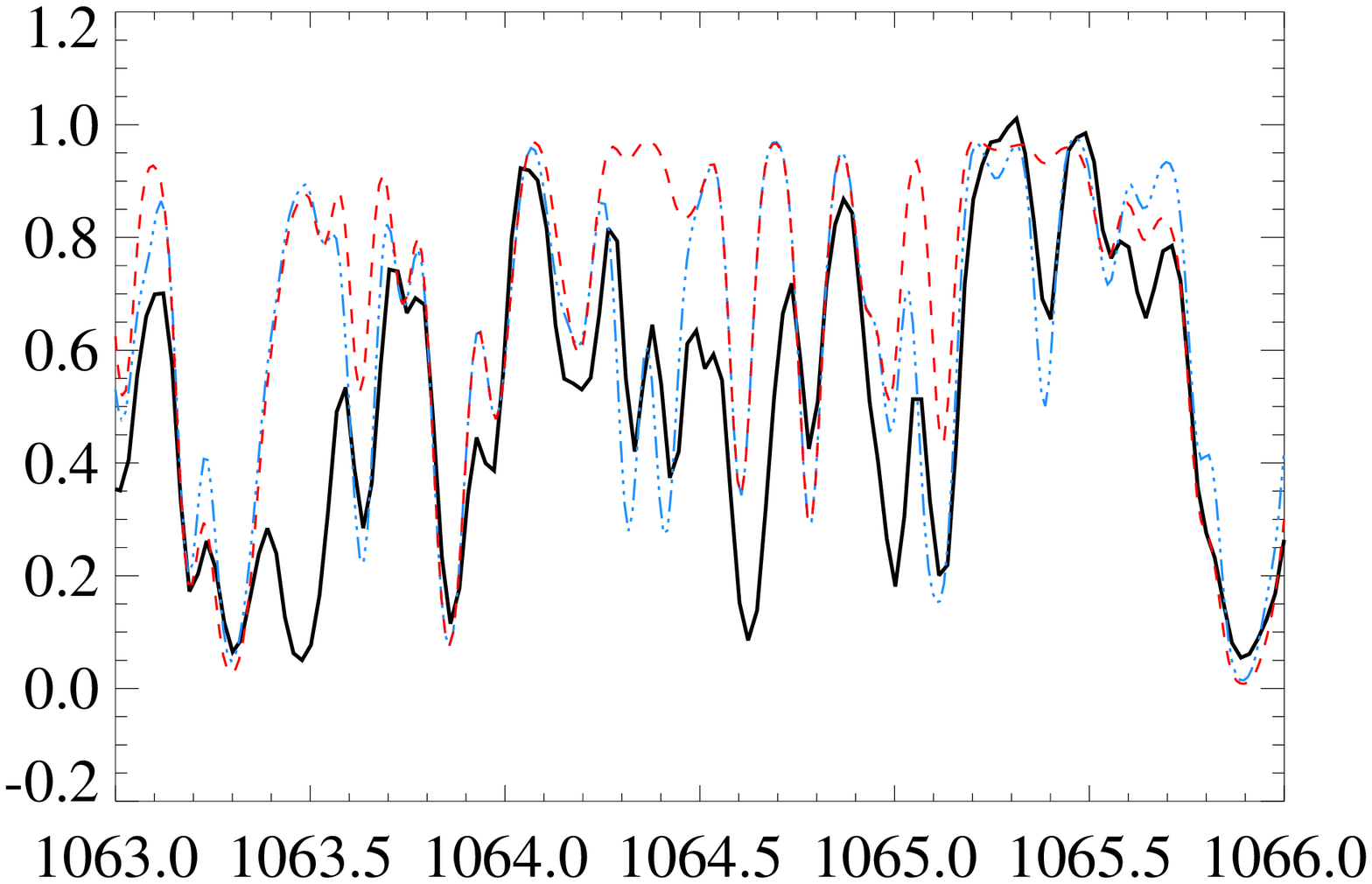}}\\
\resizebox{\hsize}{!}{\includegraphics{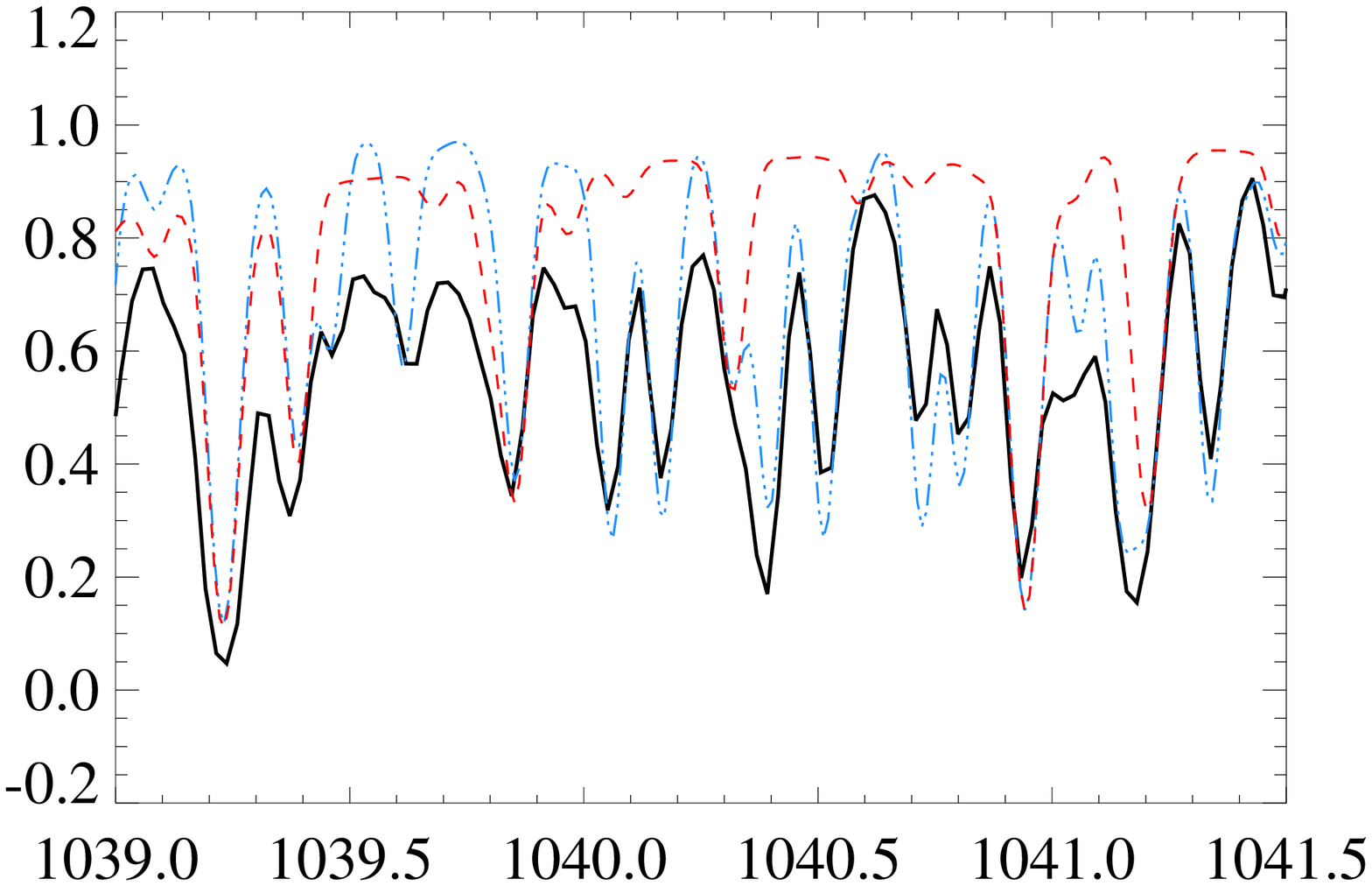}%
\includegraphics{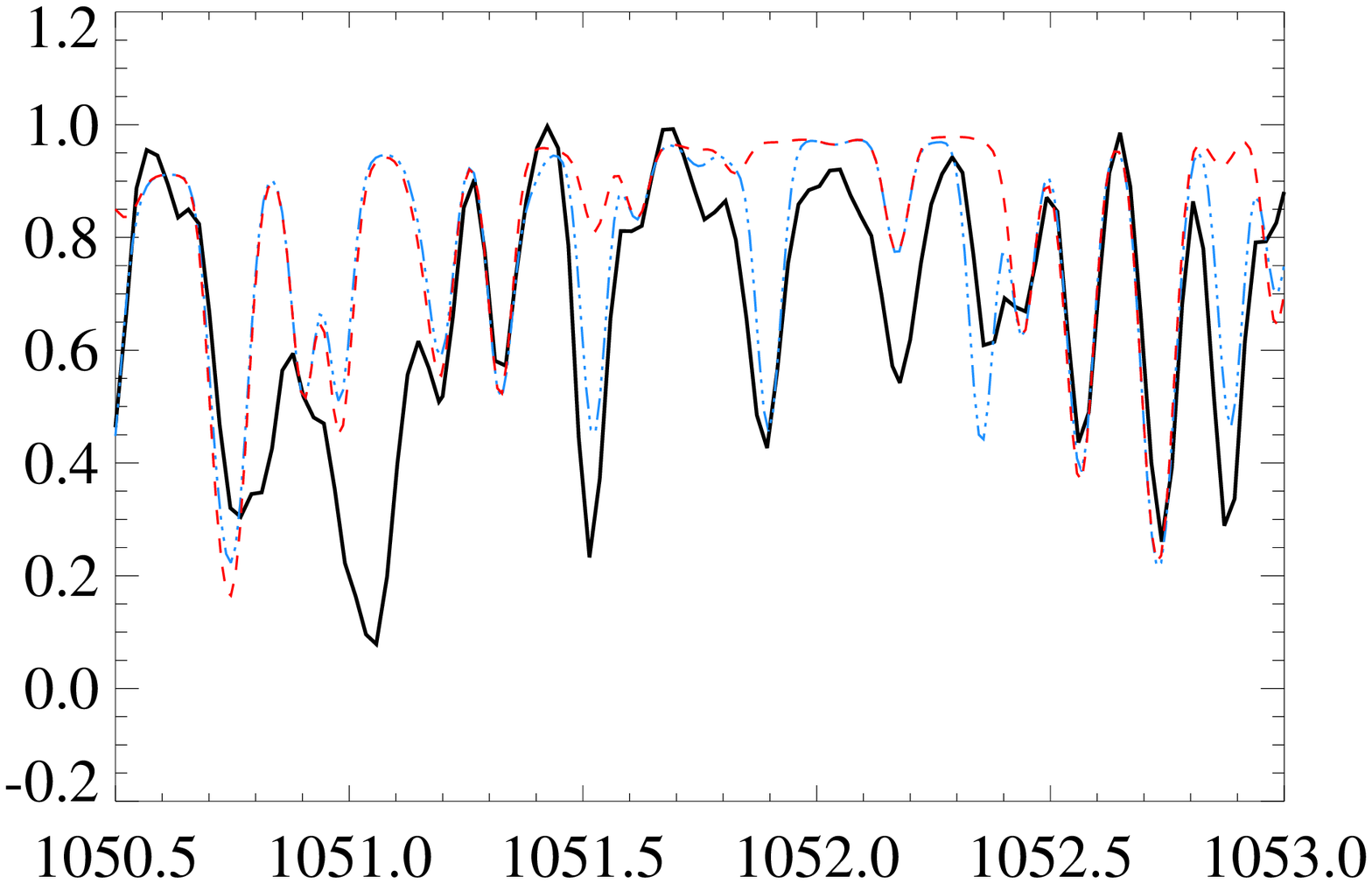}}
\caption{\label{Fig:Cr_clean_line} This figure shows the observation
  (in black), the calculated model with $\log(n_{\rm Cr}/n_{\rm
    H})$=--6.10$\pm$0.65, and without chromium (in red). The plots
  belong to the same ionization state; Cr\,{\sc iii}. Triplet lines at
  1098\,\AA\ {\it (Top left)},\ Cr\,{\sc iii} triplet lines at
  1064\,\AA\ {\it (Top right)},Cr\,{\sc iii} resonance line
  with contributions from another multiplet at
  1040\,\AA\ {\it (Bottom left)}, strong Cr\,{\sc iii} line at
  1051\,\AA\ {\it (Bottom right)} }
\end{figure}

We confirmed this value using Cr\,{\sc iii} triplet lines at 1040\AA\ and
1064\,\AA\ windows and a strong Cr\,{\sc iii} line at 1051\,\AA.\ What
they all have in common is that they arise from low energy levels and
have oscillator strengths that is between 0 and --1. Figure
\ref{Fig:Cr_clean_line} shows the observation and the best-fitting
models.

\subsection{Manganese, Z=25}
\label{manganese}
In the database selected here, there are 1448 manganese lines. Almost all
of them are in the form of Mn\,{\sc iii}, and the rest are in the form
of Mn\,{\sc ii}, with very few lines of Mn\,{\sc iv}. In the UV
spectrum of $\iota$~Herculis manganese is observed mostly in the form
of Mn~{\sc iii} lines throughout, with a few strong Mn\,{\sc ii}
lines. The contribution from Mn\,{\sc iv} is too weak to be
  discernible. The measurements from the Saha equation, indicates that
  Mn\,{\sc iii} is the dominant state of ionization at this
  temperature which is consistent with our observation.

\begin{figure}
\resizebox{\hsize}{!}{\includegraphics{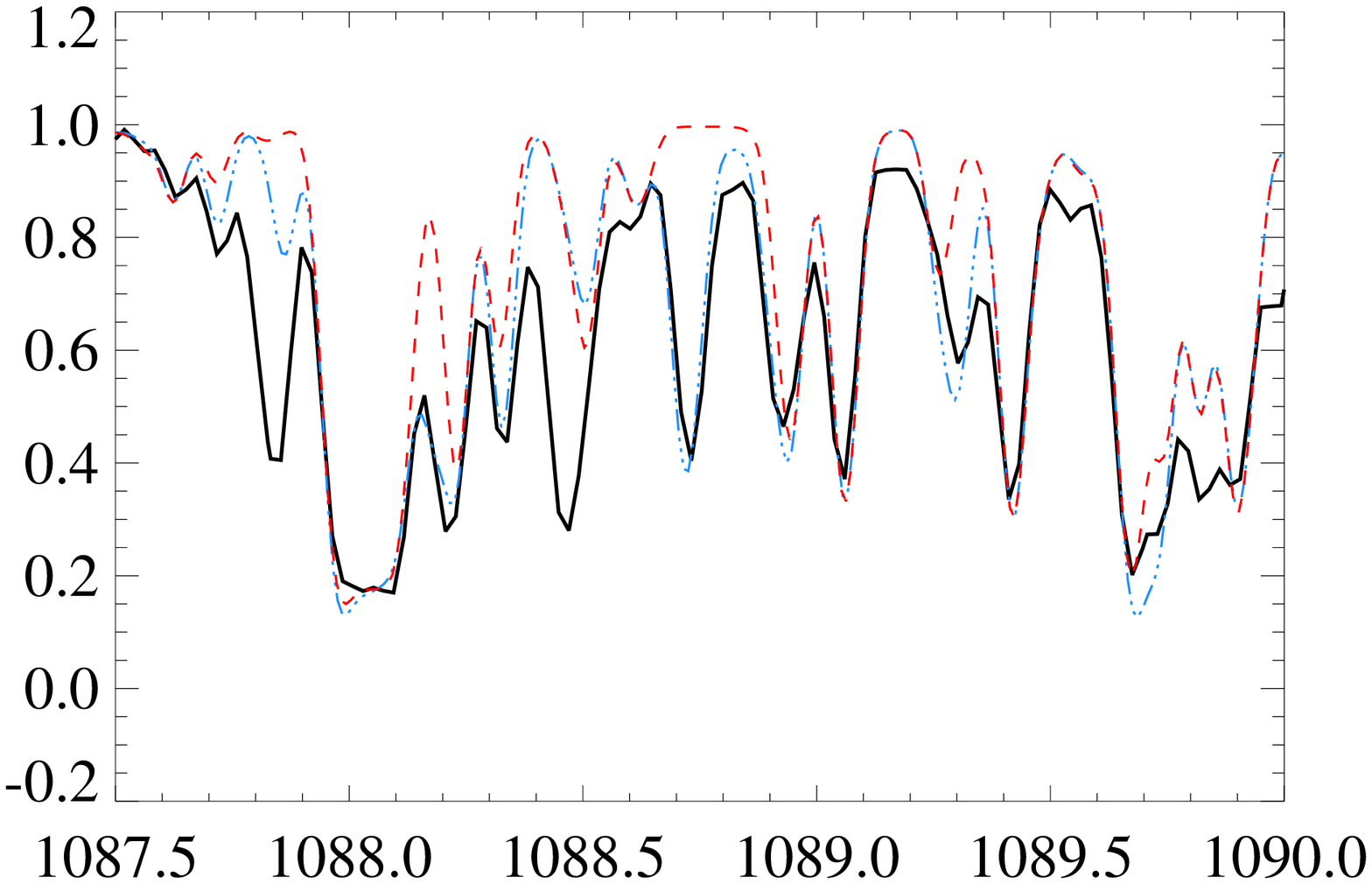}%
\includegraphics{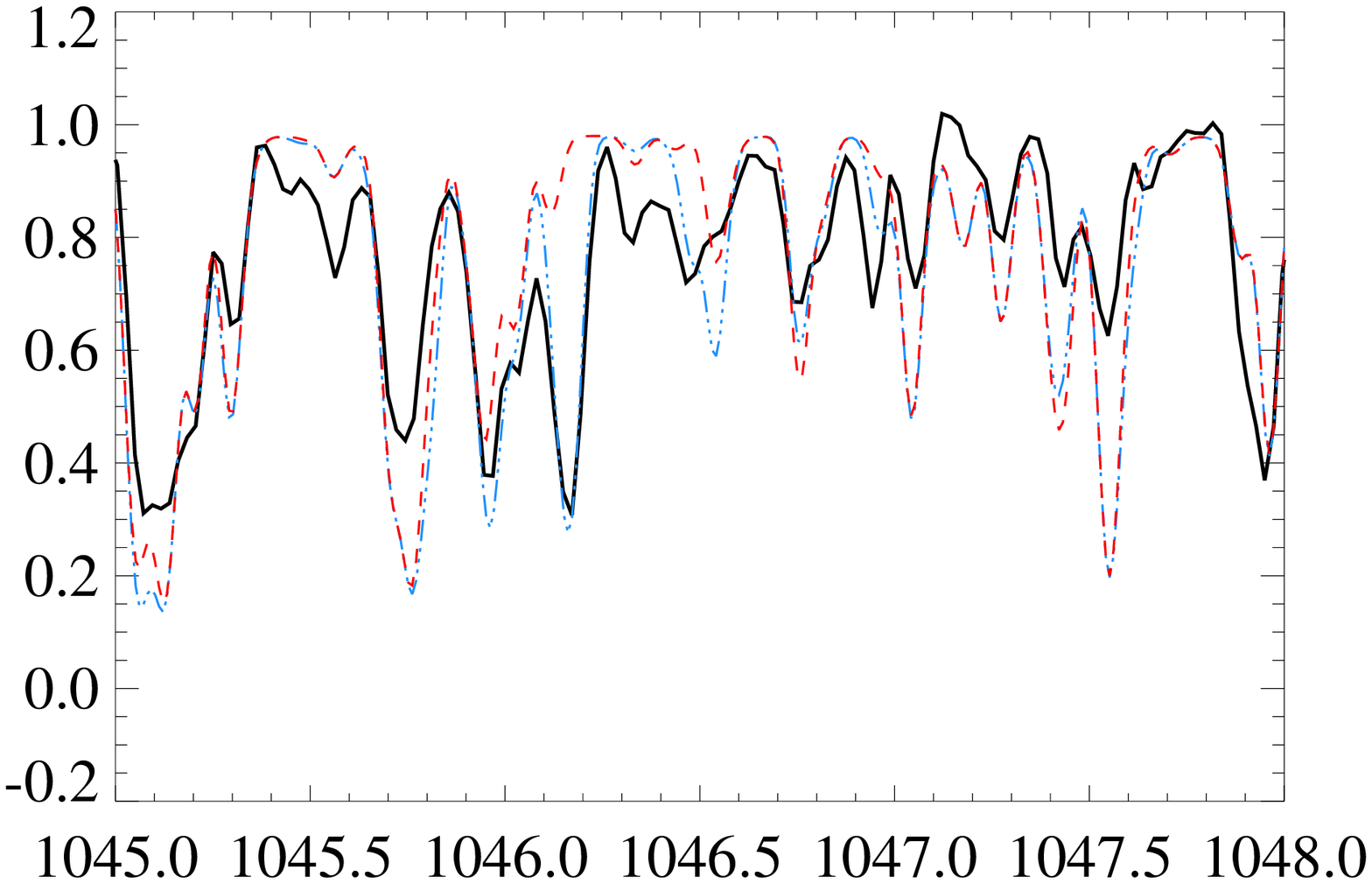}}\\
\resizebox{\hsize}{!}{\includegraphics{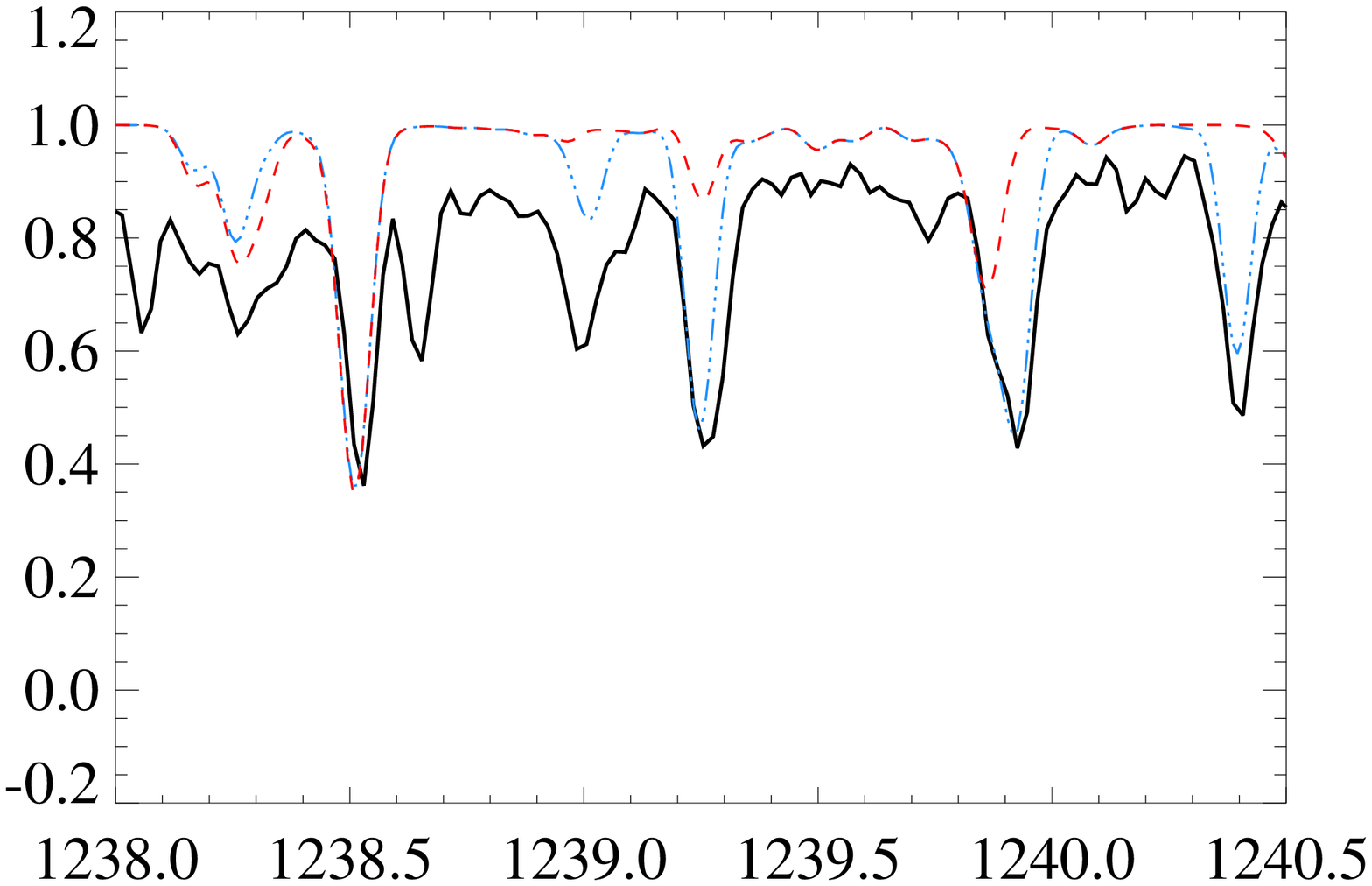}%
\includegraphics{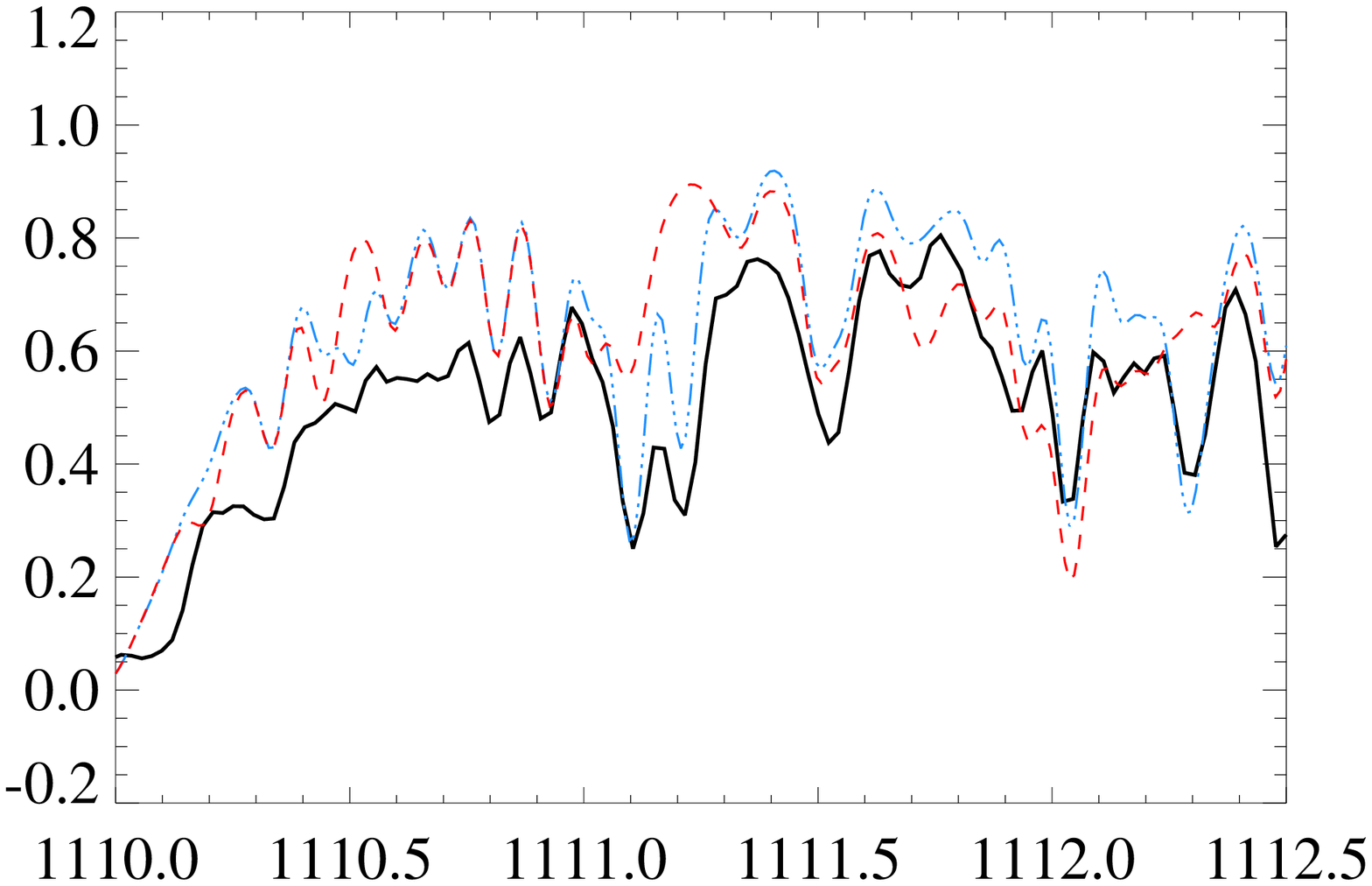}}
\caption{\label{Fig:Mn_clean_line} This figure shows the observation
  (in black), the calculated model with
  $\log(n_{\rm Mn}/n_{\rm H})$=--6.78$\pm$0.45, and without manganese (in
  red). These plots belong to the same ionization state of Mn\,{\sc
    iii}; doublet lines at 1088\,\AA\ {\it (Top left)},
  doublet lines at 1046\,\AA {\it (Top right)}, doublet
  lines at 1239\,\AA\ {\it (Bottom left)}, doublet lines at
  1111\,\AA\ {\it (Bottom right)}.}
\end{figure}

We have used the unblended strong Mn\,{\sc iii} lines in the
1088\,\AA\ window to determine the abundance. We find
$\log(n_{\rm Mn}/n_{\rm H})$=--6.78$\pm$0.45 for manganese. We confirmed this
value by modeling the Mn\,{\sc iii} doublet lines in the  1046,
1239 and 1111\,\AA\ windows. All the lines chosen here have $\log gf$
values around -1 and arise from low lying energy levels
($\sim$4~eV). The observed spectrum and models are shown in Figure
\ref{Fig:Mn_clean_line}.

\subsection{Iron, Z=26}
\label{iron}
Iron lines are prominent in almost every region of the UV spectrum of
$\iota$~Herculis. Our selection of the VALD database contains 8165
iron lines. More than half of them are in the form of Fe\,{\sc
  ii}. The Fe\,{\sc iii} lines, even though fewer in number, seem to
be stronger and deeper than the lines of Fe\,{\sc iv} which are very
weak.

In the UV spectrum of $\iota$~Herculis the majority of iron lines are
observed in the form of moderately strong Fe\,{\sc ii} lines and fewer
Fe\,{\sc iii} lines. Because the lowest excited state is $\sim$19~eV above the
ground state, Fe\,{\sc iv} does not appear in in our
observed spectral window. The calculations from the Saha equation (see
Table 3) indicate that the dominant ionization
state at this temperature is Fe\,{\sc iii}.  Nevertheless, Fe\,{\sc iii} lines do
not appear strongly since for most such lines the lower level is on
average 10~eV or more above the ground state. In contrast the lower
levels of strong Fe\,{\sc ii} lines are only slightly (about ~4eV)
above the ground state which strengthens lines from this state.

Despite the frequency of Fe\,{\sc ii} lines in the spectrum, they are
heavily blended and are not ideal for abundance measurements. We
instead used the strong pair of Fe\,{\sc iii} lines in the
1130\,\AA\ window with a slight overlapping blend from Fe\,{\sc ii}
lines. This set of lines is suited for our purpose because they, not
only are unblended, but also arise from very low lying energy state
(~0.1eV) with $\log gf$ of -1 (VALD3). The best-fitting model results
in $\log(n_{\rm Fe}/n_{\rm H})=$--4.9$\pm$0.4.

\begin{figure}
\resizebox{\hsize}{!}{\includegraphics{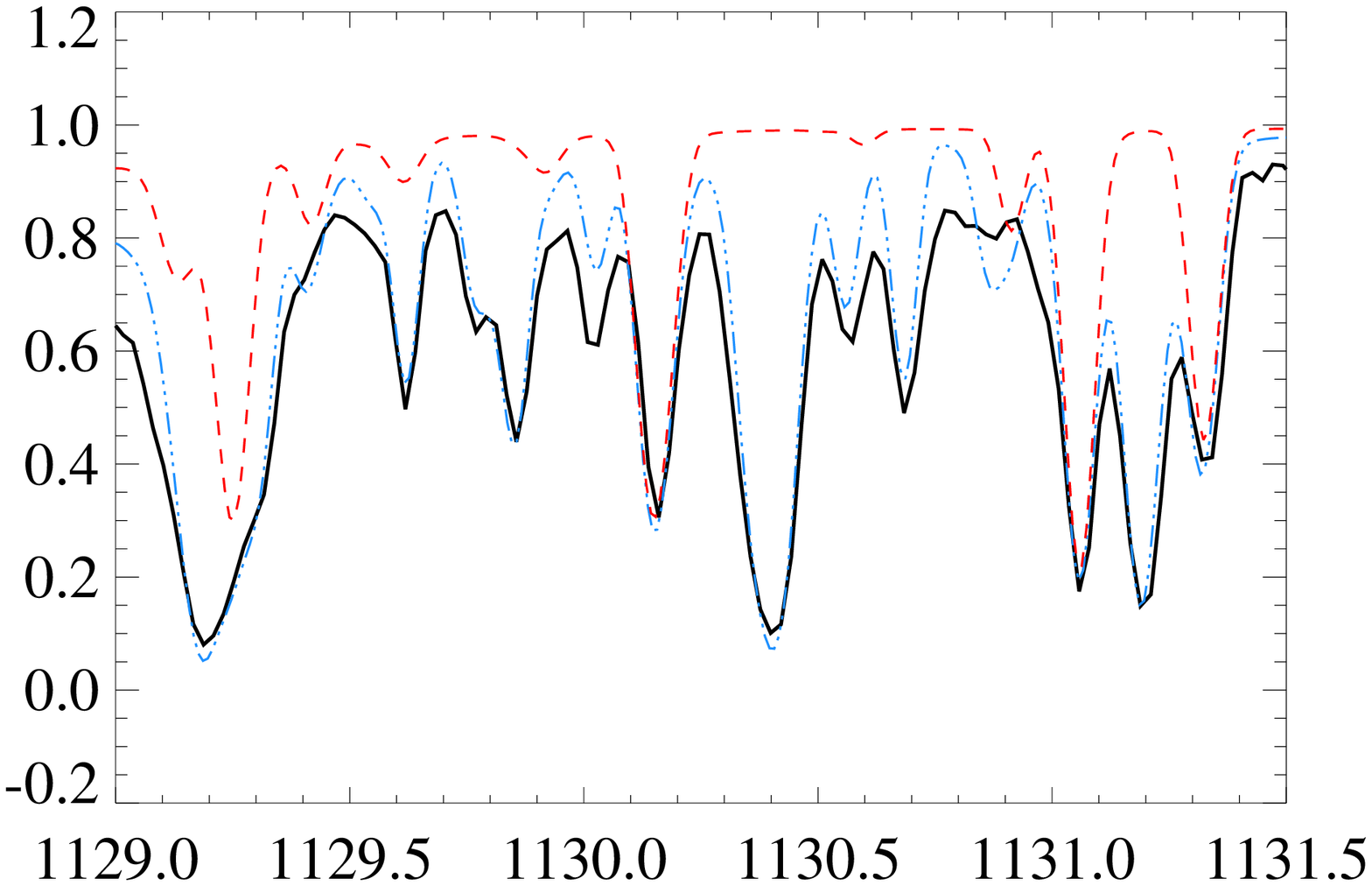}%
\includegraphics{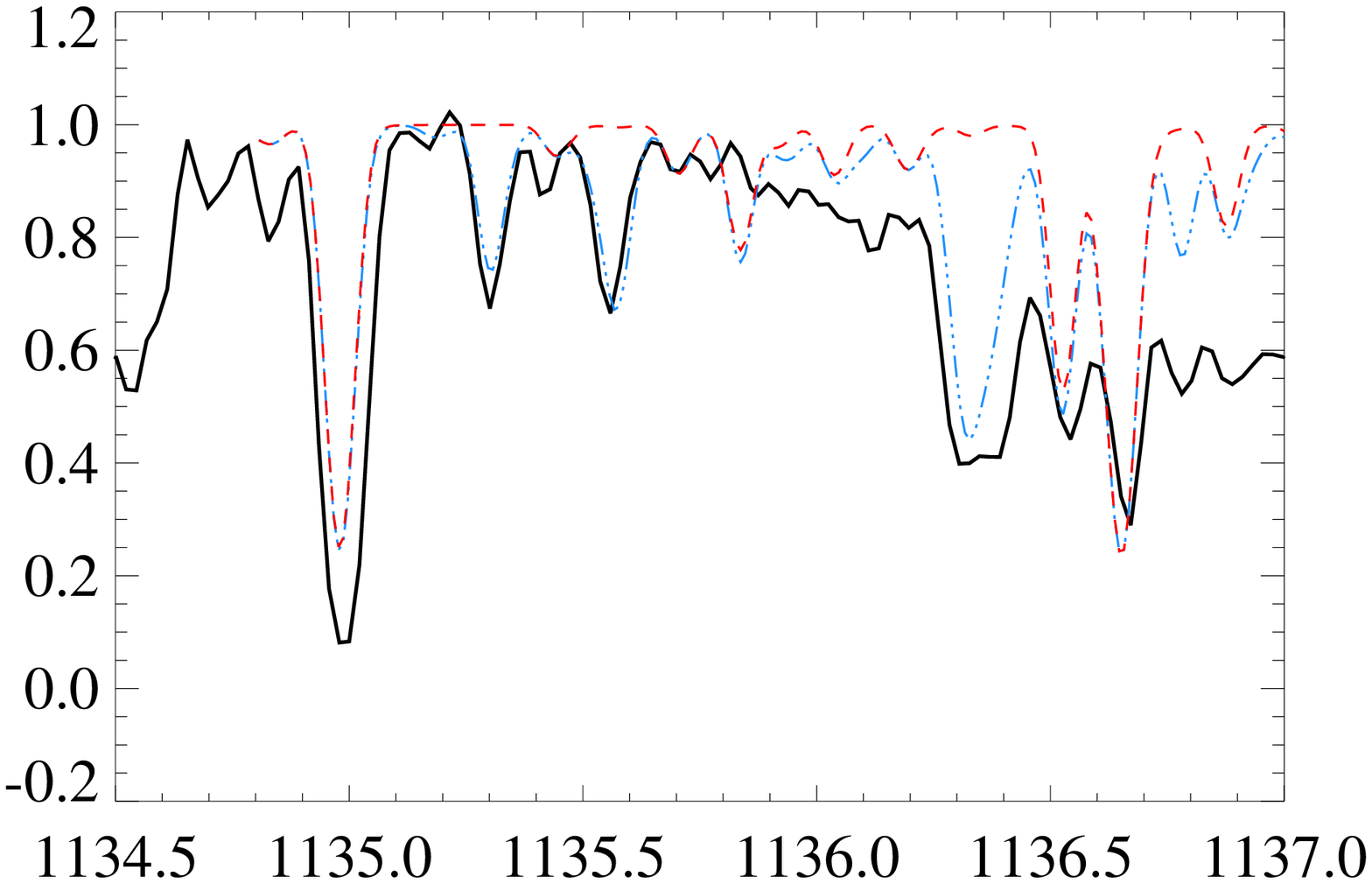}}
\resizebox{\hsize}{!}{\includegraphics{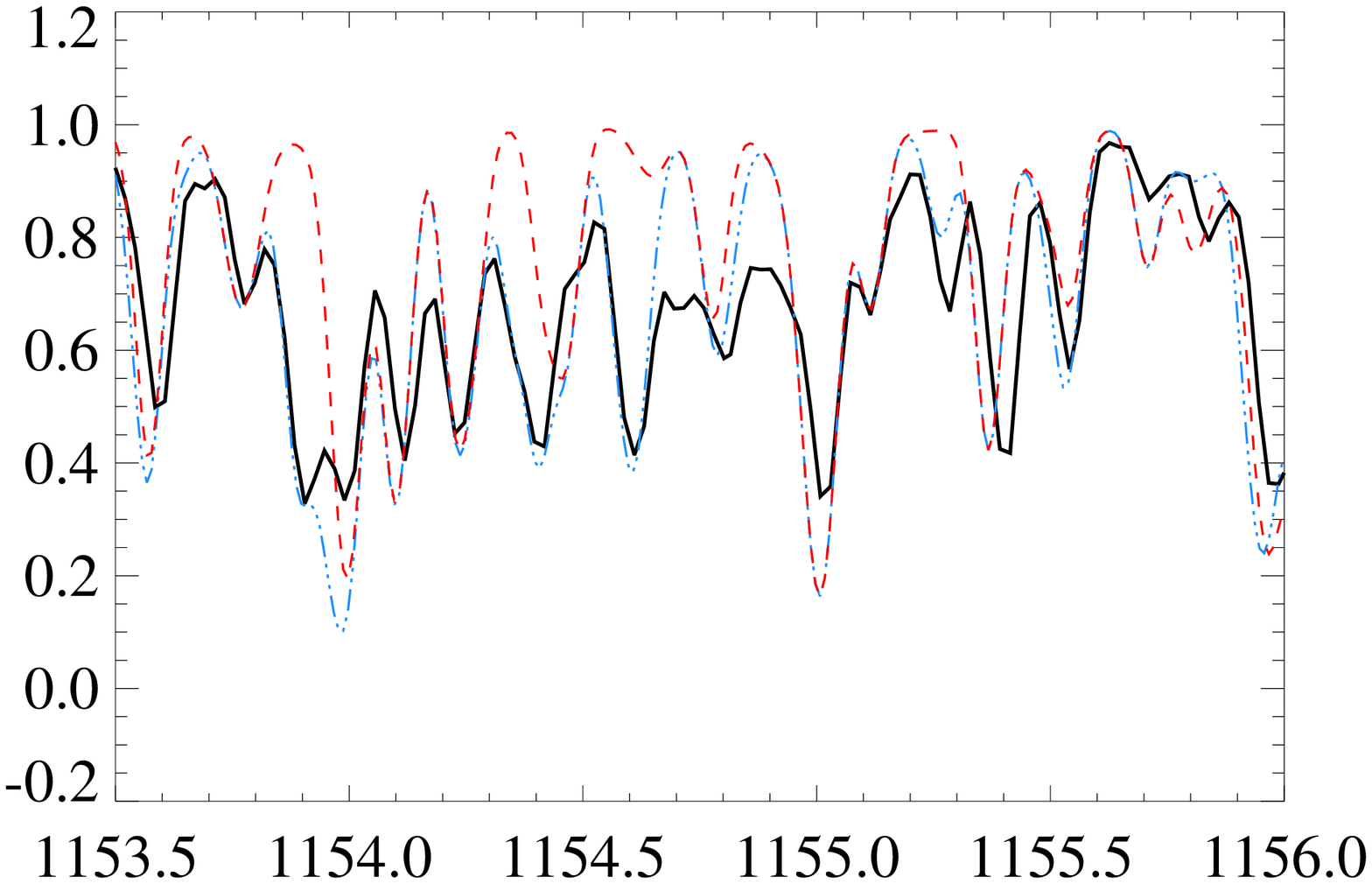}%
\includegraphics{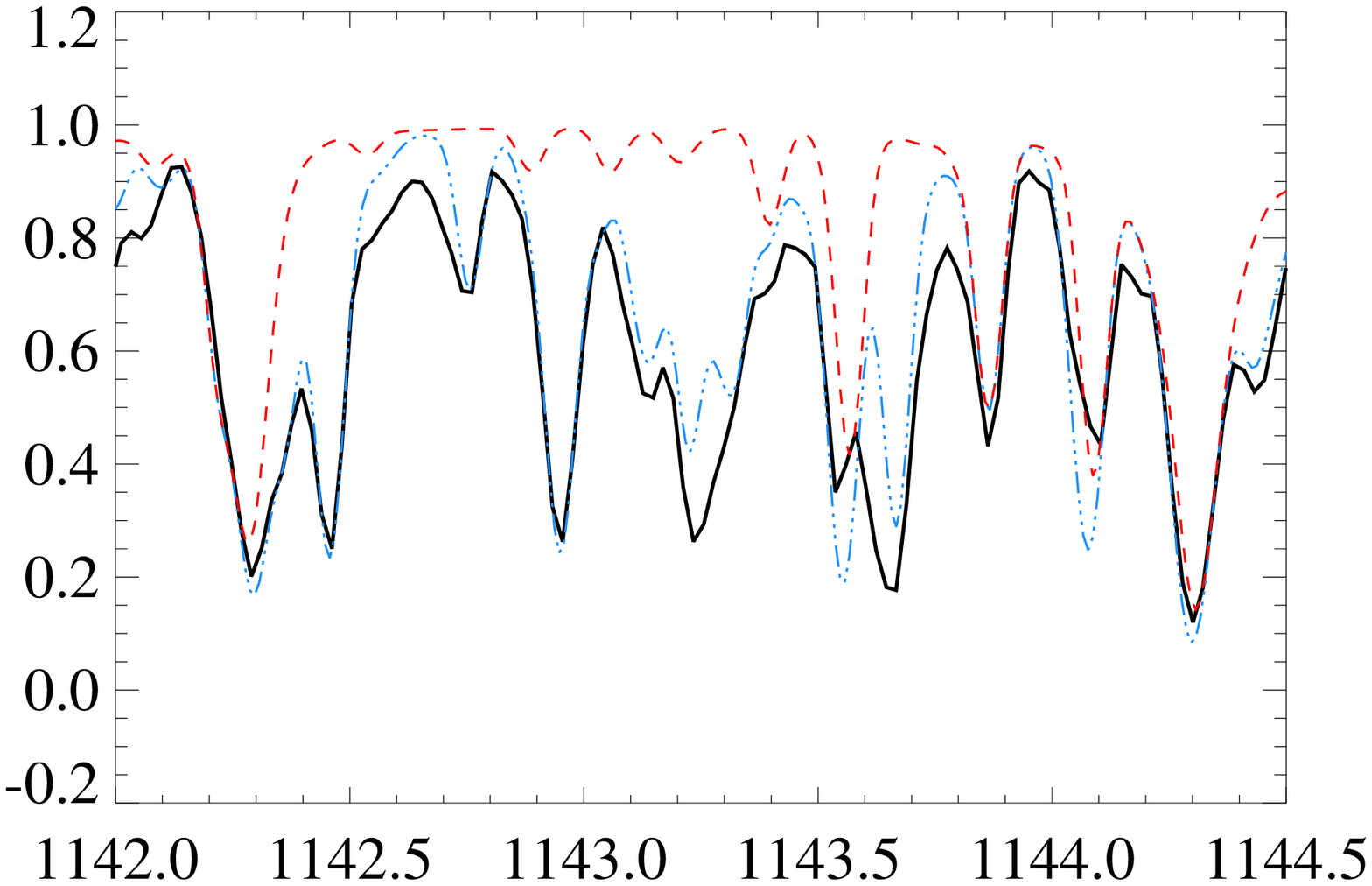}}
\caption{\label{Fig:iron_clean_line} This figure shows the observation
  (in black), the calculated model with
  $\log(n_{\rm Fe}/n_{\rm H})$=--4.9$\pm$0.4 (in blue), and without iron (in
  red). {\it Top left:} The Fe\,{\sc iii} lines at 
  1130\,\AA,\ blended with Fe\,{\sc ii} lines.
  {\it Top right:} Fe\,{\sc ii}
  doublets at 1135\,\AA.
  {\it Bottom left:} Fe\,{\sc iii} doublet at
  1154\,\AA,\ {\it Bottom right:} The strong Fe\,{\sc
    iii} line at 1142\,\AA\ blended with Fe\,{\sc ii} lines.}
\end{figure}

 We confirmed this value through modeling a cluster of iron lines in
 the 1142-43\,\AA\ window including Fe\,{\sc iii} and Fe\,{\sc ii}
 lines, Fe\,{\sc ii} lines around 1135-26\,\AA, and Fe\,{\sc iii}
 lines in the 1154\,\AA\ window. All these lines arise from very low
 lying energy levels (0-4~eV) and have $\log gf$ values around zero or
 --1. Figure~\ref{Fig:iron_clean_line} shows the observation and the
 modeled spectra.

\subsection{Cobalt, Z=27}
The selected database here contains 581 cobalt lines. In the UV
spectrum of $\iota$~Herculis, cobalt is observed over the entire
wavelength range. Almost all of the strong lines are in the form of
Co\,{\sc iii}, with a few Co\,{\sc ii} lines and only one Co\,{\sc iv}
line. The Saha equation predictions shown in Table
3 are consistent with the observed ionization
ratios.

\begin{figure}
\resizebox{\hsize}{!}{\includegraphics{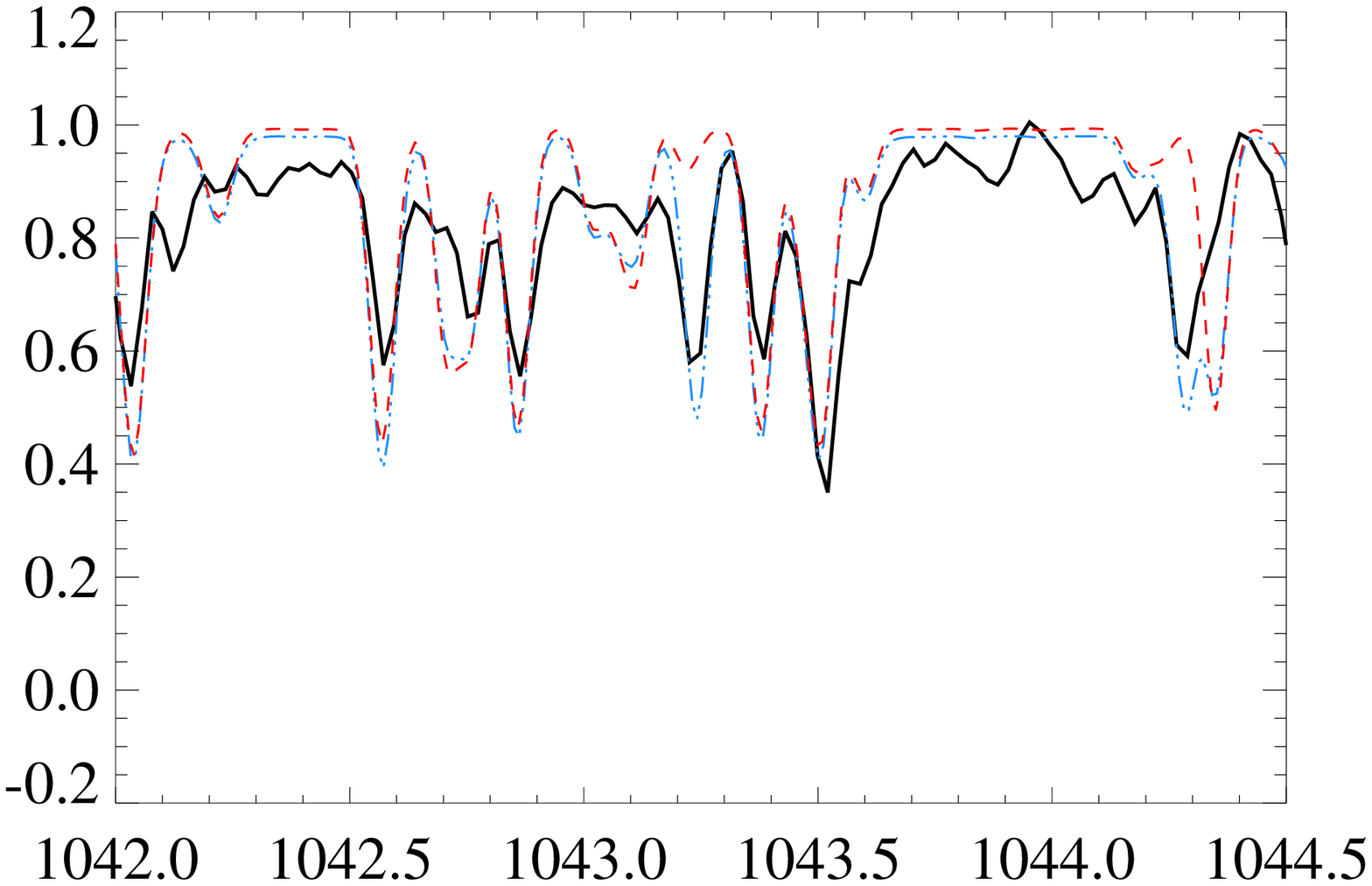}%
\includegraphics{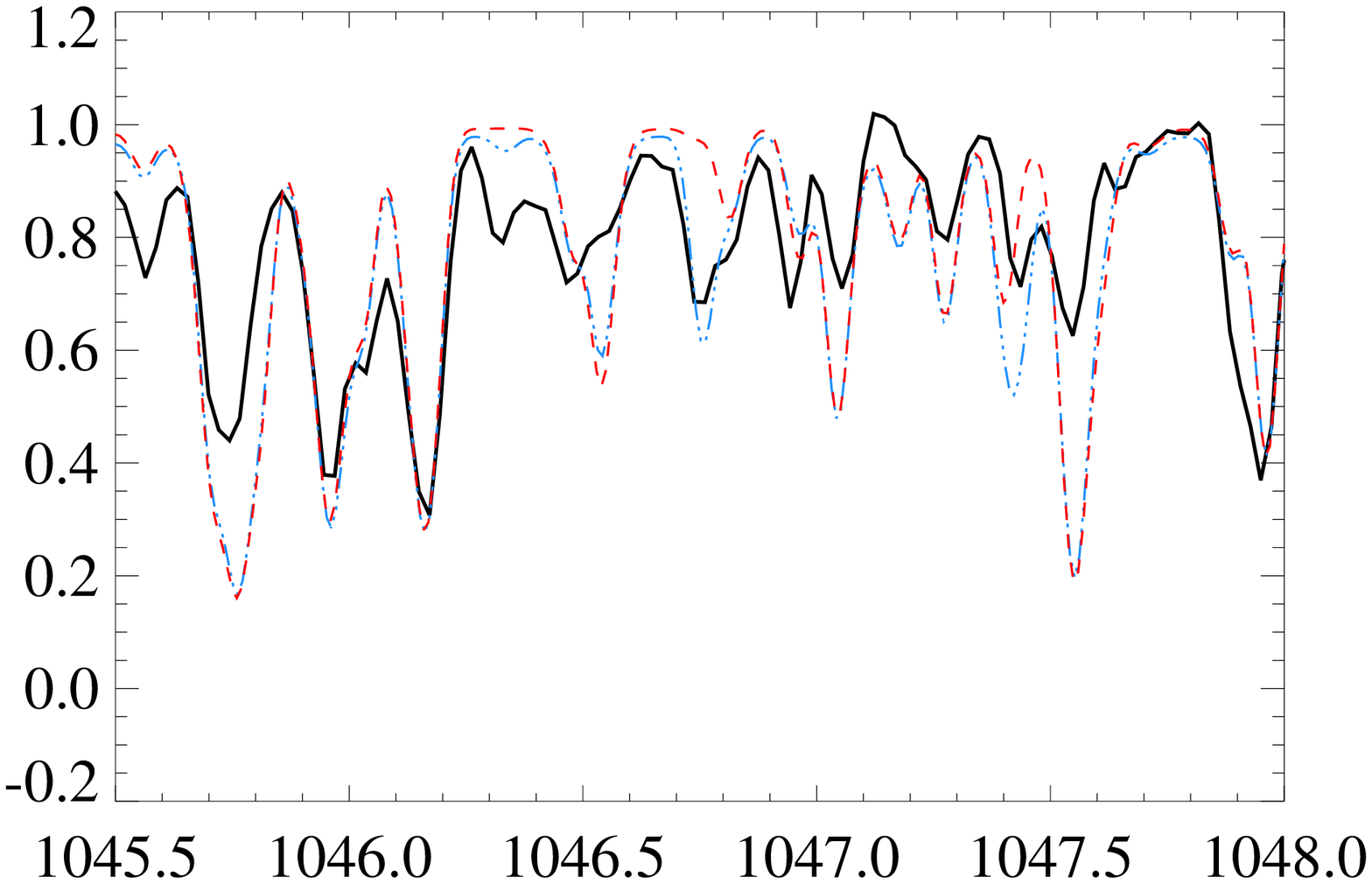}} \\
\resizebox{\hsize}{!}{\includegraphics{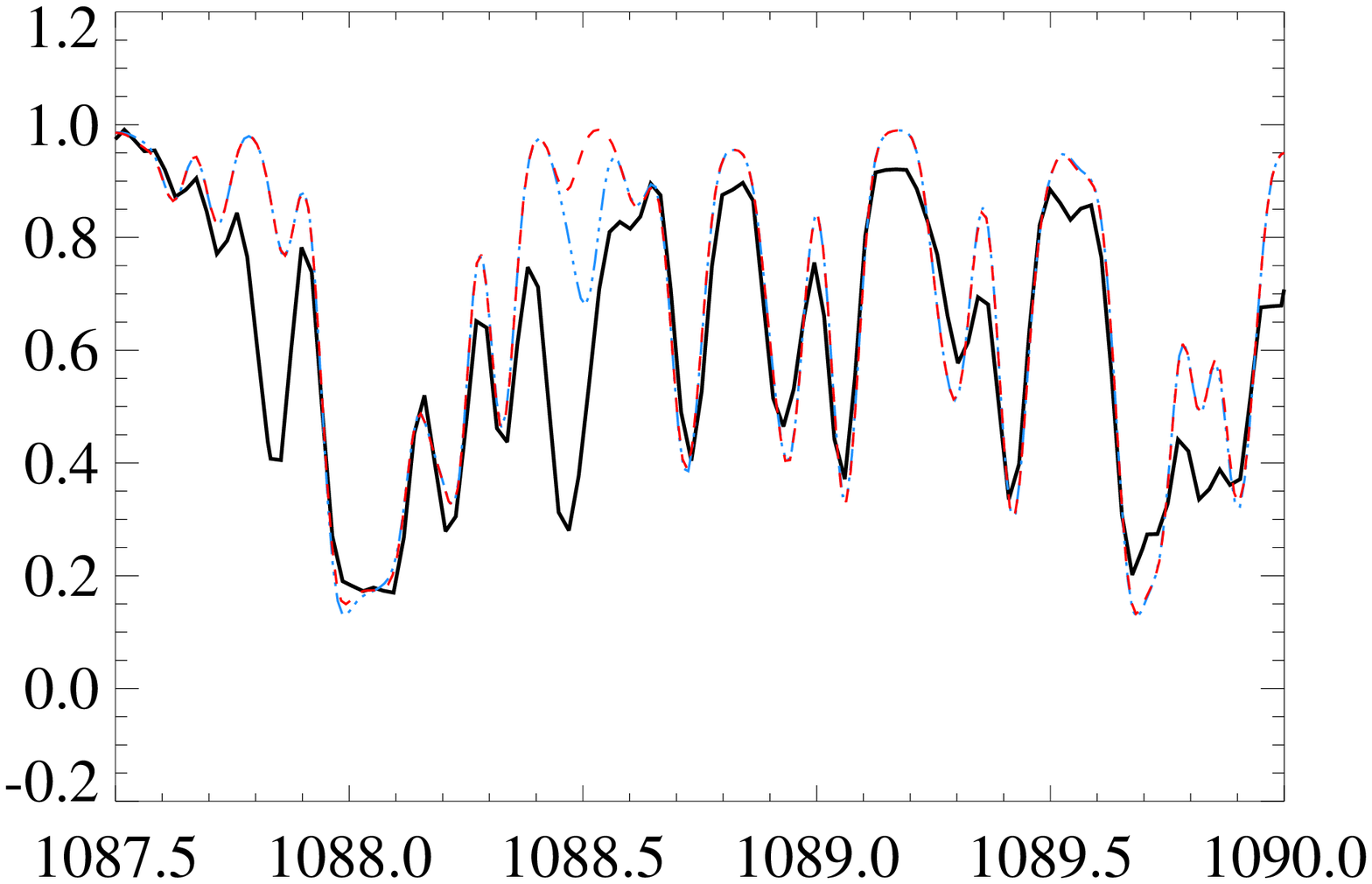}%
\includegraphics{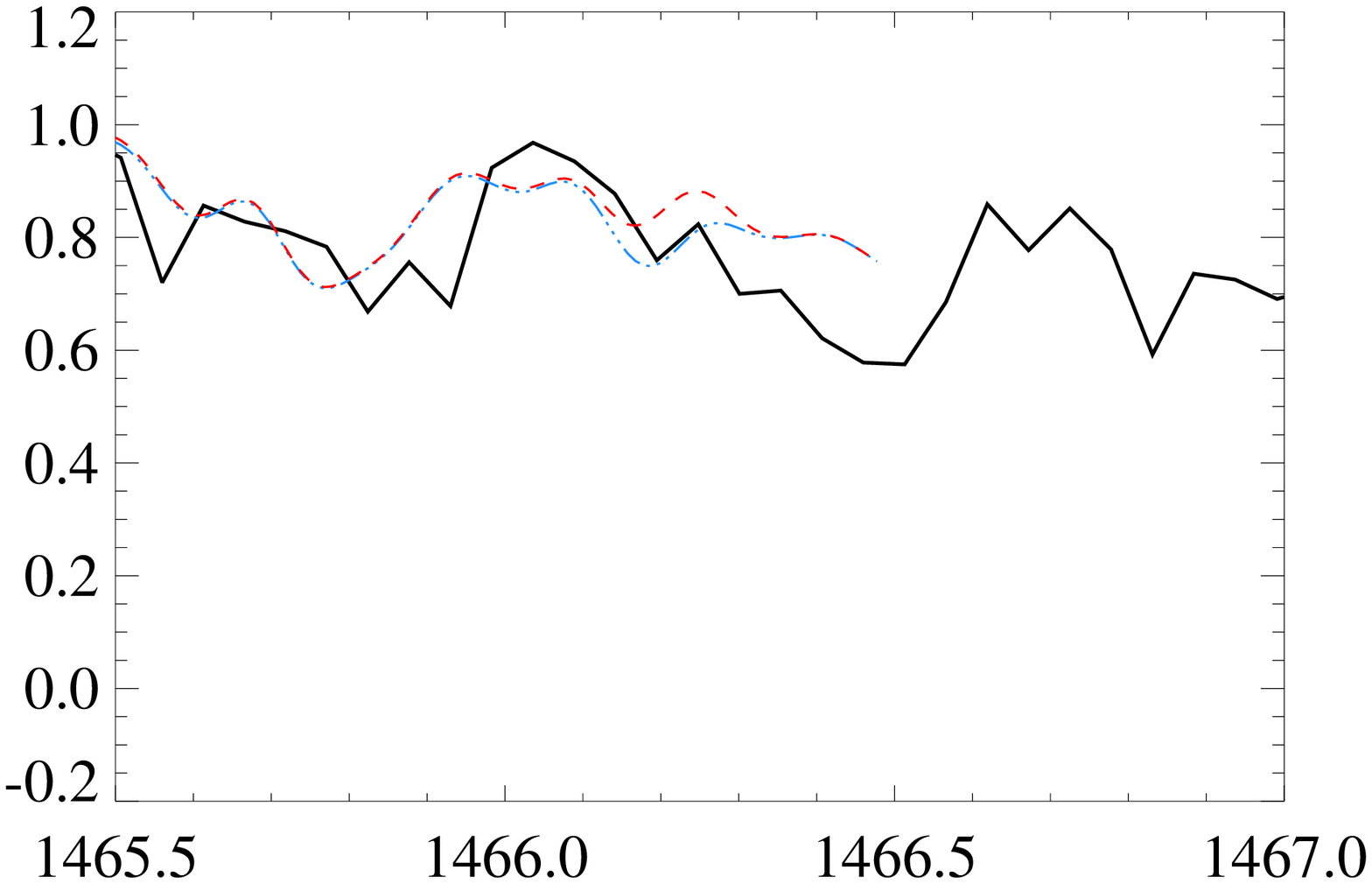}}
\caption{\label{Fig:co_clean_line} This figure shows the observation
  (in black), the calculated model with $\log(n_{\rm Co}/n_{\rm H})= $
  --7.55$\pm$0.49 (in blue), and without cobalt (in red). {\it Top
    left:} Co\,{\sc iii} line at 1043\,\AA,\ {\it Top right:}
  Co\,{\sc iii} line at 1046\,\AA,\ {\it Bottom left:} Co\,{\sc
    iii} line at 1088\,\AA,\ {\it Bottom right:} Co\,{\sc ii} line
  at 1466\,\AA.}
\end{figure}

We modeled a group of Co\,{\sc iii} in the window between 1043-88 and
found $\log(n_{\rm Co}/n_{\rm H})=$ --7.55$\pm$0.49. These lines all
arise from low energy states at 1.9~eV with well determined oscillator
strength values around --1 (VALD3). We tested this value by also
modeling the resonance Co\,{\sc ii} line at 1466\AA. Figure
\ref{Fig:co_clean_line} shows the best-fitting model and the
observation.
\subsection{Nickel, Z=28}
\label{nickel}
The line list selected for this work contains 2145 nickel lines. More
than half of the lines in the list are in the form of Ni\,{\sc iii}. A
smaller fraction is in the form of Ni\,{\sc ii}, with only a few lines
of Ni\,{\sc iv}. In the UV spectrum of $\iota$~Herculis nickel
  is observed mostly as Ni\,{\sc iii} and Ni\,{\sc ii} with very
  little contribution from Ni\,{\sc iv}. The Ni\,{\sc iv} lines hardly
  appear in the spectrum because the excited states are at least 14~eV
  above the ground state (VALD3). This distribution is consistent with
  the ratios estimated from the Saha equation (see
  Table~3).

We determined the abundance of this element using the Ni\,{\sc iii}
triplet  lines in the 1321-22\,\AA\ window and found
$\log(n_{\rm Ni}/n_{\rm H})=$--5.70$\pm$0.35. These lines arise from $\sim$6~eV
and have $\log gf$ around -2, but they are suitable for this purpose
because they are clean and unblended.

\begin{figure}
\resizebox{\hsize}{!}{\includegraphics{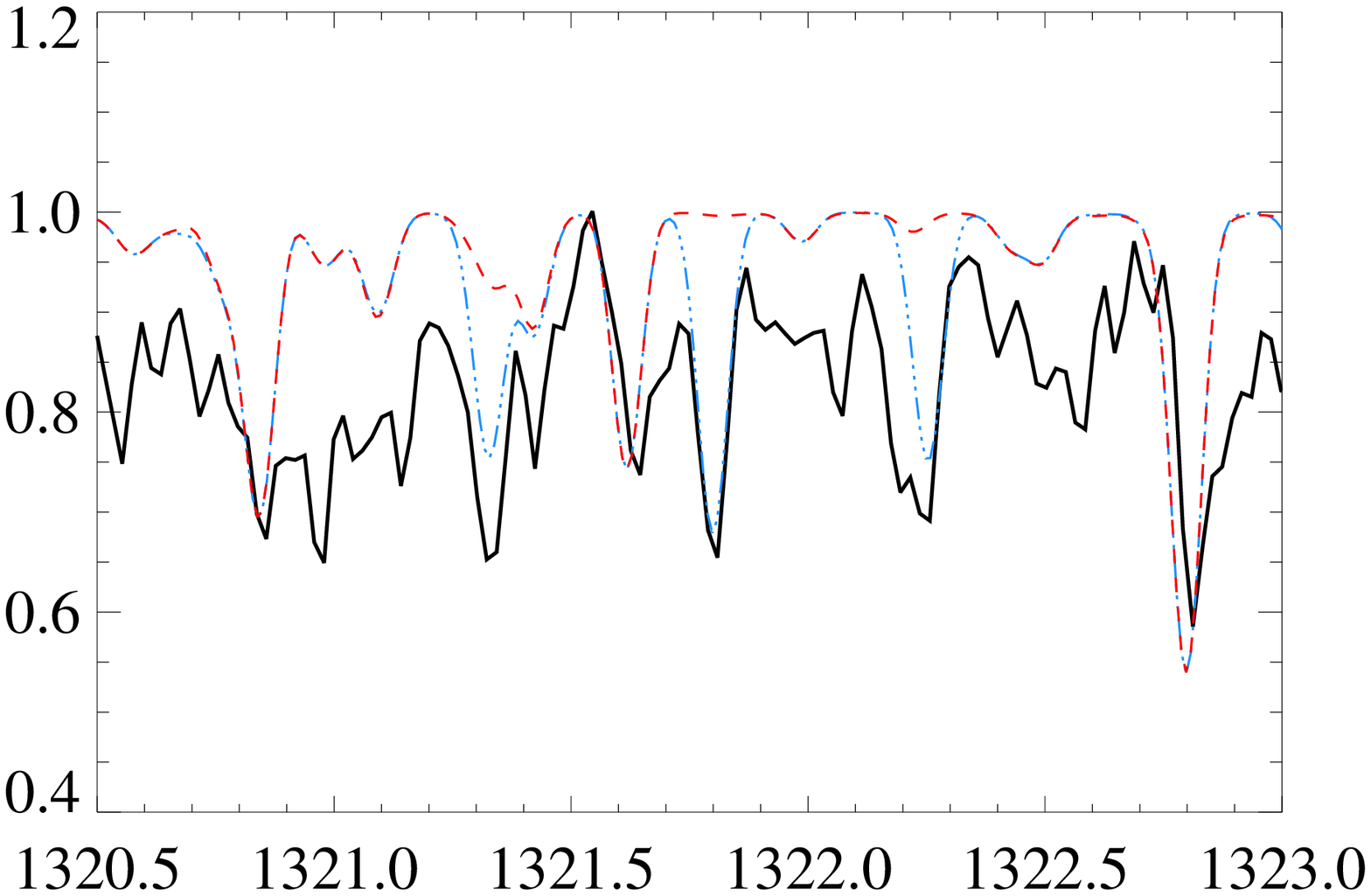}%
\includegraphics{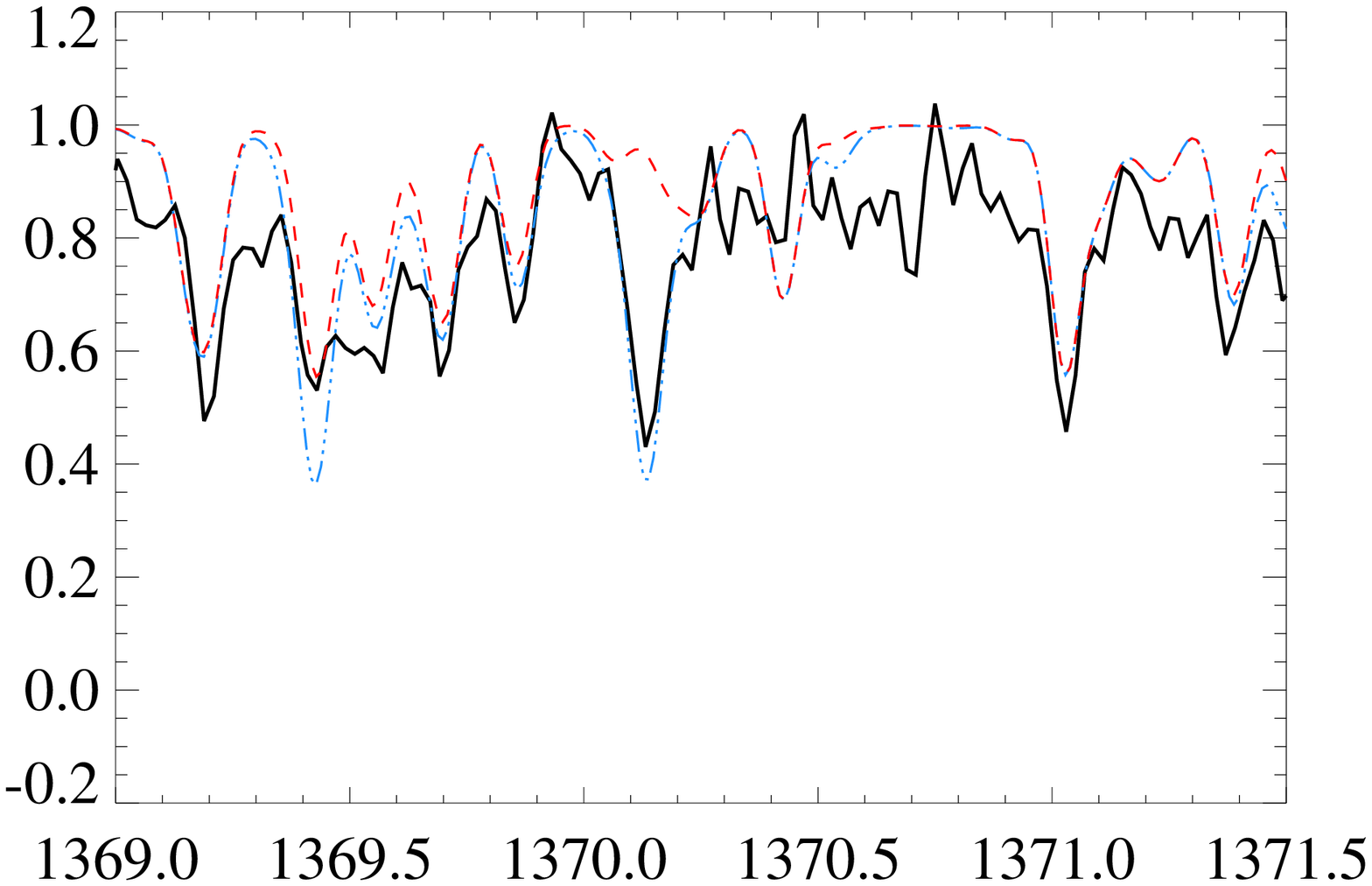}}\\
\resizebox{\hsize}{!}{\includegraphics{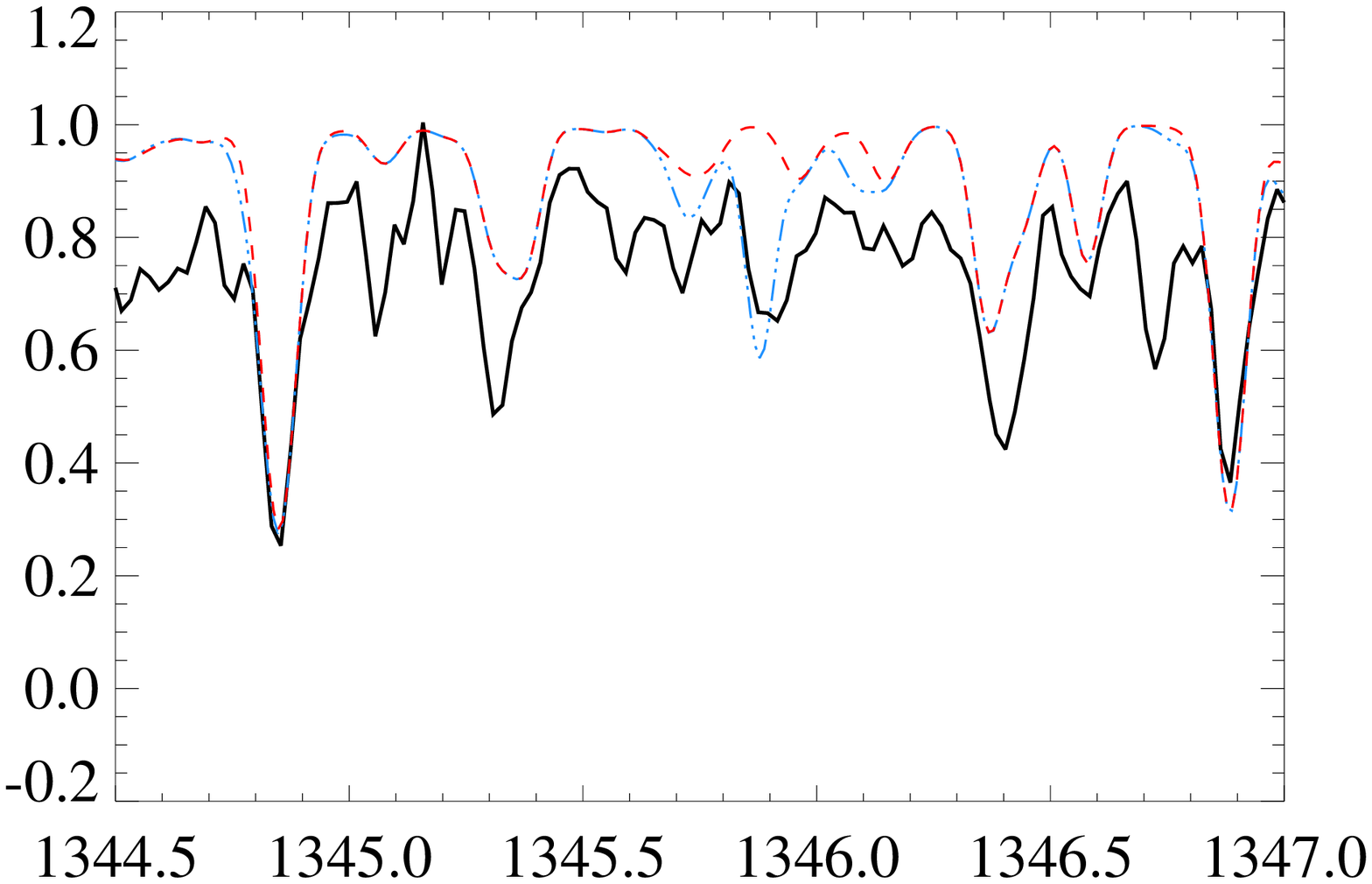}%
\includegraphics{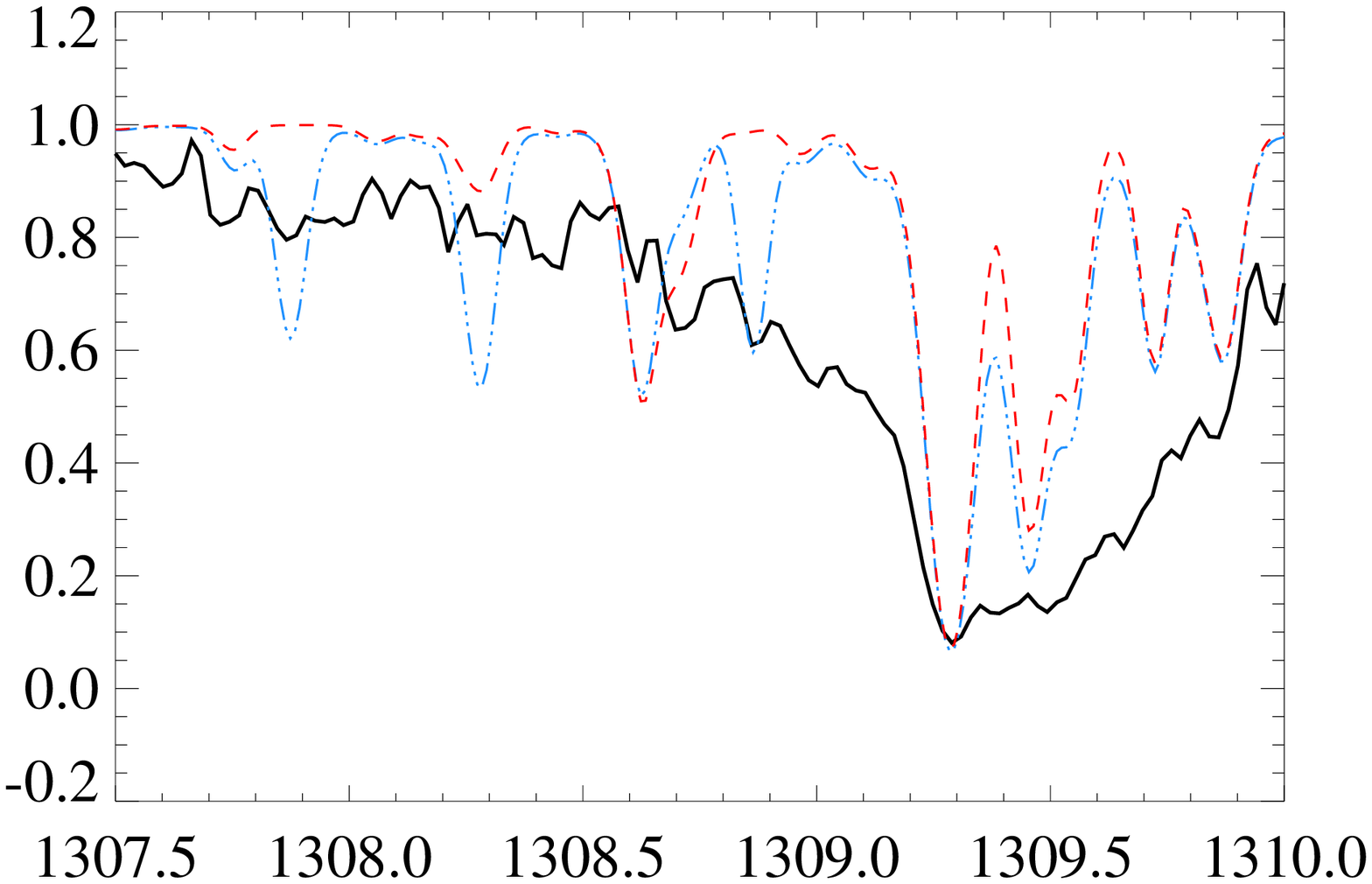}}
\caption{\label{Fig:nickel_clean_line} This figure shows the
  observation (in black), the calculated model with $\log(n_{\rm Ni}/n_{\rm H}) =
  $--5.70$\pm$0.35 (in blue), and without nickel (in red). {\it Top
    left:} Ni\,{\sc iii} triplet lines at 
  1321--22\,\AA,\ {\it Top right:} a resonance Ni\,{\sc ii} lines at
  1370\,\AA,\ {\it Bottom left:} a resonance Ni\,{\sc ii} line at
  1345\,\AA,\ {\it Bottom right:} a resonance Ni\,{\sc ii} line at
  1308\,\AA.}
\end{figure}

We confirmed this value by modeling three resonance Ni\,{\sc ii} lines
in three wavelength windows including 1308, 1345, 1370\,\AA. Figure
\ref{Fig:nickel_clean_line} shows the resulting model using this
abundance and the fit to the observation.

\subsection{Copper, Z=29}
In the subset of VALD database used here there are 62 Cu\,{\sc ii}
lines. In the UV spectrum of $\iota$~Herculis, copper is observed in
the form of Cu\,{\sc ii} throughout the entire wavelength range
studied here. The results of the Saha equation indicate that copper
must be dominated by Cu\,{\sc iii} at this temperature which is
inconsistent with our line list. The reason seems to be a lack of
atomic data about Cu~{\sc iii}.

\begin{figure}
\resizebox{\hsize}{!}{\includegraphics{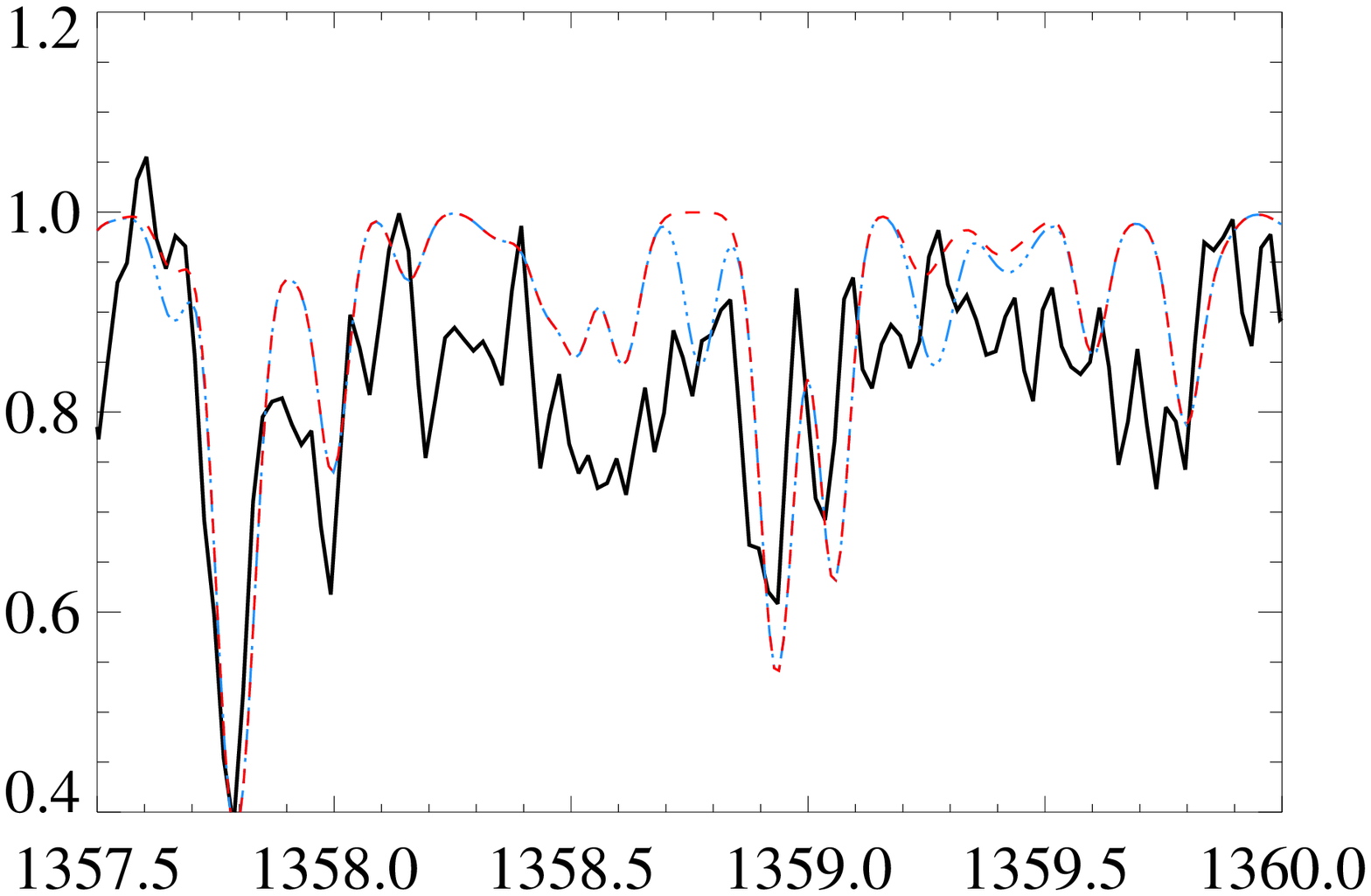}}\\
\caption{\label{Fig:cu_line} This figure shows the observation (in
  black), the calculated model with $\log(n_{\rm Cu}/n_{\rm H}) = $--9.00 (in
  blue), and without copper (in red). This is a resonance Cu\,{\sc ii}
  line at 1358\,\AA.}
\end{figure}

We modeled the strong resonance line of Cu\,{\sc ii} at
1358\AA\ (Ross, 1969) with $\log gf$ of almost 0, and found an
abundance of $\log(n_{\rm Cu}/n_{\rm H})=$--9.0. This line is the only
unblended line that we have available in our spectrum. We present this
as an upper limit for this element due to lack of other unblended
lines for comparison. The uncertainty could not be estimated on this
value since there are not enough clean lines for comparison. The least
uncertainty on this value is due to stellar parameter inaccuracy which
is of the order of 0.04~dex (see \S3.3 ). Figure~\ref{Fig:cu_line}
shows the observation and the model.

\subsection{Zinc, Z=30}
Zinc has 127 weak lines in the form of Zn\,{\sc iii} in our selected
VALD line list. In the spectrum of $\iota$~Herculis, zinc is mostly
observed in the longer wavelengths and it appears only in the form of
Zn\,{\sc iii}. This ionization distribution is consistent with the
ratios estimated from the Saha equation (see Table~3).

\begin{figure}
\resizebox{\hsize}{!}{\includegraphics{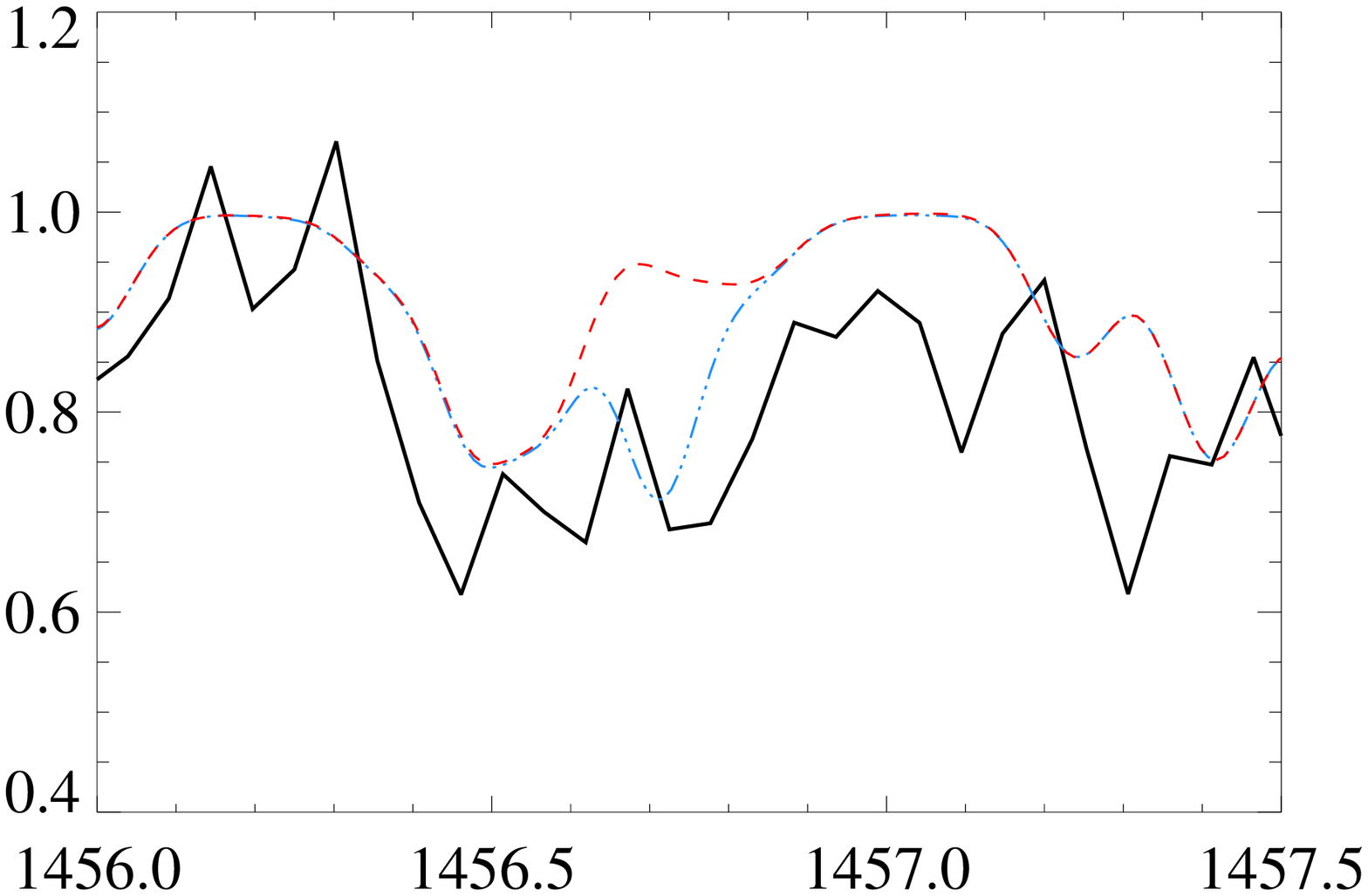}%
\includegraphics{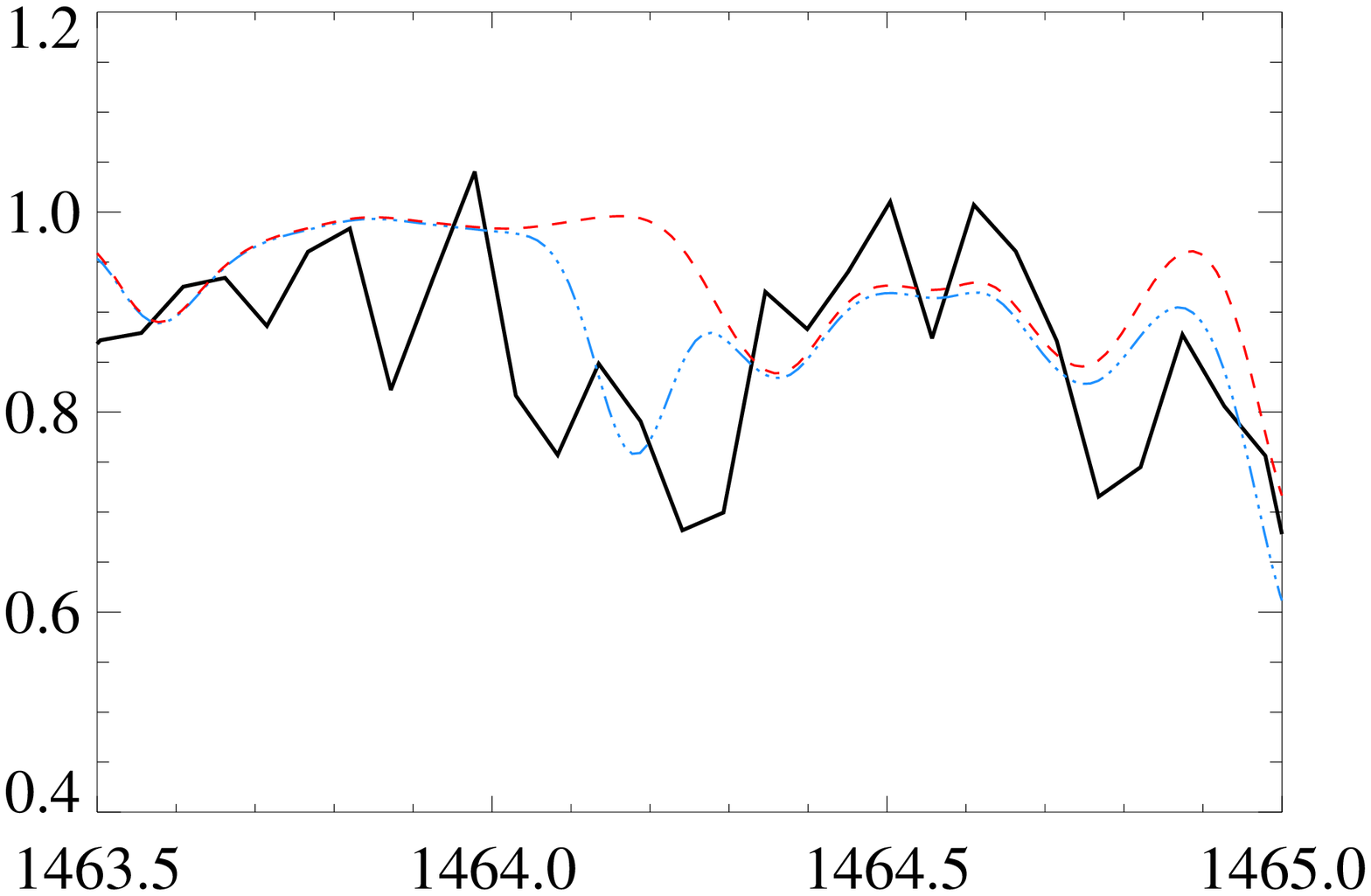}} \\
\resizebox{\hsize}{!}{\includegraphics{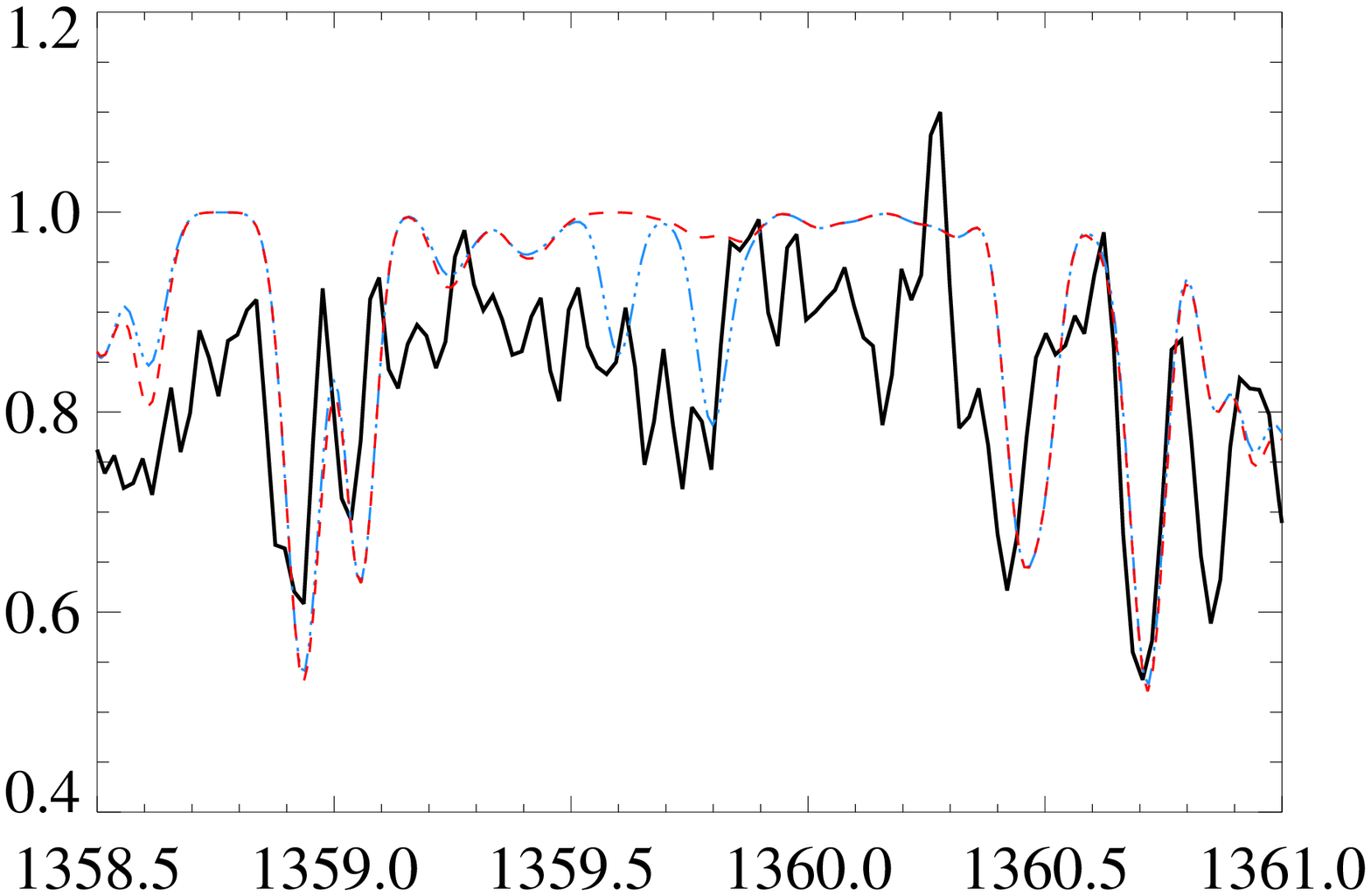}}
\caption{\label{Fig:zn_line} This figure shows the observation
  (in black), the calculated model with $\log(n_{\rm Zn}/n_{\rm H}) = $
  --6.85$\pm$0.20 (in blue), and without zinc (in red). {\it Upper:}
  Zn\,{\sc iii} lines at 1456 and 1464\,\AA. {\it Lower:}
  Zn\,{\sc iii} lines at 1359\,\AA.}
\end{figure}

The lines are weak and heavily blended and arise from at least 9~eV
above the ground state. The lowest excited energy state belongs to a
Zn\,{\sc iii} doublet line in the 1456-64\AA\ window. These lines are
weak but fairly unblended. We have used them for abundance
determination and found $\log(n_{\rm Zn}/n_{\rm H})=$--6.85$\pm$0.20.  The
uncertainty arises from the fact that each of these lines can best be
modeled with a slightly different abundance (see Figure
\ref{Fig:zn_line}). The Zn\,{\sc iii} line  in the 1359\AA\ window
can also be used to confirm this value. The observation and models are
shown in Figure~\ref{Fig:zn_line}.

\subsection{Germanium, Z=32}
\begin{figure}
\resizebox{\hsize}{!}{\includegraphics{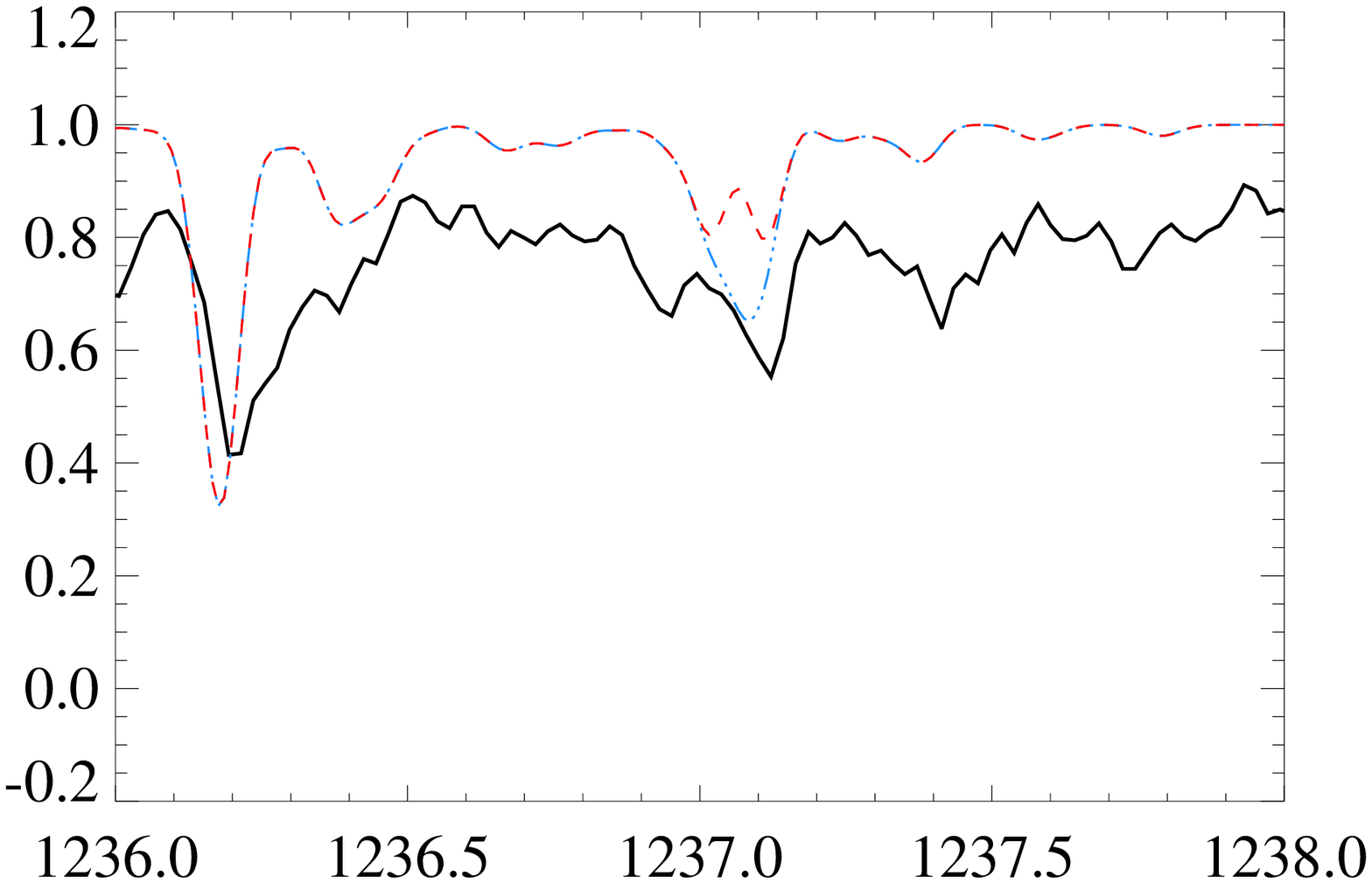}%
\includegraphics{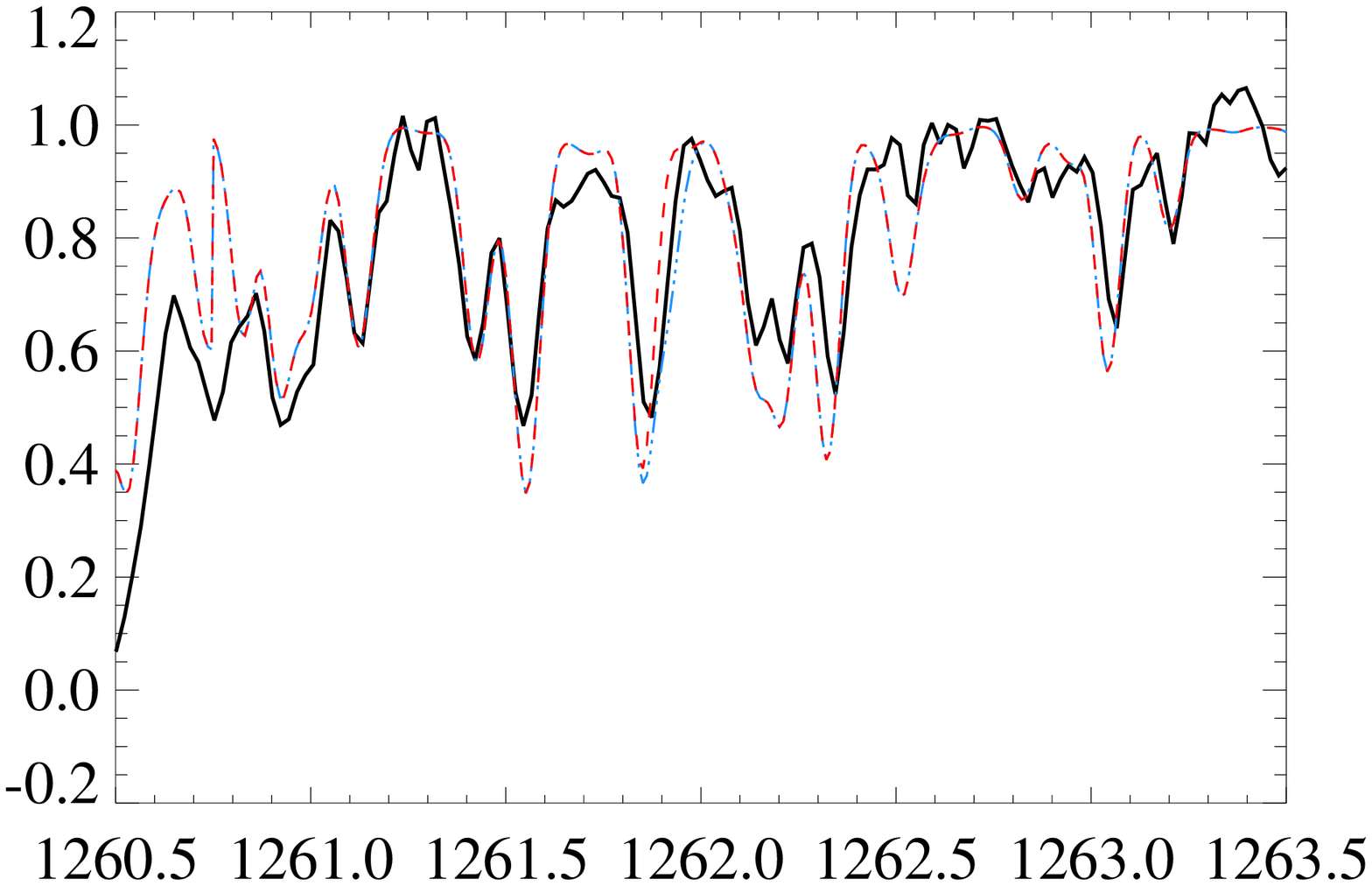}}
\caption{\label{Fig:Ge_line} This figure shows the observation (in
  black), the calculated model with $\log(n_{\rm Ge}/n_{\rm
  H})=$--8.50 (in blue), and without germanium (in red). {\it left:} A
  strong Ge\,{\sc ii} line at 1237\AA,\ {\it right:} A part of a
  Ge\,{\sc ii} doublet at 1261\AA.}
\end{figure}

In the VALD line list selected here, there are 8 Ge\,{\sc ii}
lines.  All of these lines arise from very low lying energy states
(0--0.2~eV) with $\log gf$ values between 0 to -1. However, most of
these lines are heavily blended which greatly complicates the task of
abundance determination. We have used a relatively strong resonance
Ge\,{\sc ii} line at 1237\AA\ to find $\log(n_{\rm Ge}/n_{\rm H})=$--8.50. We do
no have any other clean lines to confirm this value but a Ge\,{\sc ii}
doublet in 1261--64\AA\ region shows that this value can provide an
upper limit. Figure~\ref{Fig:Ge_line} shows the model and observation.

\subsection{Platinum, Z=78}
The selection of VALD list used here contains 148 platinum lines. All
of them are in the form of Pt\,{\sc iii}. There is only one Pt\,{\sc
  ii} line in the selected list. The dominant state of
  ionization as estimated by the Saha equation (see
  Table~3), is consistent with what is observed in
  the spectrum of $\iota$~Herculis.

\begin{figure}
\resizebox{\hsize}{!}{\includegraphics{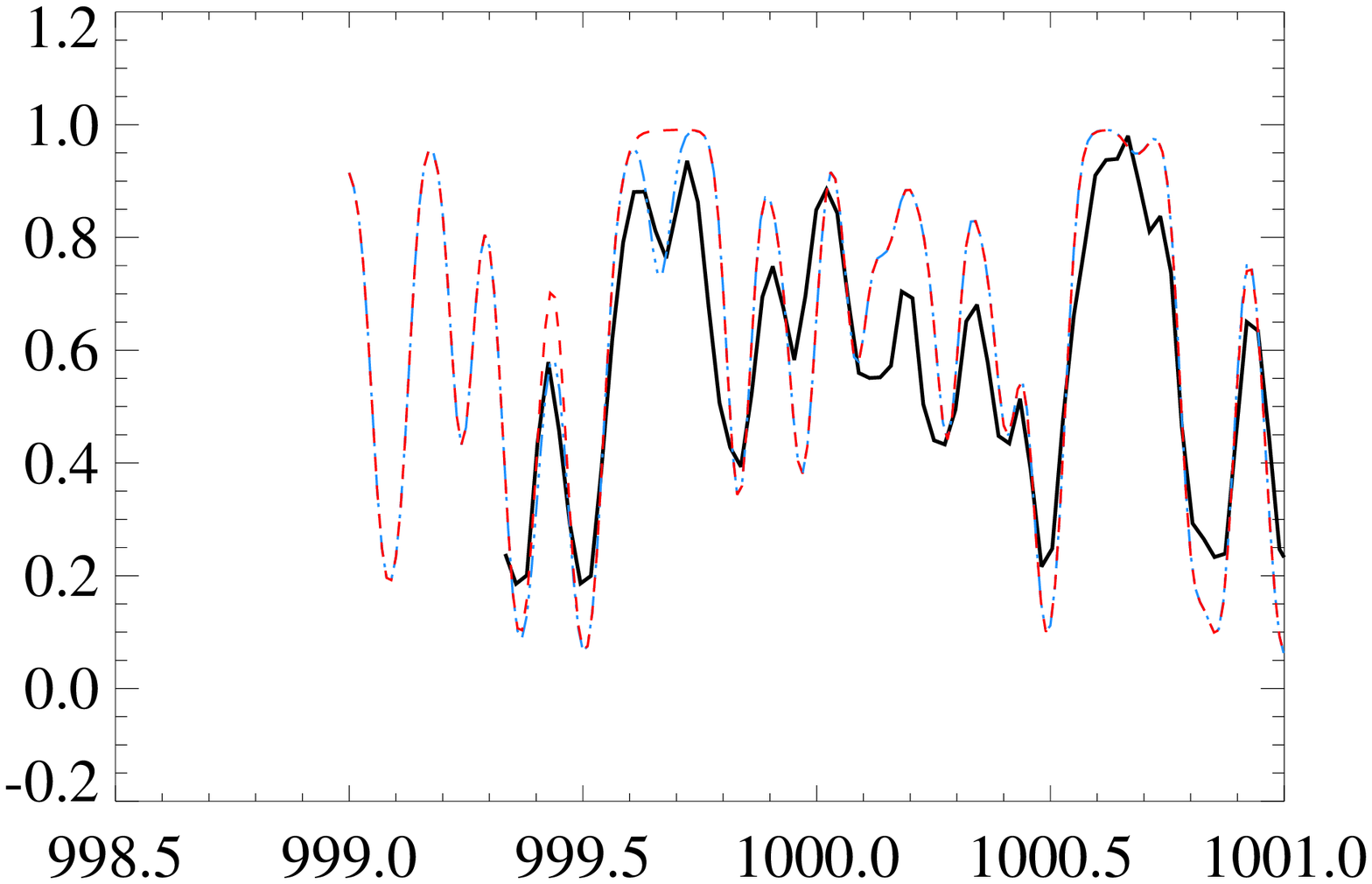}}
\caption{\label{Fig:pt_clean_line} This figure shows the observation
  (in black), the calculated model with $\log(n_{\rm Pt}/n_{\rm H}) = $ --8.825
  (in blue), and without platinum (in red). This is the Pt\,{\sc iii}
  line at 999\,\AA.}
\end{figure}

There are only a few weak lines throughout the spectrum with a minor
concentration at shorter wavelengths. We have used a fairly
clean Pt\,{\sc iii} line around 999\AA.\ Our calculation yields
$\log(n_{\rm Pt}/_{\rm H})=$--8.825. Results are shown in Figure
\ref{Fig:pt_clean_line}. This value was not confirmed with any other
line since there are no other clean and unblended lines available
throughout our data. Therefore, we present this value as an upper
limit for this element for which we can not determine an uncertainty.

\subsection{Mercury, Z=80}
The subset of the VALD database selected here contains 13 mercury
lines. In the observed spectrum of $\iota$~Herculis, the majority of
mercury lines are observed as Hg\,{\sc iii} with a minor
contribution from Hg\,{\sc ii}. In our wavelength range, apart from 
two resonance lines, the rest of Hg\,{\sc ii} lines arise from higher
energy states of around $\sim$7~eV. At this temperature,
the Saha equation also predicts the same ionization distribution. 

\begin{figure}
\resizebox{\hsize}{!}{\includegraphics{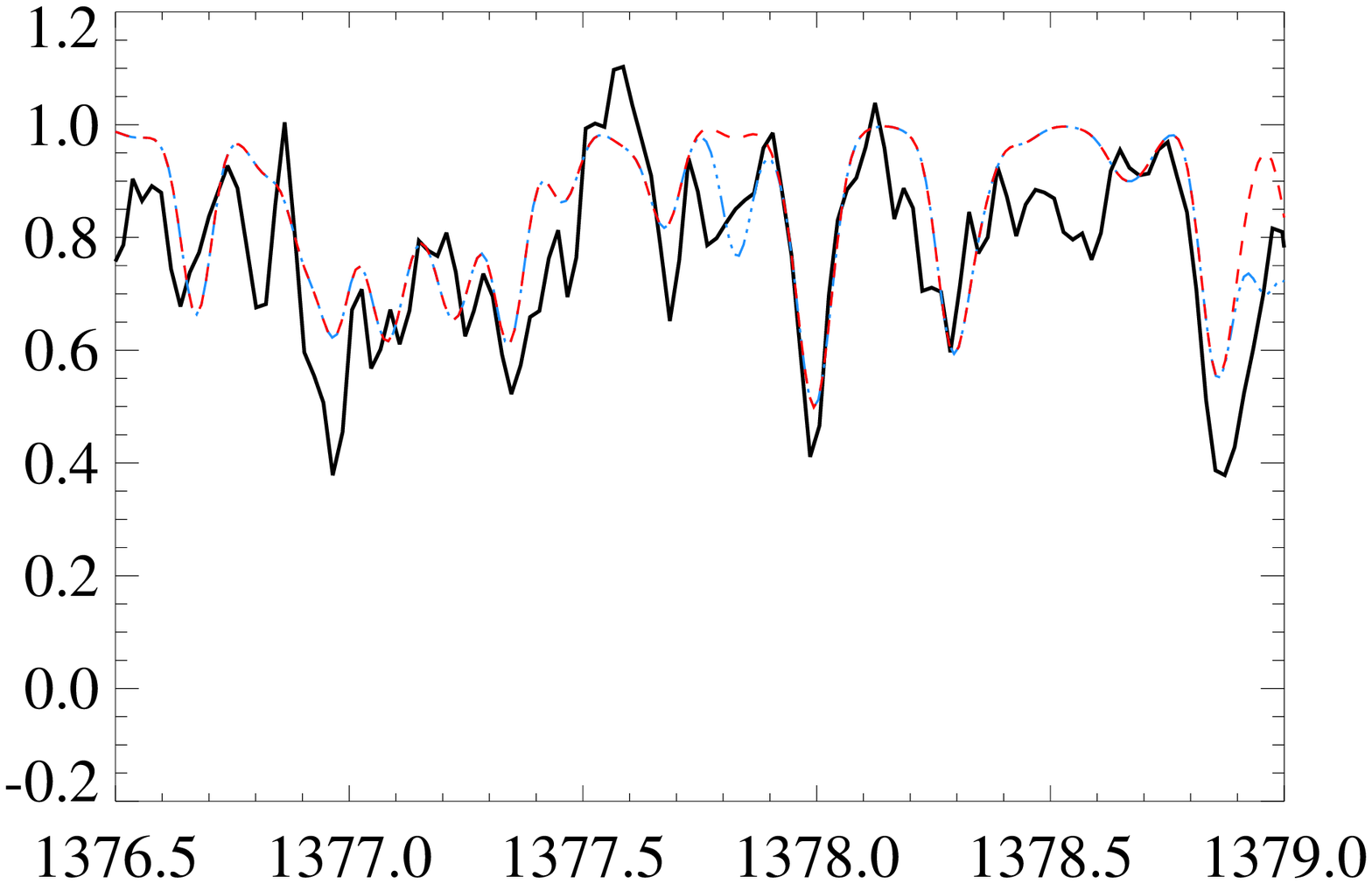}%
  \includegraphics{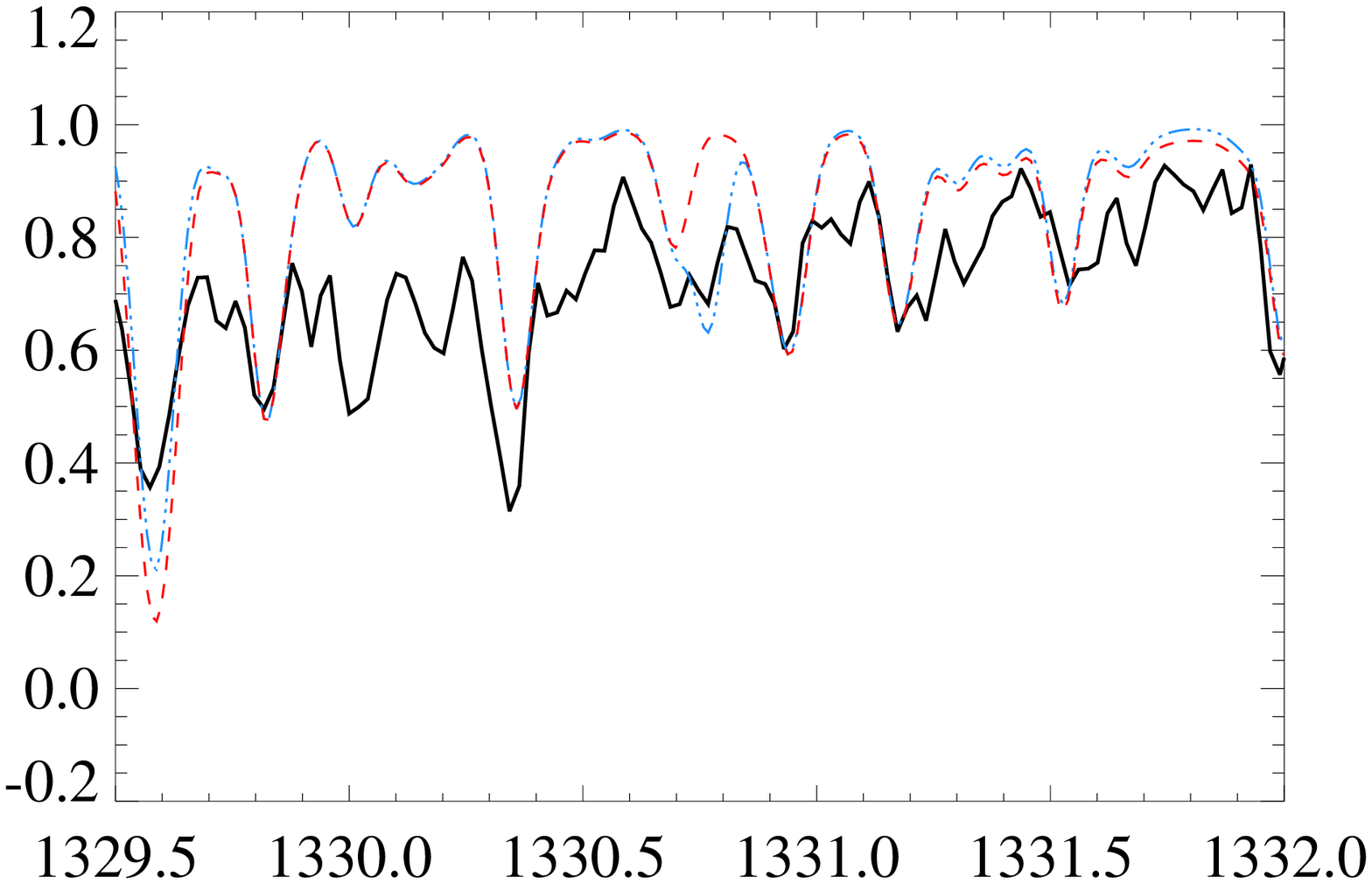}}
\caption{\label{Fig:hg_clean_line} This figure shows the observation
  (in black), the calculated model with $\log(n_{\rm Hg}/n_{\rm H}) = $
  --8.95$\pm$0.13 (in blue), and without mercury (in red). {\it Left:
  } Hg\,{\sc iii} at 1377\,\AA,\ {\it Right: } Hg\,{\sc iii} at
  1330\,\AA}
\end{figure}

There are only very few mercury lines suitable for our purpose and
they are mostly blended. We used the Hg\,{\sc iii} line at
1377\,\AA\ and 1330\,\AA\ and we find $\log(n_{\rm Hg}/n_{\rm
  H})=$--8.95$\pm$0.13. Figure~\ref{Fig:hg_clean_line} shows the
resulting model and the observation.

\begin{onecolumn}
\section{Appendix B}

In this appendix, we present the results of the spectrum
synthesis. The spectrum in black is the observation and the model
spectrum calculated using the final abudances given in the text, is
shown in blue.
\label{Appendix_B}

\begin{figure*}
\resizebox{\hsize}{!}{\includegraphics{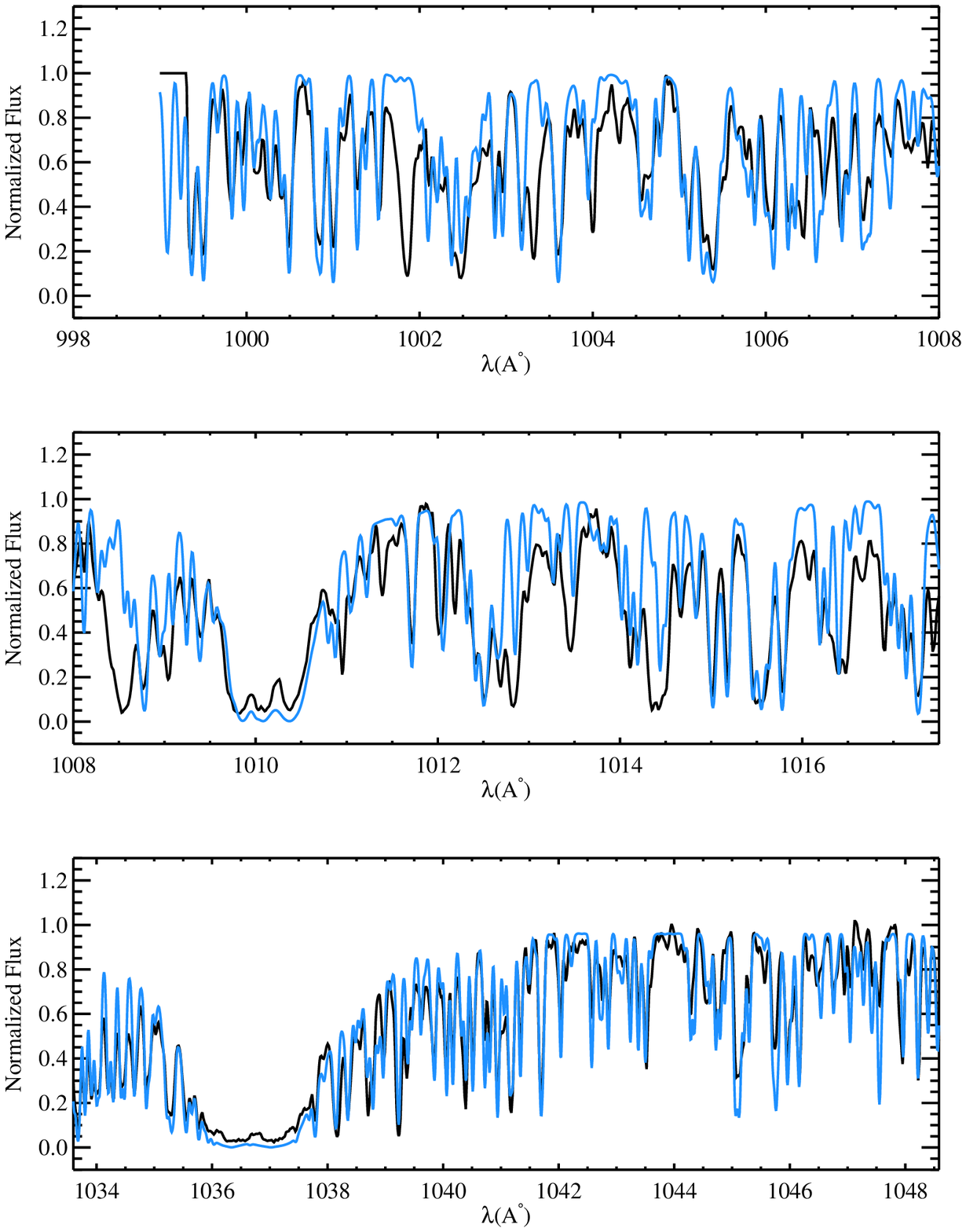}}
\end{figure*}
\begin{figure*}
\resizebox{\hsize}{!}{\includegraphics{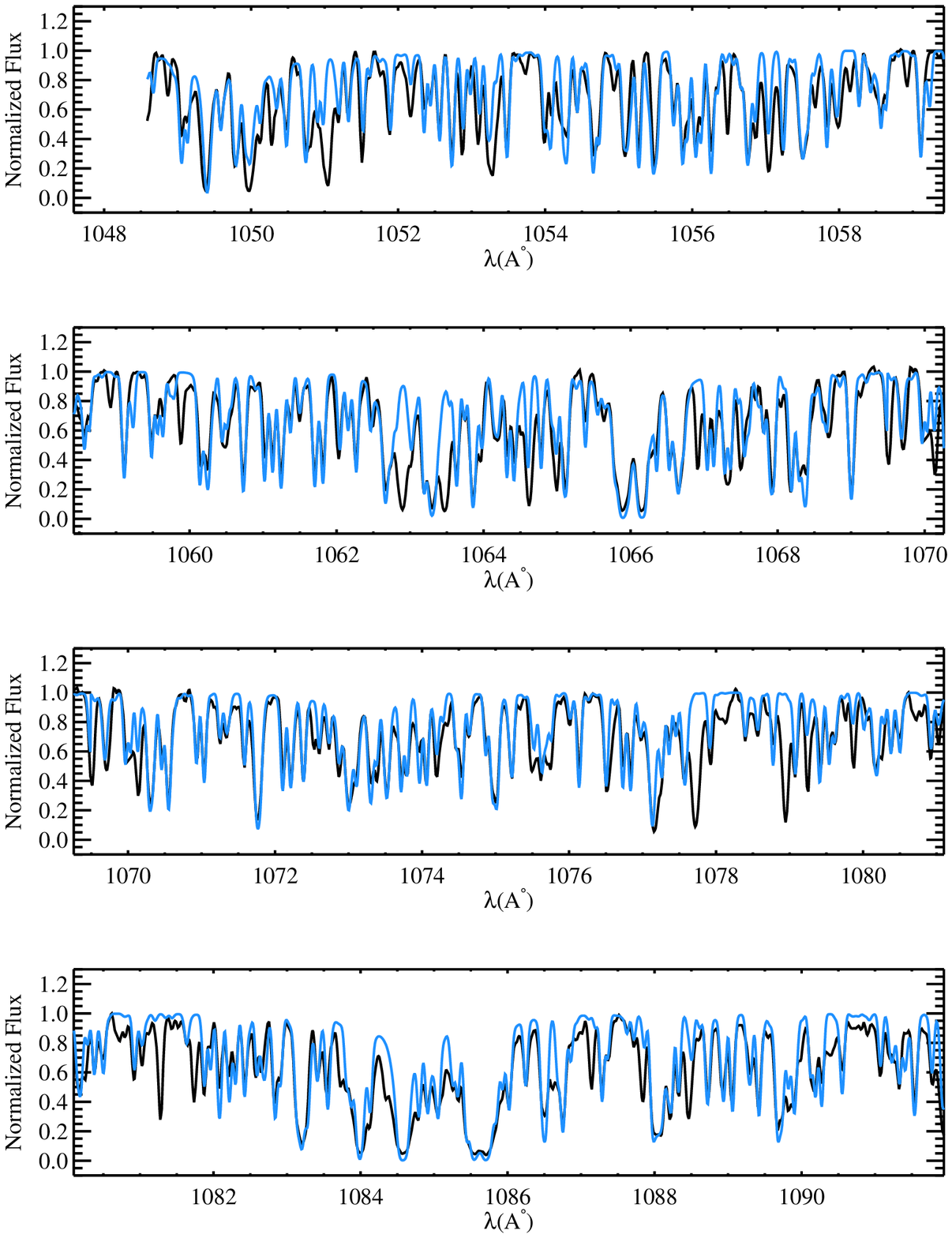}}
\end{figure*}
\begin{figure*}
\resizebox{\hsize}{!}{\includegraphics{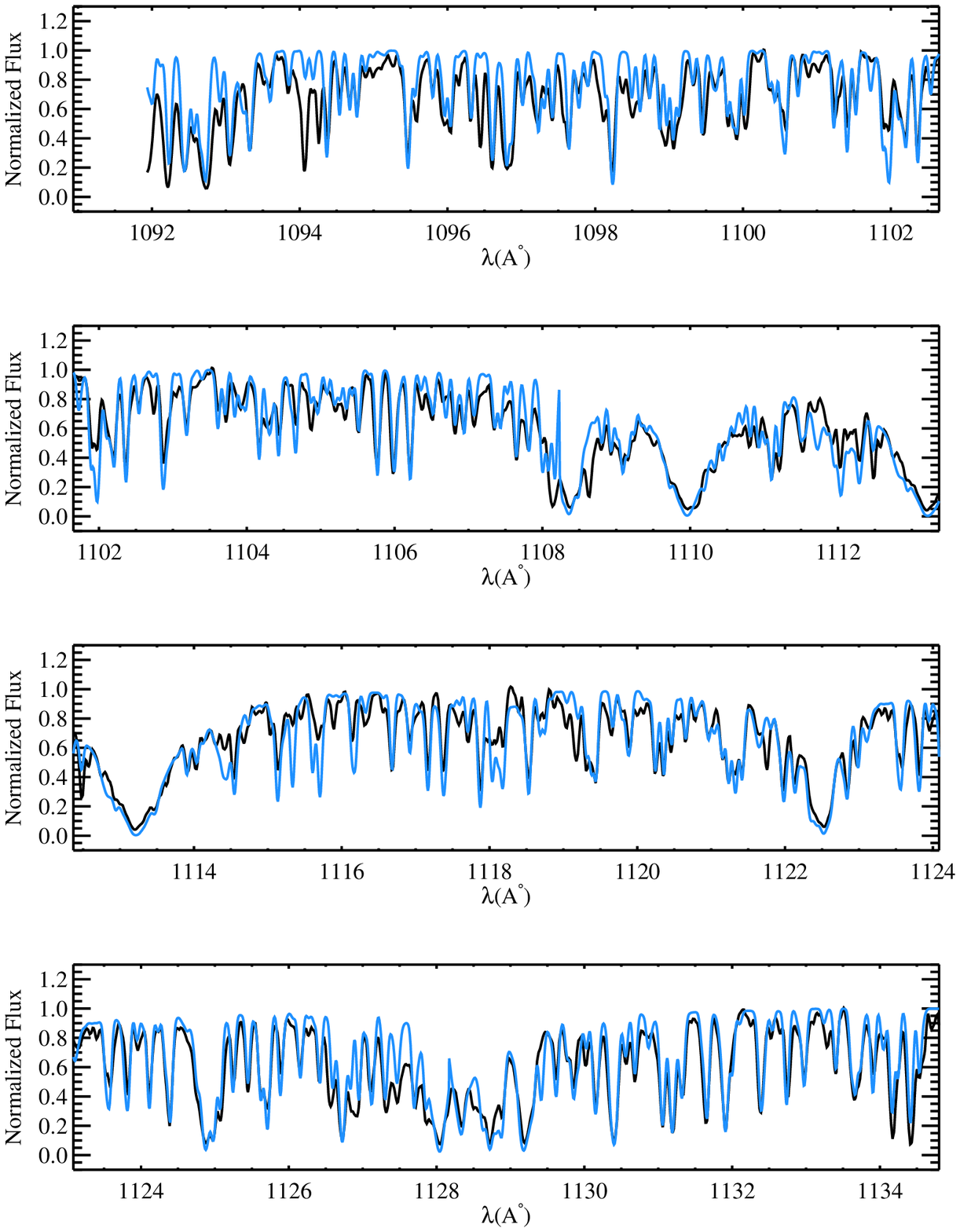}}
\end{figure*}
\begin{figure*}
\resizebox{\hsize}{!}{\includegraphics{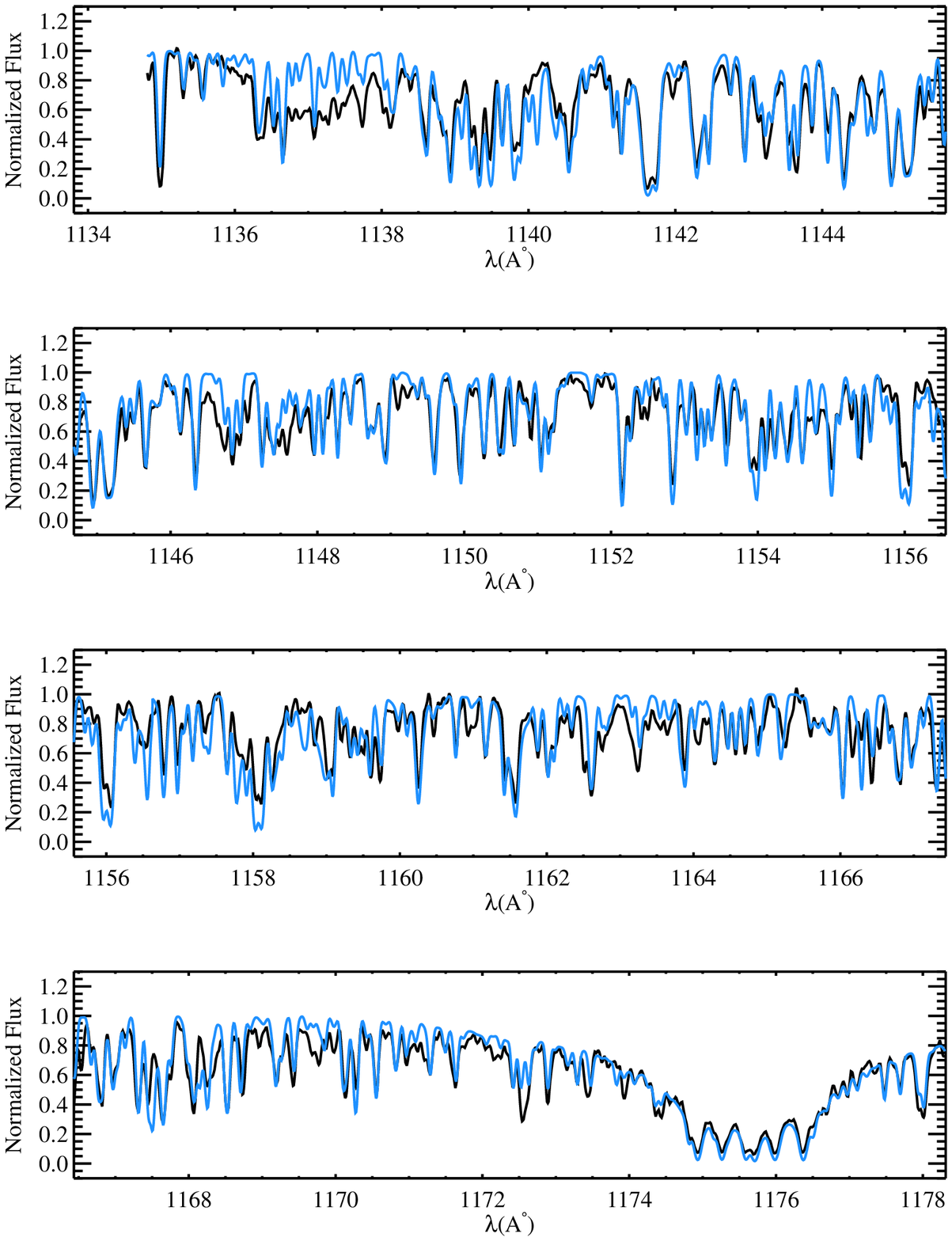}}
\end{figure*}
\begin{figure*}
\resizebox{\hsize}{!}{\includegraphics{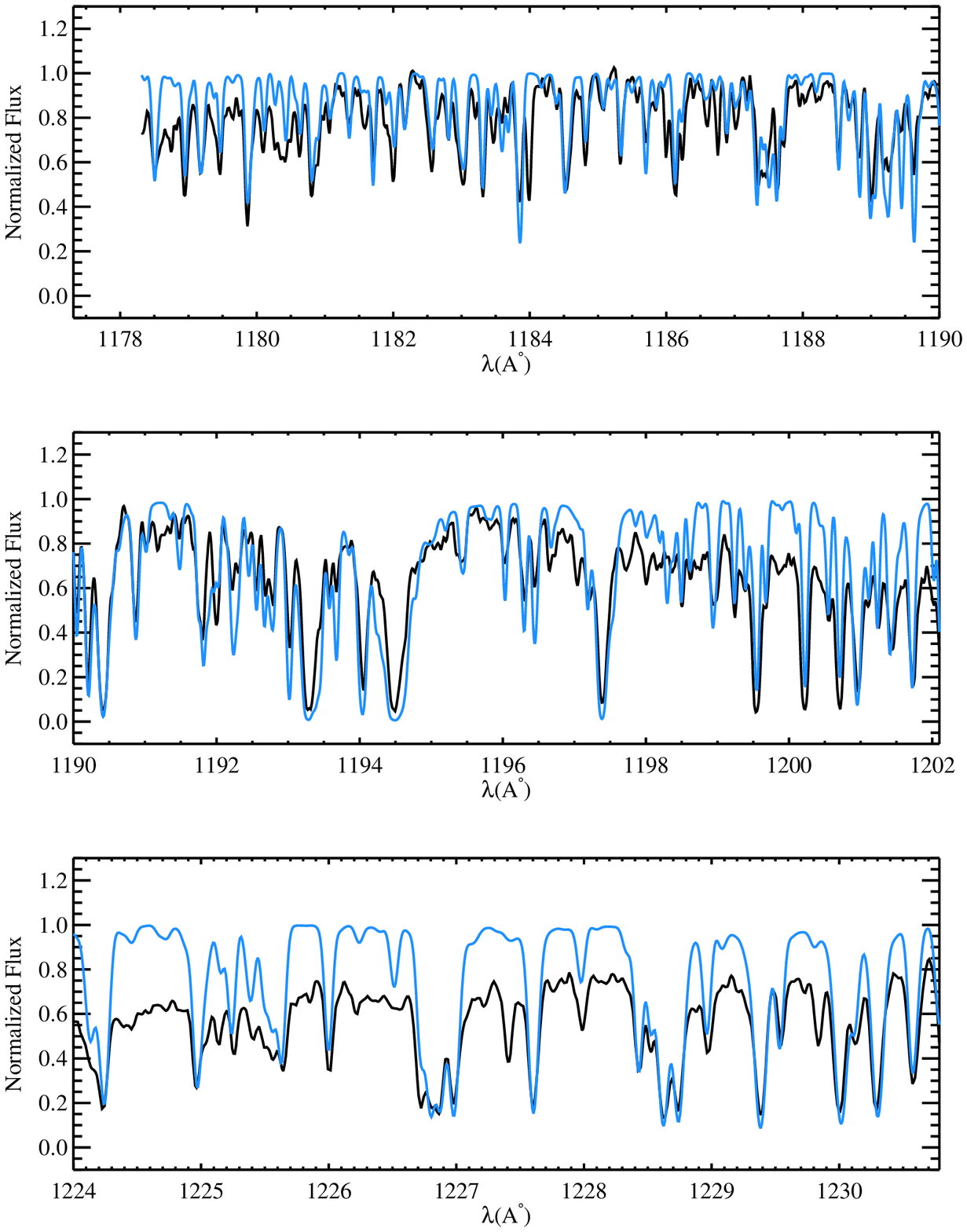}}
\end{figure*}
\begin{figure*}
\resizebox{\hsize}{!}{\includegraphics{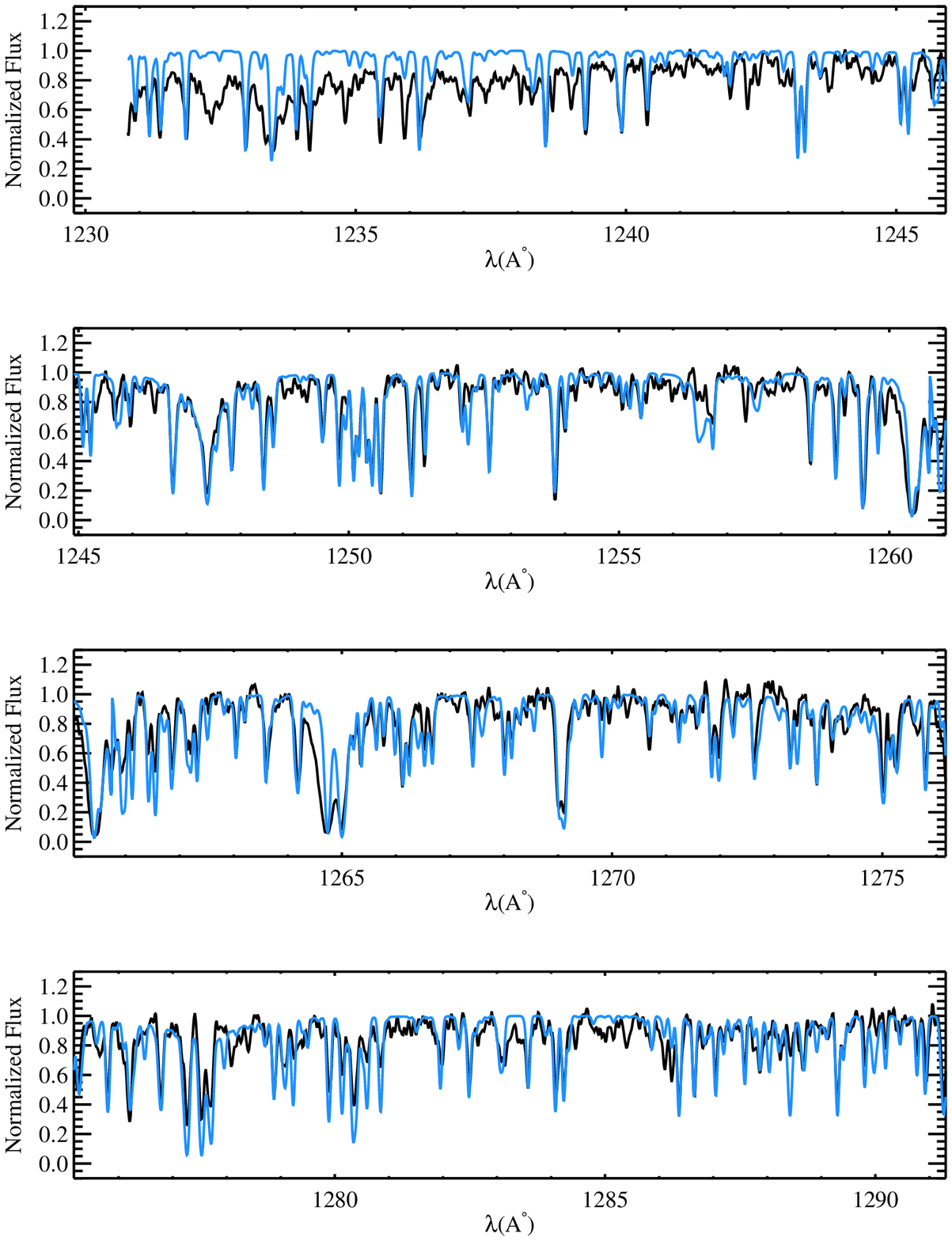}}
\end{figure*}
\begin{figure*}
\resizebox{\hsize}{!}{\includegraphics{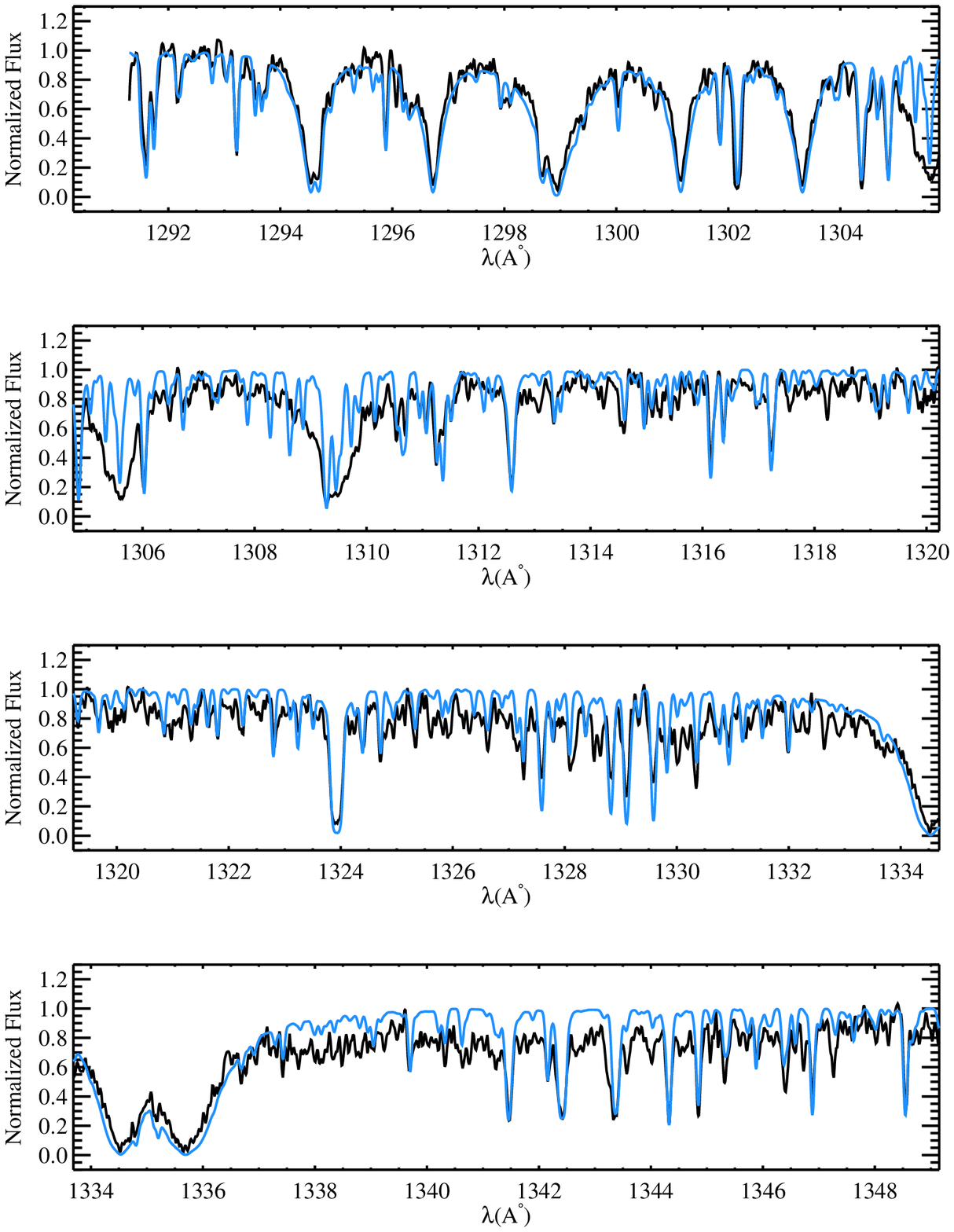}}
\end{figure*}
\begin{figure*}
\resizebox{\hsize}{!}{\includegraphics{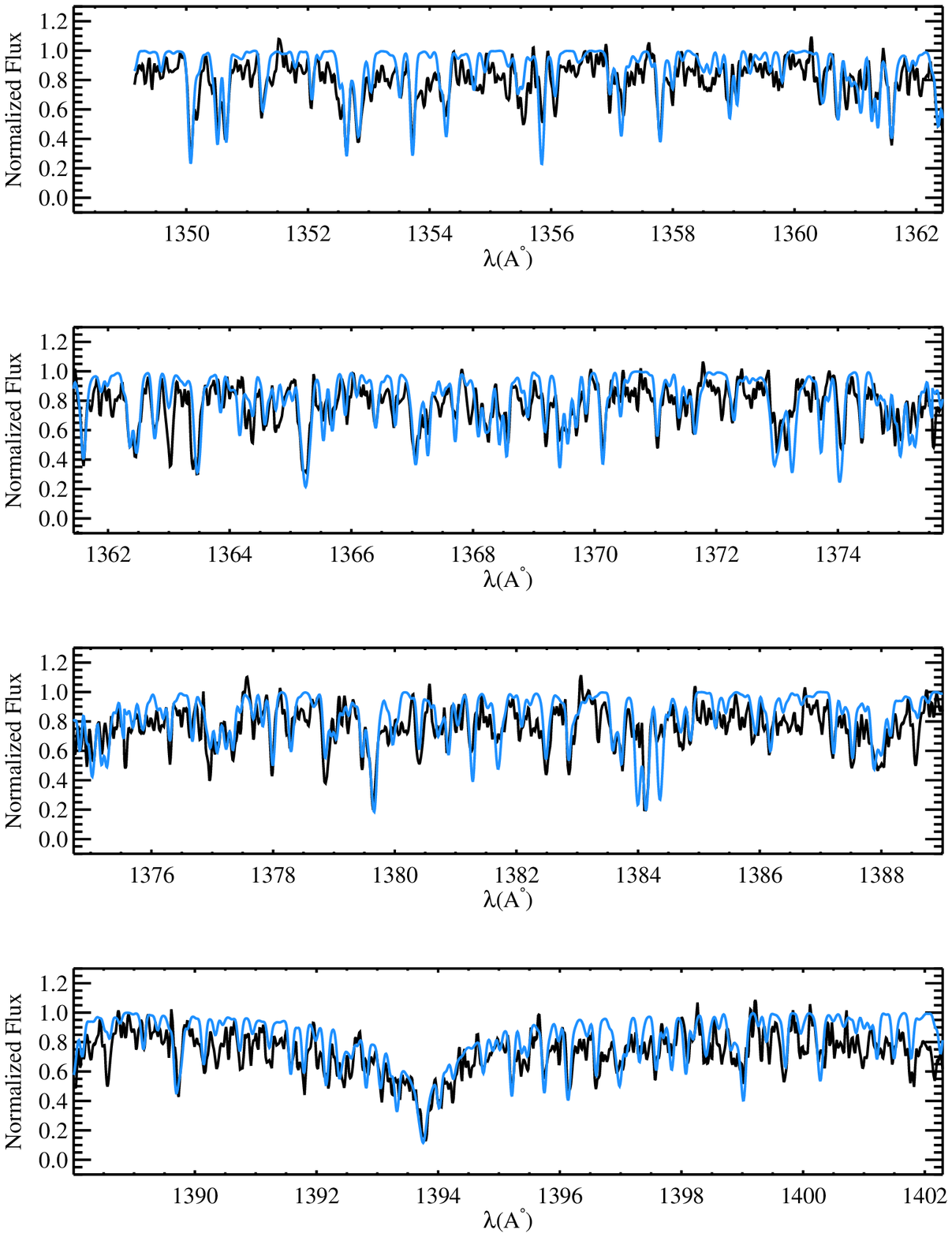}}
\end{figure*}
\begin{figure*}
\resizebox{\hsize}{!}{\includegraphics{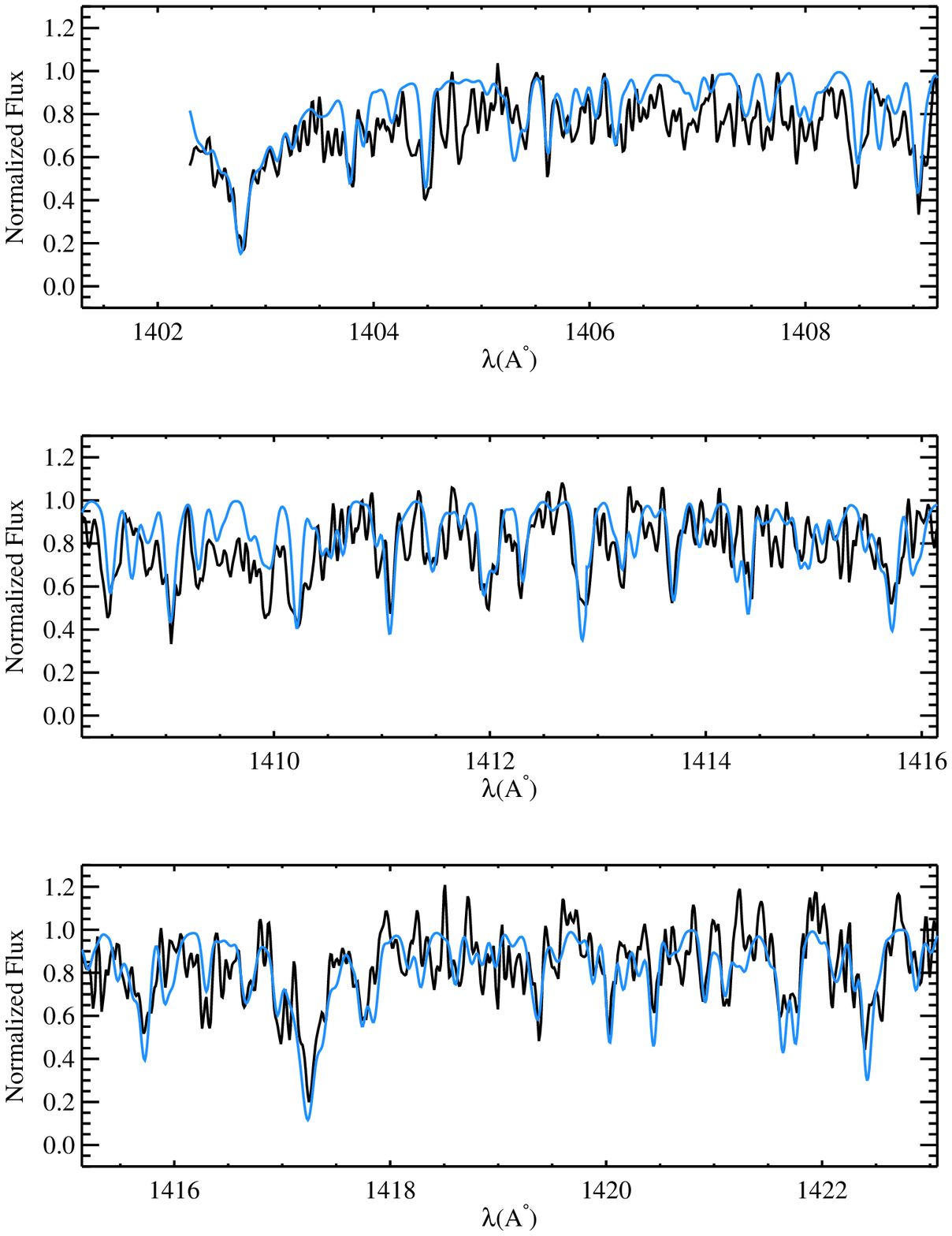}}
\end{figure*}
\begin{figure*}
\resizebox{\hsize}{!}{\includegraphics{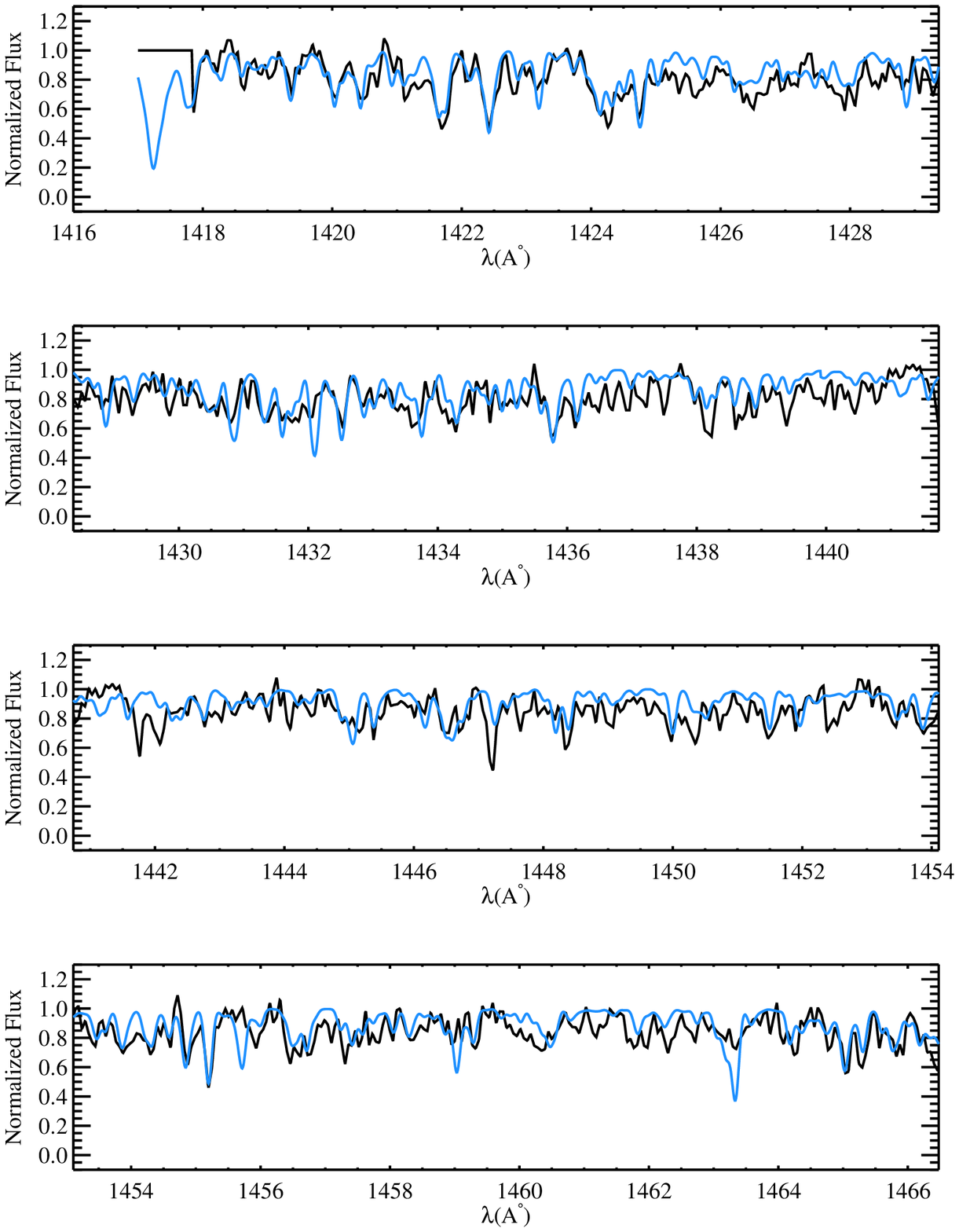}}
\end{figure*}

\end{onecolumn}